# Immersive virtual worlds:

## Multi-sensory virtual environments for health and safety training

Dr Glyn Lawson, Emily Shaw, Dr Tessa Roper, Tommy Nilsson, Laura Bajorunaite, Ayesha Batool
University of Nottingham, University Park, Nottingham, NG7 2RD

The University of Nottingham
UNITED KINGDOM · CHINA · MALAYSIA

**IOSH, the Chartered body for health and safety professionals, is committed to evidence-based practice in workplace safety and health. We maintain a Research and Development Fund to support research and inspire innovation as part of our work as a thought leader in health and safety.**

All recipients of funding from our Research and Development Fund are asked to compile a comprehensive research report of their findings, which is subject to peer review.

For more information on how to apply for grants from the Fund, visit **www.iosh.com/getfunding**, or contact:

Duncan Spencer
Head of Advice and Practice
**duncan.spencer@iosh.com**

Mary Ogungbeje
OSH Research Manager
**mary.ogungbeje@iosh.com**

Ivan Williams
OSH Research Adviser
**ivan.williams@iosh.com**

**Acknowledgement:** IOSH would like to thank the peer reviewers of this report.

# Contents

# List of tables

# List of figures

# List of publications

The following publications have been produced based on the research conducted for this project:

Nilsson T, Roper T, Shaw E, Lawson G, Cobb SV, Meng-Ko H, Miller D, Khan J. Multisensory Virtual Environment for Fire Evacuation Training. In CHI EA'19 Extended Abstracts of the 2019 CHI Conference on Human Factors in Computing Systems. Glasgow, Scotland, May 4–9, 2019. ACM New York. Paper no. INT044.

Shaw E, Roper T, Nilsson T, Lawson G, Cobb SV, Miller D. The Heat is On: Exploring User Behaviour in a Multisensory Virtual Environment for Fire Evacuation. In CHI'19 Proceedings of the 2019 CHI Conference on Human Factors in Computing Systems. Glasgow, Scotland, May 4–9, 2019. ACM New York. Paper no. 626.

Wareing J, Lawson G, Abdullah C, Roper T. User Perception of Heat Source Location for a Multisensory Fire Training Simulation. In Proceedings of the 2018 10th Computer Science and Electronic Engineering (CEEC), Sept 19–21 2018. IEEE New York, pp. 214–218.

Tahsiri M, Lawson G, Abdullah C, Roper T. A multisensory virtual environment for OSH training. In 2018 IEEE Conference on Virtual Reality and 3D User Interfaces (VR) 2018 Mar 18–22 2018, IEEE New York, pp. 699–700.

Other dissemination:

IOSH magazine. What's the drill? Available from: https://www.ioshmagazine.com/article/whats-drill [accessed 27th June, 2019]

Demonstration of the multimodal VE during a roundtable session at the IOSH 2017 annual conference, Birmingham, 20–21 November 2017.

Workshop session accepted for the IOSH 2019 annual conference Birmingham, September 2019.

Demonstration at the 15th annual EuroVR Association conference, EuroVR 2018, Savoy Place, London 22–23 October 2018. Abstract: http://eurovr2018.org/Docs/Posters/EuroVR_2018_paper_18.pdf

# Acknowledgements

The authors would like to thank the Institution of Occupational Safety and Health (IOSH) for supporting this project through its Research and Development Fund. We are also grateful to all the experts who participated in this research. The contribution of the following experts is particularly acknowledged:

**Advisers**

Peter Caines, Rolls Royce plc
Brian Waterfield, Jaguar Land Rover
Dr Pete Bibby, University of Nottingham
Dr Harriet Allen, University of Nottingham
Dr Andrea Venn, University of Nottingham
Dr Aled Jones, University of Nottingham
Professor Monique Smeets, University of Utrecht
Paul Speight, Leicestershire Fire and Rescue Service

**Virtual environment development**

Tommy Nilsson, University of Nottingham
Tony Glover, DRT Software Ltd
Che Zulkhairi Abdullah, University of Nottingham
Daniel Miller, University of Nottingham

**Other contributors**

Meng-Ko (Morgan) Hsieh, University of Nottingham
Dr Sue Cobb, University of Nottingham
Professor Sarah Sharples, University of Nottingham
Joe Wareing, University of Nottingham
Mina Tahsiri, University of Nottingham
James Khan, University of Nottingham

# Abstract

Virtual environments (VEs) offer potential benefits to health and safety training: exposure to dangerous (virtual) environments; the opportunity for experiential learning; and a high level of control over the training, in that aspects can be repeated or reviewed based on the trainee's performance. However, VEs are typically presented as audio-visual (AV) systems, whereas engagement of other senses could increase the immersion in the virtual experience. Moreover, other senses play a key role in certain health and safety contexts, for example the feel of heat and smell in a fire or smell in a fuel leak.

A multi-sensory (MS) VE was developed, which provided simulated heat and smell in accordance with events in a virtual world. As users approached a virtual fire, they felt heat from three 2 kW heaters and smelled smoke from a scent diffuser. Behaviours in the MS VE demonstrated higher validity than those in a comparable AV VE, which ratings and verbatim responses indicated was down to a greater belief that participants were in a real fire. However, a study of the effectiveness of the MS VE as a training tool demonstrated that it did not offer benefits over AV as measured by a written knowledge test and subjective ratings of engagement, attitude towards health and safety and desire to repeat. However, the study found further evidence for the use of AV VEs in health and safety training, particularly as the subjective ratings were generally better than for PowerPoint-based training.

Despite the lack of evidence for MS simulation on traditional measures of training, the different attitudes and experiences of users suggest that it may have value as a system for changing trainees' attitudes towards their personal safety and awareness. This view was supported by feedback from industrial partners.

# Executive summary

The project investigated the use of multi-sensory (MS) virtual environments (VEs) for health and safety training. The rationale behind the project was that traditional health and safety training can fail to motivate and engage employees, and can lack relevance to real-life contexts. Health and safety training also often takes employees away from their primary jobs, reducing productivity. VEs offer potential solutions by increasing trainees' engagement and understanding by affording them experiential learning in safety-related scenarios; customising VEs to trainees' own workplaces; and reducing costs by allowing trainees to train locally, and at their convenience, rather than travelling to dedicated training facilities. Thus, the overall aim of this research was to produce evidence-based guidance for the development and use of VEs in engaging and effective training using cost-effective and accessible solutions.

The development work conducted for the project including integrating heat and smell simulation with a virtual environment. The heat was provided through three 2 kW infrared patio heaters, with the heat being controlled by fins acting as heat shields. These provided adjustments in the level of heat that corresponded to the distance from the virtual heat source (e.g. fire) in the VE. The fins were mechanically attached to servo motors, which were controlled by Arduino. Scent was delivered through a SensoryScent 200 fragrance diffuser, which was again rewired for Arduino control. The fragrances were purchased off the shelf from a supply company.

To simplify the process of creating bespoke custom virtual workplaces, an investigation was conducted into the use of scanning technologies, including Google Tango. While in principle this would enable a company to scan a real environment, then use the scan data to create a virtual training scenario, in reality this is currently not feasible for several reasons. Firstly, technical skills were required to manipulate the mesh, which would not typically be available within an organisation. Secondly, and related to the first point, the scanned file sizes were large, meaning that they had to be broken into sections during the scan and joined together in the VE. Finally, obtaining the scan posed challenges such as capturing scan data from all round an object, and treating doors as static data (whereas they would need to move in the VE). However, a protocol was created to help address these issues.

Given these difficulties, the project created virtual worlds in Unity, with guidance from two industrial partners. Two scenarios were developed: a building fire/evacuation and an engine disassembly task with fluid leak and response actions. The former was used to investigate any differences in the validity in the behaviour people demonstrated when multi-modality was added. Fifty-two participants were recruited to a between-subjects study (audio-visual (AV) vs multi-sensory (MS)). Participants were instructed to navigate towards a meeting room in the VE, where they began a series of tasks. A virtual fire was started, and participants' behaviours were recorded. A post-trial questionnaire captured subjective ratings and verbal feedback. The results gave evidence for a greater level of belief that they were participating in a real emergency in the MS. For example, ratings were higher in the MS for "level of time pressure", "the building is on fire" and "I need to find the exit nearest to me". Behaviours such as avoidance of directly passing through the fire and comments such as "I wanted to get out fast" support the notion that participants' experiences in the MS condition were closer to that of a real fire.

A second study looked at the training effectiveness of the MS VE. Fifty students were recruited to a mixed design, in which they would experience either VE or PowerPoint training (between-subjects factor) with or without multi-sensory simulation (within-subjects factor). Training was given in two scenarios: fire safety and engine disassembly/fluid leak. The VE training performed better than PowerPoint for several measures, including knowledge retention and ratings of engagement, attitude towards health and safety and motivation to participate in future training. However, there was no evidence to support the addition of multi-sensory feedback.

Despite the results from the second study, the data from the study of behavioural validity and feedback from industrial partners suggest that MS VE can contribute to improved health and safety outcomes. This benefit lies in the different experience afforded by the MS VE, which is closer to a real emergency. Thus, while Study 2 did not find evidence for knowledge uptake, as tested in a question-and-answer-type quiz, it may be that the more realistic experience of the MS VE better prepares people for an emergency.

# 1 General introduction

Fatal fires occur worldwide and often in buildings of high occupancy (1), and in the 12 months to March 2018 fire and rescue services in England reported attending 167,150 fires and over 300 fire-related fatalities, including 71 related to the high-profile Grenfell tower incident (2). Unsurprisingly, the implementation and monitoring of safety regulations and the interest in fire safety systems and equipment are increasing (3). Fires are unexpected, unpredictable and often dangerous situations requiring immediate, appropriate action (4). Minimising loss of life is critically dependent on safety training (1,5).

Research on human behaviour in fire during real-world fire incidents has suggested that a lack of knowledge relating to the spread and movement of fire often means that occupants are unprepared and misjudge appropriate actions (6,7). Effective training in fire safety behaviours is a vital means to improve occupant survival rates by reducing evacuation times and potentially fatal errors (7). Fire safety forms part of occupational safety and health (OSH) training. In safety-critical contexts it is not usually practical, cost-effective or ethical to conduct training in the real situation. Traditionally, OSH training has been largely theoretical and delivered through training lectures or written materials, which often fail to motivate and engage target audiences (8,9). For example, the use of pre-planned emergency evacuation fire drills of specific incidents can prove disruptive to the workforce and costly to the employer and are often carried out inconsistently, limiting educational value. Additionally, fire drill simulations typically lack critical psychological and affective elements associated with a fire emergency, e.g. smoke-filled corridors resulting in unrealistic behavioural responses, for example lack of urgency (7,10,11).

Virtual environments (VEs) are an effective tool for research and training in environments and scenarios that would be too dangerous in the real world (12). They allow multiple rehearsals on a wide variety of scenarios in a cost-effective and safe way (5). For example, Tate et al. (13) found using immersive VE was an effective training tool for shipboard firefighting, while Mól et al. (14) demonstrated similar behaviours in VE simulated emergency evacuations as in the real world. VEs could provide the opportunity for a more experiential learning approach, which may improve relevance, engagement, transfer/application of learned knowledge to real situations, and employee motivation to train. However, research has uncovered potential inconsistencies that highlight the need to assess how user behaviour in the VE corresponds to or deviates from the real world (15). For example, Smith and Trenholme (10) investigated fire evacuation behaviour using a VE and found that although evacuation patterns were similar to real life, time to evacuate tended to be longer and was influenced by gaming experience. They also raised concerns about some of the observed behaviours within the VE, e.g. willingness to open doors with smoke coming from underneath them, which they attributed to missing sensory elements that would be present in a real fire, such as heat. They suggest further work to increase the realism of the VE. Chalmers and Ferko (16) echo this by stressing the importance of multi-sensory simulation within VEs, explaining that, in order to achieve valid user behaviour, it is necessary to go beyond traditional audio-visual experiences and deliver simulation for all senses.

Despite these recommendations, there remains a lack of research on the impact of multi-sensory simulation on user behaviour in the context of safety-related VEs. In particular, research has not yet assessed whether, or how, the validity of user behaviour could benefit from additional modalities. Exploring this topic would help developers produce more effective VE safety solutions, and in turn contribute to the mitigation of any damage that might result from future real-world incidents. The aim of this research was to investigate whether the addition of multi-sensory feedback (namely heat and smell) offered benefits over audio-visual VEs. We developed a multi-sensory VE and used it to compare the outcomes of OSH training, such as effectiveness of the training and trainee attitudes, to those of traditional training methods. We also compared the multi-sensory VE to a VE without heat or smell, to see whether the additional sensory feedback made a significant difference to behaviours demonstrated in the VE and thus whether multi-sensory systems were worth the extra investment.

# 2 Literature review

## 2.1 Human behaviour in real-world emergency situations

Extensive research has been carried out over the last few decades investigating human behaviour during emergency evacuation (15,17) in recognition of the impact human behaviour has on the outcome of the situation. Studies have used survivor interviews from real-world incidents (6), fire drills (18) and, more recently, VEs in an attempt to establish valid data on human behaviour in emergencies for use in research and training (15). Advances in the versatility of VEs have enabled high-fidelity visualisation and simulation of practically any environment (12). However, due to the complex nature of the behaviour being studied and the limited access to direct behavioural observations for ethical and safety reasons, proving the validity of behavioural responses from a simulated experience presents a unique challenge (5,19). Various methods have been suggested for assessing the validity of behavioural responses from research studying human behaviour in emergencies. Perhaps most notably, work by Canter et al. (17) developed various taxonomies of human behaviour in fire for different types of buildings, which can provide a basis for investigations of validity. This work has since been used and built upon by Lawson (15) to include emergent behaviours from research using VEs. Key findings from these bodies of work highlight behavioural elements seen in real-world settings that can be used to compare behavioural responses produced by participants exposed to fire emergency scenarios in VEs.

Contrary to popular assumptions regarding panic in fire emergencies, irrational behaviour during evacuation is rare (20). Humans have been shown to use rational decisions about exit routes out of buildings (20–22). Decision making is limited during emergencies; therefore, people tend to use heuristics: making choices using information that is easiest to recall or most available in their immediate environment. Satisficing, a heuristic method that chooses the first available workable option, even if it is not optimal, is more likely to be used in time-pressured situations, where conditions are dynamic (23–25). This would explain why occupants are more likely to use familiar exit routes or retrace their steps to evacuate the building (1,26). Research by Edelman (27) on the real-life evacuation of a nursing home demonstrated the importance of understanding human behaviour in fire in different contexts to inform areas of fire safety. Nearly all residents reported using the main stairwell for evacuation. Reasons cited for this were a lack of fire drills and, therefore, lack of experience using emergency exits; following or being assisted by care workers; or concerns for being reprimanded for using exits not normally permitted (27).

Research suggests that occupants often perceive alarms as ambiguous. For example, Purser and Bensilum (28) found that occupants were slow to respond to alarms, and often continued with non-evacuation activities before exiting. They concluded that building occupants need to recognise the importance of an event to break their commitment to current activities. Proulx (29) also found that occupants are likely to conduct non-evacuation activities, such as warning others and fighting the fire before evacuating. Wood (6) noted that the more serious a person considered a fire to be, the less likely they would be to attempt to fight it. The level of seriousness is linked to the perceived risk felt by the occupant and reflects objective measures such as high levels of smoke spread and density. This is supported by Proulx (29) who suggested that occupants were more likely to respond to an alarm if it was accompanied by other cues such as smell. Additionally, both Wood (6) and Purser and Bensilum (28) reported that occupants often make judgements of seriousness based on a lack of knowledge about fire and its potential speed of development, which further highlights the importance of research and training in this area.

Kuligowski (21) suggested that people are more likely to define the situation as an emergency when there are a higher number of consistent and unambiguous sensory cues. This can contribute to faster initiation of evacuation and therefore overall egress time. For example, following the World Trade Center disaster, survivor reports highlighted cues such as the smell of burning fuel, hearing the explosion and feeling the building sway as causing rapid evacuation (30). These findings are of particular importance when considering the role of VE in research and training for fire evacuation scenarios and the case for using multi-sensory (MS) VEs in order to increase the sense of realism and ecological validity of this research tool. The subject of MS simulation in VEs for health and safety applications is revisited in section 2.5 following a review of the advantages and challenges of virtual environment training.

## 2.2 The advantages of virtual environment training (VET)

Virtual environment training (VET) offers several advantages over conventional training and other technology-based training. For example, VET can provide experience-based learning that enables trainees to observe the consequences of their actions (31). According to McGuire (32), this active learning process helps the learner to achieve understanding of the real world via an "ongoing process" that produces a sense of new information through their own version of reality, compared to conventional training that is based on the trainer's perspective. Trainees in VET move autonomously and engage in self-directed activities among their learning contexts, or experience the consequences of their actions, learning through these processes (33). As a result, this can lead to greater effectiveness of the training (34).

VEs are capable of offering interaction, observation, examination and other experiences that are impractical or impossible to achieve via other means (34,35). For example, VR has been used to study a variety of emergencies, such as aircraft crashes (36), mass-casualty incidents in emergency medicine departments (37) and subway evacuation (38). Cha, Han, Lee and Choi (39) developed a training simulator for a vehicle fire in the Jukryeong Tunnel in Korea to provide safe and convenient training for firefighters. In all of these examples, the real-life incidents these simulated scenarios are recreating are fortunately rare. Therefore, there are few training opportunities in the real world, and VET can provide an alternative method of gaining and maintaining the necessary skills and knowledge.

VET can be motivating and engaging, particularly when the VE is personalised and individuals seek to achieve their own goals within them (35). This can be further enhanced by using a game format to make the learning more interesting and fun (34). Game-based systems often use elements such as exploration, challenge and progression through levels. While often enjoyable to interact with, such systems have a purpose (in this case learning about fire safety), hence they are often referred to as 'serious games'.

VEs can be tailored according to the characteristics, capabilities and needs of the trainees. For instance, VEs can enable learners to progress through training according to their experience and competence, and at their own pace. For example, Smith and Veitch (40) compared simulation-based training to lecture-based training in four offshore emergency response procedures. They used a mastery-based approach with their VET, in which competence needed to be achieved before progressing to the next task. Smith and Veitch (40) report task performance improvements, less time required for the training and more risk-averse behaviour following training with their simulation-based approach. They conclude that simulation-based approaches are not only useful in demonstrating competence but can also help address individual variability by customising the training to learner needs. Further to the notion of tailoring VEs, it is possible to create and conduct training that represents trainees' own physical locality (34,40). This has the potential to improve training by increasing its relevance to an evacuee's own location. Thus, VET presents the possibility to train geo-specific aspects such as the location of fire safety elements, or the layout of buildings and evacuation routes.

The ability to design scenarios is a key element in realising the benefits of VET. VEs enable developers to create scenarios in order to fulfil distinct training purposes. A scenario illustrates a hypothetical, but plausible, circumstance that may be employed for a variety of purposes. Although it may not decrease uncertainties corresponding to a future case in the real world, the scenario can generate a more concrete situation; as a result, users are able to practise proposed response strategies or crisis management systems in a self-consistent and reasonable manner (41). Furthermore, multiple varying scenarios can be used to explore different hypothetical scenarios. For example, trainees can be exposed to a fire in which different exit routes are blocked, in which different flammable materials are present, or where they are in different parts of the workplace, and can experience the differences in each. This is typically not accommodated by structured training and drills.

Due to their reliance on computational technology, VEs provide good opportunity for measuring learning and performance as the sessions can be easily monitored and recorded (42). Moreover, they provide the opportunity to implement other features in support of training, such as reviewing the training experience from a different perspective (e.g. bird's-eye view to help understand evacuation route choice), or in real-time, sped up or slowed down, or fast-forwarded to key decision points, to review with a trainer.

## 2.3 Challenges with virtual environment training (VET)

While VEs offer the potential advantages outlined above, there are challenges associated with VET. Several studies have reported that use of VEs can bring about symptoms of simulator sickness (e.g. 43–45). The sickness is associated with a wide range of symptoms, including headache, eyestrain, dizziness, fatigue and nausea (45). There are differences in the level of sickness symptoms attributed to VR display type, with head-mounted display (HMD) often causing more (e.g. 45). With regard to the reasons behind HMD sickness, some scholars (46,47) attribute it to lag in the system which leads to sensory conflict between the control operation and the resulting output seen on the display (48). Importantly, cybersickness negatively impacts attitudes towards virtual training technologies, and is associated with poorer learning (49).

There is mixed evidence for the proven effectiveness of VET in the academic literature. The success of VET depends on the nature of the training and on the instructional content, and on whether the environments grant engagement and stimulate elements such as reasoning skills, transfer of knowledge and other factors that support training (50). While there is some evidence for the effectiveness of VR-based training in different domains including automotive/manufacturing (51–56), aviation (57) and medical (58–60), there are often methodological concerns with the research. These include a focus on a limited set of performance criteria, often time and error (51), lack of validation of the evaluation methods and a tendency towards user evaluations (rather than controlled studies of training effectiveness) with positive results bias (49). There is also a lack of research on higher-level cognitive skills, with most research focused on remembering or understanding facts (49), rather than exploring higher-level educational outcomes such as analysis, creativity and evaluation, or more fundamental shifts in attitude towards a topic.

There are issues with controlling movement in a virtual environment which arise as a consequence of the interface layer between the user and the virtual world. Input devices, such as a standard mouse, joysticks or other controllers, do not afford similar feedback to that received in the real world and, consequently, users can have difficulty navigating in VEs (61). Lawson (15) found users demonstrated overshoot errors and difficulties getting through doors, and frequently got lost when navigating a fire evacuation in a VE. Smith and Trenholme (10) attribute longer evacuation times in a virtual building evacuation than in the real world to control issues.

While the costs of VR equipment have reduced considerably in recent years (49), there is still a resource associated with buying and developing VETs. Smith and Trenholme (10) report that their virtual replica of a multi-storey computer science department took a single developer three weeks to construct, which would likely take longer for someone without computer science or VE development experience. Importantly, the environment creation is only one element; considerable resource must be assigned to the instructional design if the training content is to be effective.

## 2.4 Existing studies of VR in emergency situations

Several studies have already investigated VR in fire safety training, or in the study of human behaviour in fire scenarios. These have shown performance improvements which have been attributed to the VET. For example, Tate, Sibert and King (13) evaluated the effects of employing VE to train shipboard firefighters in the navy in comparison with conventional methods of mission preparation. Navy trainees wore an HMD and used 3D joysticks to interact with the VE. Despite a limited sample size, the study demonstrated that VE-trained participants completed a navigation mission faster than the control group who had been trained using traditional approaches. In addition, the VET group committed fewer mistakes (navigational errors) than the control group (13).

Some researchers have focused on the validity of behaviours that users demonstrate in VEs, given the potential importance of user experiences that match the real world in training (62). Gamberini et al. (62) investigated participants' responses to a fire emergency in a virtual library. The participants wore an HMD and used a joystick to interact with the VE. Initially, participants were given the chance to navigate without any hazardous phenomenon so that they could orient themselves within the VE. They were then required to arrive at a predetermined point in which an emergency would be initiated. The researchers found that the participants identified the emergency and responded with behavioural adaptations, such as movement patterns more focused on evacuation. This leads the authors to conclude that VE is suitable for studying behaviour in emergencies, and posit its potential use as a training tool. Kobes et al. (1) compared evacuation behaviour in a real hotel to a virtual one, and concluded that generally wayfinding behaviour demonstrated relative validity, with the exception of a scenario in which exit signs were located near the floor; in this condition participants in the virtual environment demonstrated an unexpected tendency to not use the nearest exit route. In a related study which also used a virtual hotel, Kobes, Helsloot, de Vries and Post (26) reported the influence of smoke on evacuation behaviour, which increased the likelihood of evacuees using a closer exit. VEs have also been used to study evacuation signage (63,64). Duarte et al. (63) argue for the validity of their results based on a comparison to previous events in the real world. They also highlight the usefulness of VE as a research tool, given the need to avoid exposing study participants to hazards. However, Tang et al. (64) express concern that some of their results, in particular that construction and fire safety workers were not significantly better at wayfinding than the general population, would not be replicated in the real world. However, this concern is not rationalised, so would require further investigation.

Another area of focus in the prior literature has been on enhancing the realism of fire and smoke simulation in virtual environments, often using computation fluid dynamics (CFD) (39,65–67). Ren et al. (65,66) developed a simulation of a subway using CFD, which they propose as a safe and inexpensive tool for virtual fire drills and training. However, their focus is on the technical development and they do not present any behavioural outcomes or user tests. Cha et al. (39) developed a VET simulator providing a wide range of experiences for the general public or inexperienced trainers and commanders so that they were able to make quick decisions and safe and organised responses in real-world situations. They believed that CFD could improve the fire and smoke graphics. User feedback from firefighters was positive on the realism of the fire in their simulation, and they commented that the VET had potential for training applications, although this was not tested empirically. Their feedback also included a request for gamification elements and a diverse range of scenarios to reflect the variability of events in real life. The users also requested multi-sensory feedback such as touch and heat, given the importance of non-visual cues to firefighters, for example when finding their way in the dark and through smoke-filled environments by relying on tactile perception through hands (39).

The gamification topic has been raised by other authors in recognition of its contribution to improving training experiences or outcomes (7,10,68–70). The potential advantages of game-based evacuation training include greater motivation for building occupants to engage with the training material (which is more interesting than traditional training material) and therefore spending longer engaging with it, and the ability to customise the training according to the users' training needs (7,70). Chittaro and Ranon (7) also recognise the potential importance of VET to employers, in that it may have less of an impact on company processes than traditional training approaches, and would be safe for trainees. User feedback from their game-based system included a desire to understand more about the effects of heat on their character, and a need to increase the emotional intensity of the game-based training. Chittaro and Ranon (7) emphasise the importance of studying retention and transfer of the training knowledge to the real world in future work. Smith and Trenholme (10) used a gaming platform to develop their fire evacuation simulation, in recognition of the time advantages such technologies bring to creating a bespoke VE. While they found the patterns of evacuation times for their three scenarios were consistent between the real world and the VE, the overall times differed. This was affected by gaming experience, as also seen by Ribeiro et al. (70) who found virtual evacuation times were quicker for regular video game players. Smith and Trenholme (10) report on behaviours seen in the VE which they would not expect in real life, such as going through a door with smoke coming from underneath it, and attribute this behavioural discrepancy to the lack of multi-sensory simulation in the VE. They also report on the lack of interactivity with some salient objects such as fire extinguishers. Backlund et al. (69) provided some interactivity with salient objects by including physical artefacts (a hoze nozzle and breathing mask) within a cave automatic virtual environment (CAVE). They present an architecture for a game-based simulator, using game-based features such as progression through levels with different learning objectives. The outcome of the user testing was that generally an appropriate level of fidelity was achieved. Similar to the findings from Chittaro and Ranon (7), they report on the importance of psychological strain on the training experience. The authors propose the simulator as a useful tool to prepare students for live training, by progressing them along the learning curve before the real training (69). In an earlier study of their simulator, Backlund et al. (68) demonstrate learning effects, as participants performed better in the simulator on subsequent uses.

They also report high levels of user enjoyment, attributed to the game-based approach, but also note instances of cybersickness.

## 2.5 Benefits of multi-sensory VEs for research and training in health and safety

Chalmers et al. (71) state that if VEs are to be used as an established and valuable tool for research and training, especially in areas of OSH, VEs needs to be perceptually equivalent to the same experience in the real world. As humans perceive the world with all of their senses, it is suggested that high-fidelity VEs require multi-sensory (MS) simulation to provide this level of immersion (5,72). The use of MS VEs has been investigated in other applied domains. For example, Jiang, Girotra, Cutkosky and Ullrich (73) found the use of haptic technology demonstrated the effectiveness of 'feel' in the VE with participants making fewer procedural errors and completing some tasks more rapidly with the addition of haptic (vibration) feedback. Thermal stress has been shown to impair endurance and vigilance in real-life fire emergencies (74), suggesting that the addition of thermal simulation to the VE could increase its realism. A recent user perception study of heat source location has provided support for this with participants rating thermal simulation from infrared heaters as realistic during a simulated emergency (9). Additionally, empirical studies, e.g. (75), have shown that the addition of olfactory simulation increases the sense of presence within the VE. Previous research has explicitly evaluated the validity of user behaviour in virtual fire scenarios (15) but not with MS feedback. Researchers such as Ren, Chen and Luo (65) have used simulated models of flames and smoke to increase the realism within the VE, but this does not provide the experiential simulation of feeling the heat or smelling the smoke. MS VEs are purported to be able to efficiently combine sensory information from several channels to provide a greater extent of sensory information to the user. This, in turn, increases environmental richness resulting in a more complete, coherent experience and greater consistency of cues (76). There is evidence to suggest that multi-sensory VEs may improve learner motivation and engagement via immersion (77) For example, Chittaro and Buttussi (8) reported that negative emotional arousal can be especially effective in knowledge retention in VR safety education.

# 3 Aims and objectives

This project aimed to produce evidence-based guidance for the development and use of VE in engaging and effective training using cost-effective and accessible solutions, with the following specific objectives:

## Objective 1

Develop and test a low-cost multi-sensory VE to establish the benefits of MS VEs in OSH training. To produce a functional, low-cost, virtual training simulator integrating off-the-shelf technologies to minimise development costs both during this project and to companies who will ultimately benefit from the training system. The simulation was to include:

a) visual and audio simulation
b) thermal simulation
c) olfactory (smell) simulation.

We aimed to take a user-centred and agile approach to the development of the simulator, involving the industrial partners, Jaguar Land Rover and Rolls Royce, to define requirements for the training simulator and to evaluate a prototype. The simulator was to be built using a suitable 3D development platform, such as Unity.

## Objective 2

To study the validity of behaviours demonstrated within the VE and the effectiveness of the training with and without MS simulation in the fire evacuation and dangerous chemical use cases. To look for differences in suitable measures, for example evacuation times, route choices, response to/interaction with fire cues, between the conditions (audio-visual vs. multi-sensory) to statistically compare the results. Additionally, to compare the observed behaviours to those reported in academic literature and incident reports to understand their validity, as done by Lawson (15).

To test the benefits of MS simulation on training effectiveness, including knowledge uptake and time to achieve proficiency, by comparing traditional training e.g. PowerPoint presentation with single- and multi-sensory VE training and conducting post-trial tests of OSH knowledge to determine whether the addition of MS information improves OSH outcomes.

## Objective 3

To prototype the VE in industrial use cases including fire evacuation training and response to chemical hazards. To create and build VEs based on salient aspects of scenarios provided by industrial partners.

## Objective 4

To test low-cost scanning techniques for creating virtual representations of factory premises to demonstrate the feasibility of wide-scale adoption of VE training approaches in the short-term (2–5 years) future.

To develop a quick-scan 3D VE, with the functionality to support quick scanning of a site, including capturing the physical geometry and identification of 'classifier-based features', for example common objects such as signage. To provide the ability to auto-generate the classified objects and integrate these into the VE to reduce the impact on resource required to build the VE, e.g. hiring costs associate with external consultants or impact on internal resource management. To develop the system such that scenario customisation is possible, for example instigating a virtual fire in specific locations. We will also implement recording and playback functionalities. The advantages of the auto-generated 3D VE include a reduction in 3D modelling cost, specificity to a training site and low technical requirements for use by the local OSH officer. This objective will be achieved once we successfully implement a scan of a real building in a virtual training scenario.

## Objective 5

Refine the hardware and software to optimise wearability and usability for company employees. To have the prototype system assessed by industrial partners and determine acceptability based on this feedback.

## Objective 6

To evaluate the benefits to business of the virtual training system, including:

1. cost and time savings
2. OSH compliance
3. motivation to train
4. knowledge management.

To give a clear demonstration of potential impact of VE-based training approach through a powerful combination of quantitative and qualitative metrics, for example through interviews with the industrial partners, giving consideration of any improvements in time to proficiency and calculation of possible reductions in company downtime, and subjective ratings from participants in the studies of the VE. To discuss potential barriers to adoption and use of the VE system with industrial partners, making recommendations to overcome them.

# 4 Overall approach

The first stage of the project focused on exploration of different technical solutions (Objective 1) and developing a multi-sensory prototype system informed by potential use cases with industry partners (Objective 3). In line with Objective 1, we researched user needs to ensure that the options we investigated would be suitable for users and contexts of use. We conducted a number of meetings, discussions and visits with our partners at Jaguar Land Rover and Rolls Royce, as well as meeting with the university's own health and safety advisers to understand fire safety and safe handling of hazardous chemicals. The focus for the first stage was the development and testing of a functional MS VE:

- developing a Unity 3D virtual environment for simulating emergency scenarios
- designing the hardware for simulating heat, and designing a control system using Arduino
- evaluating scent systems
- experimenting with fans and extraction to present the fragrances to the user
- rewiring the above heat systems and scent diffuser for Arduino
- developing Python control software for managing the Arduino hardware trigger, which is synchronised with the Unity 3D virtual environment
- preparing the system to meet the university safety requirements.

Selected technologies and VE development needed to be suitable for intended purpose based on desired use cases.

The second stage of the project focused on use-case development and developing user studies to research training material and delivery of training for the chosen cases, as well as iteratively developing the VE in line with the cases (Objective 2). Initial discussions with industry partners explored the use of training both for emergency situations and for safe operations in the workplace to reduce the risk, or mitigate the effects, of major incidents. In particular the identification of hazards relying on multiple sensory cues (such as gas leaking from a pipe) was noted as a potentially useful case for the application of MS VEs. Two user studies were outlined. An initial study investigated whether feedback modalities affected validity of behaviours in the VE, comparing a range of qualitative and quantitative measures, including evacuation times, qualitative interview data, analysis of movement data and subjective ratings, with and without thermal and olfactory simulation. A second study compared the effectiveness of (1) traditional training (slide presentation), (2) a vision-only VE and (3) an MS VE for two training scenarios developed with industry partner input. Knowledge uptake, motivation to train and attitude towards OSH were compared.

Over the course of the project, the VR prototype was refined with industry and stakeholder feedback to optimise practicality and usability for workplace training contexts, and acceptability of the technology was evaluated (Objective 5). This included analysis of barriers to adoption of VR-based training solutions and recommendations to overcome these issues. An overarching aim of the research was to produce evidence-based guidance on the use of VR in OSH training.

Further detail is provided on key stages of the overall approach in the following sections.

## 4.1 Development of the use cases

During exploration of possible use cases with Rolls Royce, a number of training scenarios that would be highly beneficial to the company were suggested. However, it was decided that these were too specialised to the company's unique situation. While it would be valuable to work on a use case with a particular need identified, to work on these specialist cases would present additional challenges as we would be highly dependent on Rolls Royce to provide details about the training subject, existing training materials, methods of measuring success, etc.; furthermore, it may not be possible to run studies with 'lay' participants, which could make these use cases unsuitable for experimental study due to limited access to suitable participants.

We tailored our selection of use cases to ensure that we have one to two use cases for experimental study.

A large number of potential training scenarios were generated and discussed in collaboration with Jaguar Land Rover (JLR) and Rolls Royce teams, who demonstrated a clear interest in the application of the technology to a broad range of H&S training areas. Our criteria for shortlisting options for user study development included the following:

- training was replicable in a VE
- met a relevant (current or potential) training need
- incorporated activities, environments or objects that involved heat and/or smell
- had measurable learning objectives
- was sufficiently challenging to find differences in performance
- was suitable for testing with available participant groups
- had sufficient (non-confidential) material available to deliver training.

Consequently, we decided on two uses: fire safety training and vehicle disassembly, described in the following sections.

### 4.1.1 Fire safety training use case

The first use case concerned fire safety training within a specific workplace context. It was necessary for the training to have sufficient complexity for comparison of effectiveness across the different conditions. The fire safety training covered both preventative and emergency protocols, and the use case comprised the following learning objectives:

- principles of fire prevention
- raising alarms
- maintaining exit routes
- what happens after an alarm is activated
- evacuation procedure
- evacuation checking
- different extinguisher types
- local building plans.

For all training conditions, the same topics were covered and the learning objectives were the same. The traditional training condition was delivered as a self-administered PowerPoint presentation, drawn from a real fire safety training presentation (adapted for suitability for this study). For the VE condition, the same material was taught, but 'in situ' in the VE, with experiential learning. For example, participants would be taught about maintaining exit routes for safe evacuation; in the VE condition they would encounter an obstructed exit route to illustrate this point.

#### 4.1.2 Vehicle disassembly use case

This use case explored whether a VE is suitable for training scenarios involving chemical smells. It involved procedural training in a disassembly task, in which an incorrect procedure led to an unsafe situation and had the potential to cause a critical incident.

The user was required to disassemble a vehicle, based on JLR's interest in end-of-life vehicle disassembly. During the procedure, the user should make safety checks and follow the correct (safe) procedure. The task was designed such that a fuel leak from the car caused a safety risk if the procedure was not correctly followed. Participants learned the procedure either with or without the addition of olfactory and thermal cues.

## 4.2 Experimental studies

In order to assess both validity of behaviours and effectiveness of VE-based training, we conducted two separate studies. The outcomes of the validity study were expected to influence the design of the training effectiveness study, and therefore the validity study was run first.

#### 4.2.1 Study 2 – Effectiveness of training

For the effectiveness of the training study there was no requirement for real-world behavioural data, but it was necessary to have access to, or to be able to produce, equivalent training through traditional training approaches such as lectures or written materials.

To assess training effectiveness, a pre-, post- and retention-test design was used. A knowledge test was be completed directly before, directly after and one week after training (the same test in all conditions) to ascertain knowledge uptake and compare participants' results. We also measured and compared motivation to train and engagement, completed post-training, based on subjective rating scales.

## 4.3 Specialist input

To support this research, we sought specialist input to complement the skills within the project team. This included:

### 4.3.1 Study 1 – Behavioural validity

For validity of behaviour, it was necessary to compare behaviours in a VE with those in the real world, and therefore the selected use case required sufficient real-world data to be available for comparison. From the list of possible use cases, the most comprehensive data for these purposes was available in relation to behaviours in a fire evacuation context. We therefore developed the VE and study design based around a fire evacuation use case, including heat and smell (burning wood).

### 4.3.2 Statistics

As agreed with the Institute of Occupational Safety and Health, we approached experts in statistics and research methods for studies of human behaviour, Drs Harriet Allen and Pete Bibby, Associate Professors in the School of Psychology, University of Nottingham. Drs Bibby and Allen supported the planning and analyses of studies conducted as part of this project. We also consulted with Dr Andrea Venn, medical statistician and epidemiologist, for support with the statistical analysis.

### 4.3.3 Fire/health and safety

We met with Paul Speight from Leicestershire Fire and Rescue Service, who has been looking at the application of virtual reality for improving safety awareness, to discuss project ideas. Paul is an experienced firefighter and was supportive of the use of virtual reality for fire safety training, thus validating our development approach.

We also received technical input related to health and safety from Dr Aled Jones, Faculty of Engineering Health and Safety Adviser, University of Nottingham.

### 4.3.4 Olfaction

We had support from Professor Monique Smeets, Chair of Sensory Systems at the University of Utrecht. Professor Smeets provided advice on the delivery of the scent, including establishing safe levels of fragrance and defining a protocol for olfactory simulation (Section 4.5.2).

## 4.4 Technical solutions

### 4.4.1 Project VR prototype

The VR prototype was based on training requirements specified by industry partners and was developed iteratively with user feedback. Formative usability studies in combination with studies of smell and heat perception informed the design of the system, including approximately 40 user study sessions in total.

The configuration of the prototype system is shown below in Figure 4-1 and Figure 4-2:

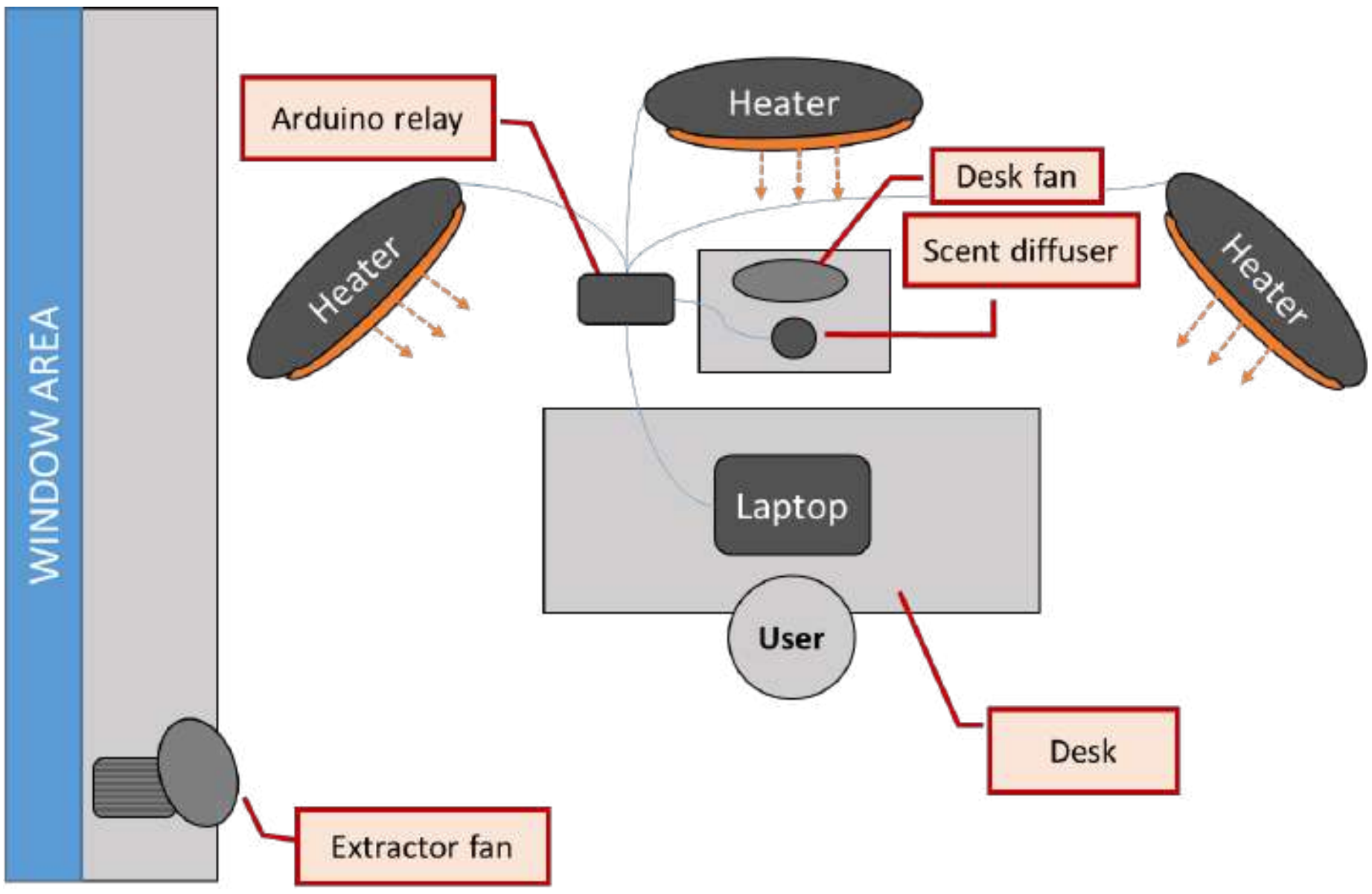

*Figure 4-1 Schematic of the prototype VE*

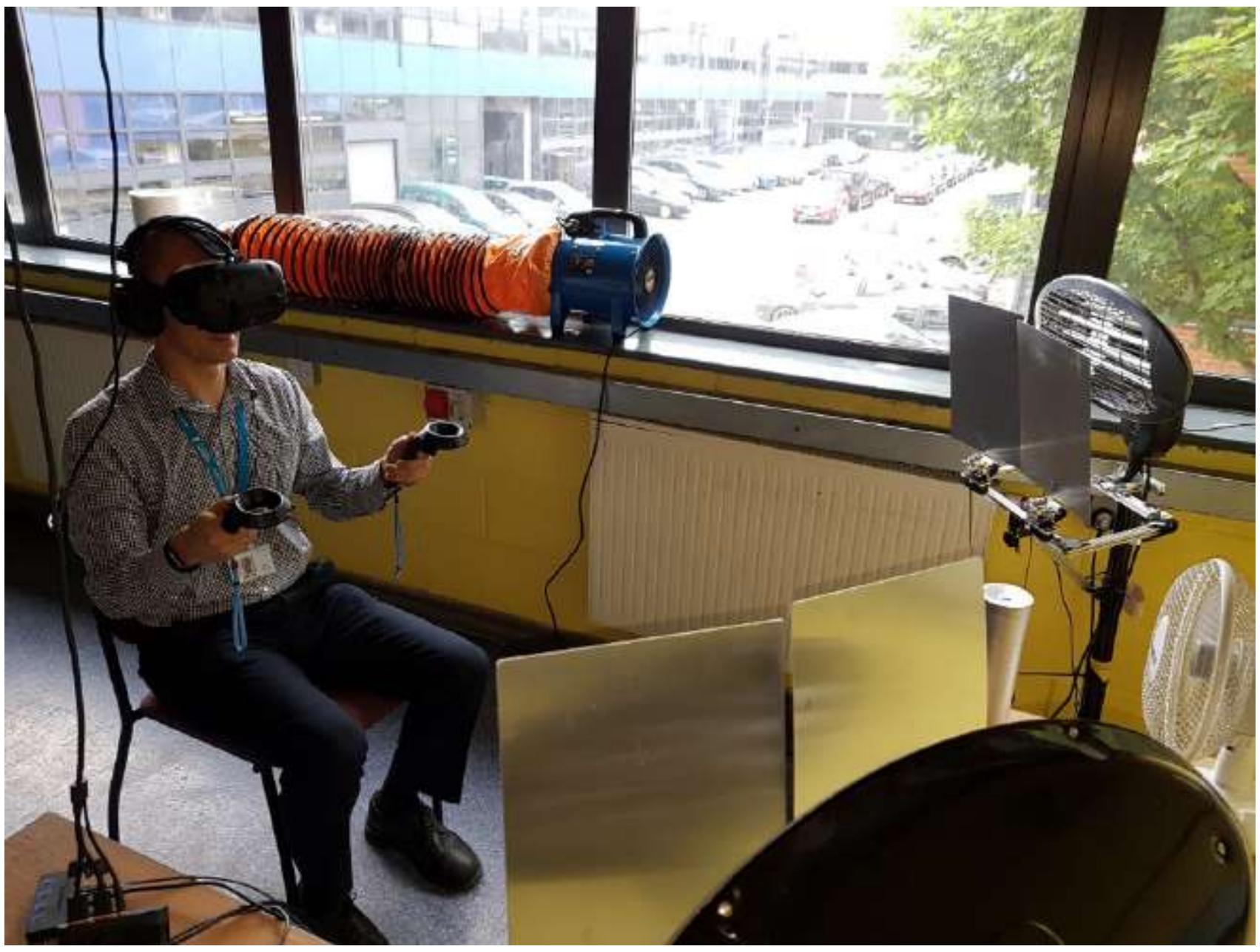

*Figure 4-2 User interacting with prototype VE set up*

The VE was run on an Alienware laptop with external Geforce GTX 1060 graphics card. A Vive (78) HMD provided eye-separated 3D stereoscopic visualisation of Unity Technologies' (79) VE and tracked user head orientation to allow the participant to visually explore the environment, enhancing the immersive experience (8,39). User direction of movement was defined by head orientation in the VE and two wireless controllers with directional touchpads – one for cross-axis movement, one for pivoting – were used for interaction and navigational movement. Noise-cancelling headphones delivered audio simulation that masked the noise of apparatus during use. This prevented possible extraneous effects of, for example, hearing the scent diffuser in advance of smelling the fragrance, while also preventing external distractions, ensuring that the user is immersed in the VE. The VE had audio integrated as appropriate for each use case. Smell and heat hardware devices were controlled by Arduino relay. Multi-sensory simulation was delivered as described in the proceeding section. Audio, thermal and olfactory simulated fire cues, e.g. crackling or burnt wood scent, were activated by participant movement over trigger points within the VE set in proximity to the fire.

## 4.5 Multi-sensory simulation development

### 4.5.1 Thermal simulation

Thermal feedback systems have been successfully implemented outside the OSH sector with Peltier devices (80) being the most common approach. Wearable devices that apply different temperatures directly to the body are often used to achieve heat simulation. Traditional interfaces often require thermal feedback to the hand or finger, and this is reflected in the literature reports on thermal interfaces. However, application of thermal simulation to immersive environments may require feedback to broader and more numerous areas of the body. Recent work by Garcia-Valle et al. (81) explored this using a haptic vest with a built-in thermal feedback system. They found that thermal feedback increased users' sense presence and realism, but control and perceived synchronisation needed to be improved.

In the context of fire evacuation scenarios, thermal feedback needs to provide the sensation of radiant heat. Thermal feedback systems in direct contact with the user's body do not provide this and there are few reports in the literature on the effective application of this sensation. One example, described by Lécuyer et al. (82), is the HOMERE system, which simulated environmental heat by surrounding the user with twelve 150 W infrared lamps. However, they reported that users did not adequately perceive the heat and therefore implementation of thermal feedback requires further development.

In the current project, heat was initially delivered through a 400–800 W halogen heater, but our initial testing showed that we needed better control over duration and perceived direction of the heat source, and a stronger heat with variable temperature. The heater was therefore replaced with two 2 kW infrared heating devices, which were placed to the left and right in front of the user such that either or both can be activated depending on the angle of their approach to the heat source in the VE. The IR heaters had the advantage of simulating heat in immersive environments where thermal feedback is not only required on contact with a single body part but should be felt as a radiant heat applied to the whole body. Servo-controlled fin-type apparatus controls the heat from the devices. This novel development was introduced to overcome the time-lag of the heaters warming up as the avatar approached the source of the heat in the VE (Figure 4-3).

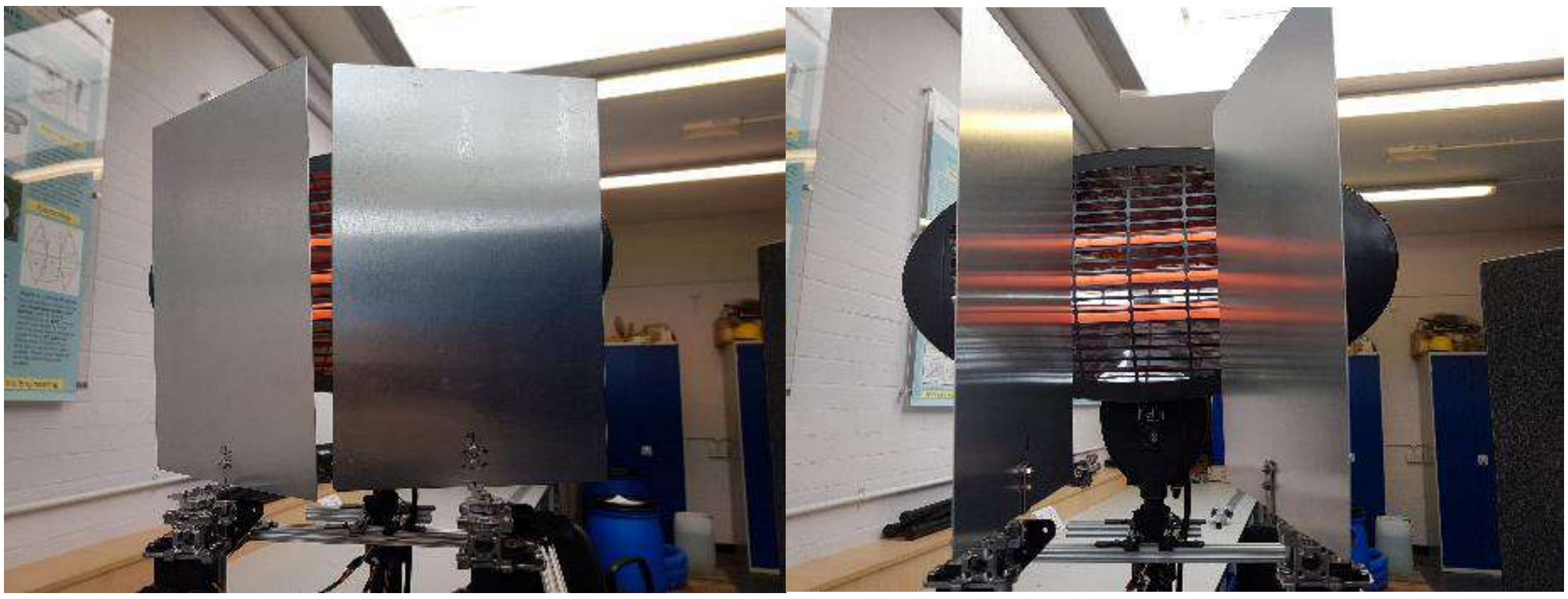

*Figure 4-3 Control of heat from 2 kW infrared heater: with fins closed by default [left] and opening with Arduino control to direct heat towards user [right]*

A student project supervised by Dr Lawson investigated the perception of heat sources in the VE (9). The project explored what extent the user's experience was affected by positioning of the infrared heaters and whether this affected subjective experiences, for example the realism of the simulation. Results from the student project showed that identification of heat source locations was generally accurate, with minimal or no deviation between perceived and actual location of the heaters. They also suggest that the addition of heaters improved the subjective realism of the simulation, but that the exact configuration of symmetrically placed heaters does not affect realism. Therefore, in training simulations, exactly matching the heat source location to the visual counterpart is not necessary for realism. The timing of the heat as delivered in this simulation study was perceived as appropriate when the heaters were placed symmetrically around the participant. However, subjective data showed that overall the heat level was too low. Improving the perceived heat level to be more realistic would require higher output heating devices. Therefore, an additional 2 kW IR heater was used in the final VR prototype set up.

### 4.5.2 Olfactory simulation

Olfactory simulation was delivered using a SensoryScent 200 fragrance diffuser, an electronic nebulising delivery system able to provide a concentrated, realistic scent. As with thermal simulation, the fragrance diffuser was controlled with an Arduino relay which triggered delivery of the scent when the user approaches the source of the smell in the virtual environment.

Of the fragrances sourced, a 'burning wood' fragrance from Osborne Technologies Ltd. (https://www.osbornetechnologies.co.uk/) was selected for the fire evacuation scenario. We also selected a diesel fragrance oil for use in the vehicle disassembly training scenario. A wide range of other fragrances are available for use with the diffuser to suit different use cases. Bespoke fragrances may also be available at additional cost if required.

Our user testing showed that the olfactory simulation was most effective if the diffuser was positioned relatively close to the user (at an approximate distance of 1 metre), with the fragrance being directed towards the user's face with a fan positioned behind the diffuser, and with an extractor fan behind the user to control flow and extract fragrances quickly. Advice from experts including Monique Smeets, and review of literature, suggested that users may become 'blind' to smells after a period, and that testing delivery of fragrances in configuration of short bursts may be most effective for detection. Additionally, certain measures can improve olfactory simulation, including ensuring an odour-free environment by removing soft furnishings (for example, use the system in an uncarpeted room without curtains, use hard chairs, and cover wood with foil).

The scents sourced for the olfactory feedback were verified as safe by the suppliers and manufacturers. However, the way the scents were used within the prototype was different from the intended purpose. Therefore, as part of the design and develop process for Objective 1, we developed a protocol for the safe use of this off-the-shelf technology for use in MS VR simulation.

**Protocol for the safe use of fragrance diffusers in the VR prototype**

Using advice from Professor Monique Smeets, the following protocol was developed and used in the production of the present study's prototype, to ensure the safety of users exposed to olfactory simulation from an MS VR training simulator:

**Risk assessment**

All fragrances used must be approved by the International Fragrance Association (IFRA) to ensure they meet recognised safety standards. A control of substances hazardous to health (COSHH) assessment must be carried out on scents used in the simulation. The quantity of substance emitted from the diffuser and the level of exposure to users during simulation should be established in order to assess risk to users, both in terms of inhalation and in the flammability of the diffused scent. For the present study this was less than 0.1 g per 1 minute of active diffusion, equating to <0.1 ml of substance by volume, and users were exposed to short intervals each of approximately 3 seconds, up to a total duration of exposure of 2 minutes. To prevent build-up of excessive vapour and limit exposure levels the following safety measures should be taken:

- An extraction system should be set up to remove diffused fragrances.
- The prototype should be set up in a well-ventilated room with windows that can be opened.
- If exposure levels exceed those expected, the study must be terminated.

**User testing**

To minimise risk to user comfort and safety, fragrance oils should be used in the lowest concentration possible. User testing should take place to establish the minimum exposure required for the scent to be perceived. User testing for the present study exposed participants to low, medium and full concentrations of 3 second bursts of each of the fragrance oils under the same conditions used in the VR prototype. Results showed that the wood smoke scent was only perceived consistently at full concentration when the diffuser was placed no further than 1 metre from the user and a desk fan was used to direct scent particles. Therefore, it was concluded that the wood smoke fragrance concentration could not be reduced for the physical configuration of hardware for scent delivery should be based on the outputs of this user testing to optimise olfactory simulation. However, the diesel fragrance was reliably detected at 50% concentration, and was therefore diluted with a carrier agent.

**Participant screening**

To minimise risks to participant safety and health, a screening process should be used prior to use of the simulation and any participants with allergies, asthma, respiratory conditions and odour intolerances should be screened out of undertaking the training. This was carried out for the present study as part of the recruitment campaign, participant information and consent process.

## 4.6 VE development

### 4.6.1 Custom VEs

The use of low-cost/low-expertise scanning technologies to create bespoke workplace models was investigated in the initial stages of the project (Objective 4):

#### *4.6.1.1 Google Tango*

Replicating the real environment in which the training may be applied would draw on the well-established principles of context-dependent memory, which could improve recall in both preventative and emergency safety situations. We therefore investigated methods of creating custom environments. Custom VEs could be created by developers using pre-made or specially designed objects, but not all companies would have the resource to do this. Therefore, we investigated whether an existing scanning technology (Google Tango) could facilitate the creation of bespoke training environments with minimal resource and prerequisite expertise. This would allow a larger number of companies to potentially make use of this customised training. Google Tango is a low-cost (~£400) option which uses an Android device to create a virtual representation of a real environment (Figure 4-4). After scanning the scene, the resulting 3D mesh can be imported into Unity to create the VE (Figure 4-5).

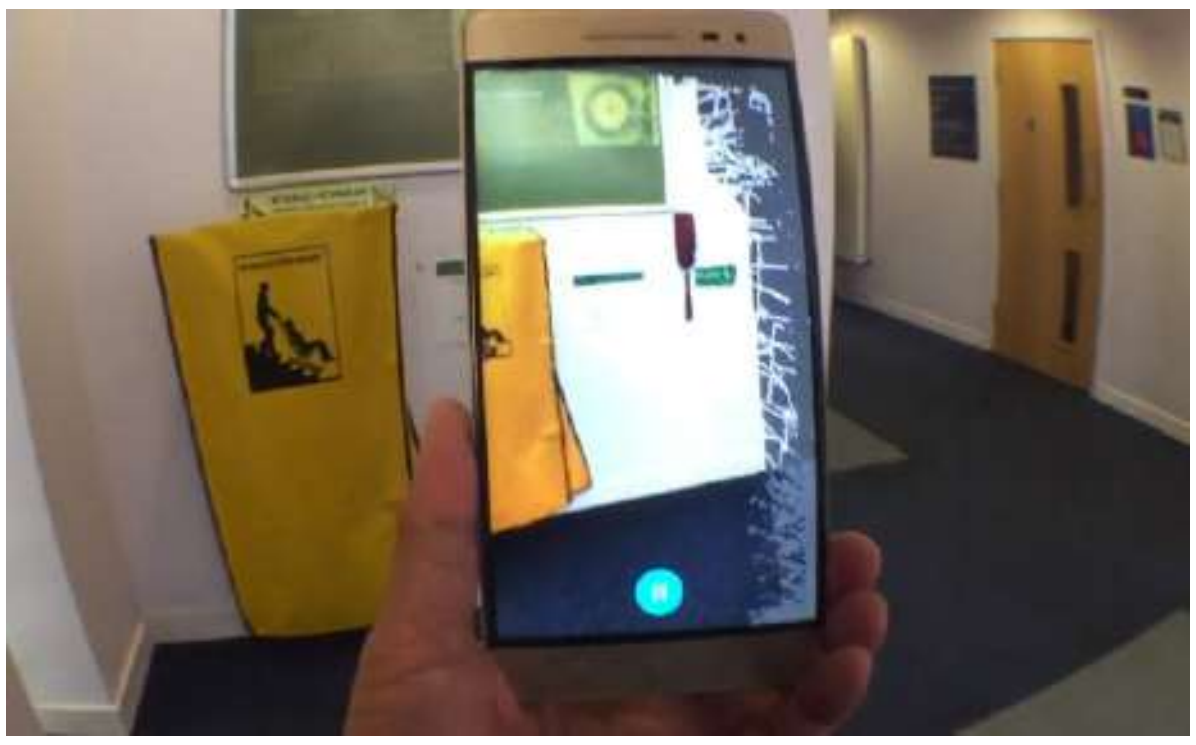

*Figure 4-4 Handheld device for real-time acquiring of 3D mesh from the built environment*

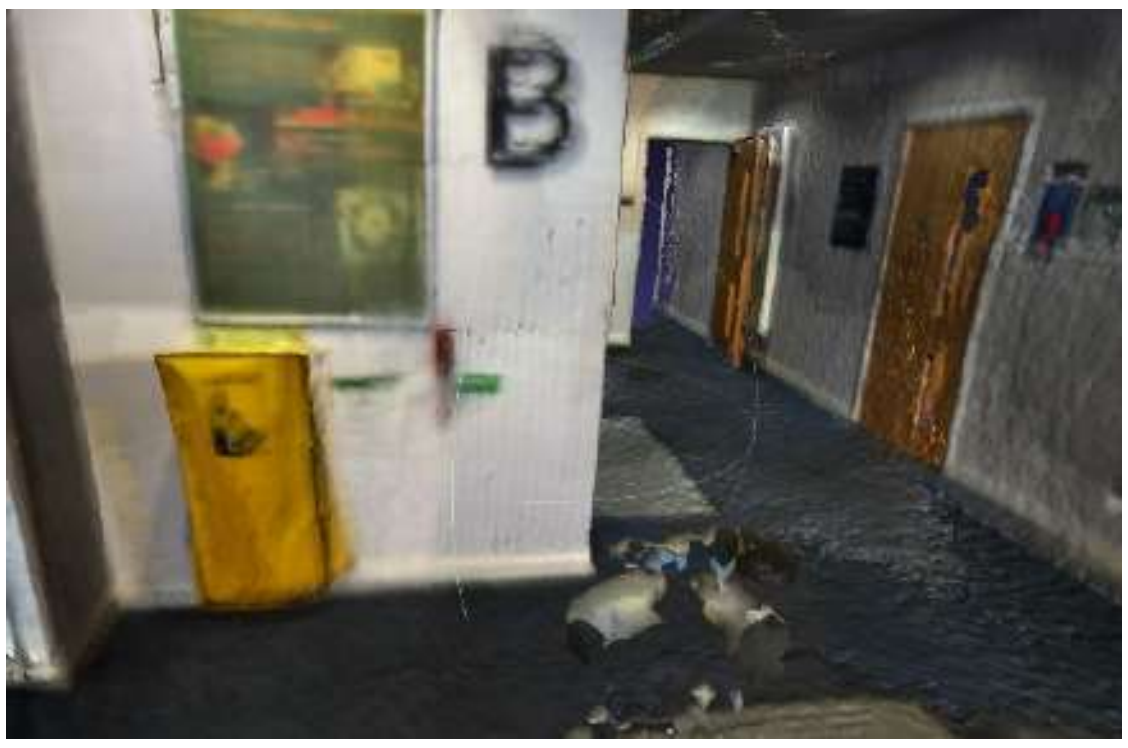

*Figure 4-5 The acquired 3D mesh is now visualised in Unity*

While we had some success with scanning small areas, we identified a number of issues with Tango that affected its suitability for these purposes. It was difficult to automate the scanning process, which reduced reliability. The data capture process was dependent on lighting, reflection and visibility of objects, which could be hidden from the scan view, and there were sometimes problems with over- and underexposure (see Figure 4-6 and Figure 4-7 for examples).

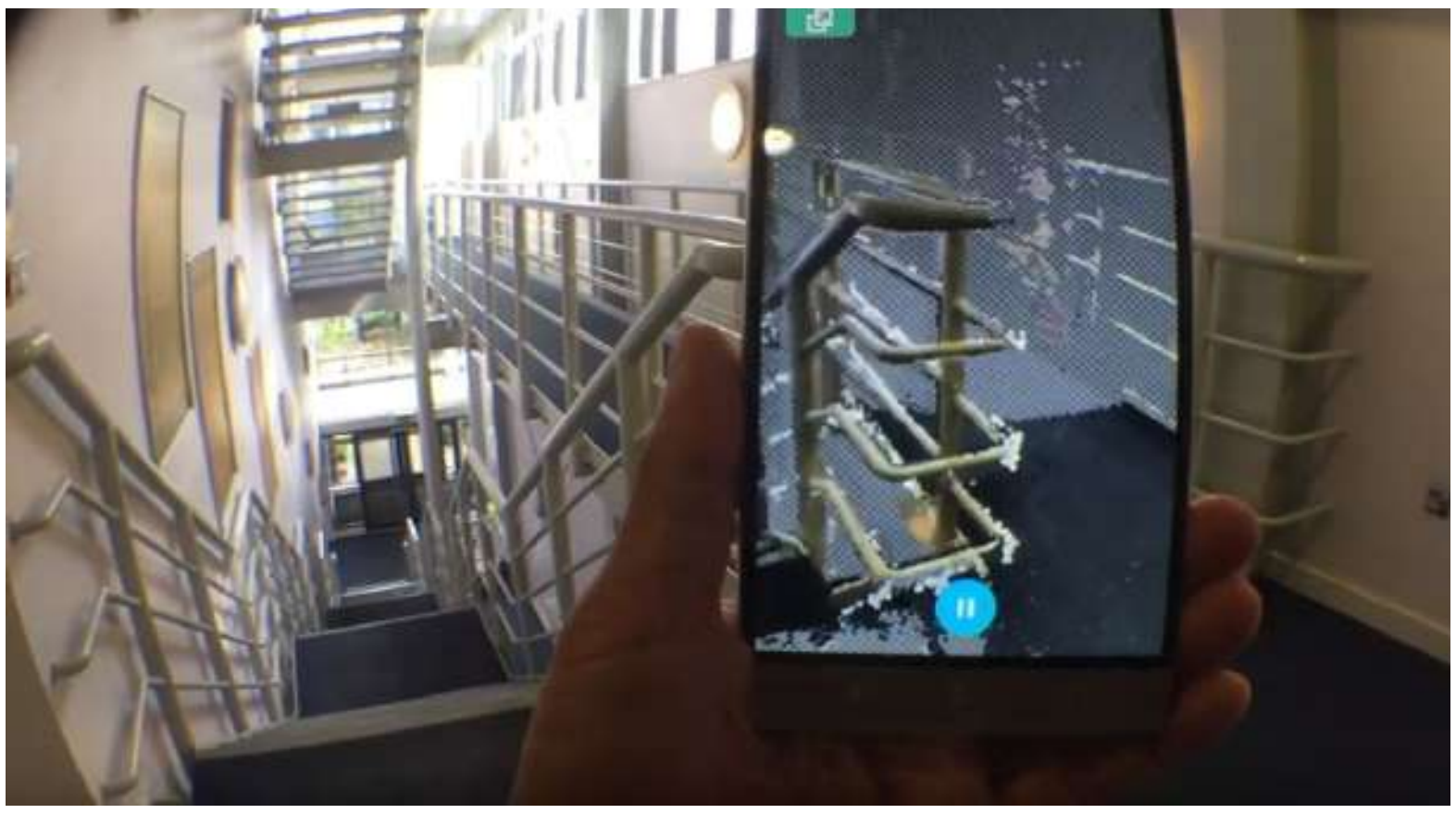

*Figure 4-6 Issues in capturing scans caused by lighting and reflections*

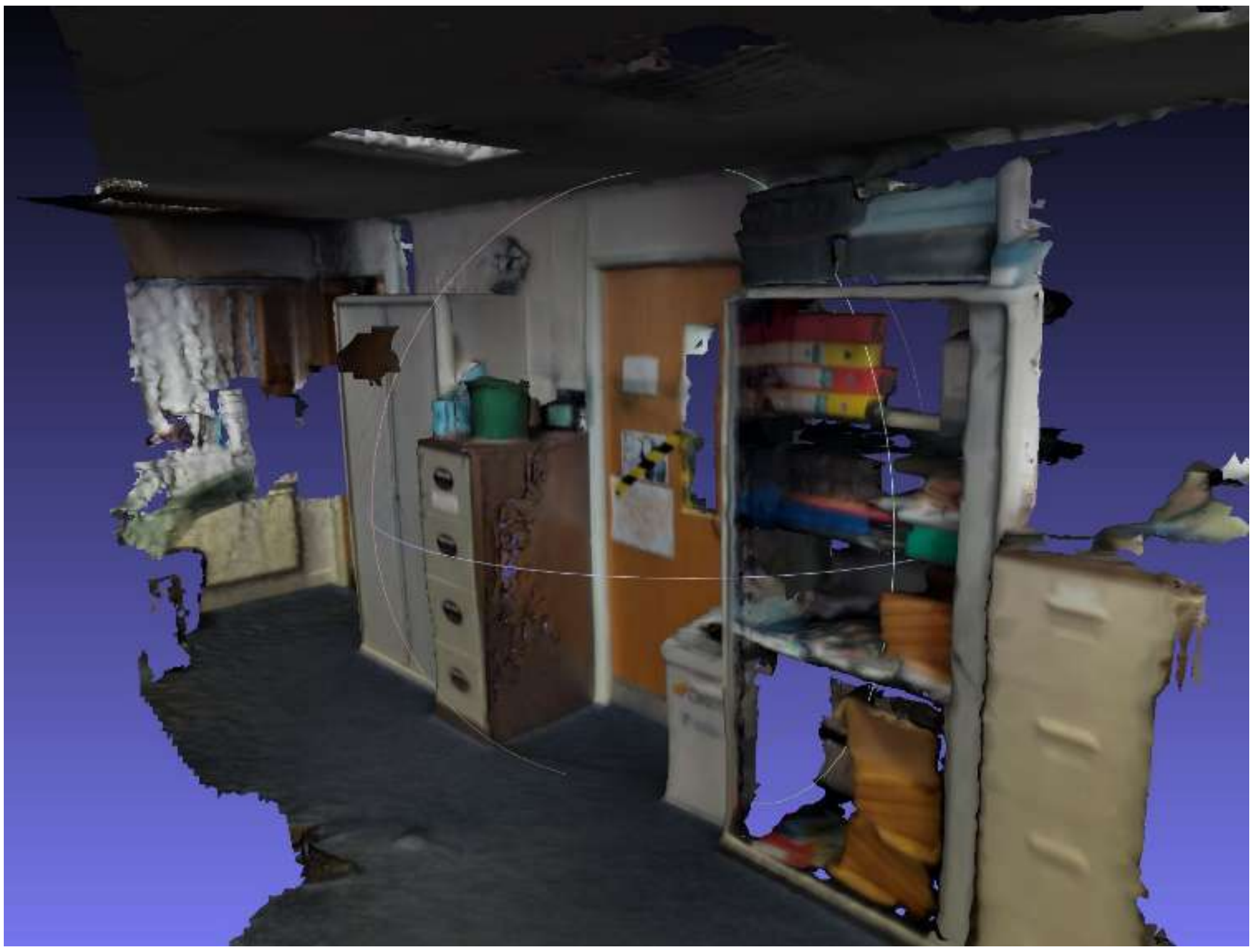

*Figure 4-7 Problems with over- and underexposure related to the amount of light within the environment and reflections from glass*

Scanning large areas was also problematic, as mesh elements can be missing (Figure 4-8), and the file sizes can become large.

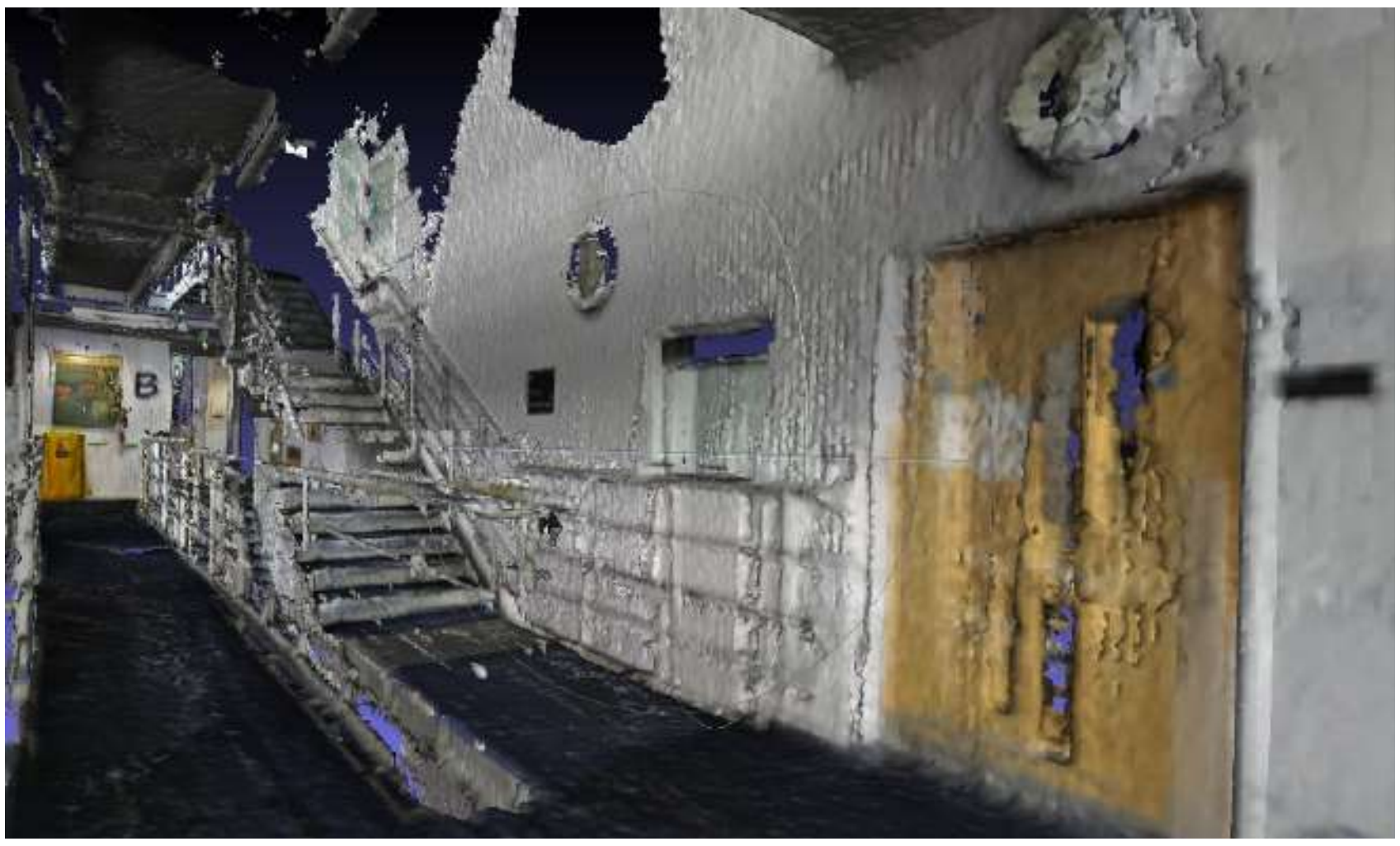

*Figure 4-8 Scanning of a large area with missing mesh elements*

Double exposure or loss of orientation experienced by the device could result in a duplicated mesh or incorrect mesh orientation (Figure 4-9).

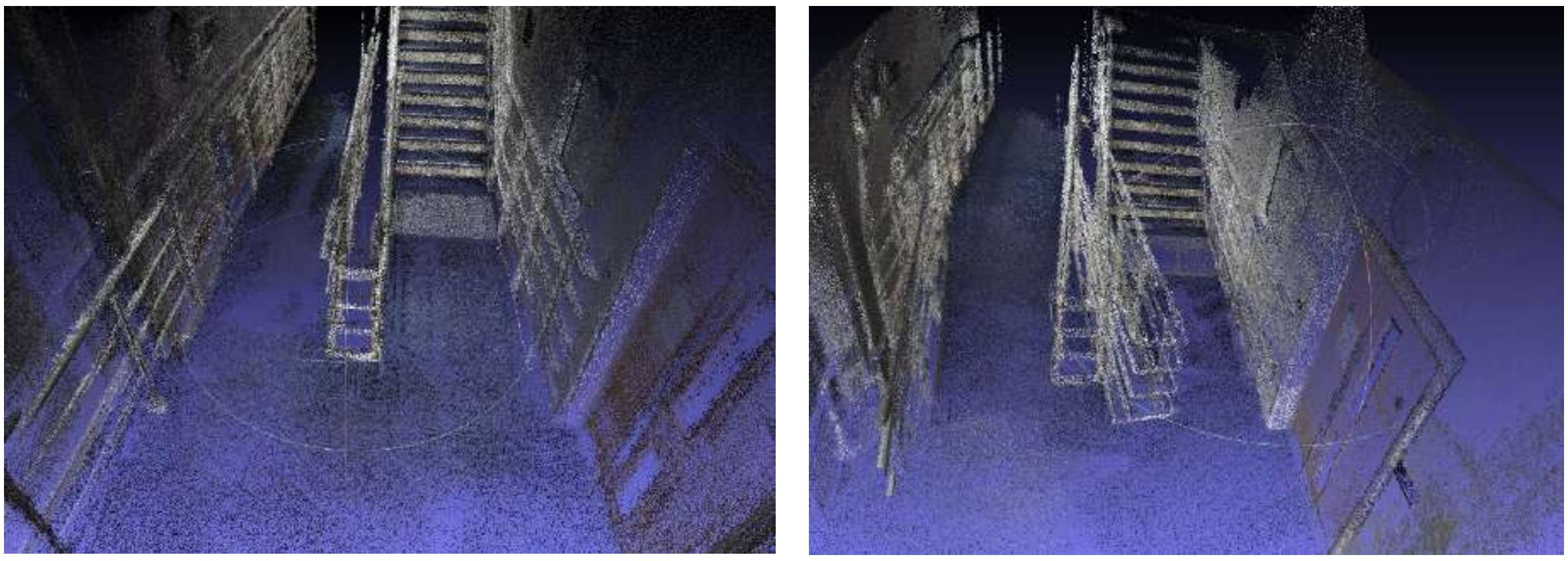

*Figure 4-9 Double exposure and orientation issues*

In typical use, the resulting Unity scene was not of a sufficient quality for the training application; see Figure 4-10 for an example of a typical scan imported to Unity. When the texture (the image applied to the surface of the 3D object) was removed from the mesh (Figure 4-11), it was apparent that the automatically reconstructed mesh failed to follow the exact shape and geometry from the built environment.

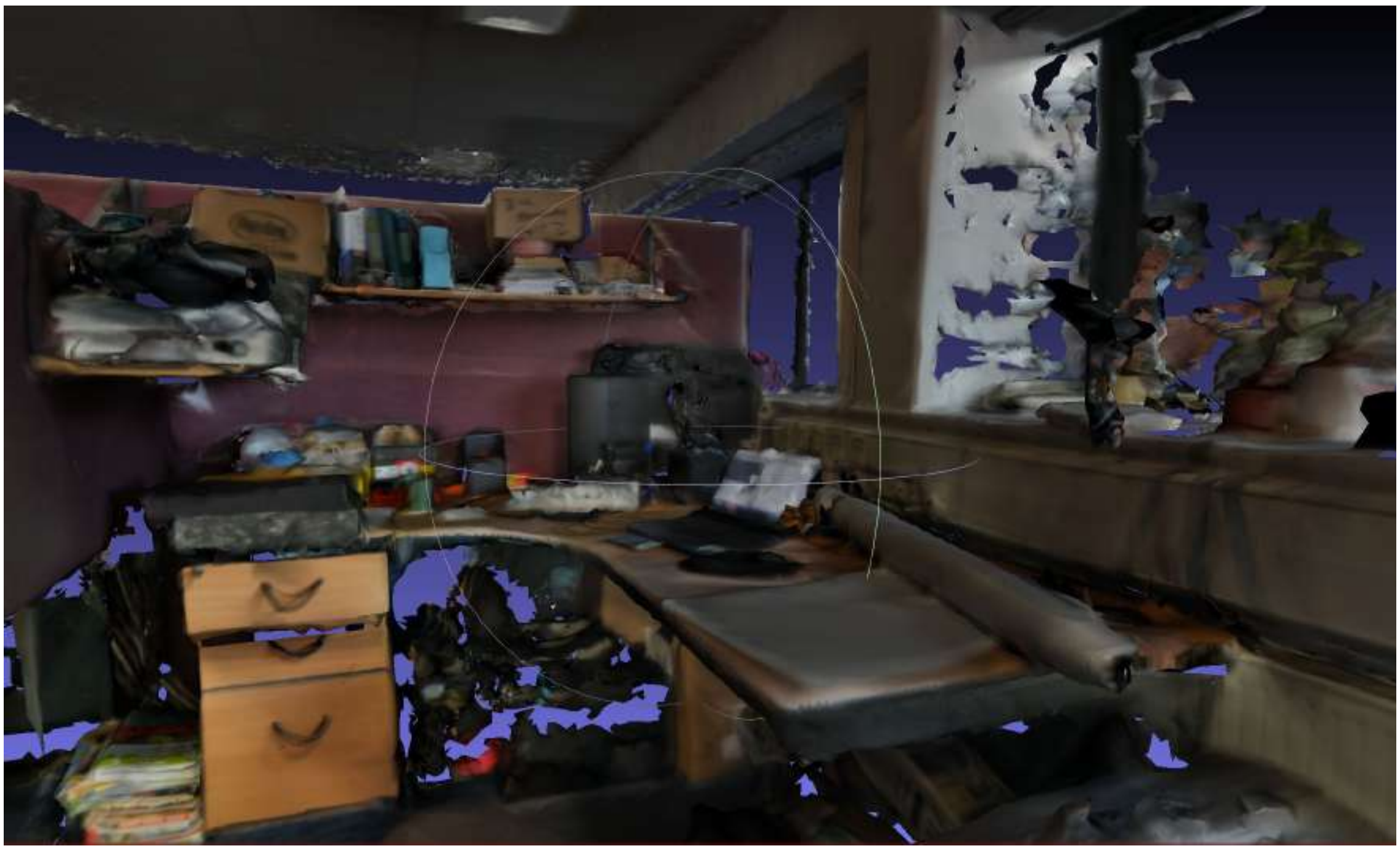

*Figure 4-10 A typical Unity 3D scene quality with the Project Tango acquired 3D mesh*

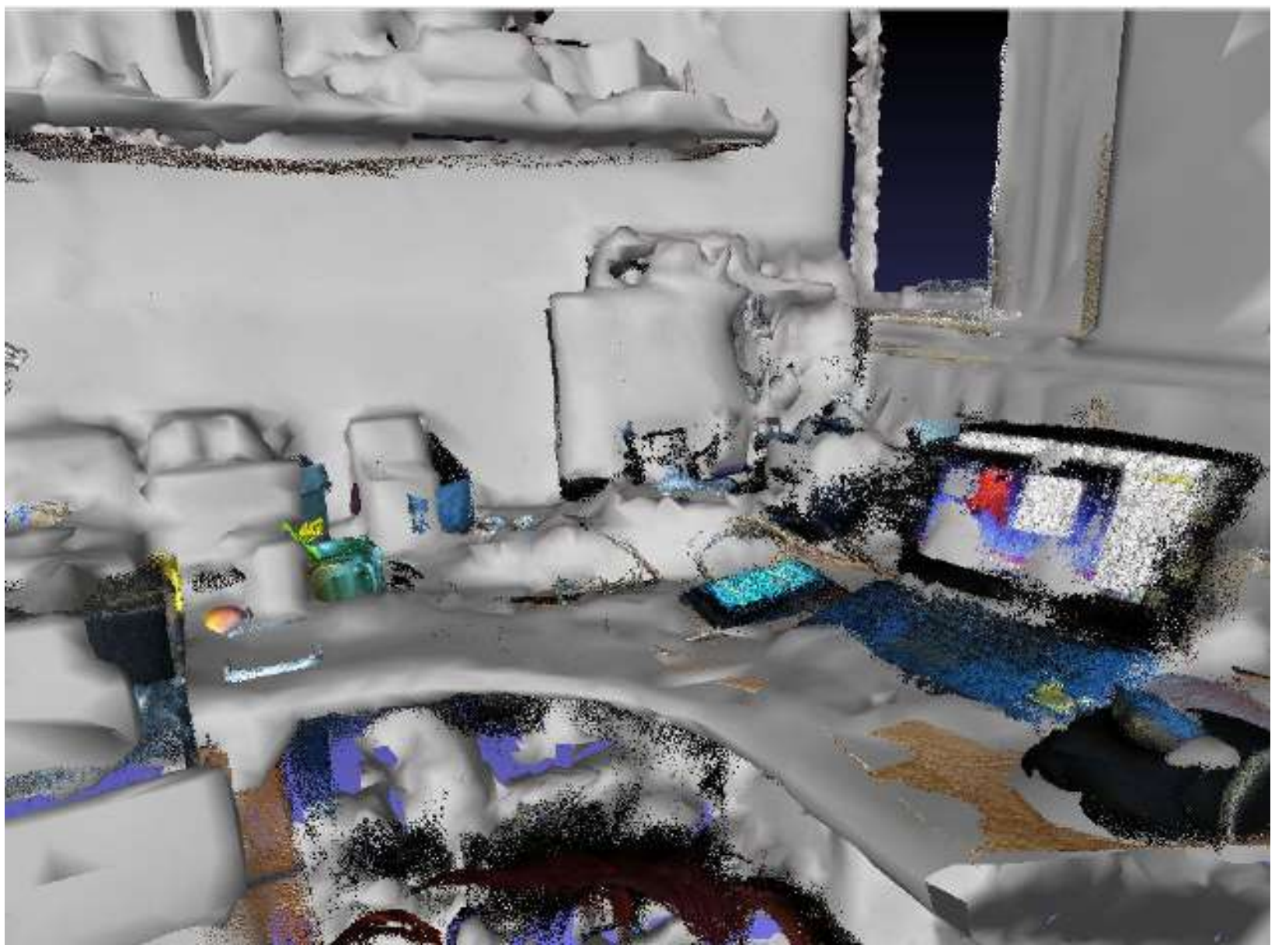

*Figure 4-11 Mesh with texture removed, showing meshing issues that have produced the poor-quality result. The automatic mesh is in grey, while the point cloud is in vertex-colour*

The algorithm for building and triangulating the 3D cloud points into a 3D mesh was not accessible via Unity and therefore tended to reduce the meshing and produce a poor-quality result. It was possible to process and clean the point cloud using an open source software called Meshlab. This involved the user selecting the mesh, deleting outliers, compacting and looking for duplicates or overlaps before exporting back to a file format that could be read correctly by Unity. Figure 4-12 shows the result of partial cleaning using Meshlab, which visibly improved the quality. However, this process required expertise and knowledge which would not typically be available within companies.

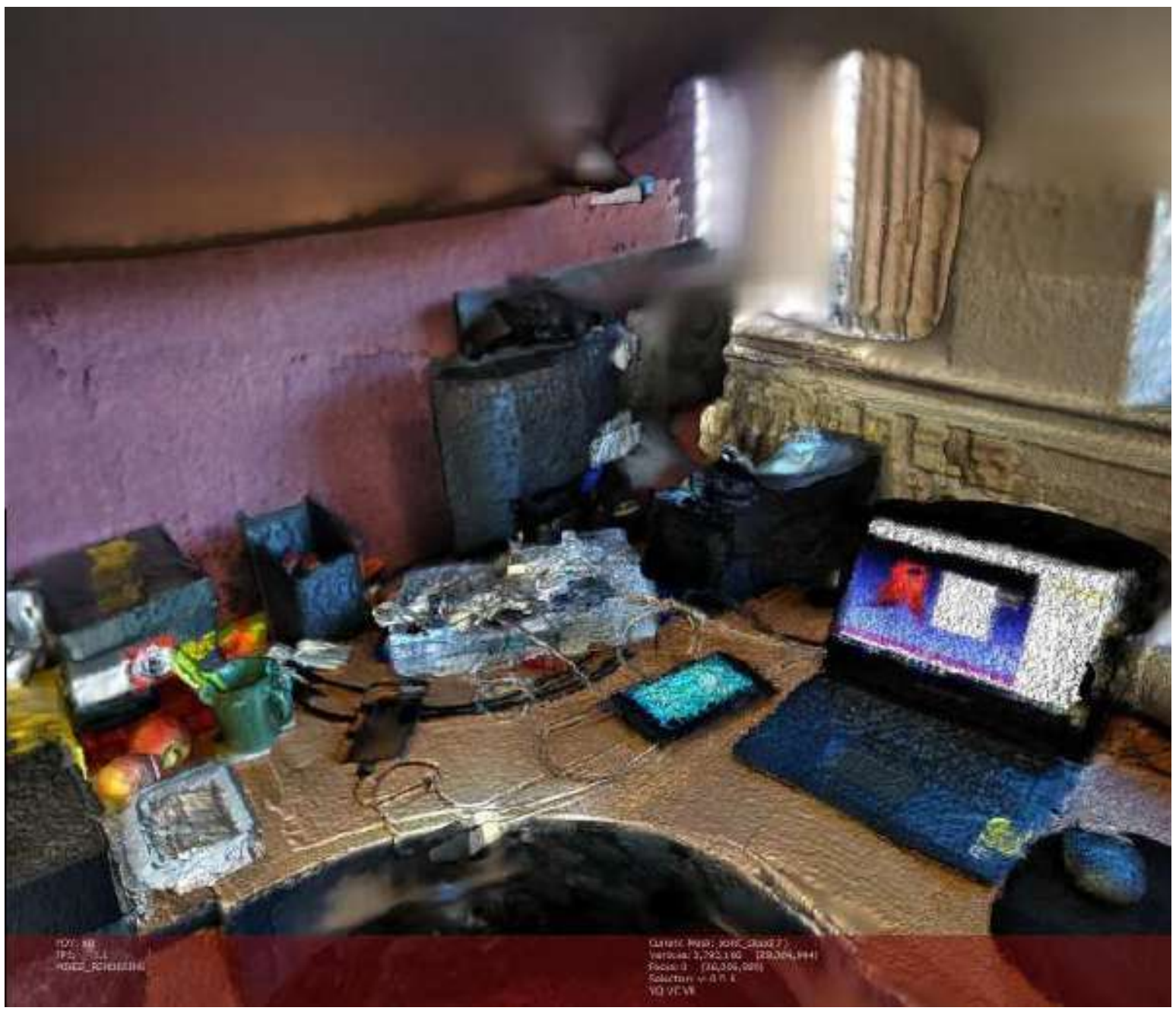

*Figure 4-12 Result from using Meshlab filter type – Screened Poisson Surface Reconstruction*

Overall, while Tango offered the potential benefits arising from creation of customised VEs at a low cost, there existed significant challenges with the use of Tango devices for a company to replicate their own premises with sufficient quality for training application. However, the quality can be improved with user guidance and limited subsequent editing. Therefore, we created a protocol for creating site scans using Tango to improve results, based on the conclusions of our testing. In brief, this comprises:

- optimising lighting in the area to be scanned (*avoid over-exposure; ensure consistency; test to establish adequate lighting levels*)
- clearing objects from rooms, especially elements not next to the wall. Any objects offset from the wall could become noise elements when captured by Tango
- deciding a forward-view path; scanning the left-side of the wall to a maximum of 5 metres distance, saving the file, then return to the starting point, and scanning the right side of the forward-view path for a maximum of 5 metres distance again
- importing into Unity3D, copying the files through USB connection to Tango
- scaling and rotating to fit the scene, and placing this near to the virtual Tango camera location inside Unity3D
- changing the material type for all imported meshes from Tango to VertexLit to get the right rendering for this mesh inside Unity3D.

Additionally, in order for the VE to provide interactive training, some modifications were required to the scanning process. For example, doors needed to be left open for scanning to enable the user to subsequently insert an interactive door element, which can be opened and closed from the object libraries. We created this solution and included the door element in the classifier section described in section 4.6.1.4 below. The object library enabled manual addition of any objects that could not be scanned.

#### *4.6.1.2 Alternative solutions*

We identified alternative options to Tango for users to scan their own environments, such as 3D laser scanning. However, they had significant drawbacks, including high hardware costs, extensive training prior to use, very large file sizes and high-end computers required to process the scans. Most systems involved placing multiple markers in the scene and performing a number of scans, which were then manually resolved using specialist software. We concluded that this was unlikely to be an accessible option for SMEs.

#### *4.6.1.3 Matterport*

An intermediate alternative identified was the Matterport capture system (http://matterport.com), which largely automated the scanning process and simplified the procedure, reducing the need for experience or training. It also used Cloud processing so that results could be obtained relatively quickly without the need for high-end hardware in-house. However, Matterport still required combining multiple scans from static locations. Matterport created a polygon planar mesh and texture based on the camera texture scanning rather than using 3D point cloud for the end result, which might affect accuracy on some scales. The Matterport was a more expensive option than Tango, starting at approximately $4,000 with recurring costs for processing the data, required for additional spaces.

#### *4.6.1.4 Classifier objects*

The training scenarios required development of case-specific classifier objects. For example, the evacuation checking procedure uses a token system, designed to monitor the evacuation status of the workplace. In this system, any person passing a fire token should collect the token, check the rooms listed on it (knocking loudly if they are unable to enter), and, once they reach the evacuation assembly point, hand the token to the fire marshal to confirm that the designated area has been evacuated.

We developed classifier objects, which were objects with predefined behaviours that could be used as part of training scenarios, for example alarms that could be activated by the user. The aim was that any VE – bespoke or generic – could be customised with elements related to the training objective. Our classifier development was based on two approaches: the first was user-insertion of classifier objects into the scene, and the second was automatic determination of important elements to a scenario design, for example 'emergency signage' which used a 2D pattern-match algorithm to automatically classify. Figure 4-13 shows an example library of selected elements that users can customise and insert into a runtime scene to create a scenario. The elements in this example were related to a fire scene. Several libraries like this could be designed to address different scenario requirements. Figure 4-14 shows an example where the user selected a door module from the library and inserted this into a runtime scenario. Upon insertion, the user could manipulate properties like the door's orientation, position and scale. This was an interactive module, where the door component had colliders placed at ground level for detecting collision with an avatar, and it could start checking for the nearest distance to the avatar for triggering door-opening or -closing animations. Alternatively, the door could have a trigger for detecting the avatar's hand before activating the door-opening animation sequence.

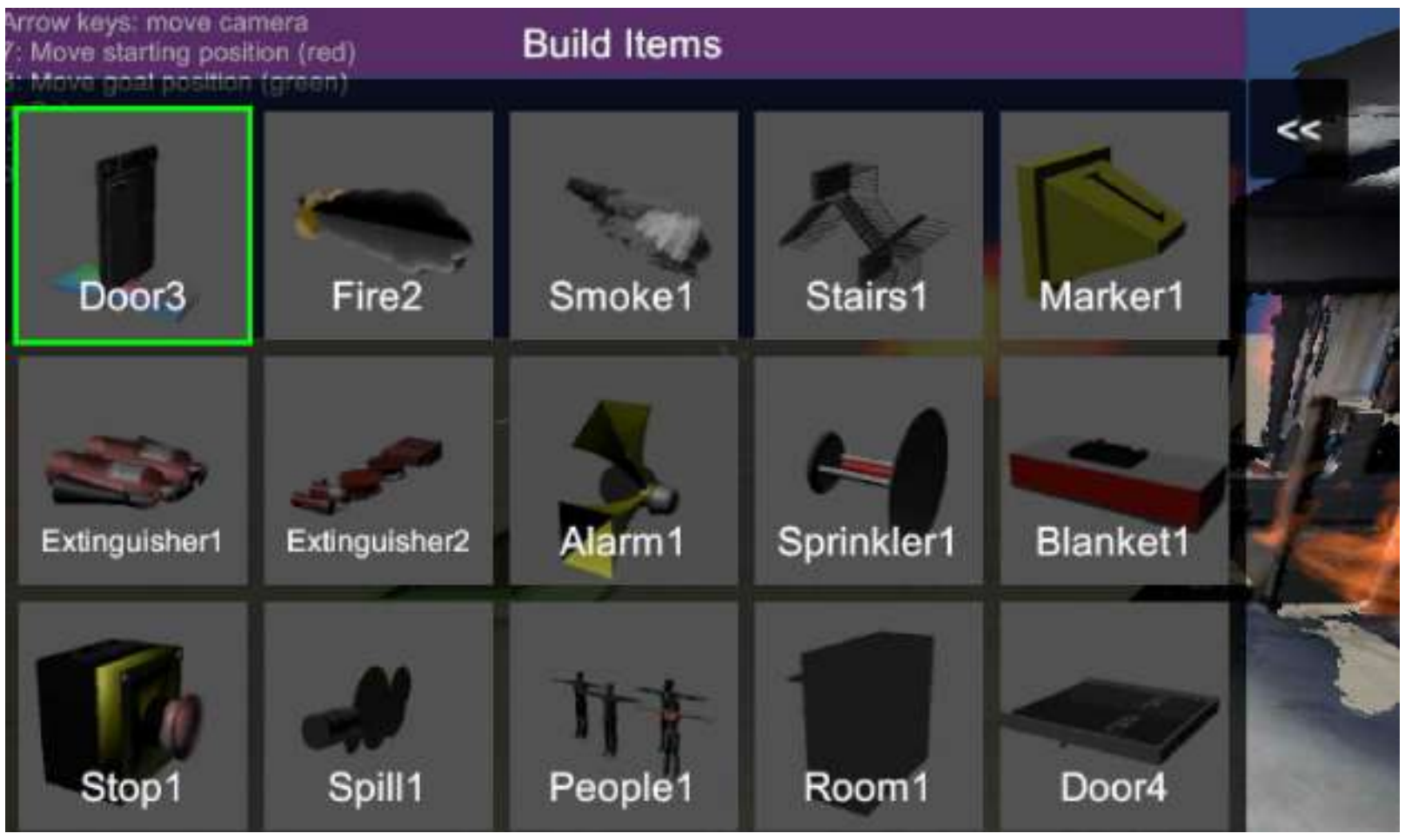

*Figure 4-13 A library of elements the user can insert into the scenario*

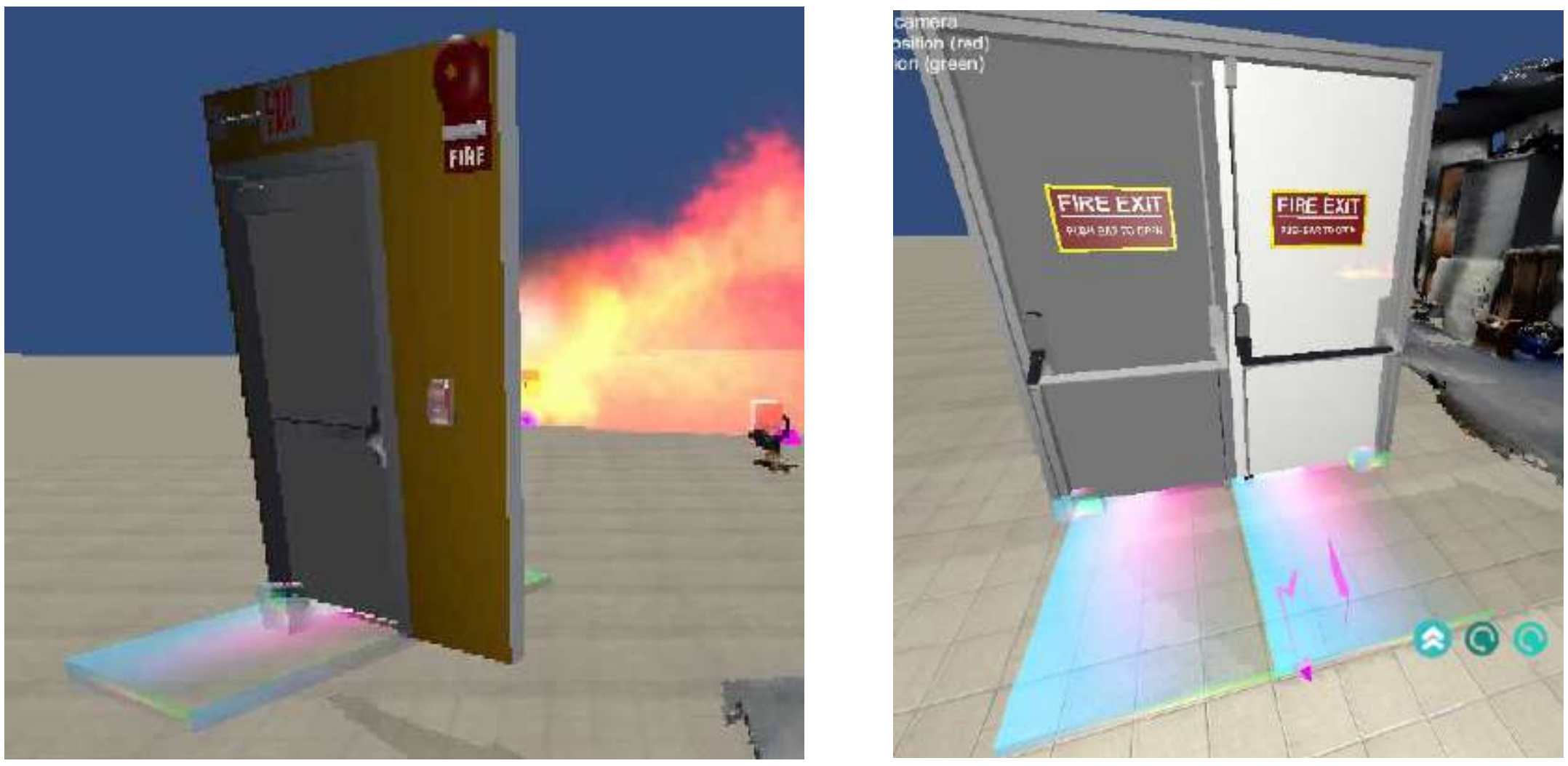

*Figure 4-14 User inserts a door module into a runtime scene, and adjusts the door's properties*

Where it was difficult to acquire a 3D mesh while using Google Tango due to limited lighting or space constraints, we enabled the use of building blocks to simulate the built environment. This building block can be customised for various sizes, textures and shapes and can be stacked (Figure 4-15).

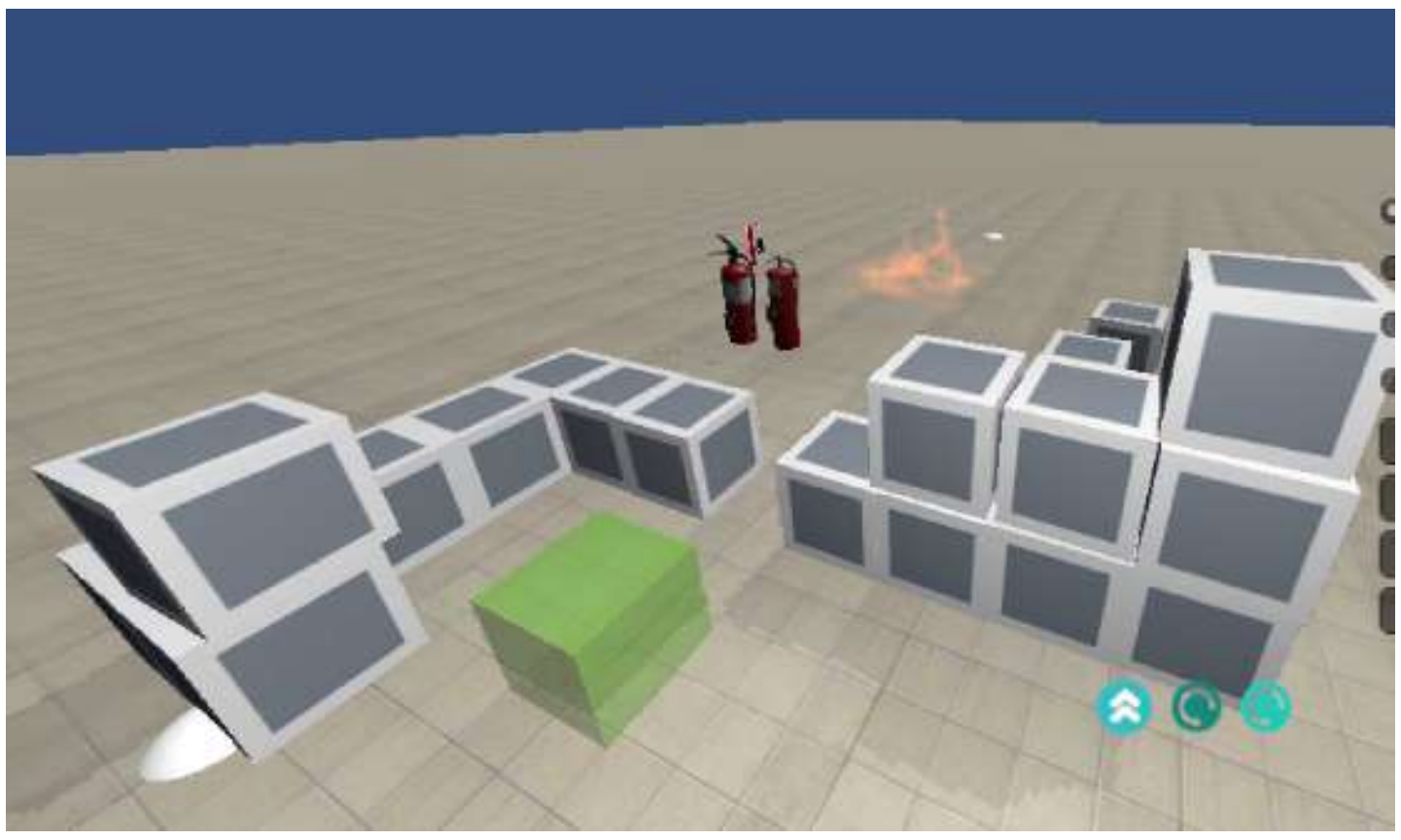

*Figure 4-15 User adding basic building blocks into the runtime scene*

Figure 4-16 shows the development of a different type of classifier based on image processing at runtime. This allowed for automatic identification of classifier objects within the Unity scene, based on image recognition. After the user imported and positioned the acquired Tango 3D mesh into the scene, they could provide the classifier library with a collection of images related to the scene. Examples given in Figure 4-16 include some fire-related images like signage, fire doors and fire extinguisher:

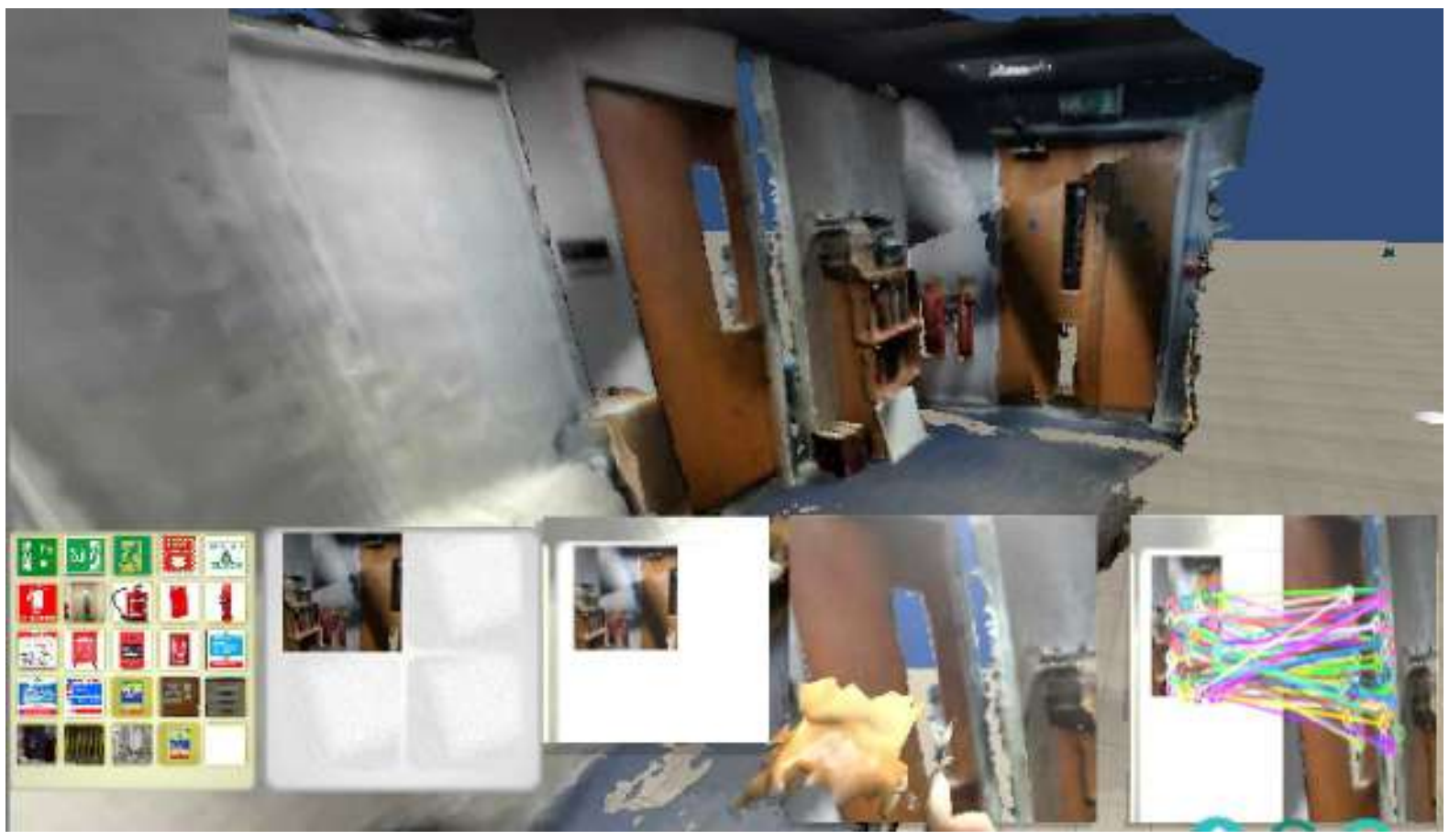

*Figure 4-16 Classifier method based on image processing for recognising certain features in the runtime scene*

The user could select a small area in the scene during runtime for this classification process, as shown in the middle section of the figure. The far-right section shows the process of 2D pattern match for similarities between the selected region and the scene as seen by the active camera.

Figure 4-17 shows a classifier image (a fire extinguisher) taken from the library via drag and drop action. The classifier looks for similar 2D patterns via the active camera in the scene. This facility allowed object behaviour to be invoked following automatic recognition of the classifier objects.

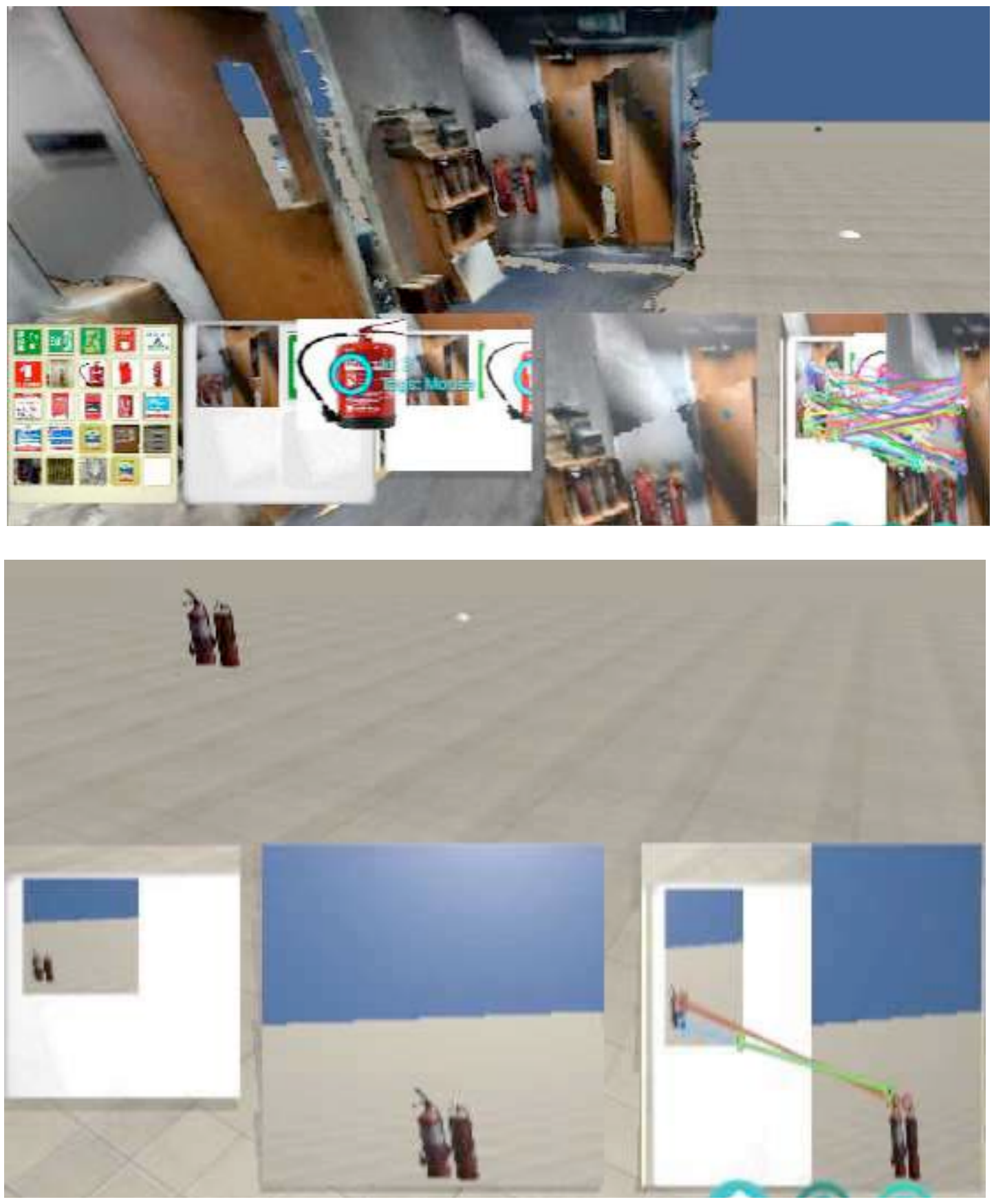

*Figure 4-17 The user process for matching items in the scene with classifier objects*

### 4.6.2 VE building design

The issues with Tango and the alternative solutions described above, in particular the additional resource and expertise required, make scanning technology an unviable option for some companies. This would likely be the case if a company sees the suggested protocol as too great a burden for the benefits of the scanning technology. It was concluded that the project VEs would be generated through other means for testing the effectiveness of training. The chosen VE development tool was the Unity 3D game engine, as this was an easy-to-obtain software which provided flexibility with licensing for current and possible future uses.

A virtual office building was chosen to serve as a backdrop for our scenarios. This was due to the fact that a host of data on validity testing of human behaviour in real-world incidents in office buildings was already available through previous research. The virtual office building was modelled based on blueprints of a geo-typical building in order to avoid the influence of familiarity on user behaviour. The modelling process was carried out using the Maxon Cinema 4D software solution and the VE was subsequently brought to life using Unity. Figure 4-18 shows the external view of the building.

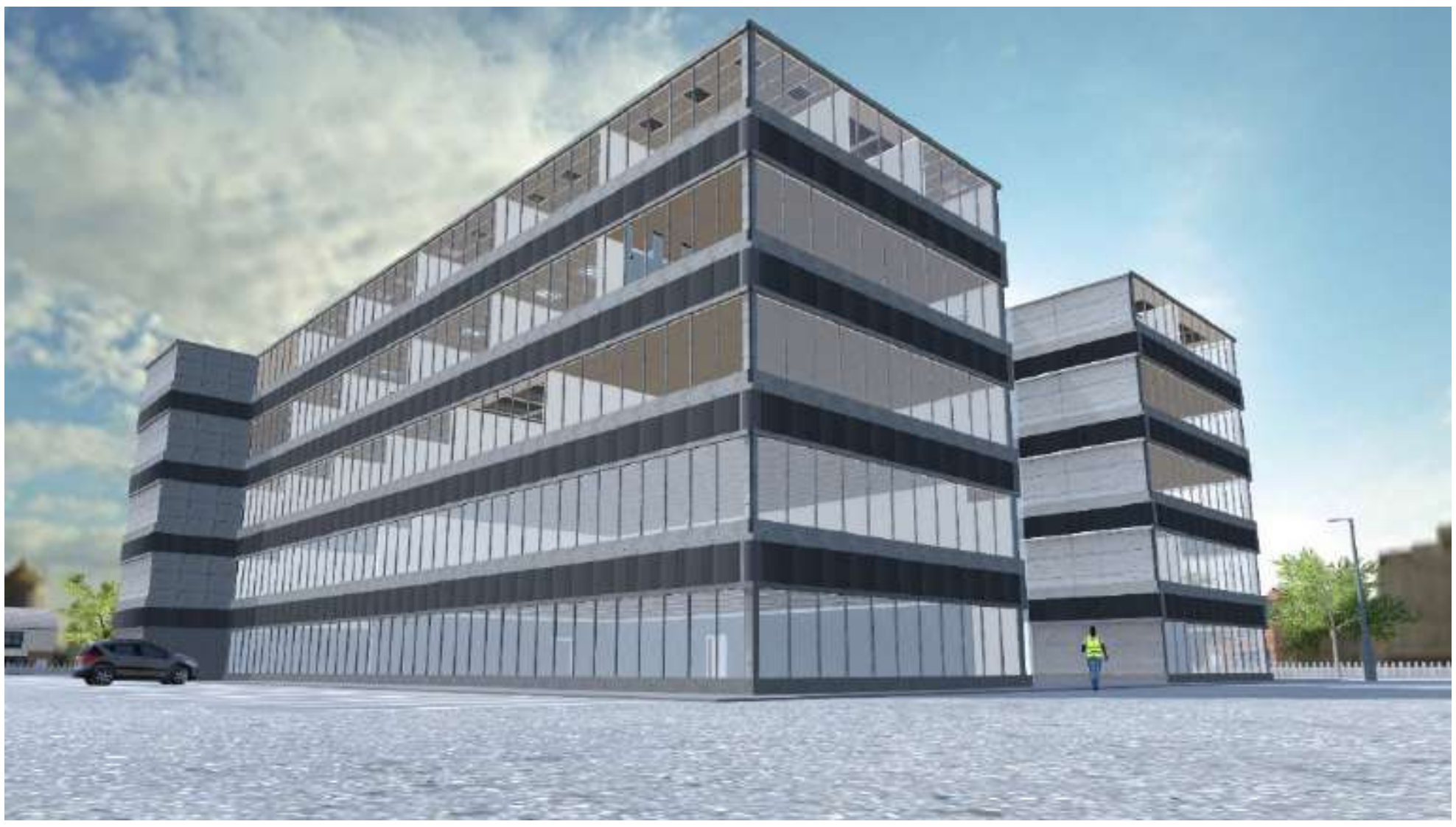

*Figure 4-18 External view of the office building model*

#### 4.6.2.1 Floor layout

The original study design for behavioural validity required users to complete the scenario four times, to study various conditions. We therefore designed and produced four alternative floor layouts. In order for exit routes to be analysed and evaluated, each of these floor layouts needed to allow for a variety of evacuation routes. In practice this meant that whenever a particular exit got blocked by a fire, there would always be one or more alternative exit routes available allowing the user to safely evacuate. Figure 4-19 shows the internal view of the building floor prior to being populated with any internal features. Figure 4-20 shows an example of a floorplan design:

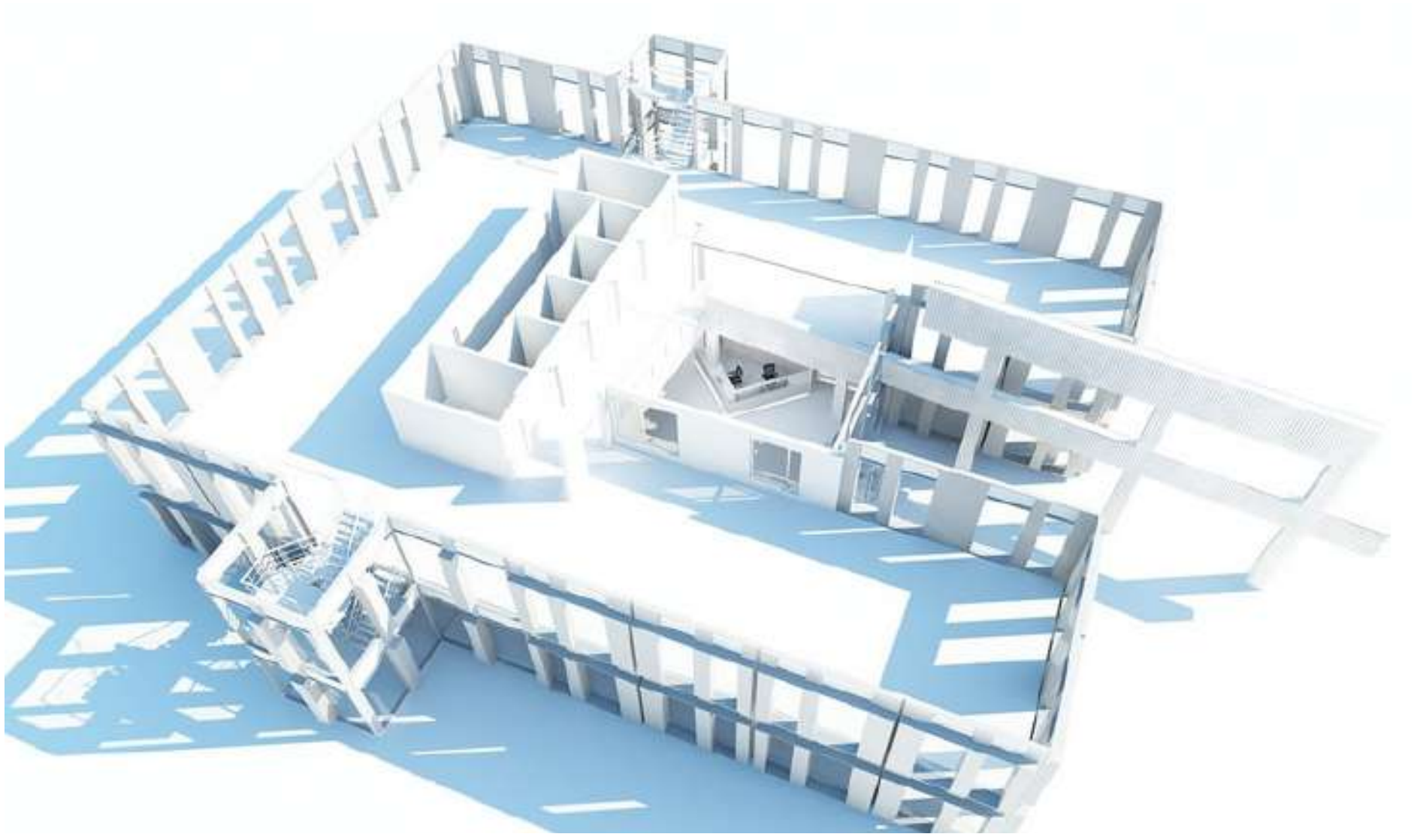

*Figure 4-19 Unpopulated internal view of the building floor*

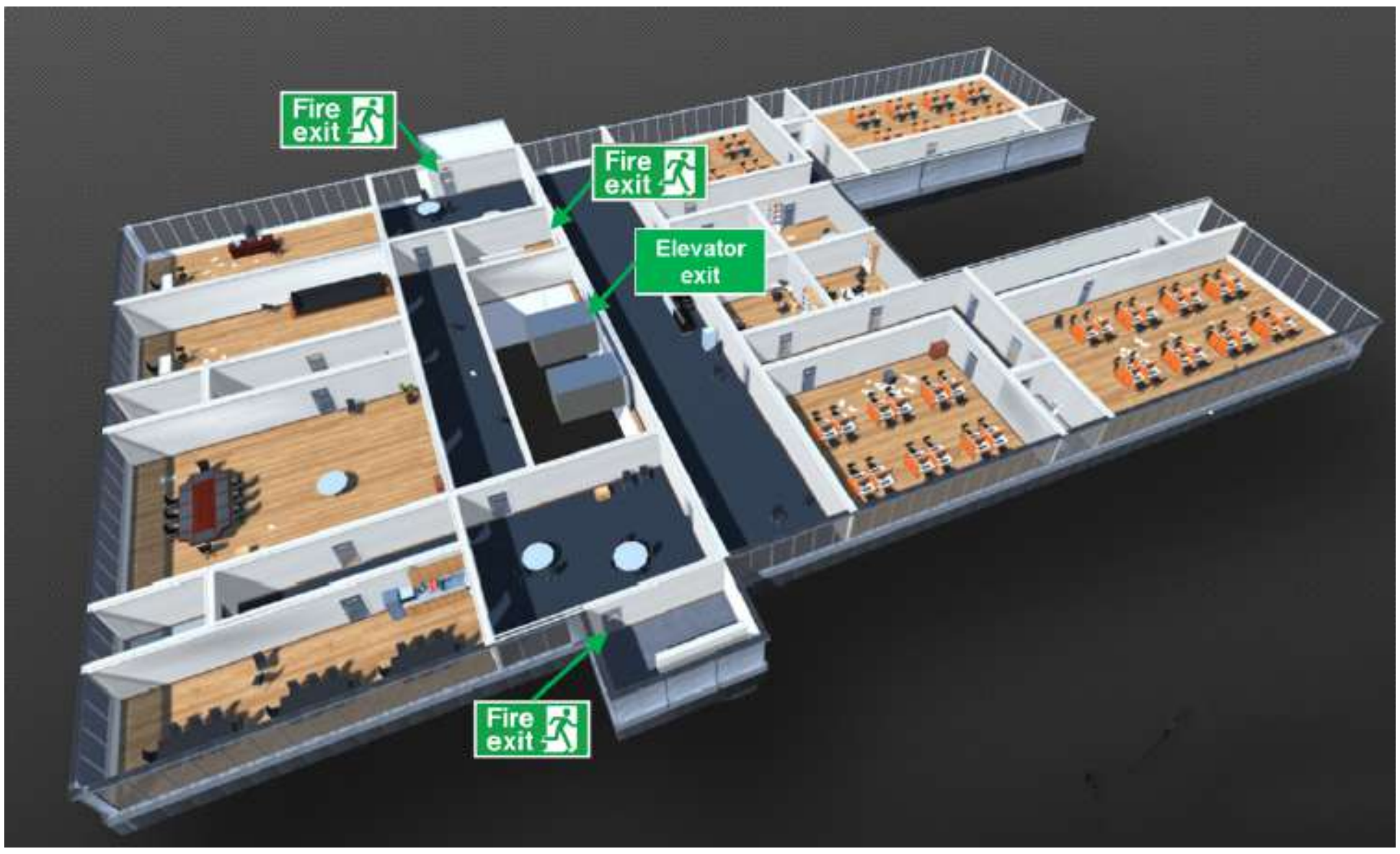

*Figure 4-20 Example of floorplan design*

### 4.6.2.2 Internal features

The virtual building was populated with features typical of an office setting, including open plan office workspaces and meeting rooms. Figure 4-21 gives an example of one of the meeting rooms:

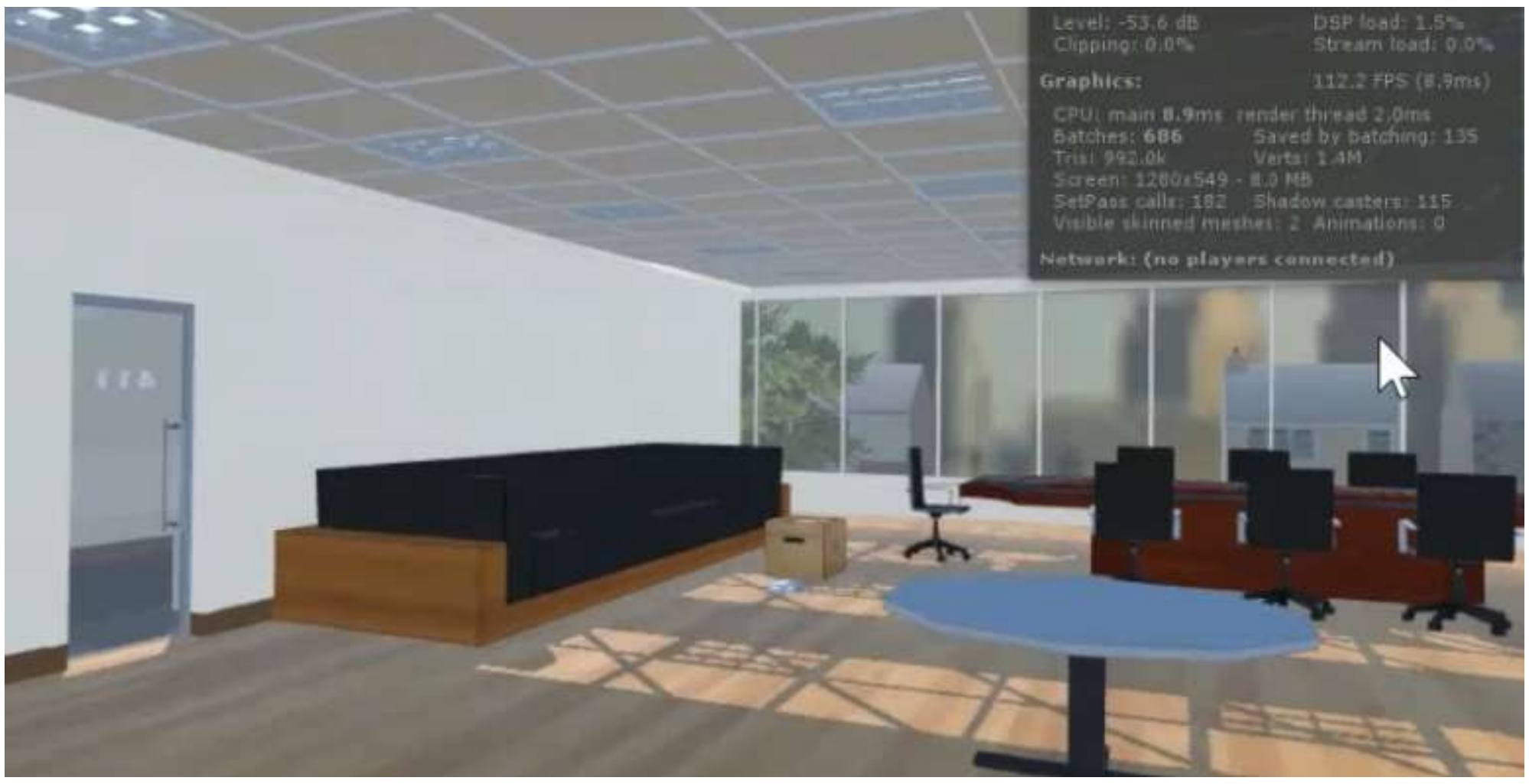

*Figure 4-21 An example meeting room*

The VE building had a reception area at the ground floor entrance. A working elevator was included in the design and the building had multiple floors so that use of the elevator for evacuation purposes could be assessed. Figure 4-22 shows the reception area on the ground floor and elevators to the upper floors in the building:

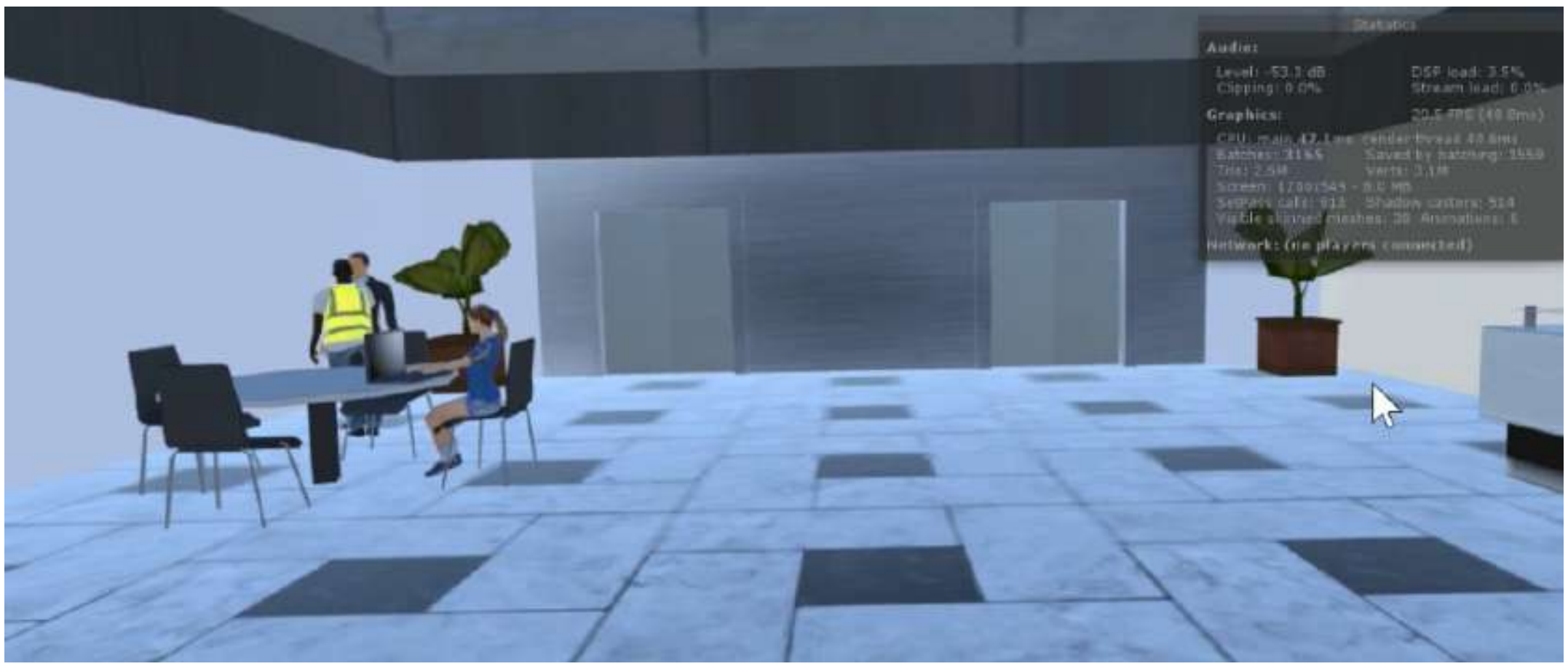

*Figure 4-22 Reception area and elevators on the ground floor*

Doors were transparent, typical of office building design. This would also allow users to make action decisions based on visual cues received prior to entering a room. Visible signage was placed throughout the building to allow for logical wayfinding, for example Figure 4-23.

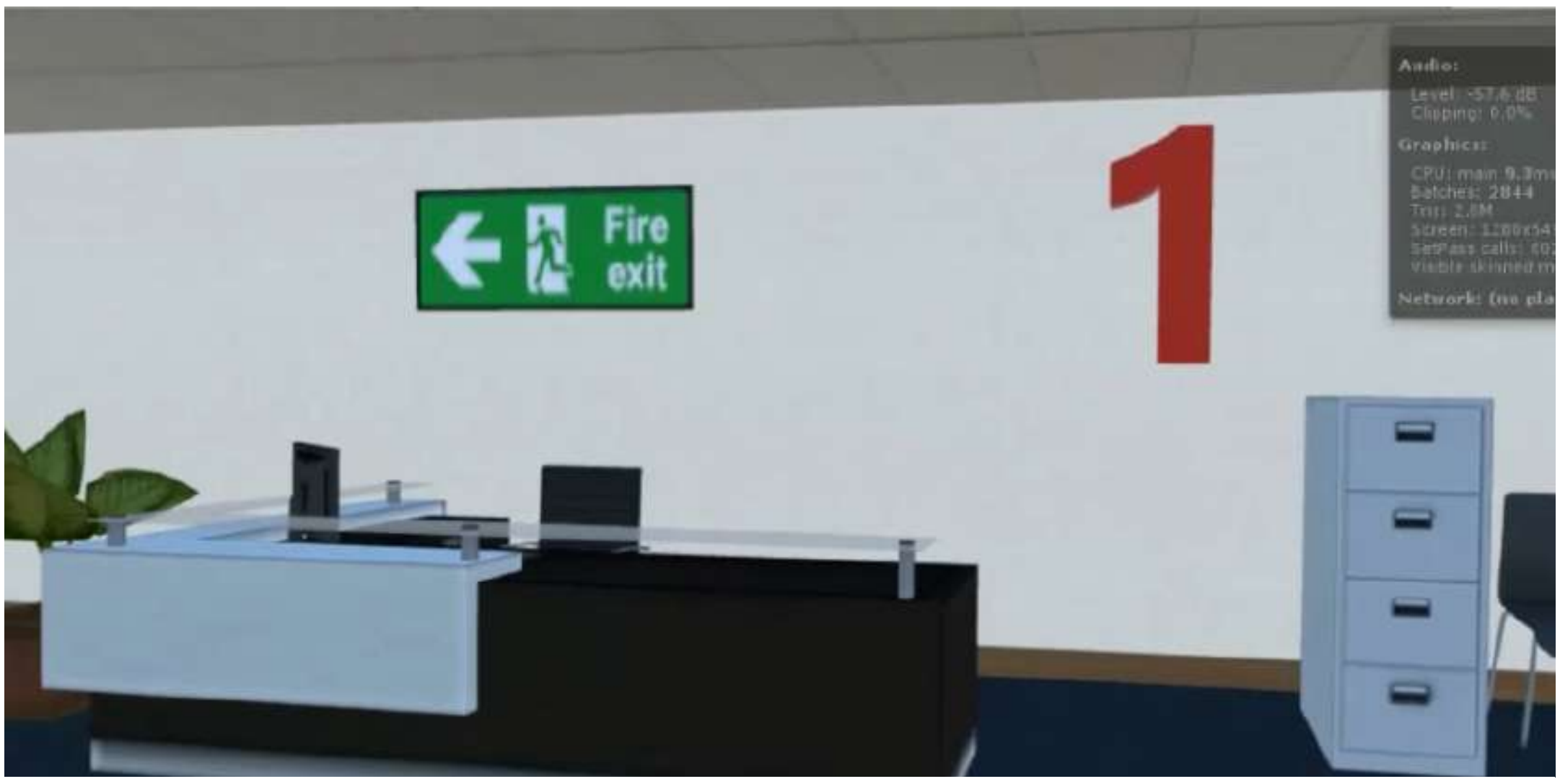

*Figure 4-23 Example of fire exit signs used throughout the building to aid logical wayfinding*

Interaction with elements within the VE was minimal. However, doors opened automatically when approached by users.

#### 4.6.2.3 Fire simulation

The fire simulation was triggered manually using button 'A' on the laptop keyboard: this triggered the visual fire cues, which consisted of smoke and flames (see Figure 4-24), along with audio cues, which consisted of a soundbite of a building fire as well as a fire alarm.

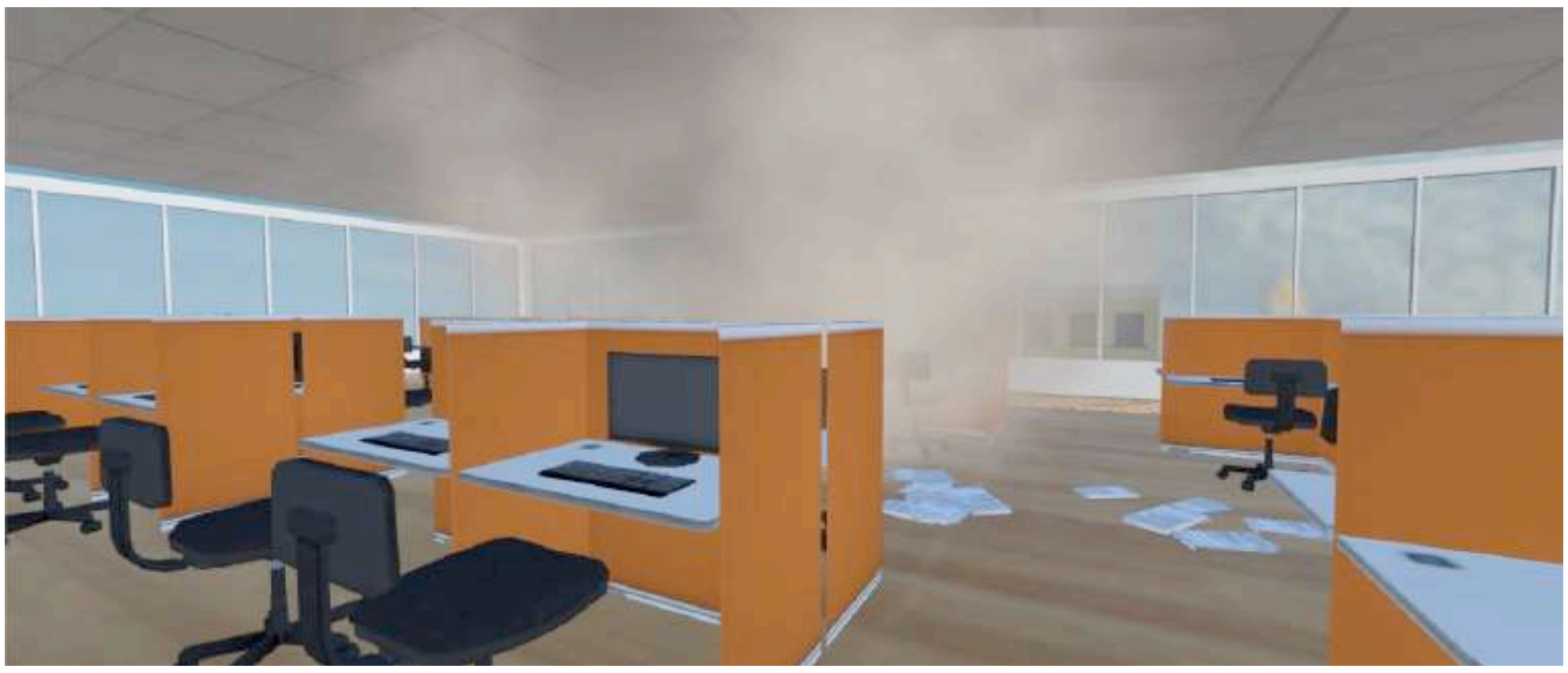

*Figure 4-24 Smoke and flame visuals were used to simulate a fire*

A set of colliders placed around the location of the fire were spawned concurrently with the audio-visual cues. These were used to control the magnitude of olfactory and thermal cues based on the user's proximity to smoke and fire. This was made possible using an Arduino controller which mediated communication between the VE on one end and the scent diffuser and heater fins on the other end.

The designed multi-sensory virtual environment then served as testbed enabling a set of usability studies which in turn fuelled the development of further iterations featuring refinements to the layout, simulation and study design.

## 4.7 Usability studies

A series of usability studies were conducted with lay users and experts from health and safety and training disciplines to refine the MS VE and to define requirements for the training scenarios. Outputs of this work were used to refine both empirical studies and for the assessment and iteration of the VE. These were loosely divided into three rounds of tests (five users in each) with ongoing minor iteration, but more substantial revisions in between each round. The first two rounds used think-aloud protocol while users navigated the prototype VE, thus obtaining feedback to refine various elements of the user experience, such as optimisation of movement. Users were asked to say aloud their thoughts as they used the system, including what they were trying to do, how they were trying to do it, what happened, whether it was what they expected.

The study used the latest iteration of the VE virtual office building being developed for Study 1, using the fire safety use case, and participants were asked to complete the following tasks:

- activate an emergency alarm
- call emergency services
- locate and use a fire extinguisher
- approach a fire to assess the situation
- exit the building safely.

Users were asked to follow the think-aloud protocol during completion of tasks, with prompting from the investigator when necessary. The investigator took notes focusing on any issues encountered while using the system to navigate and interact with the VE and virtual objects. A video from behind the user's head, showing the screen, and their use of the Vive controllers, was captured during their use of the VE for reference in subsequently analysing and reporting usability issues. No quantitative measures were taken, as this study was formative. Following completion of the VE tasks, users participated in a short, semi-structured interview to qualitatively explore their experience with the VE and prototype system.

For the third round of tests we did not use a concurrent think-aloud protocol, as we wanted to test the entire fire evacuation scenario with a view to studying behavioural validity, and concurrent verbal protocol risks interference with task performance during verbalisation, which could have implications on the behaviours exhibited in this case. Instead, participants were interviewed following use of the system. This resulted in further improvements to usability as well as to shaping the design of the behaviour study.

## 4.8 VE iteration – Study 1

Outputs from the usability studies were used to develop the VE, simulated cues and the experimental design. Key changes were as follows:

#### *4.8.1.1 Study design*

Following the usability studies it was agreed that the experimental design would change to between subjects where each participant would experience the same scenario and VE, either in audio-visual modality or multi-modality with the addition of heat and smell modal simulation. This change meant that only one floor design was required for the simulation. Therefore, the preferred floorplan design was chosen from the four available and iterations were made to this design for the final version.

#### *4.8.1.2 Number of floors*

There were a number of issues with simulator sickness with users interacting with the VE, in particular when they were using the stairwell to evacuate the building from the 4th floor. Therefore, it was decided that the number of 'floors' in the building would be reduced to one, but that no cues would be given to disclose the floor level users were on, in order to avoid influencing evacuation behaviour, e.g. users may be more likely to take the stairs if there is only one flight of stairs to take to reach the exit.

#### *4.8.1.3 Tasks and context*

During usability testing it was noted that users did not explore the building extensively and it was not guaranteed that the user would encounter the fire. When the fire was triggered users tended to exit immediately via their entry route, which was made more obvious by the doors remaining open. As users were given free movement within the VE, there was large variability in the complexity of their evacuation, with some users being able to exit very quickly without experiencing any of the fire simulation with the exception of the alarm. To counter this, a task was introduced in order to direct users along a specific and consistent entry route that would ensure the user would encounter the fire. In addition, doors would automatically close once users had passed through them. The task also provided context for users to increase a sense of immersion and distraction as it provided a purpose to the user's visit to the building, i.e. users were there to attend an assessment and interview. Additionally, it allowed for a period of familiarity prior to the evacuation where the user interacted with the VE rather than the experimenter, thereby increasing their level of immersion. The task was split into three sections. Section 1, seen in Figure 4-25, was a wayfinding task directing users to a breakout area.

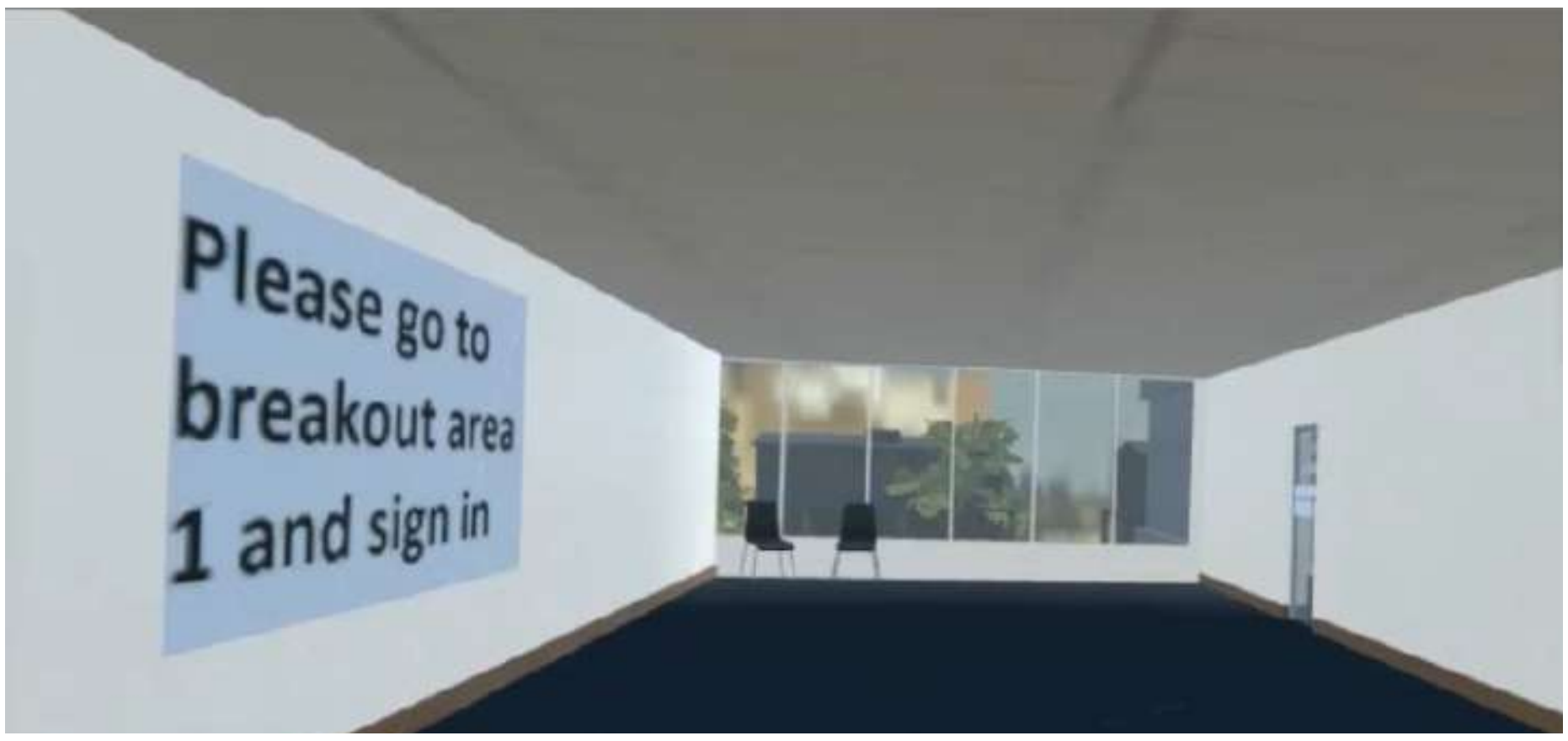

*Figure 4-25 Wayfinding task*

The second part of the task had the users walk into a reception room and enter their signature into a sign-in sheet. This action was largely figurative and involved each user looking into the sheet while standing in its proximity, whereupon a signature simply appeared automatically accompanied by a writing sound effect. Wayfaring instructions for the final section of the task were also provided with instructions to the sign-in task. Figure 4-26 and Figure 4-27 show the instructions for part two of the task and the position of the sign-in sheet and the appearance of the signature:

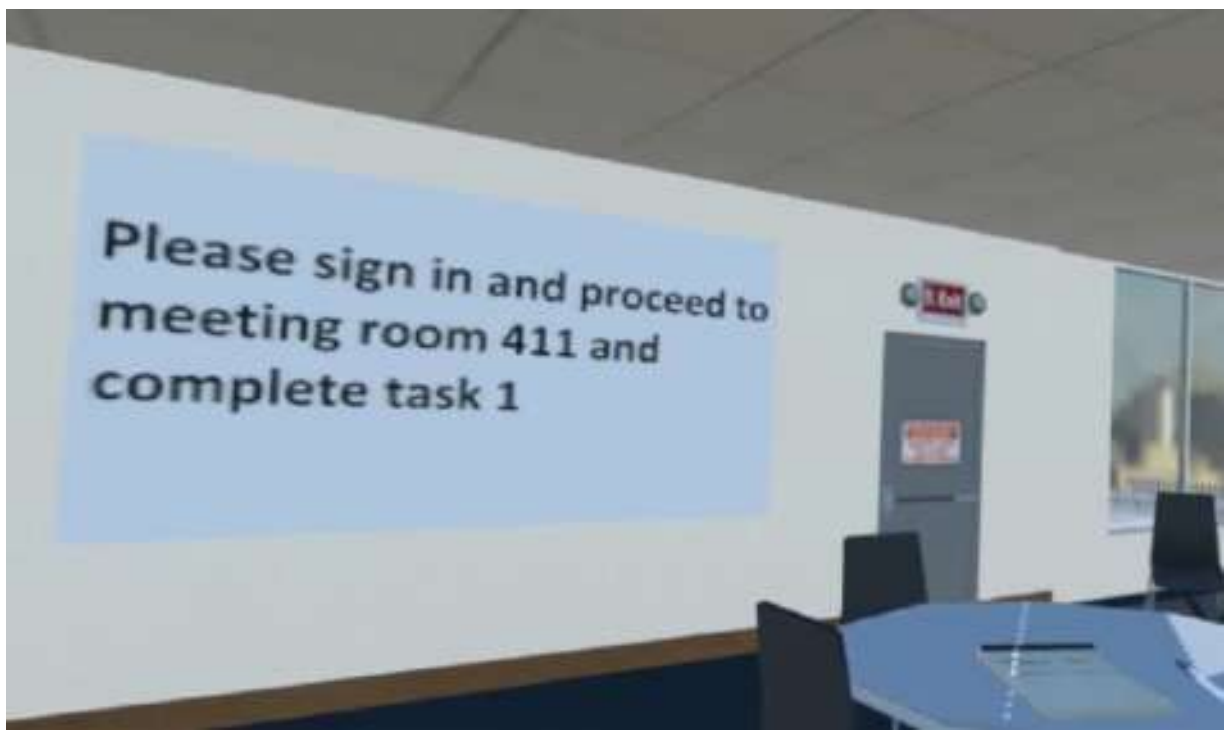

*Figure 4-26 Sign-in task and instructions for the next task*

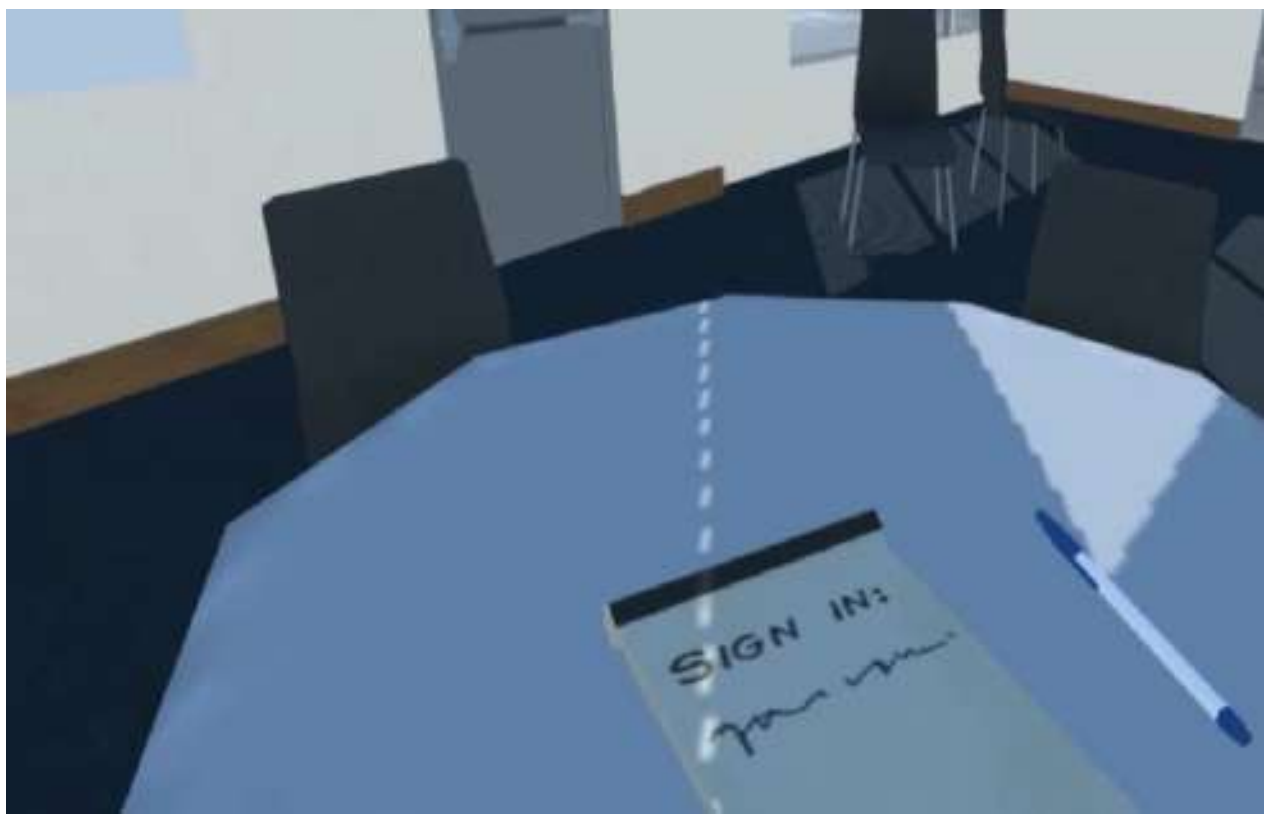

*Figure 4-27 Signature appearing on sign-in sheet*

The final part of the task was an IQ style task. Once the user was observed engaging with this task, the fire scenario was manually triggered. Figure 4-28 shows the IQ task.

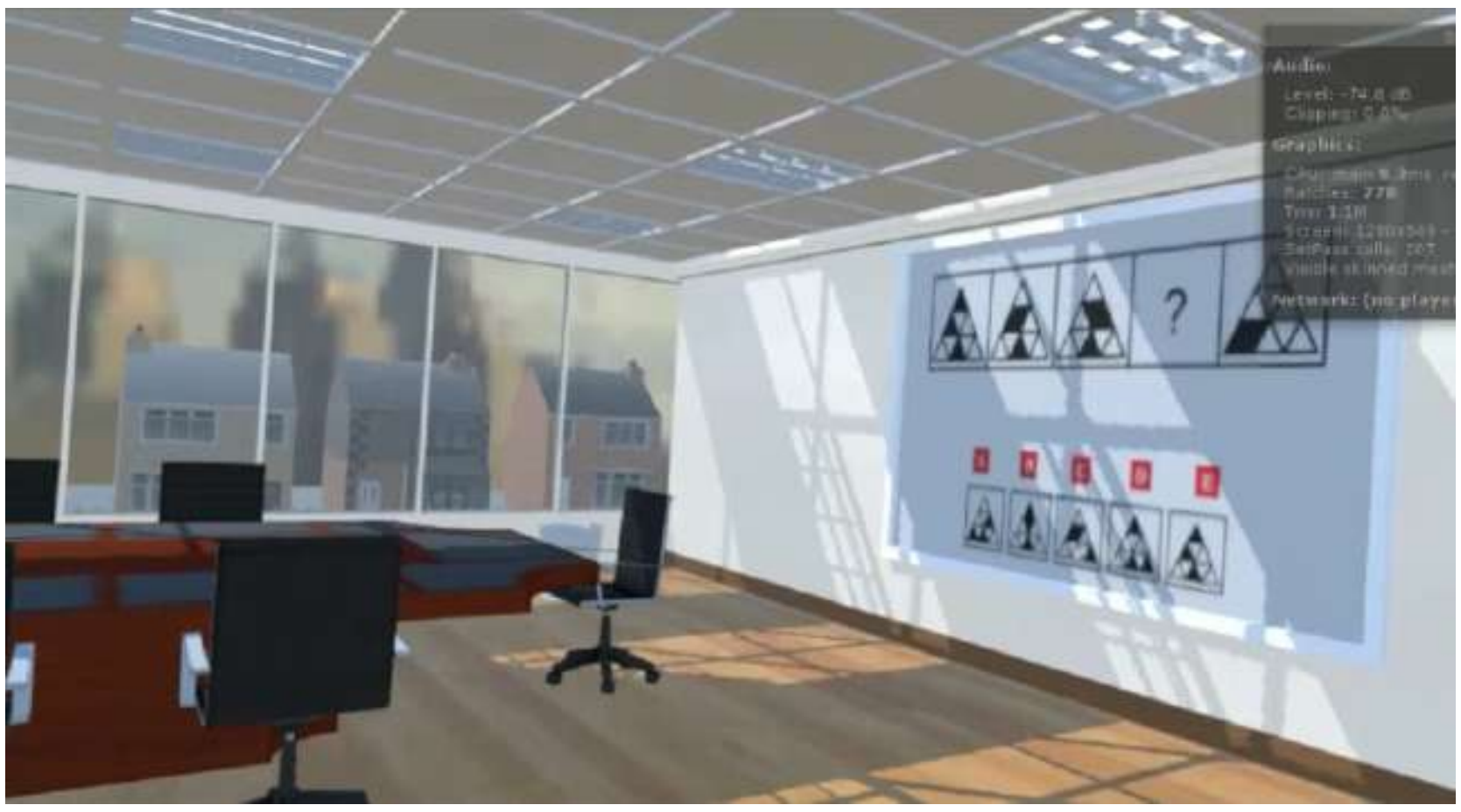

*Figure 4-28 IQ task*

#### 4.8.1.4 Fire simulation development

Feedback from users was that they were not clear that the building was on fire from the visual cues available, for example the smoke was too light in colour and there were no obvious visible flames. In response, we took a number of steps to increase the visibility and perceived strength of the fire. Specifically, the density and lifetime of smoke particles emitted in the VE was increased, resulting in thick clouds of smoke accumulating in the building corridors as the fire progressed. Flame effects were likewise enlarged and the volume of burning sound effects was increased, making fires audible even across multiple rooms.

During usability testing a number of participants either misconstrued the meaning of the alarm or failed to take notice of it at all and therefore did not evacuate the building until prompted by the experimenter. Failing to evacuate on hearing a fire alarm is well known within literature on human behaviour in fire (15); therefore, to increase the likelihood of evacuation, a small fire was strategically placed so that it could be seen by the user from the position of the final task. This fire was triggered concurrently with the fire alarm and a second larger fire elsewhere in the building. Figure 4-29 shows the position of the small fire in a wastepaper bin next to the final IQ task.

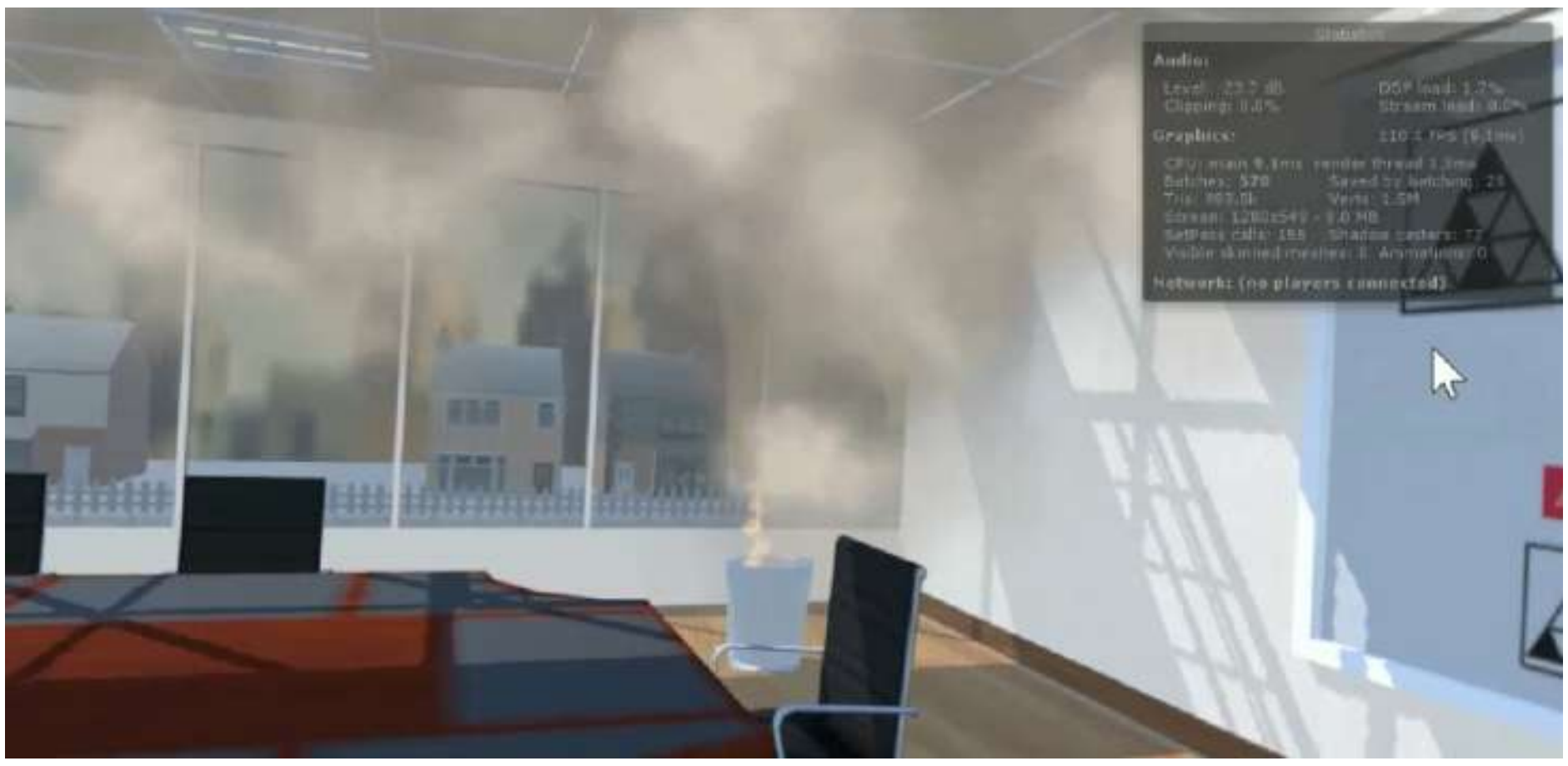

*Figure 4-29 Small fire in wastepaper bin near to the IQ task*

The inclusion of a smaller, less serious fire also allowed for analysis of pre-evacuation behaviours, for example attempting to extinguish the fire. In addition, in order to maximise the chances of the user encountering the fire, provide a consistent scenario to all participants and allow for the analysis of retracing steps for evacuation route choice, the second larger fire was positioned so that it blocked the route users took into the building. The audio-visual presentation of this fire was likewise changed in the manner described above, i.e. we amplified its perceived strength by enlarging the flame effects and increased the audibility of the accompanying burning sound effect. Figure 4-30 shows the second larger fire blocking the route users took into the building.

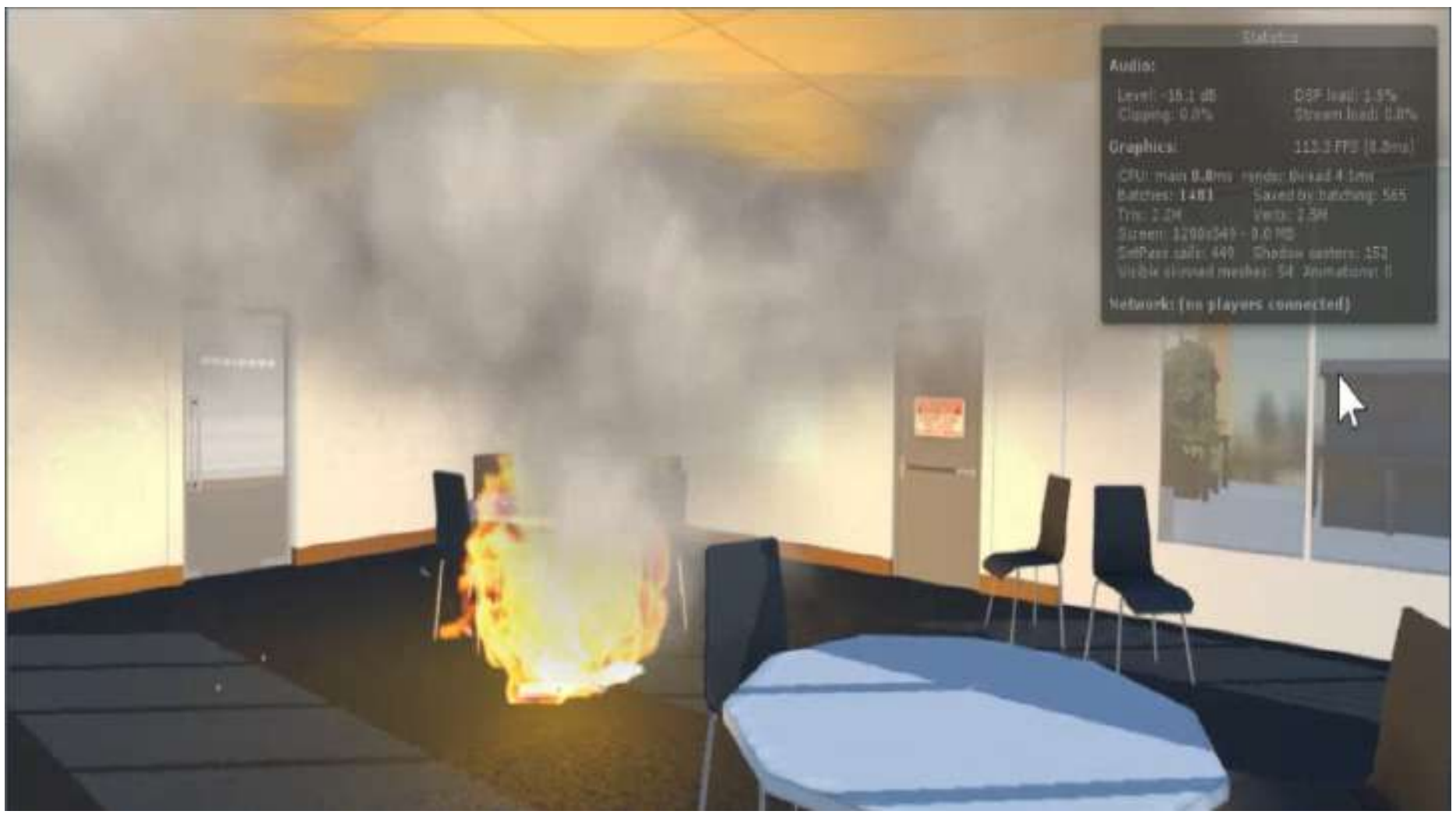

*Figure 4-30 Second larger fire placed in the breakout room where users signed the sign-in sheet*

Control of the fire alarm was changed so that it was no longer triggered by the colliders and continued to sound throughout the building during evacuation as this was felt to be more realistic. Fire colliders were positioned strategically, such as at door thresholds, in order to better align the MS feedback with a user's movement within the VE. This resulted, for example, in sudden increases in heat and smell when entering the room containing the larger fire.

#### 4.8.1.5 Building occupancy

Feedback from the usability studies was that the building felt unoccupied, and therefore the number of non-player characters (NPCs) within the VE was increased. All NPCs disappeared from the VE when the fire simulation was triggered and reappeared gathered in the fire assembly point outside the building. This was done to avoid issues relating to users interacting with NPCs, for example influencing user behaviour or evacuation strategies, or actions arising from concern over potential danger to other characters in the simulation. Ambient noise, typical of office buildings, was also added to increase the presence of others within the building.

## 4.9 VE development for Study 2

For Study 2 the same VE building was used for the fire emergency scenario but developed to fit the design of our training module. A VE for the engine disassembly use case was developed specifically for Study 2. It was important for both use cases that the learning outcomes and key concepts covered were matched between the PowerPoint and VE media in terms of content and repetition. As with Study 1 colliders were used throughout the VE to trigger olfactory and thermal simulation at points within the VE that aligned with the user's proximity to hazards within the simulated scenario, for example smoke smell with fire and diesel smell with fuel leak. A crosshair was added to the VE to assist with directional navigation and VE object interaction due to the change to using mouse and keyboard rather than the HMD and Vive controllers (see Figure 4-31).

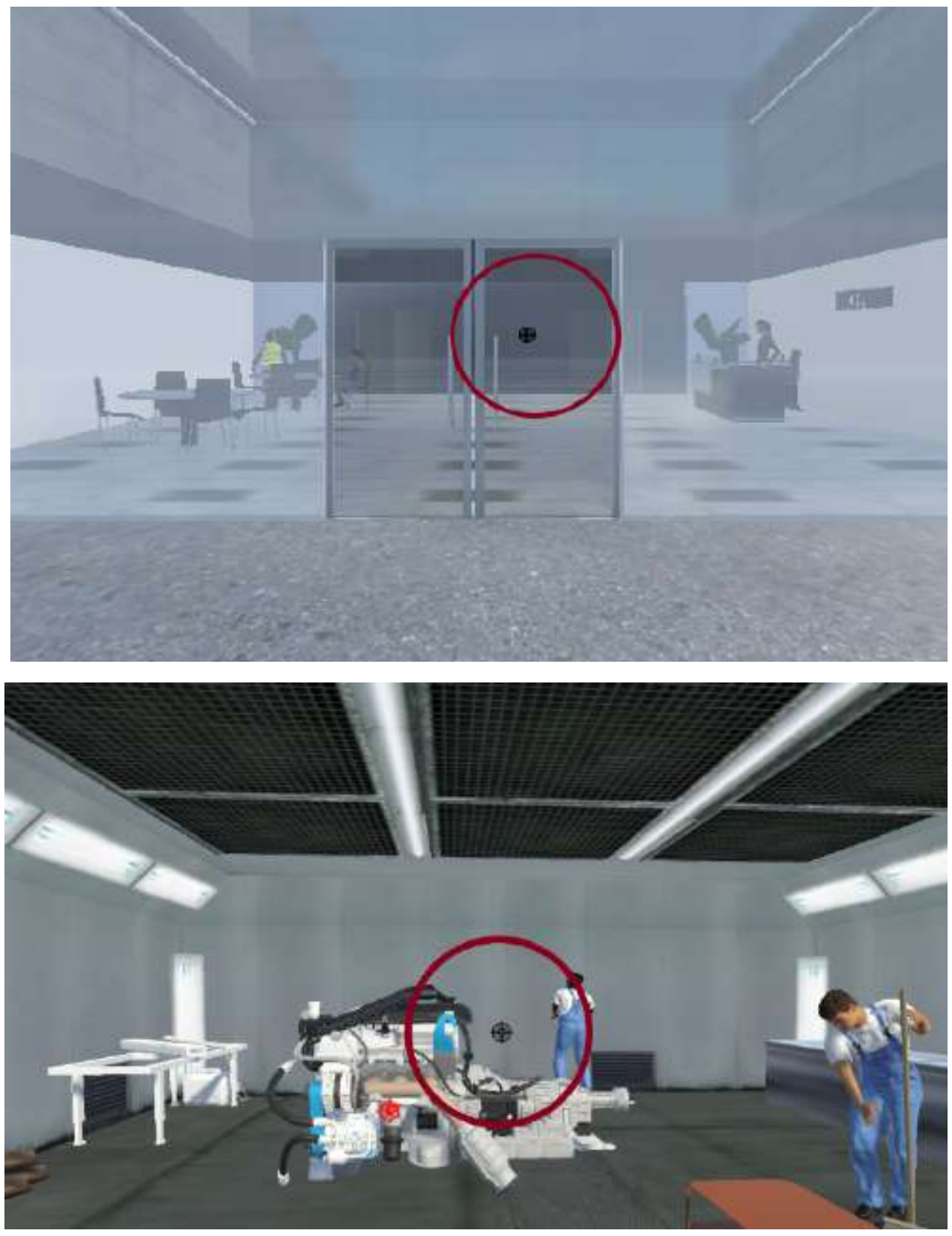

*Figure 4-31 Crosshair added to VE to assist with navigation*

The training module was designed so that users could complete all tasks without input from the experimenter. The aim was that the training module could be implemented in the workplace without the need for human instructional input, which would reduce the necessary training resources.

### 4.9.1 Fire safety use-case training VE

The VE for the fire safety use case was designed as a training module that would take users through three tasks of increasing complexity and autonomy:

1. a familiarisation task, to allow users to practise using the controls and navigate around the virtual building before beginning the training blocks
2. an instructional task, to provide users with fire safety information in the form of a game within the VE
3. a fire emergency scenario task, to allow users to put fire safety knowledge into practice and experience the consequences of their decisions and actions in a virtual emergency fire scenario within the VE.

The start screen of the training module is shown in Figure 4-32. Users were instructed to select each of the tasks in turn by clicking on the appropriate button.

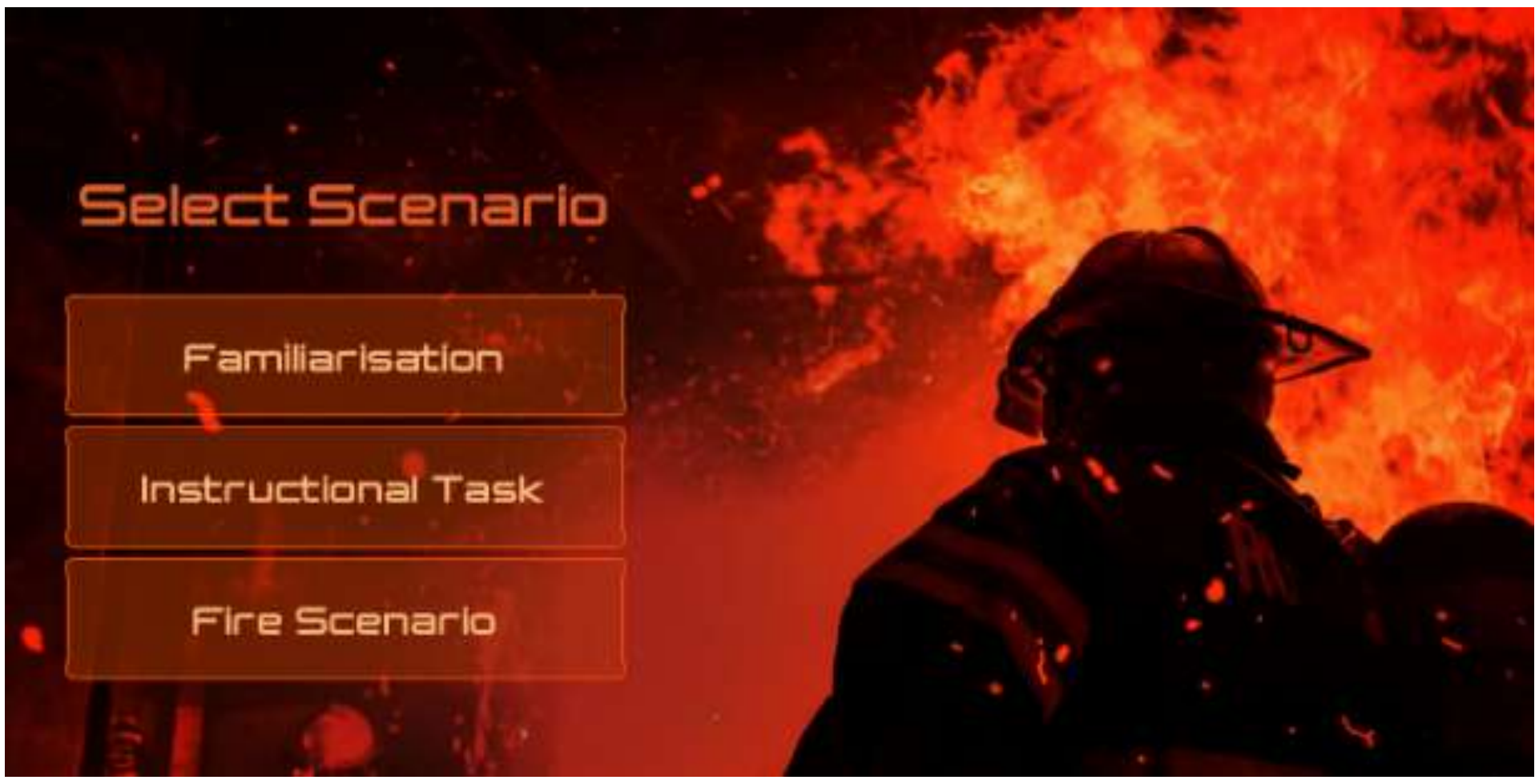

*Figure 4-32 Opening screen of the training module*

#### 4.9.1.1 Familiarisation

The purpose of this task was to give users the opportunity to experience the VE and to practise using the controls (keyboard and mouse) to move throughout the virtual building. Users were given a series of visual pop-ups with concurrent audio instructions to introduce the task and guide users through initial actions required, for example how to use the controls, shown in Figure 4-33 and Figure 4-34.

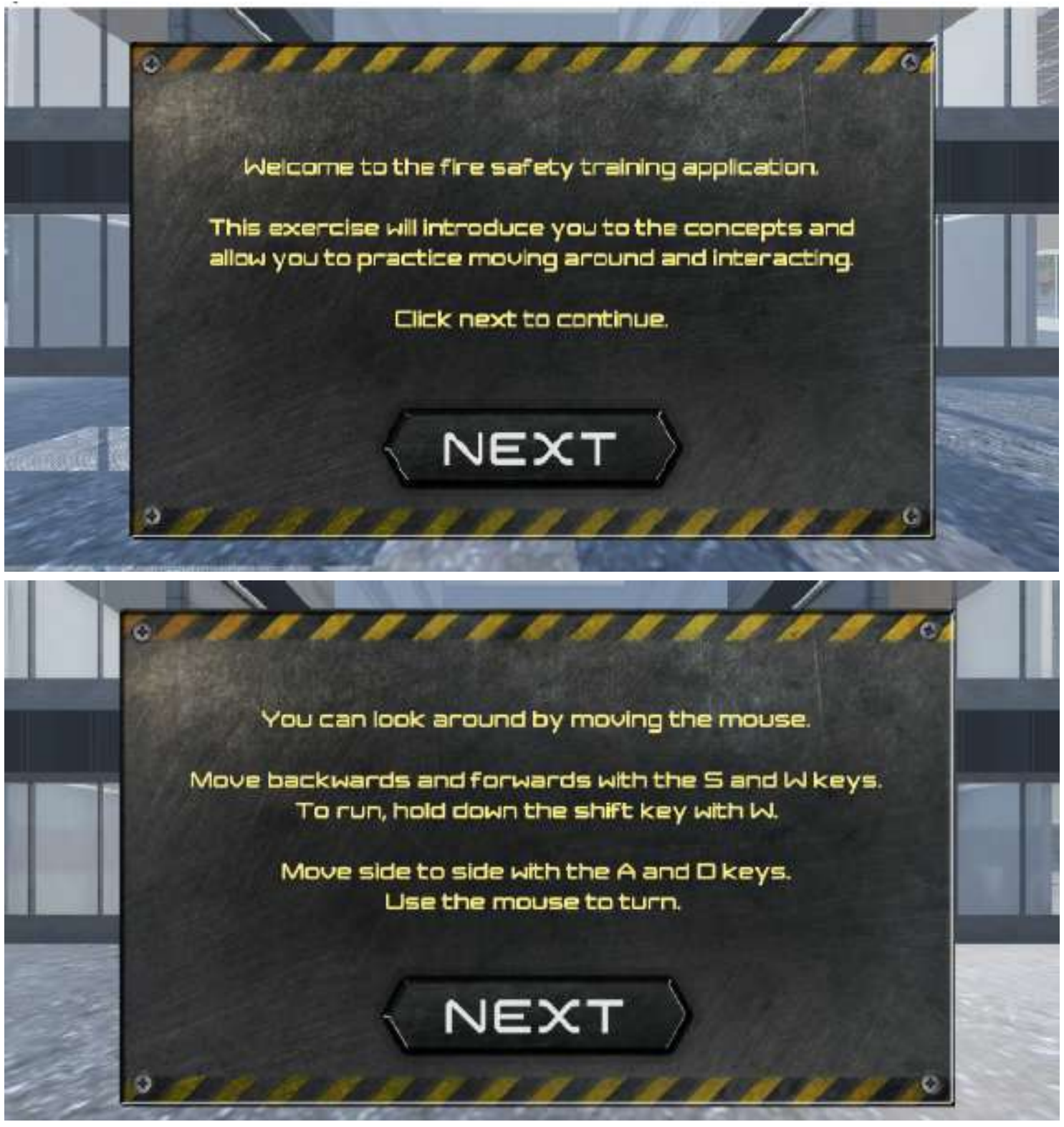

*Figure 4-33 Initial introduction to familiarisation task requirements and control instructions*

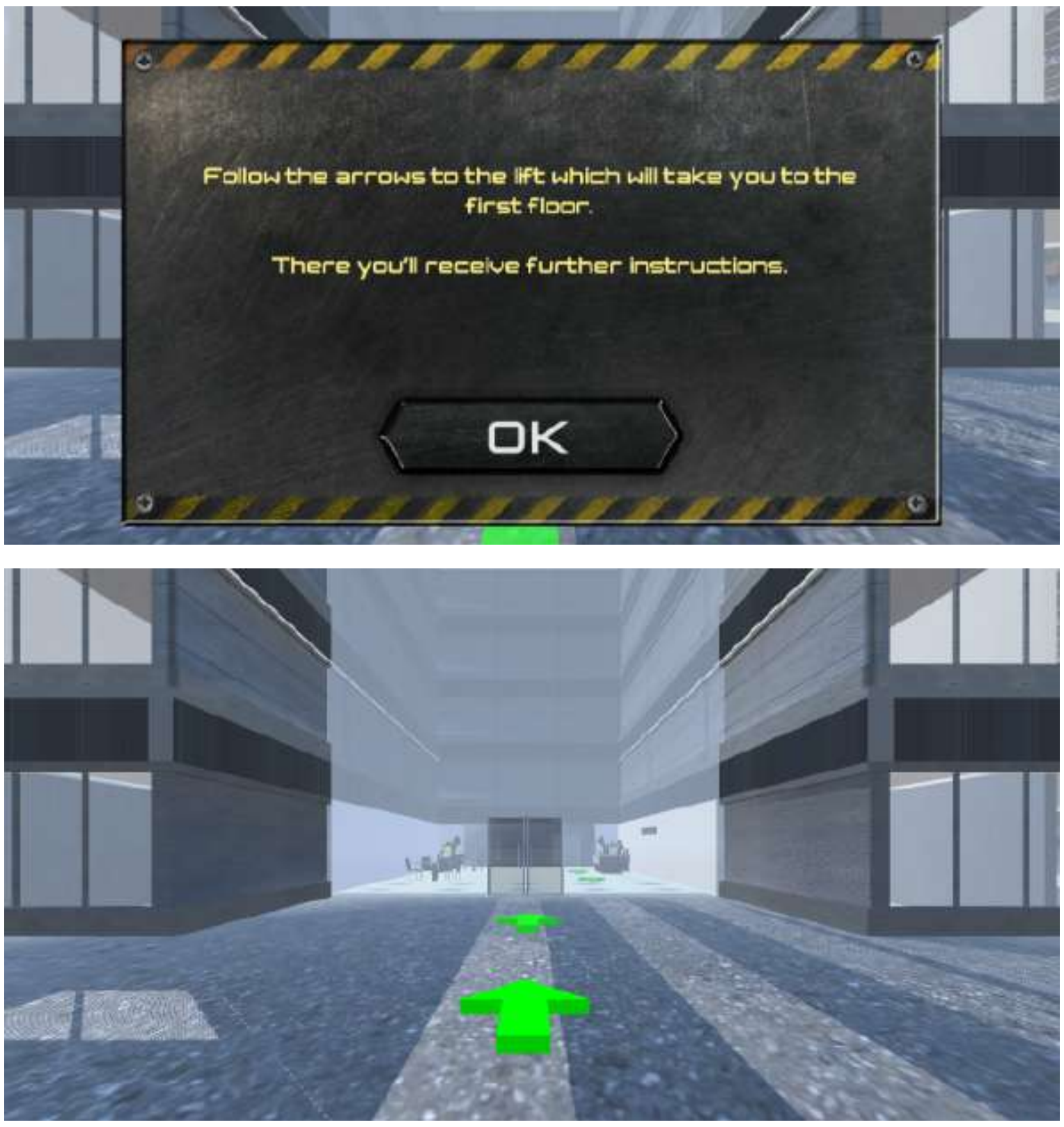

*Figure 4-34 Guidance to get users into position in the building to complete the main task*

To complete the familiarisation task users were required to activate a fire alarm and evacuate the building on encountering a fire within the VE building. Once users reached the first floor where the fire simulation would be initiated, they were given guidelines on safety detectors and alarm systems, which mirrored the guidelines given in the PowerPoint training condition, seen in Figure 4-35.

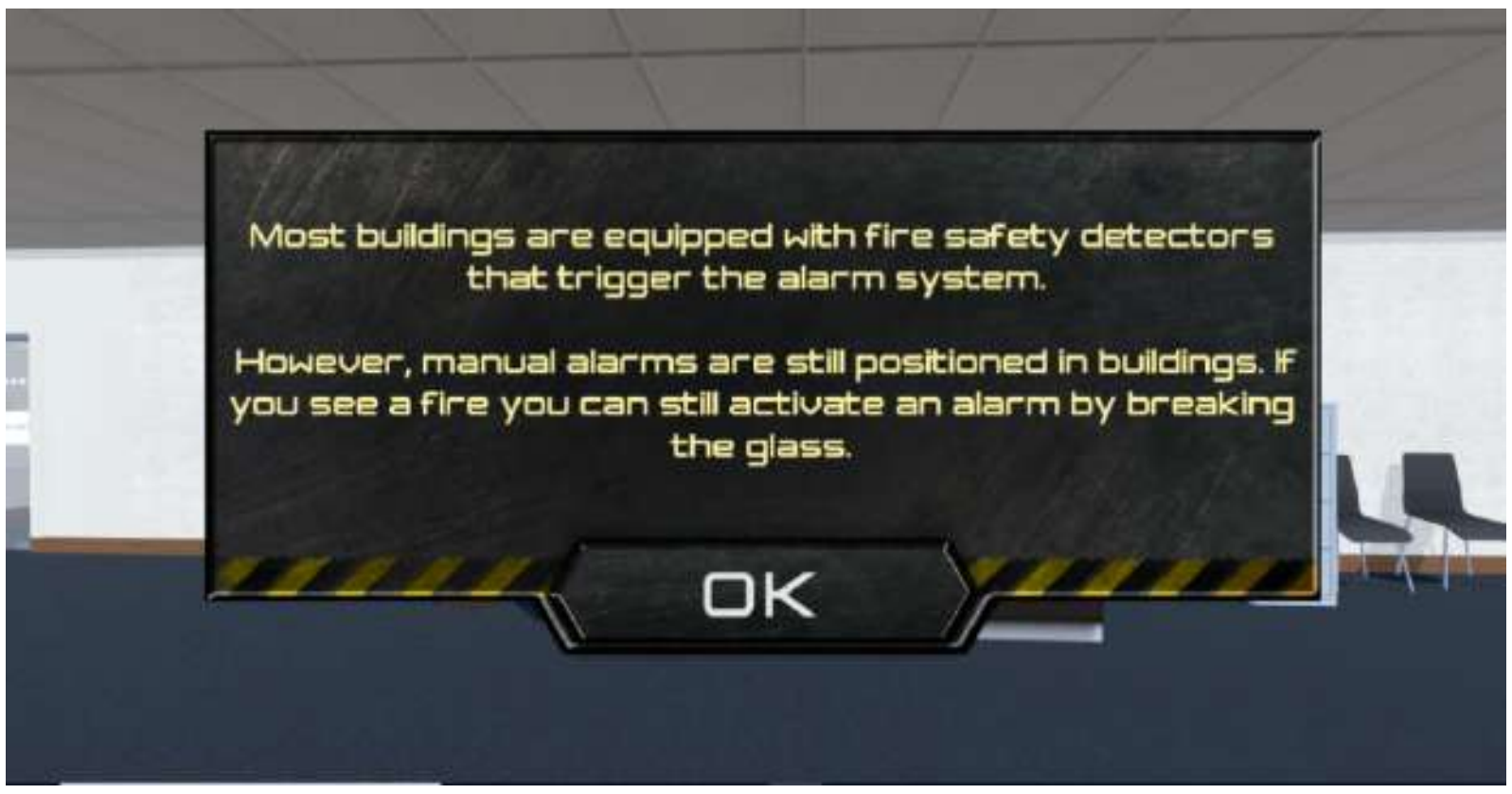

*Figure 4-35 Fire safety guidelines provided to users on safety detectors and alarm systems*

Users were guided through the VE into the required position to encounter the fire, using dynamic green arrows to highlight a path through the building as seen in Figure 4-36.

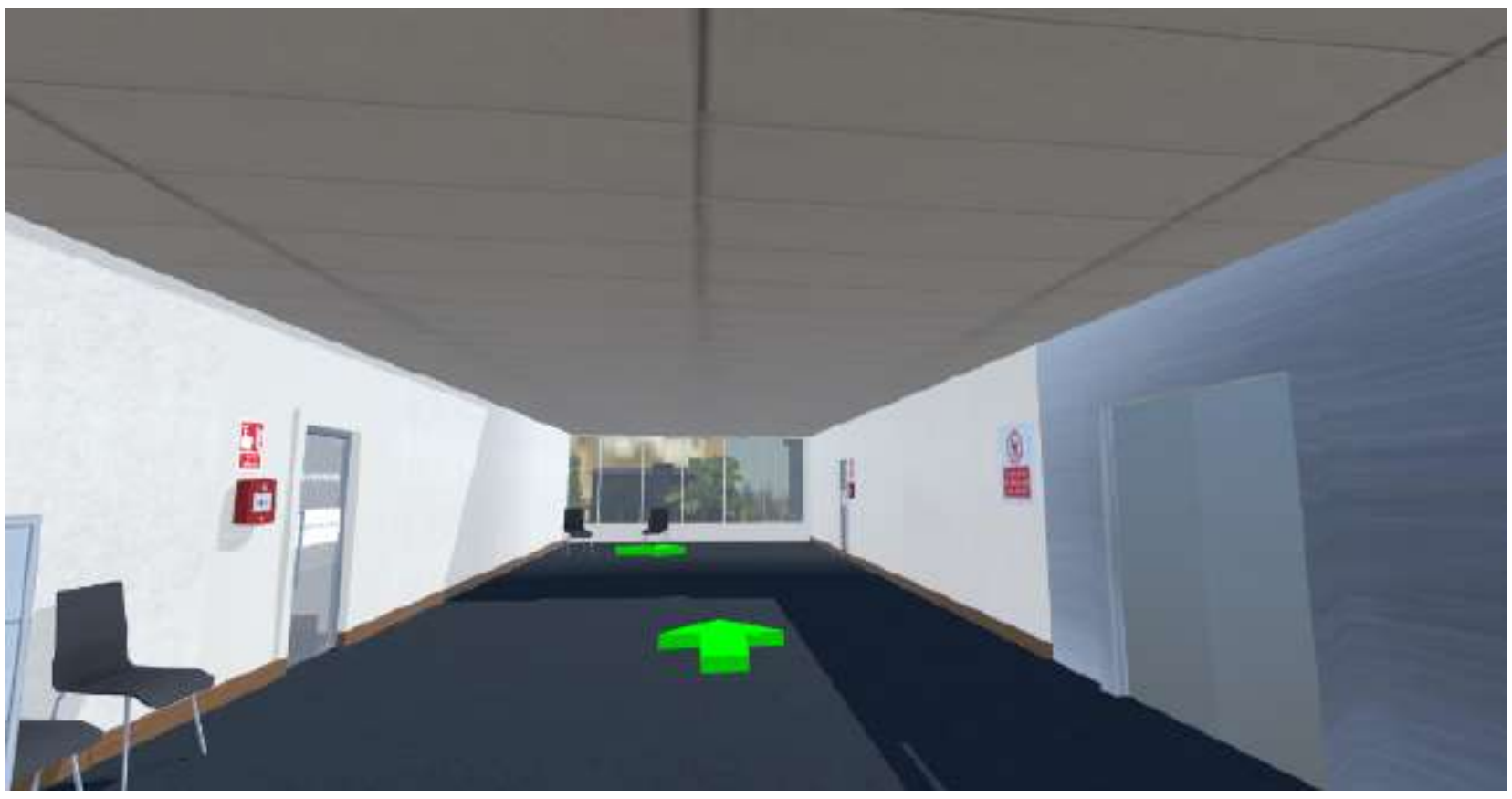

*Figure 4-36 Dynamic green arrows guiding user to a room in the VE*

Users were guided to a specific room within the VE. As they reached the entrance to the room, users received instructions on how to interact with doors; see Figure 4-37.

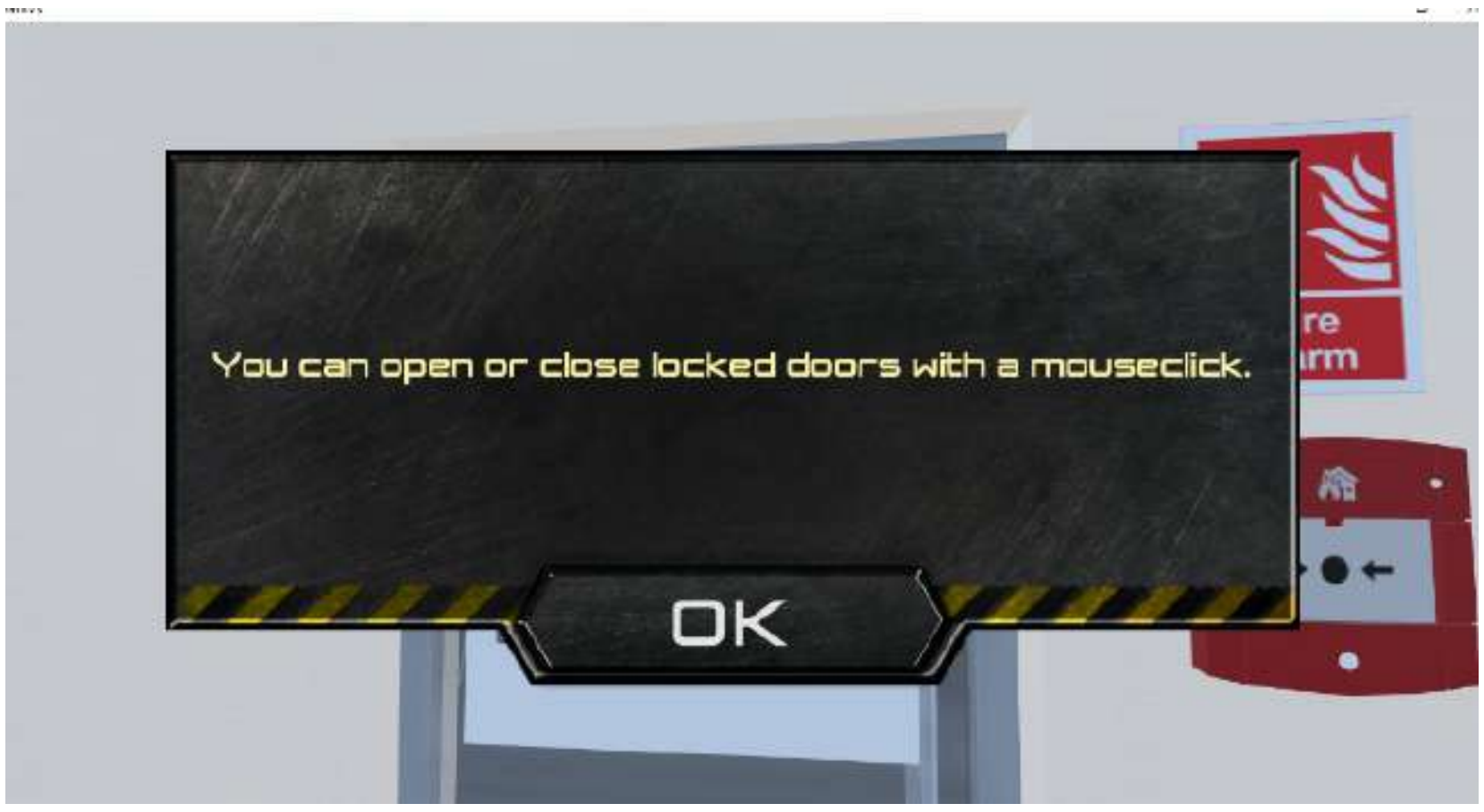

*Figure 4-37 Instructions on interacting with doors within the VE*

As the user entered the room, the fire simulation was triggered, and the user received a pop-up alerting them to the fire and instructing them to activate the nearest alarm and evacuate the building, shown in Figure 4-38 and Figure 4-39.

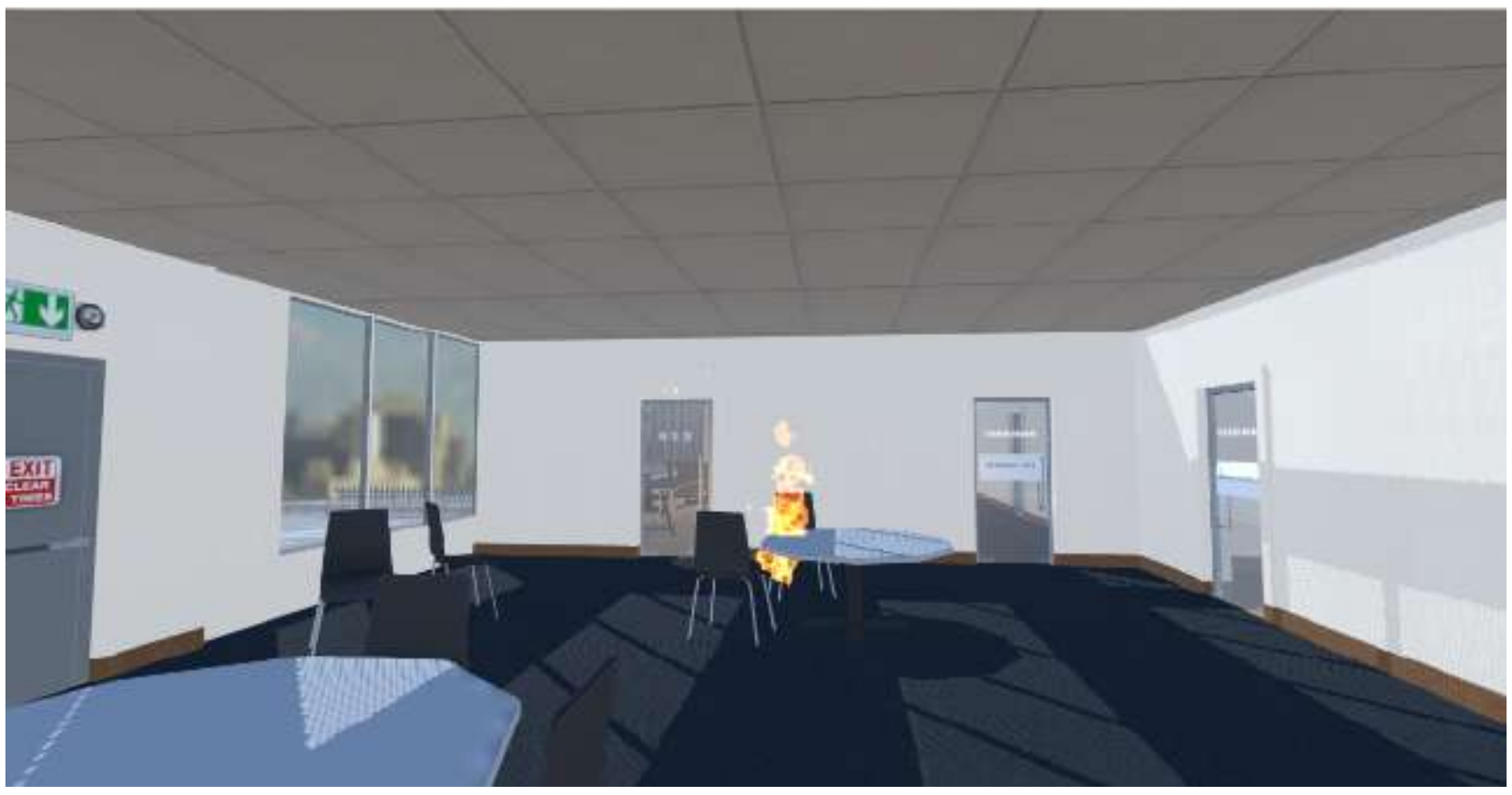

*Figure 4-38 The simulated fire*

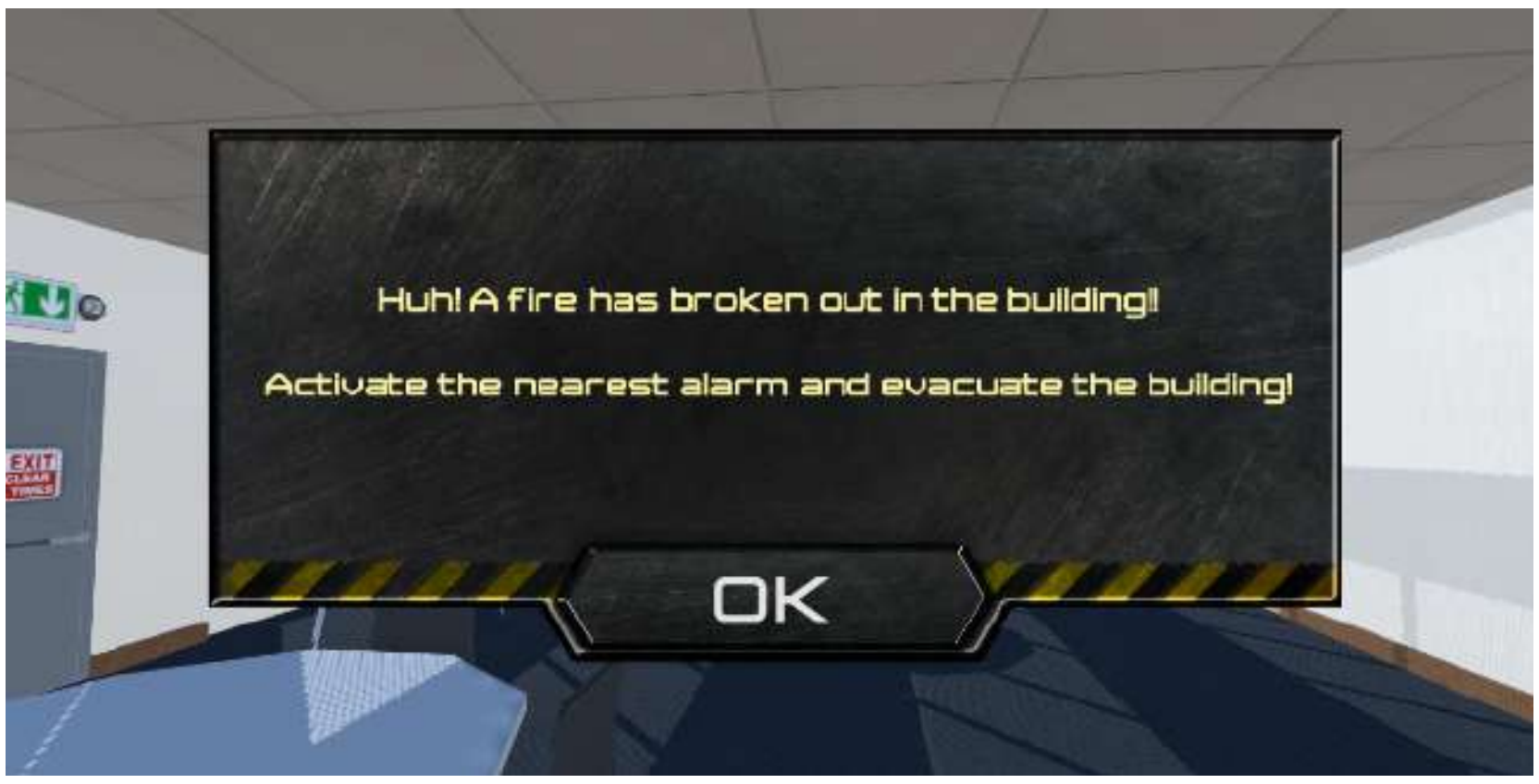

*Figure 4-39 Pop-up alert to the fire and instructions on what action to take*

Red arrows guided the user out of the room, and there were additional pop-up labels to highlight the position of fire alarms; see Figure 4-40.

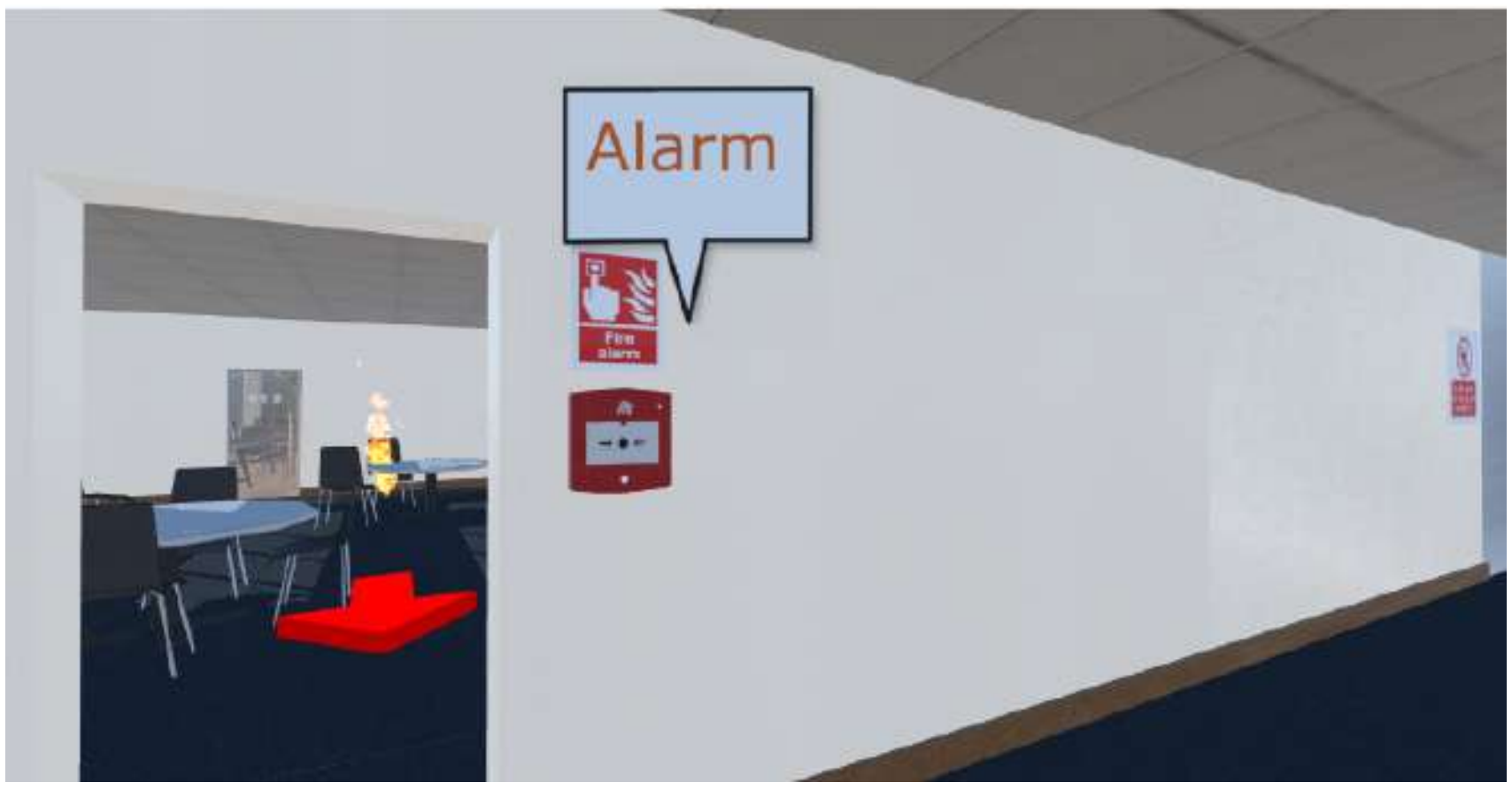

*Figure 4-40 Red arrows guiding users out of the room and pop-up label to highlight the location of an alarm system*

Once the user had activated the alarm and evacuated the building, they received a pop-up message confirming successful completion of the familiarisation task and taking them back to the initial screen to select the next task, shown in Figure 4-41.

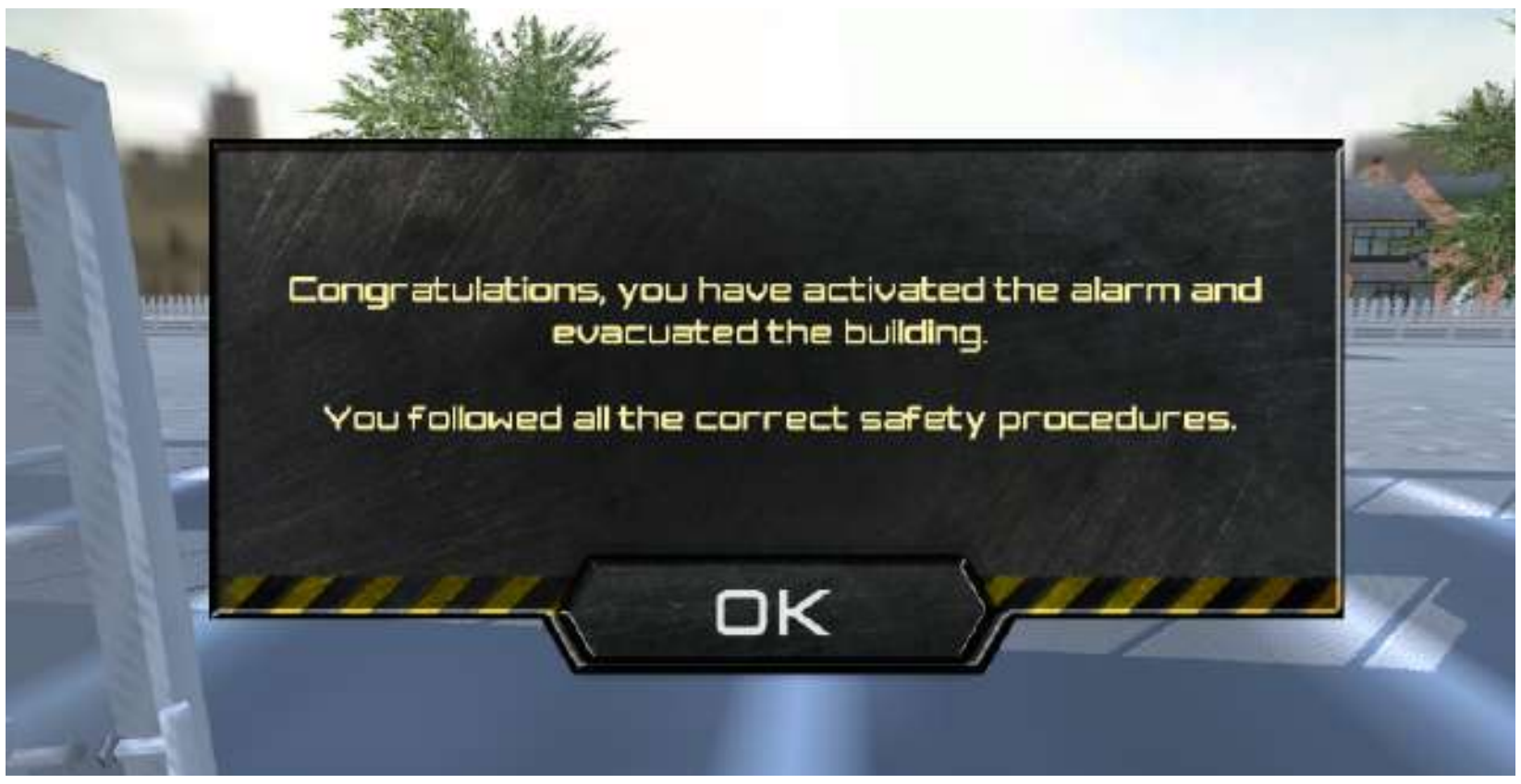

*Figure 4-41 Pop-up confirming successful completion of the familiarisation task*

#### 4.9.1.2 Instructional task

The purpose of this task was to provide participants with fire safety information consistent with that provided in the PowerPoint training condition and prior to users applying this knowledge in the emergency fire simulation task. A gamified version of the VE was used to present information to users during this training section via three sub-tasks: an information-gathering task, a spatial-awareness task and a fire token task. Users were given a visual pop-up with concurrent audio instructions to introduce the first subtask and guide users through initial actions required; see Figure 4-42.

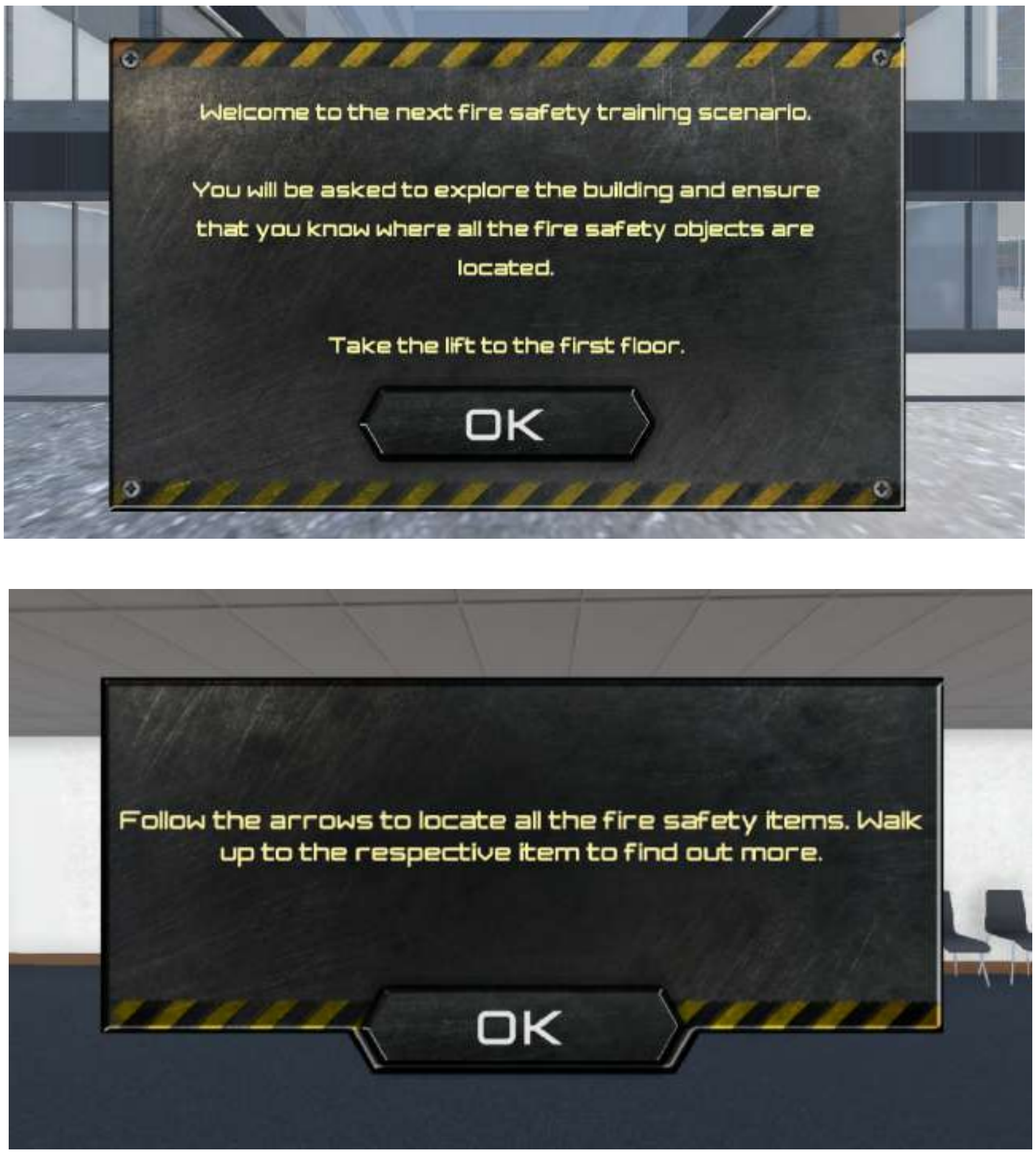

*Figure 4-42 Initial introduction to instructional task requirements and control instructions*

Users received core fire safety information by walking over floating, revolving 3D objects placed throughout the upper floor corridor in the VE building. Users were guided to each object via dynamic green arrows, which updated automatically as each action was completed. Core fire safety information included details on:

- fire exits
- automated safety systems and fire alarms
- personal belongings
- fire extinguishers

- a fire token evacuation process.

Figure 4-43 shows examples of fire exit and fire extinguisher objects.

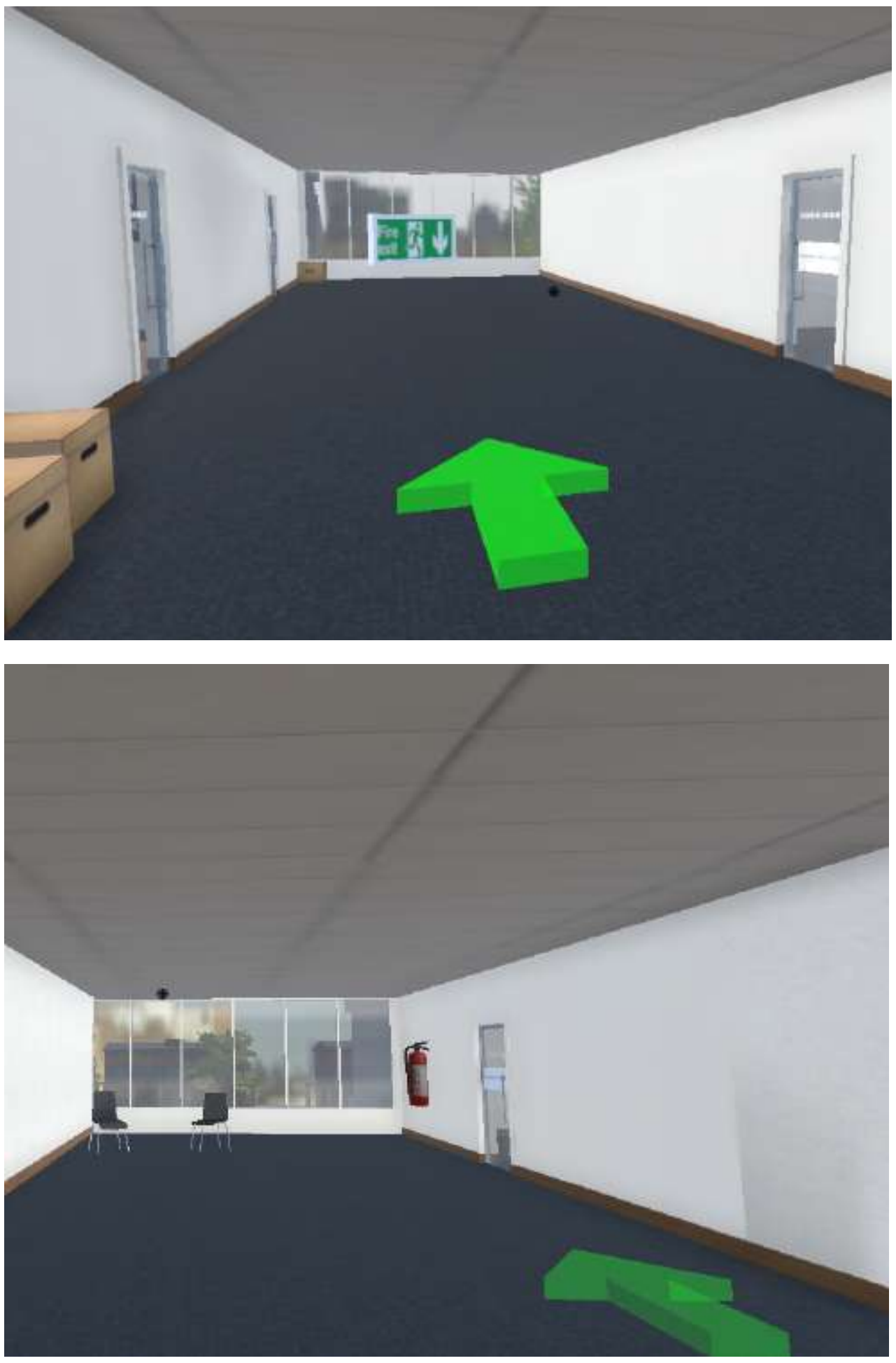

*Figure 4-43 Dynamic arrows guiding users to fire safety objects in the VE*

On passing through each floating object, users were presented with a pop-up text box providing fire safety rules relating to that object. Figure 4-44 shows the pop-up for the fire extinguisher.

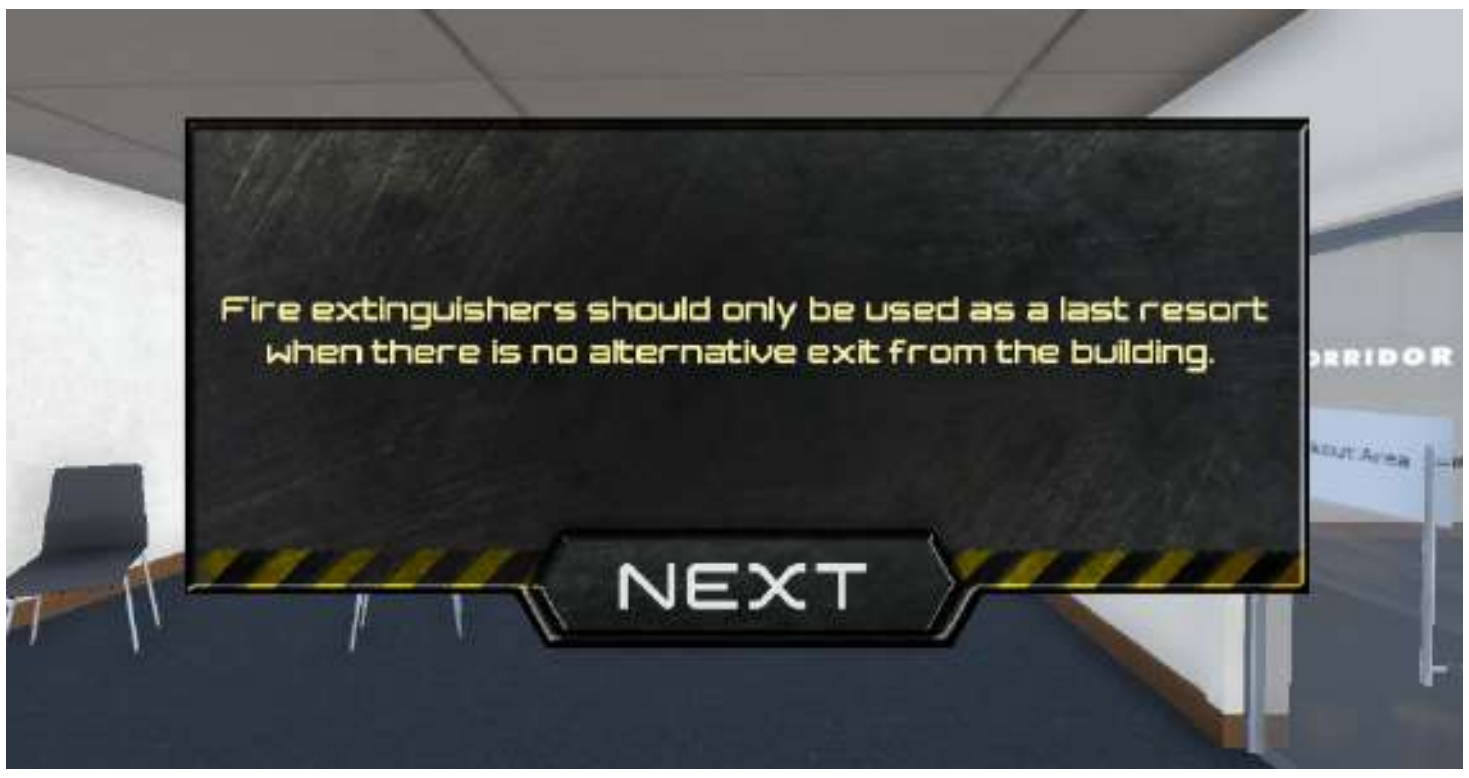

*Figure 4-44 Fire safety rules relating to the fire extinguisher*

For interactive objects, such as the fire extinguishers and doors, additional instructions were provided on how to interact with the object in the VE. Figure 4-45 shows the instructions for the fire extinguisher.

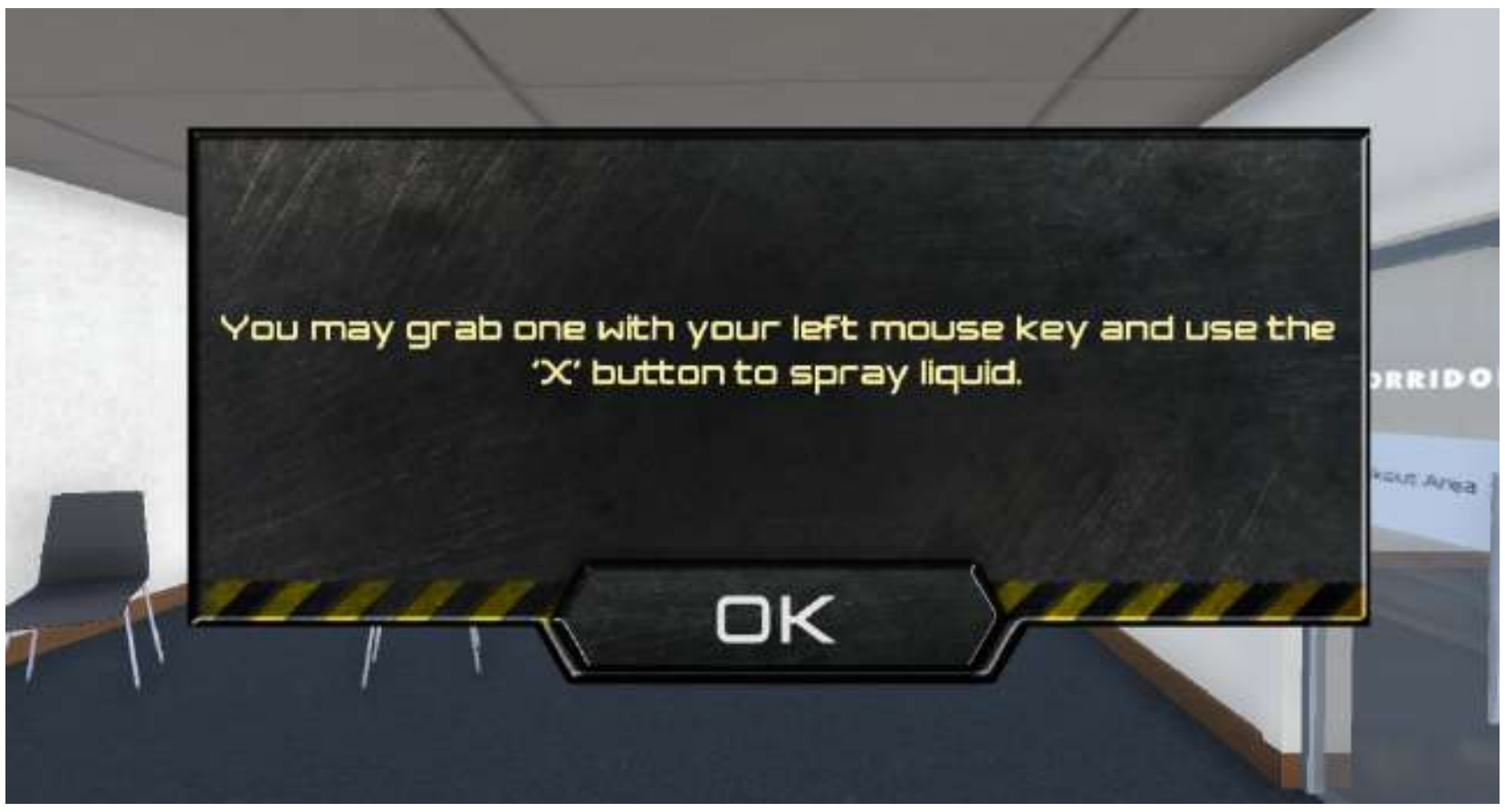

*Figure 4-45 Instructions for interacting with fire extinguishers in the VE*

Once users click 'OK', their attention was be drawn to the location of the object within the VE, as per Figure 4-46 below.

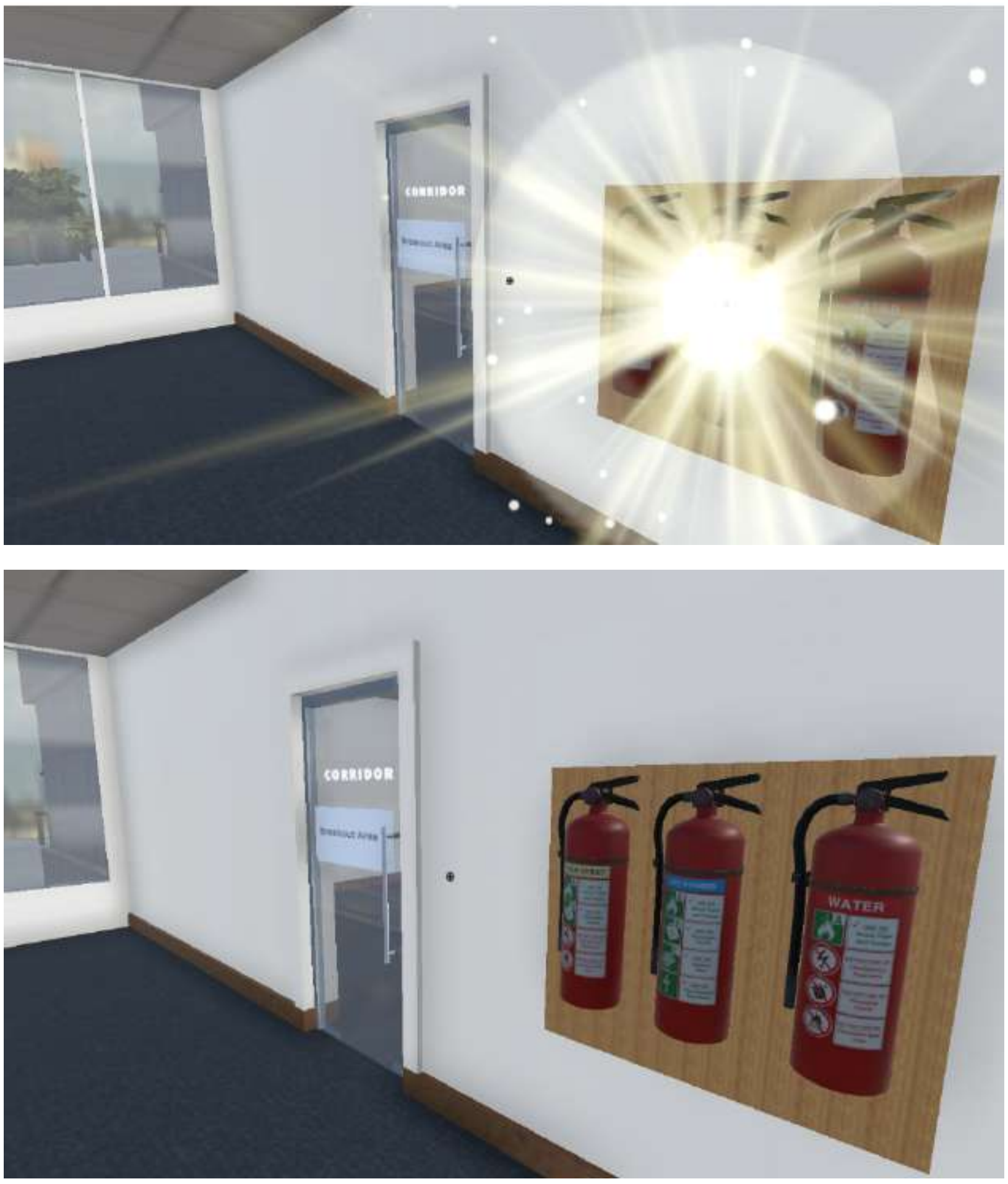

*Figure 4-46 Visuals of the fire extinguisher 'appearing' in their VE location*

When users passed through the fire token object, they were provided with both information on the fire token and how to carry out the fire token task in preparation for the final subtask of the instructional section of the training module. Figure 4-47 shows the 3D fire token within the VE and Figure 4-48 shows information provided about the fire token.

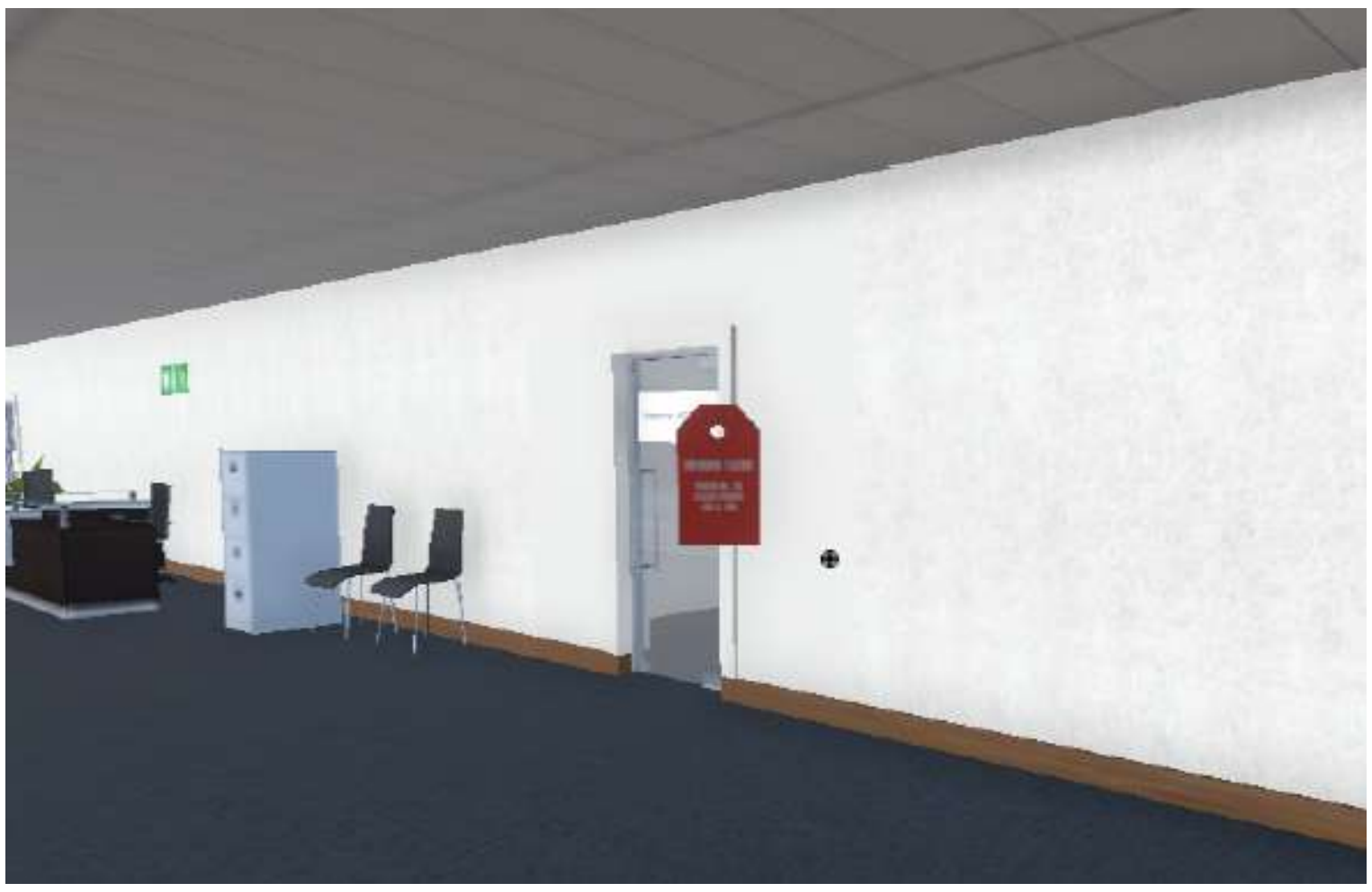

*Figure 4-47 Fire token as a 3D object*

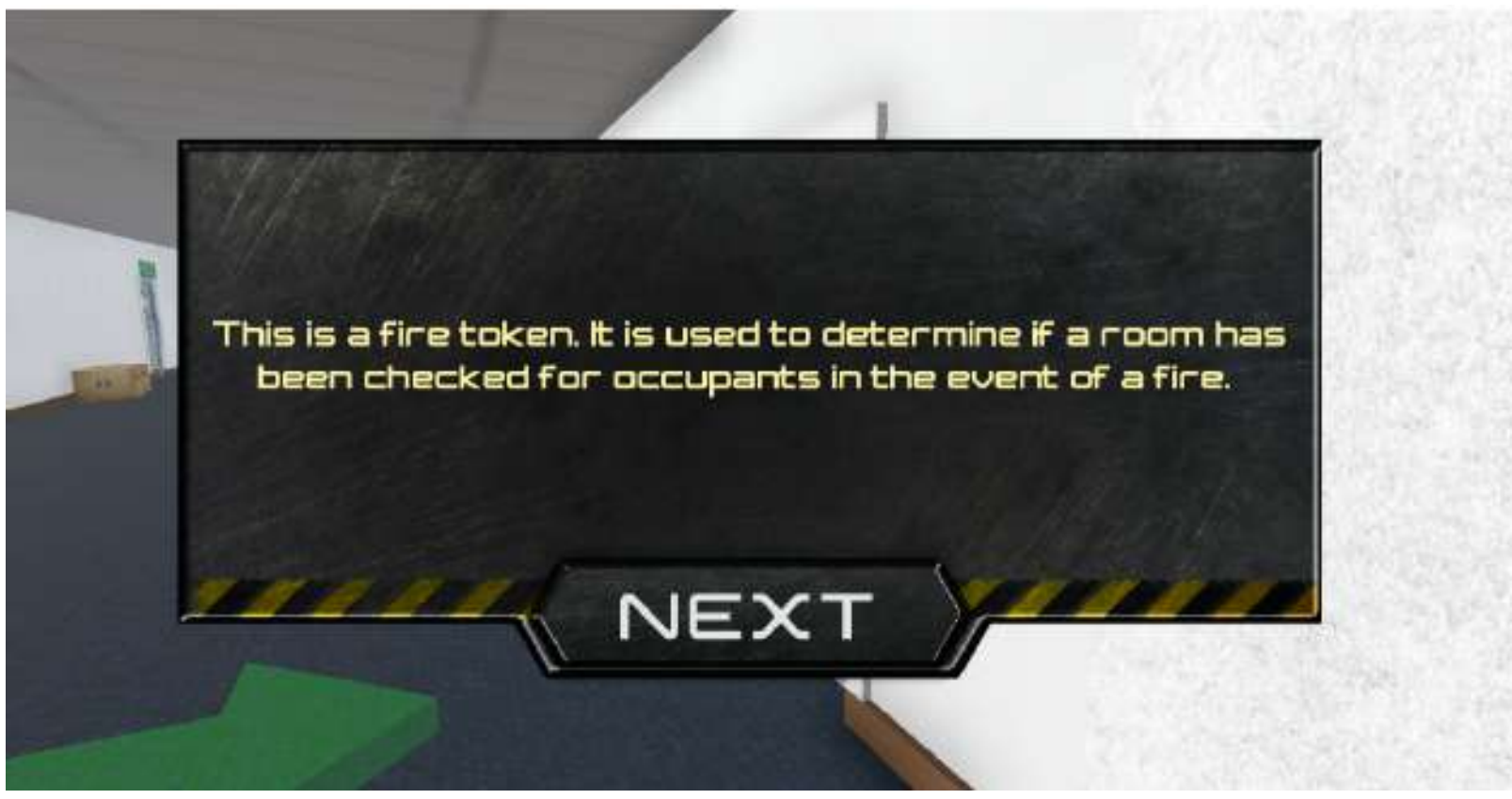

*Figure 4-48 Fire safety information for the fire token*

Users were then guided through a small series of actions to demonstrate the actions required to complete the fire token task. Figure 4-49 shows the information provided on the fire token task and the first set of action instructions:

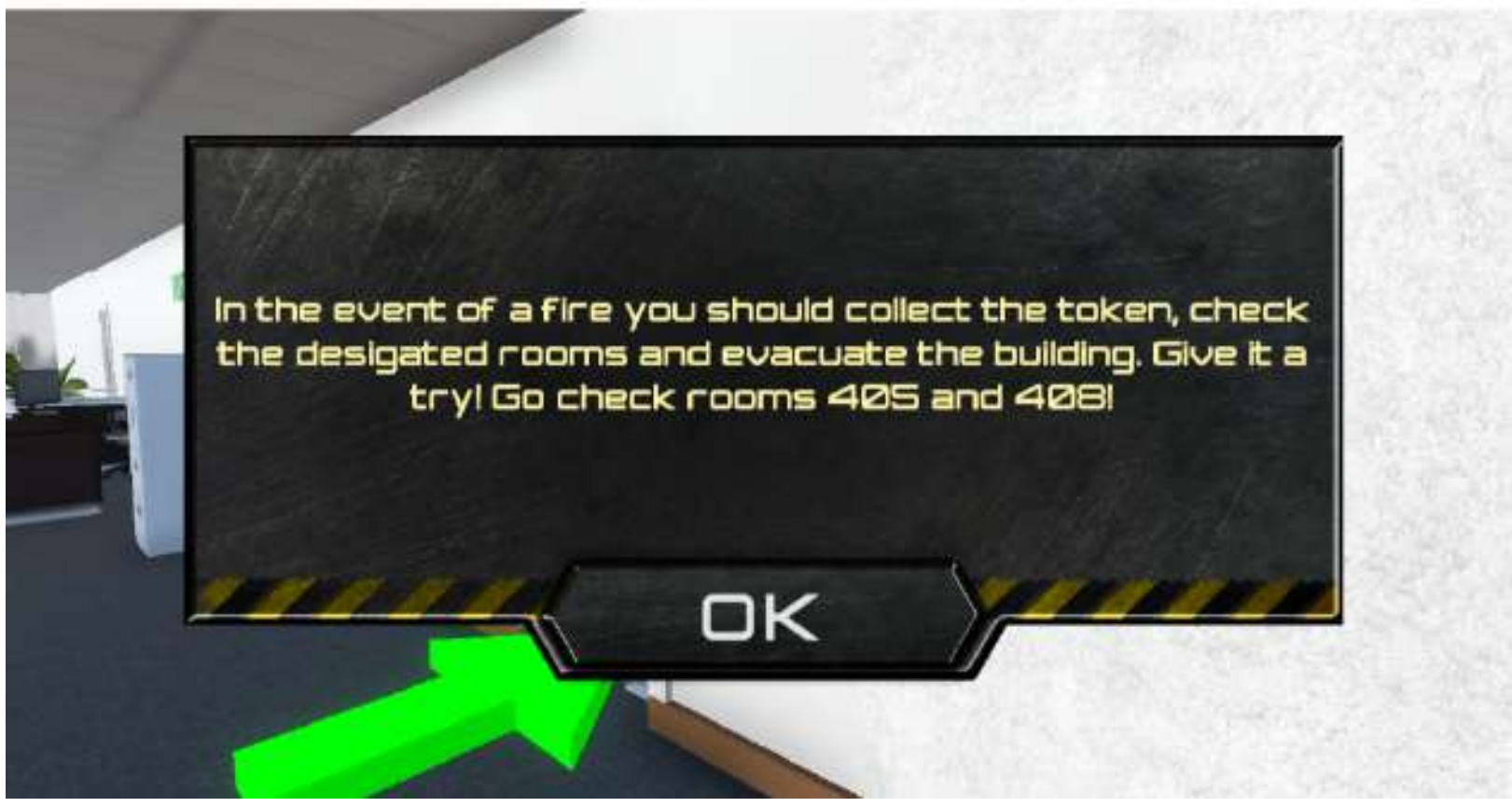

*Figure 4-49 Instructions to complete the fire token sub-task*

To complete this action the user needed to go through a doorway. This event was used to provide users with fire safety information relating to closing doors and provide a reminder on how to interact with doors in the VE. Figure 4-50 shows the information pop-up users received.

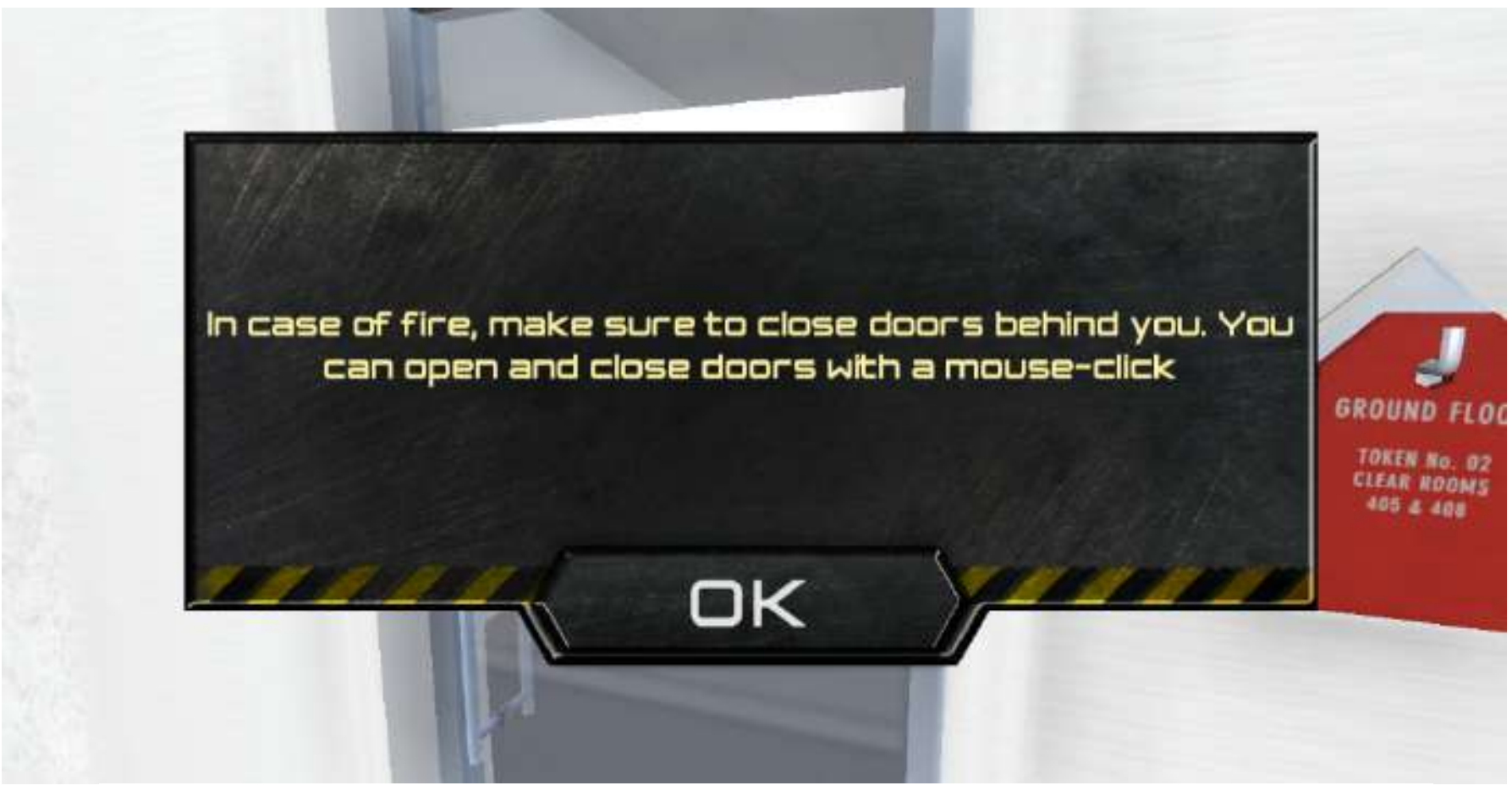

*Figure 4-50 Fire safety information and interaction instruction for doors*

On checking room 408, the user encountered an NPC and a pop-up instructing them on what action they should take; see Figure 4-51.

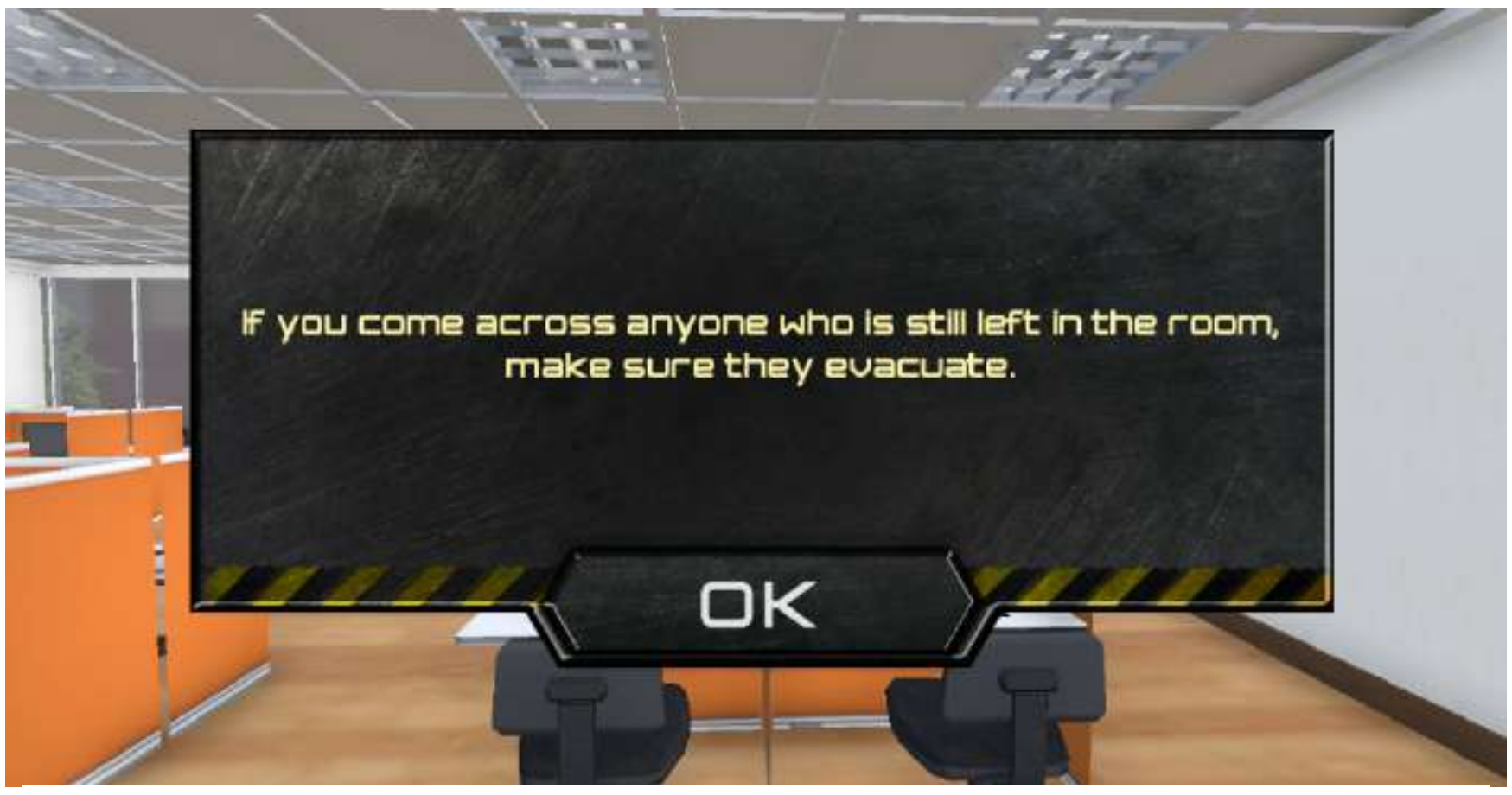

*Figure 4-51 Instructions regarding finding other occupants in the building*

This pop-up marked the end of the information-gathering subtask. Next the user was provided with a series of pop-ups guiding them through the spatial-awareness task, which required the user to collect fire safety objects located throughout the VE; see Figure 4-52.

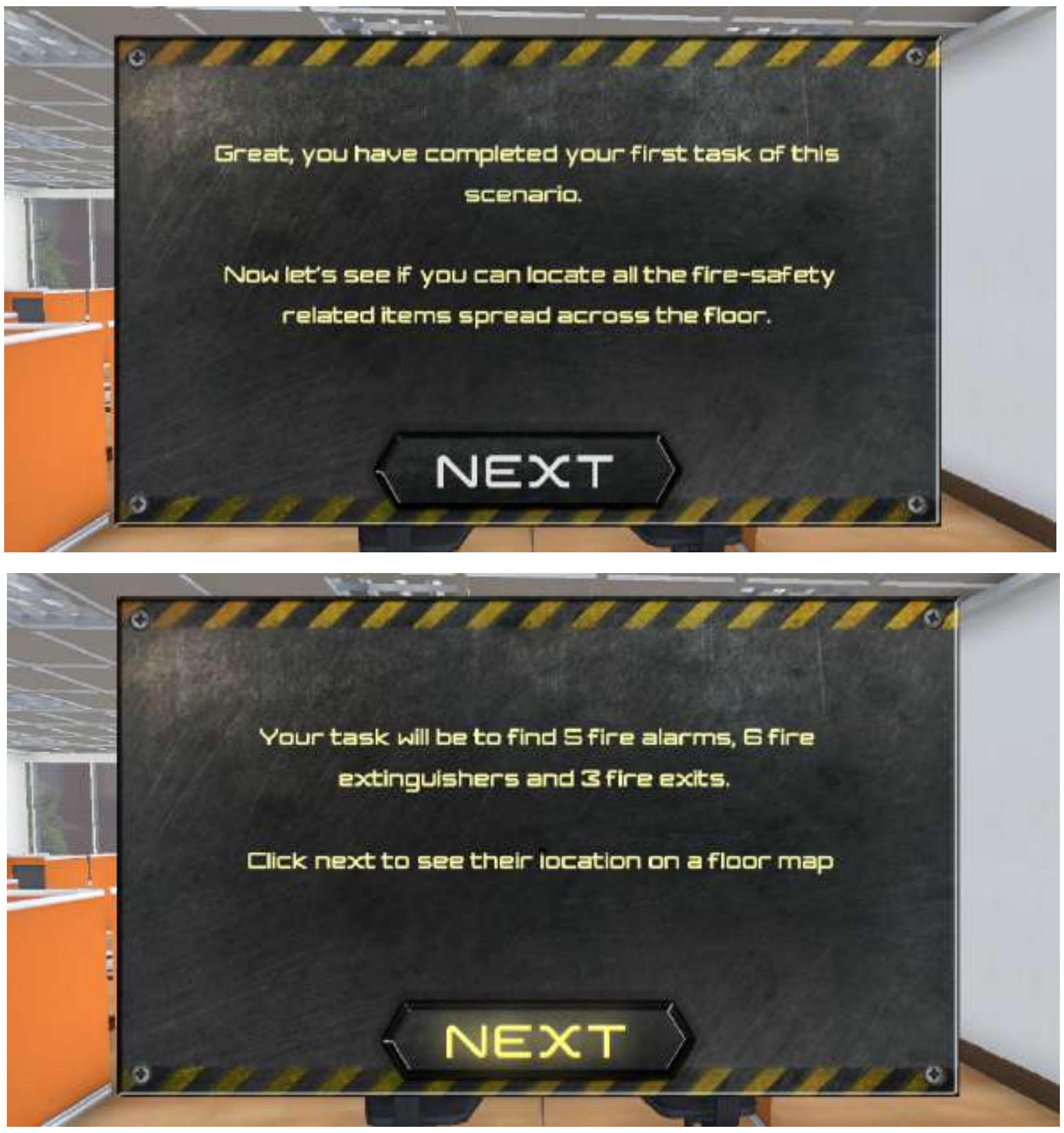

*Figure 4-52 Guidance for spatial-awareness task*

The fire alarms, fire extinguishers and fire exits were displayed as 3D floating, rotating objects throughout the building consistent with the information-gathering task. To collect the objects, users were required to walk over the 3D object. Once successfully collected, the object would move to its fixed location within the VE as in the information-gathering subtask. To assist users in this task, a toolbar was used to keep track of objects collected (Figure 4-53), which updated as objects were successfully collected.

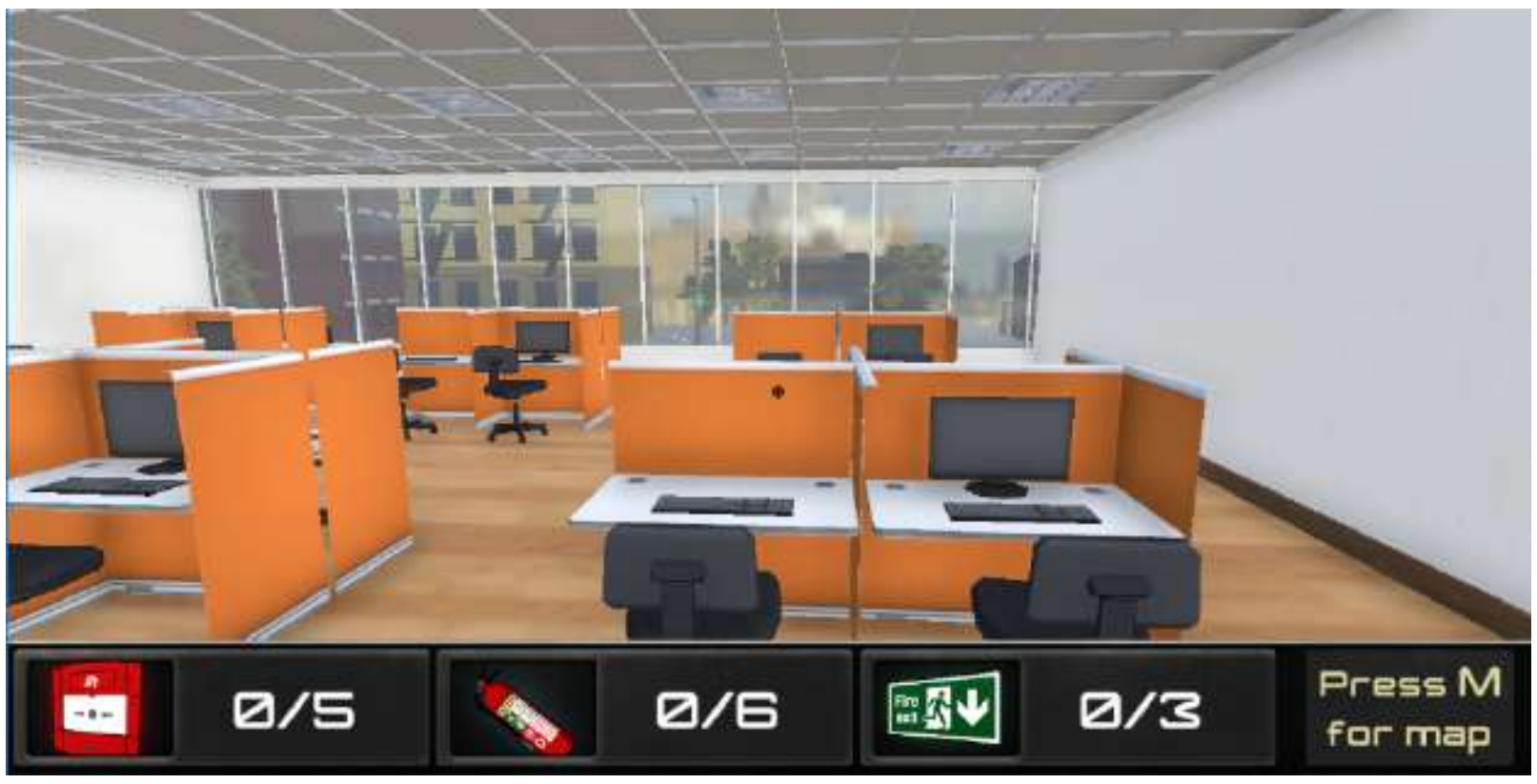

*Figure 4-53 Toolbar for spatial-awareness task*

Additionally, a live interactive map could be accessed at any point by pressing 'M' on the keyboard (Figure 4-54). The map showed the live position of the user within the VE and what direction they were facing to assist users in making a mental map of their position within the VE in relation to the objects they needed to locate.

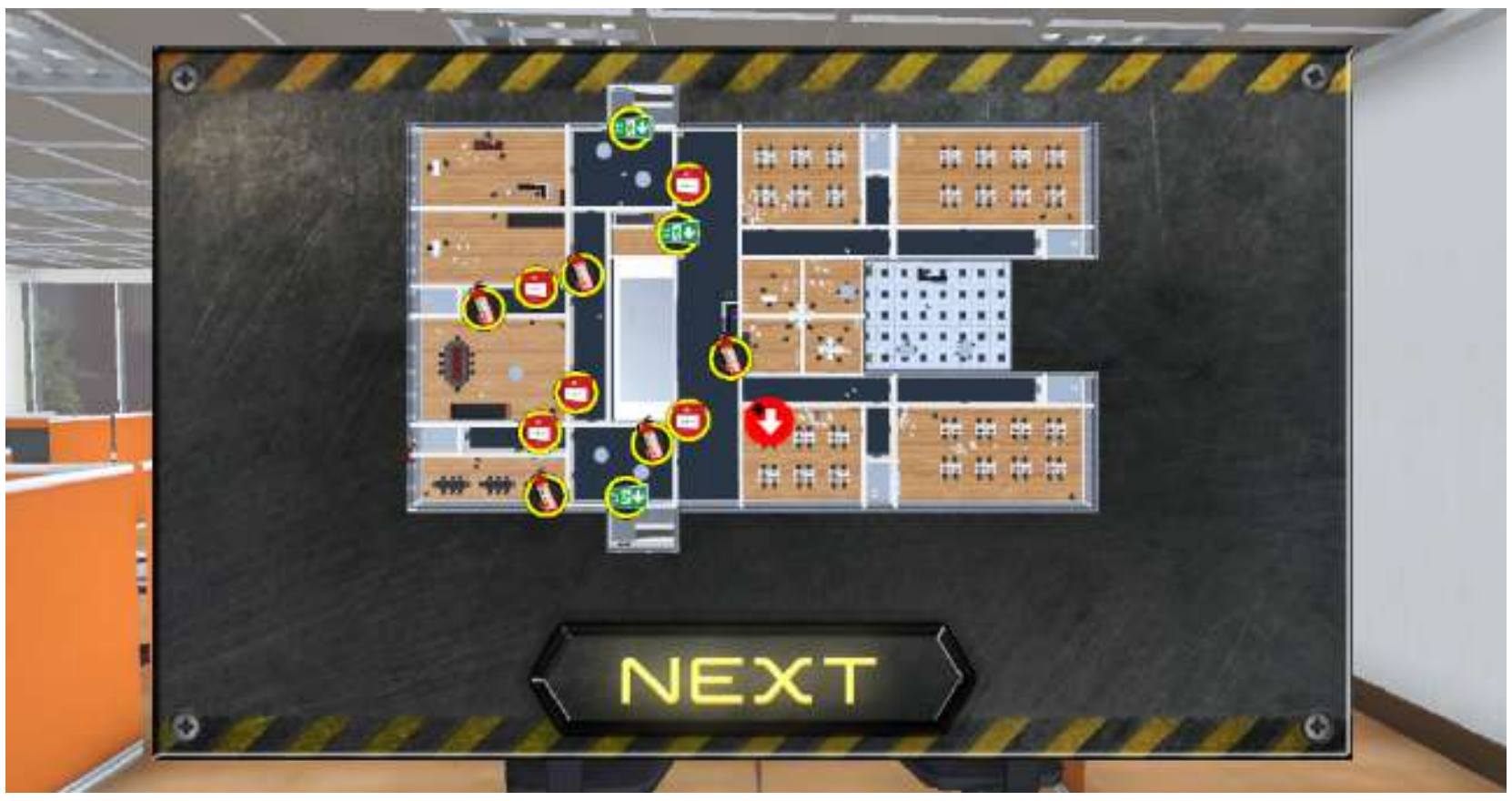

*Figure 4-54 Interactive map showing the location of the fire safety objects and the user*

Once the user had successfully collected all objects, they received a pop-up asking them to move to a specific room within the VE to collect a ‘prize’. The user was shown a map highlighting the location of this room and the interactive map was updated so that the user could see a live version of their location in relation to the room; see Figure 4-55.

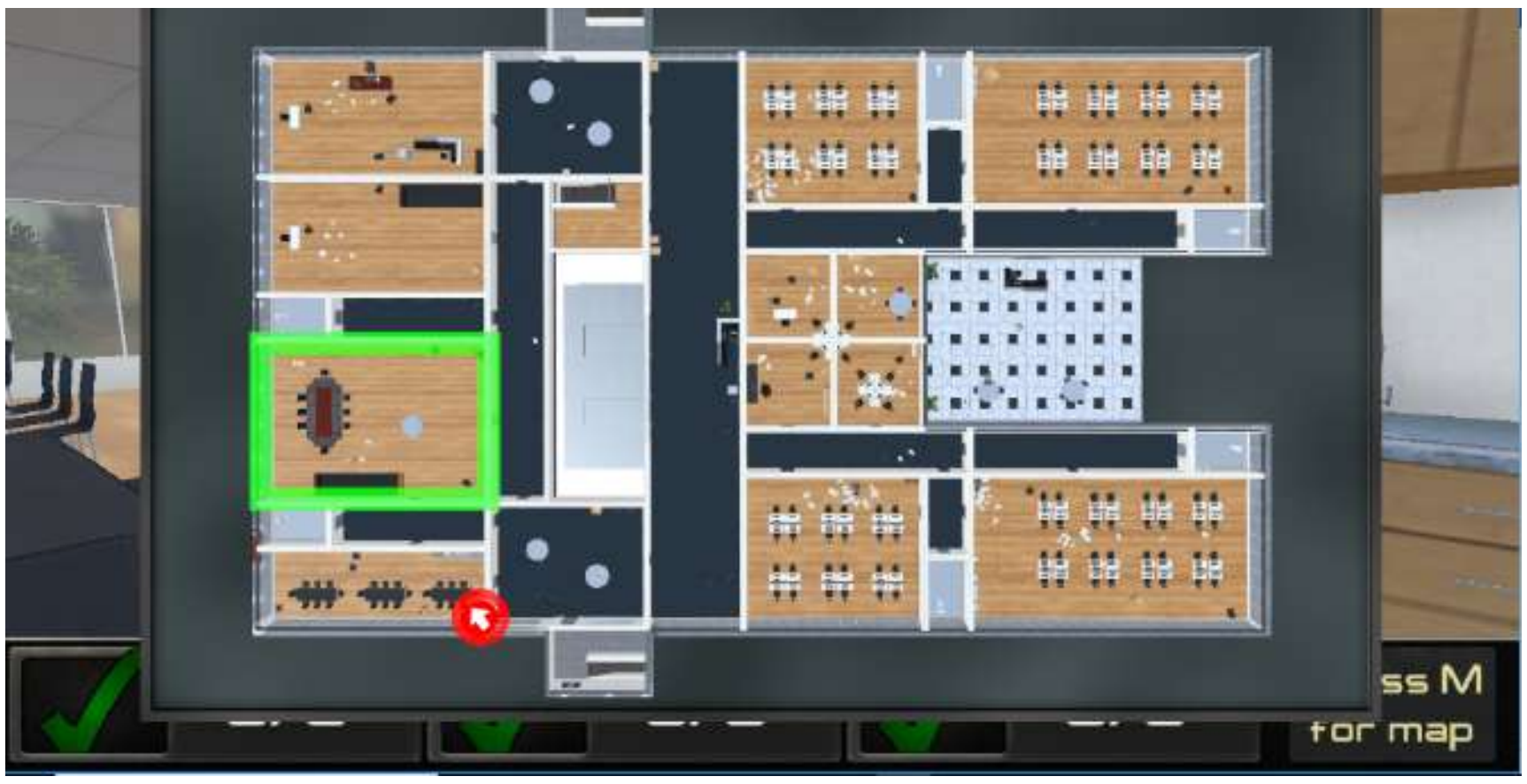

*Figure 4-55 Interactive map showing the location of the room to collect the 'prize' and the user*

As the user entered the room containing the ‘prize’, they crossed a collider, which triggered a fire simulation and simultaneously presented the user with a pop-up telling them it was a trap (Figure 4-56):

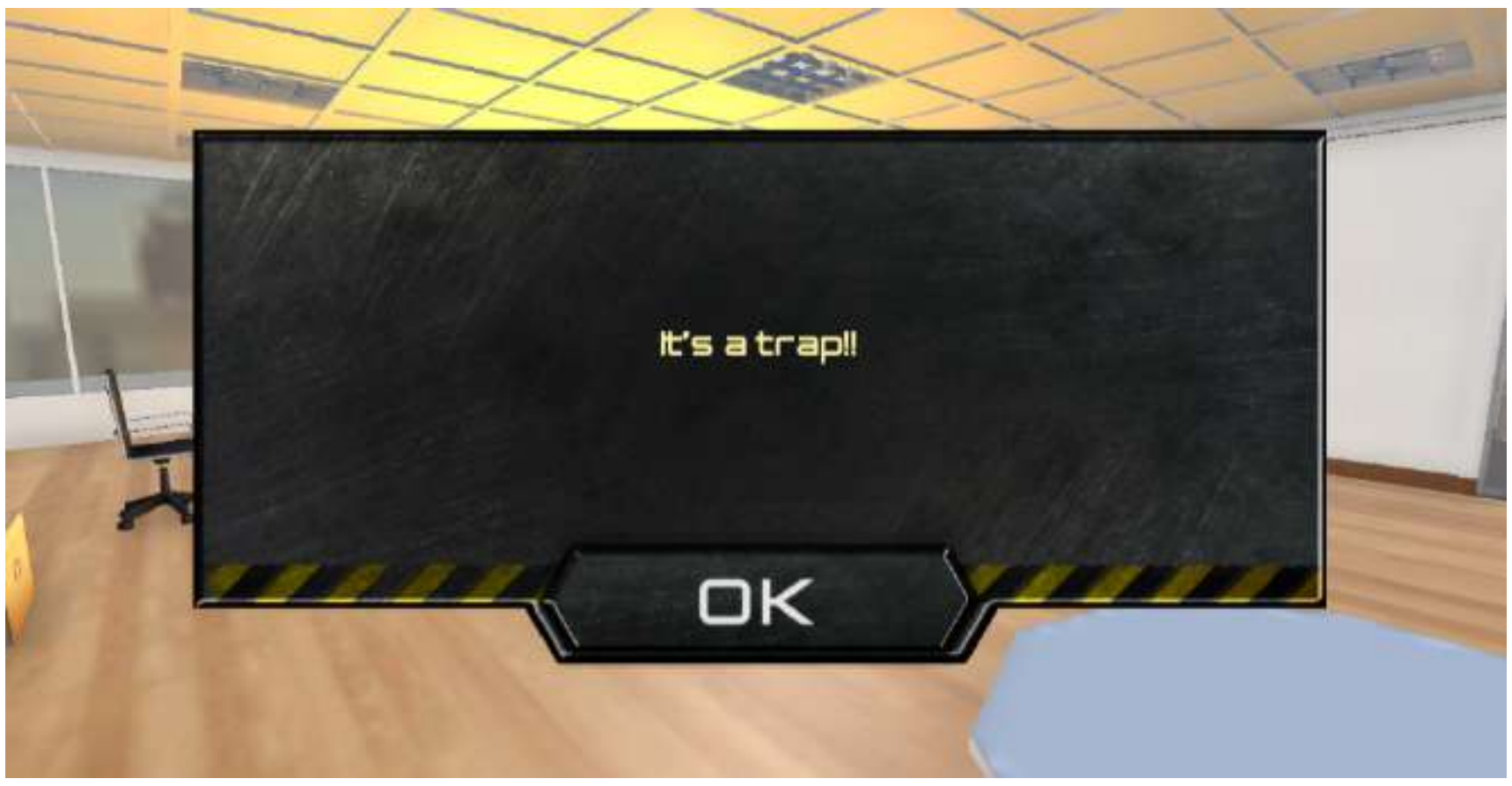

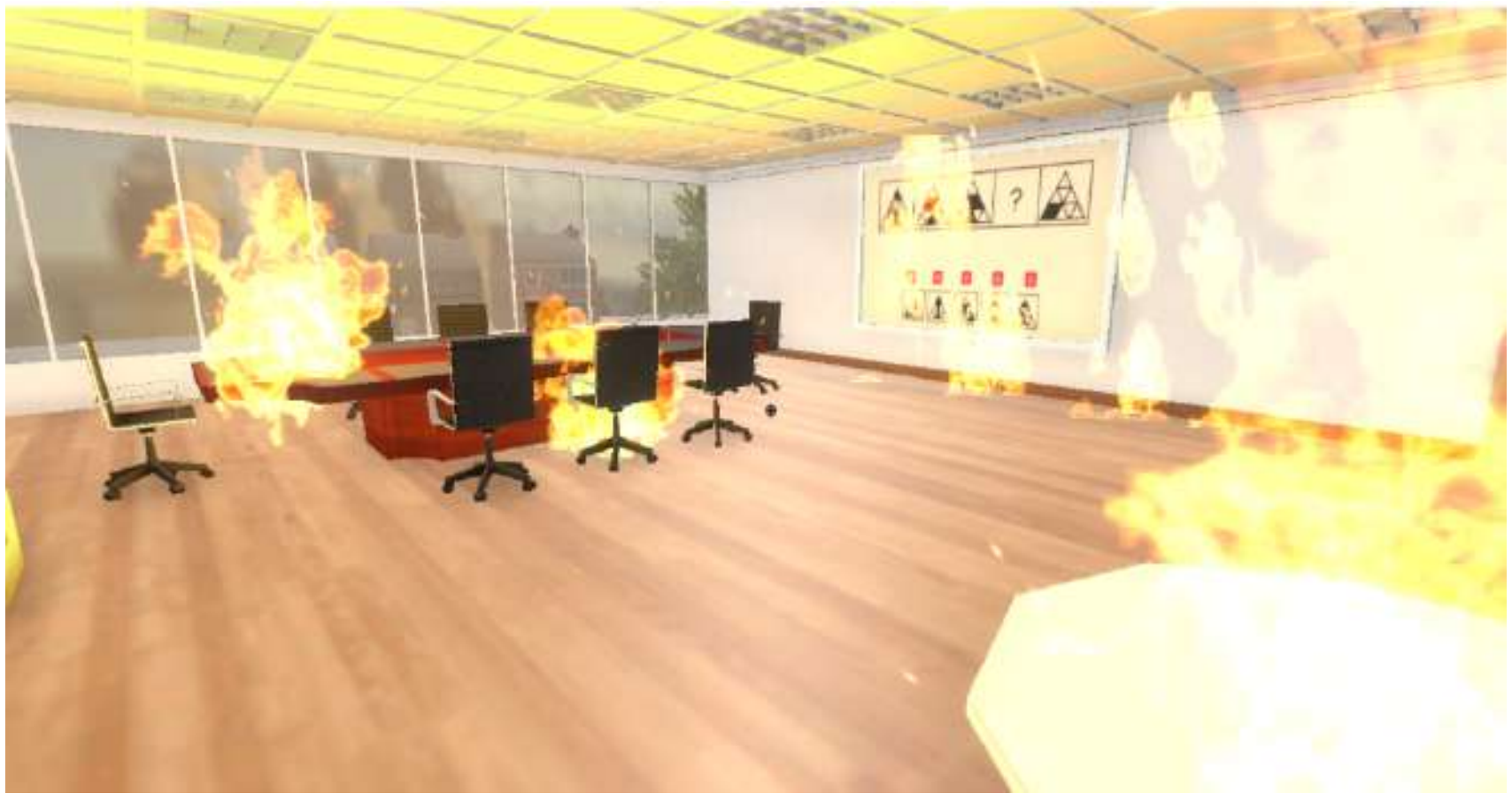

*Figure 4-56 Fire simulation and 'it's a trap' pop-up*

On evacuating the building, the user was presented with a number of pop-up yes/no options relating to fire safety points covered in the instructional task: closing doors, collecting personal belongings, collecting fire tokens and checking rooms. The answer chosen by the user would pull up a further pop-up confirming whether the chosen answer was correct and reiterating the correct action. For example, Figure 4-57 shows a pop-up for collecting personal belongings and the pop-up if the user selects ‘yes’.

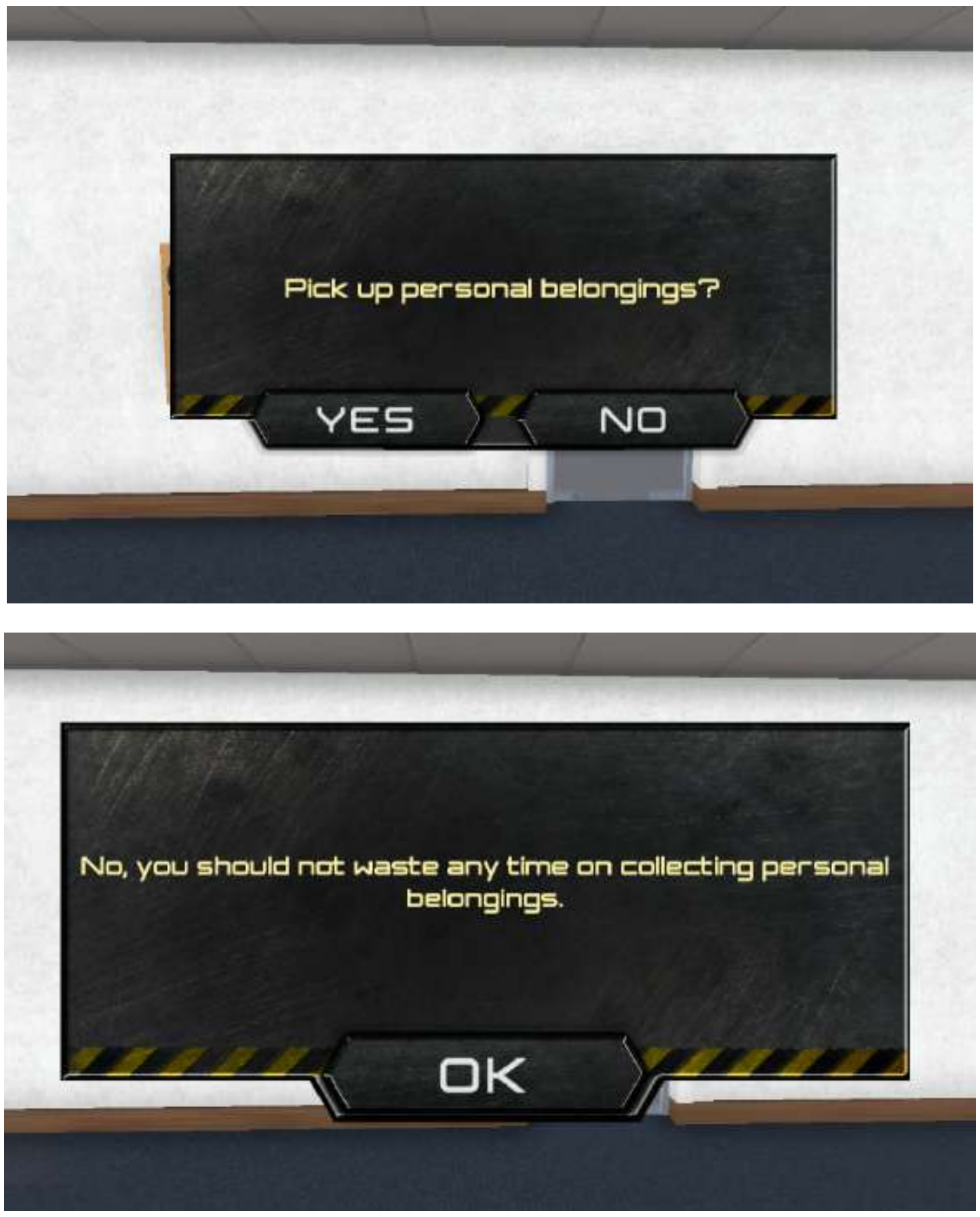

*Figure 4-57 Action option pop-up and answer if user clicked 'yes'*

On exiting the building, the user was presented with a pop-up letting them know that they have successfully escaped, and when the user pressed ‘OK’ they were taken back to the main menu to select the final ‘fire scenario’ task.

#### 4.9.1.3 Fire scenario task

The purpose of this task was to provide users with the opportunity to apply the fire safety knowledge to a virtual fire emergency scenario with no instructional input or guidance. The scenario was designed so that if the user acts in a way that contravenes the fire safety knowledge they have been given, they will experience a consequence within the VE simulation. Users were given a visual pop-up with concurrent audio instructions to introduce the scenario and guide users through initial actions required.

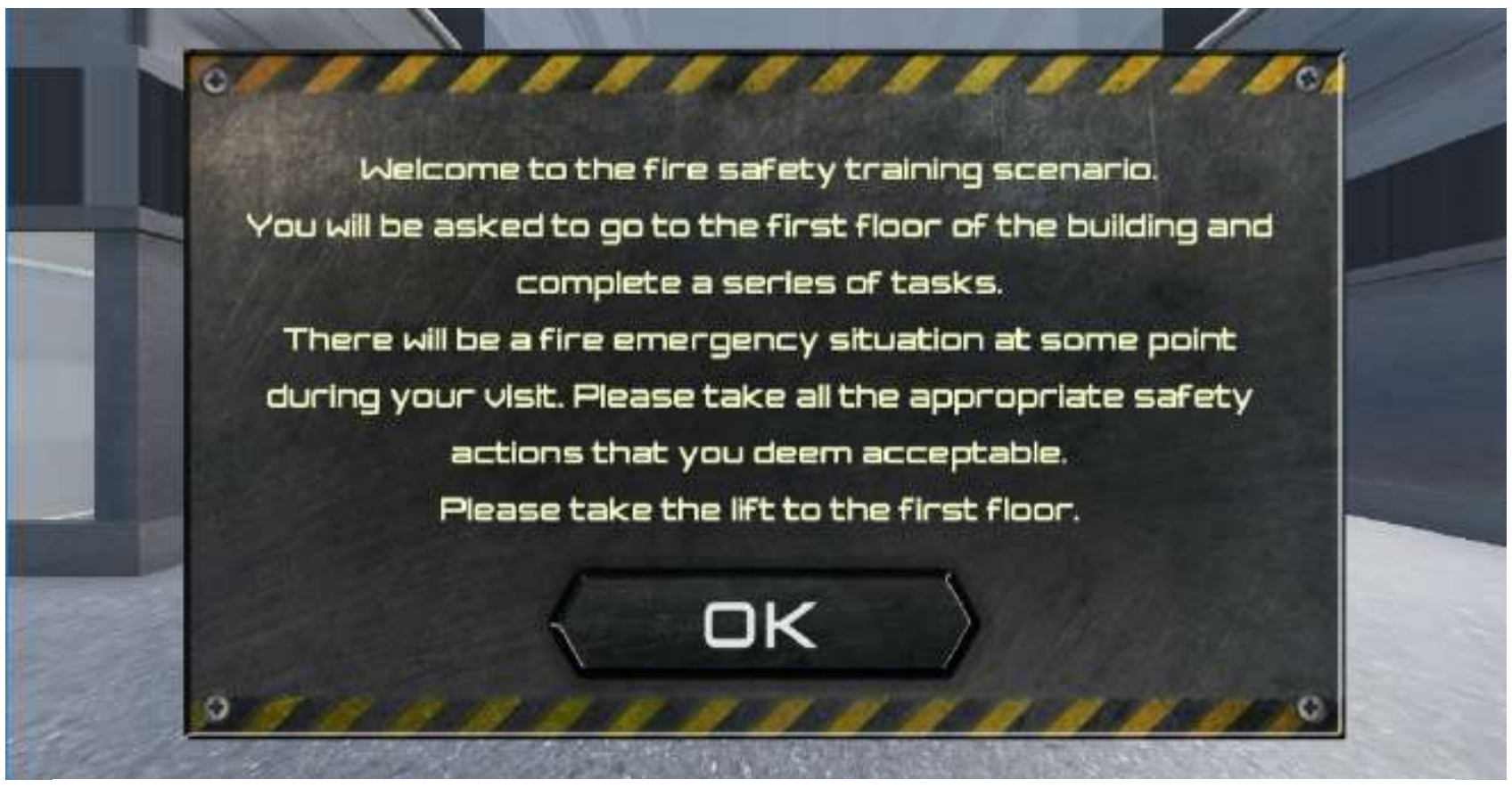

*Figure 4-58 Initial introduction to fire scenario task*

A series of tasks were used to guide users along a consistent route to the location where the fire was triggered. Instructions for these tasks were presented either via visual pop-ups with concurrent audio instructions or via audio instruction alone, for example a welcome speech made by the receptionist. Users were instructed to approach the receptionist, sign in and then follow the sign to Room 412, as shown in Figure 4-59 below.

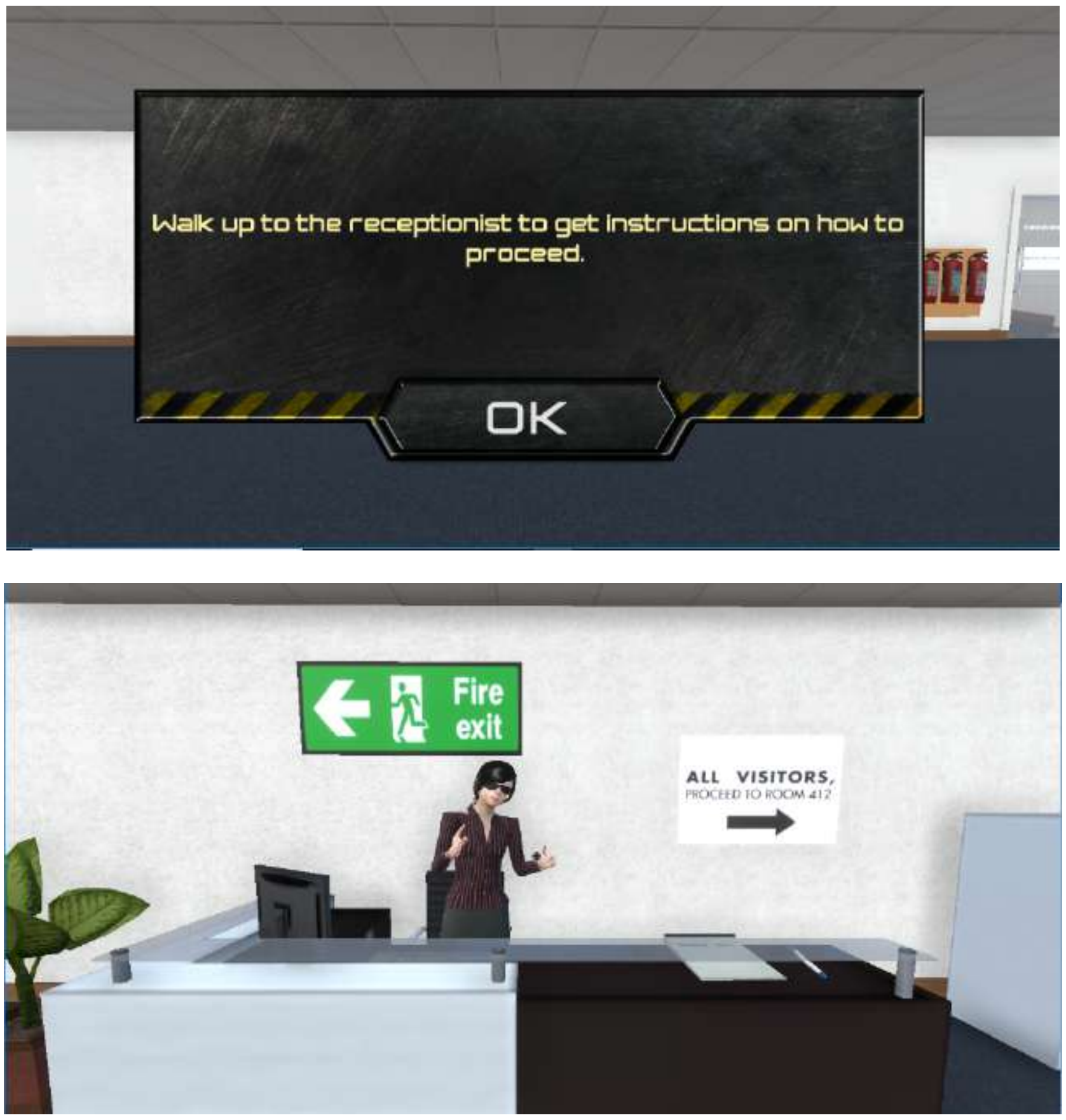

*Figure 4-59 Instructions to complete sub-tasks by visual pop-up and via audio speech made by an animated receptionist character*

The sign-in task mirrored the one used in Study 1 with the signature automatically populating once the user approaches and looks down at the sheet (see Figure 4-60).

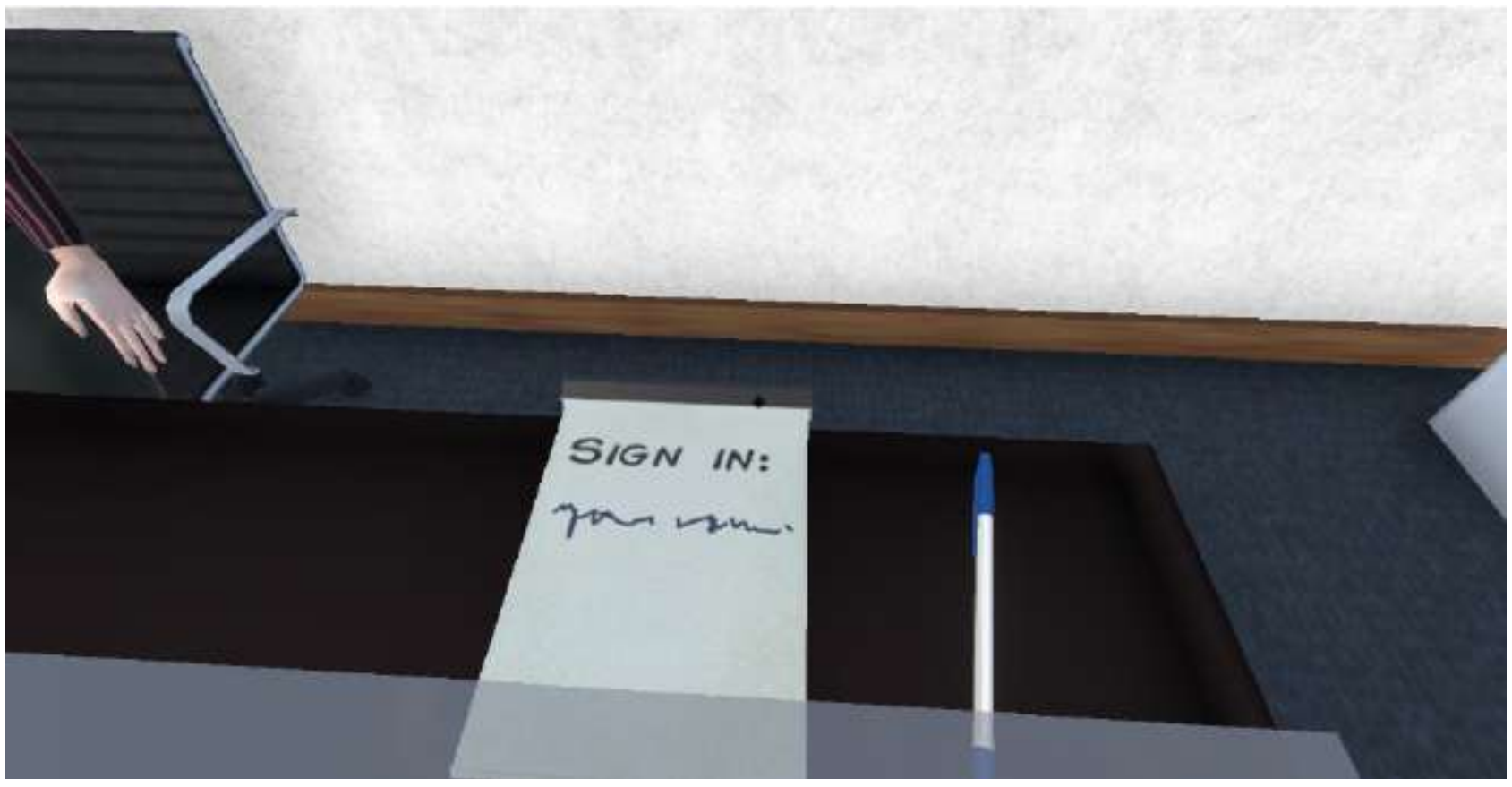

*Figure 4-60 Image showing automated population of sign-in sheet*

A collider was placed in the entrance of room 412, so that when the user entered the room, the simulated fire was triggered automatically. Once again, the simulated fire consisted of visual and auditory effects, but the user had to face a greater number of fires than in the previous study. To further amplify the perceived pressure, the fire was also programmed to gradually spread until the point where flames engulfed much of the building interior. To increase the realism of the scenario, the fire was set to start at the microwave in the small kitchen contained in this room. The fire was also positioned close to a set of fire extinguishers to present users with an opportunity to use them and to test fire safety knowledge related to the use of fire extinguishers.

Users were required to use knowledge of safety features and procedures gained from the instructional task to evacuate the building safely. In order to escape the fire and successfully evacuate the building, users had to rely on knowledge of safety procedures gained from the previous instructional task. The degree to which users followed these procedures had a direct impact on the rate at which the simulated fire spread throughout the building. For instance, by closing doors behind themselves while escaping through the building corridors, users could slow down the spread of the fire substantially and thus increase their chances of 'survival'. Incorrect behaviours, such as the use of fire extinguishers, on the other hand caused the fire to spread more rapidly and consequently hampered the user's chances of escaping in time. Additionally, the task was designed so that the first fire exit the user approached to evacuate the building was blocked (see Figure 4-61). All fire exits were initially blocked to ensure users encountered a blocked fire exit. Once the user had encountered one blocked fire exit, the experimenter could manually unblock them by pressing 'X' on the keyboard.

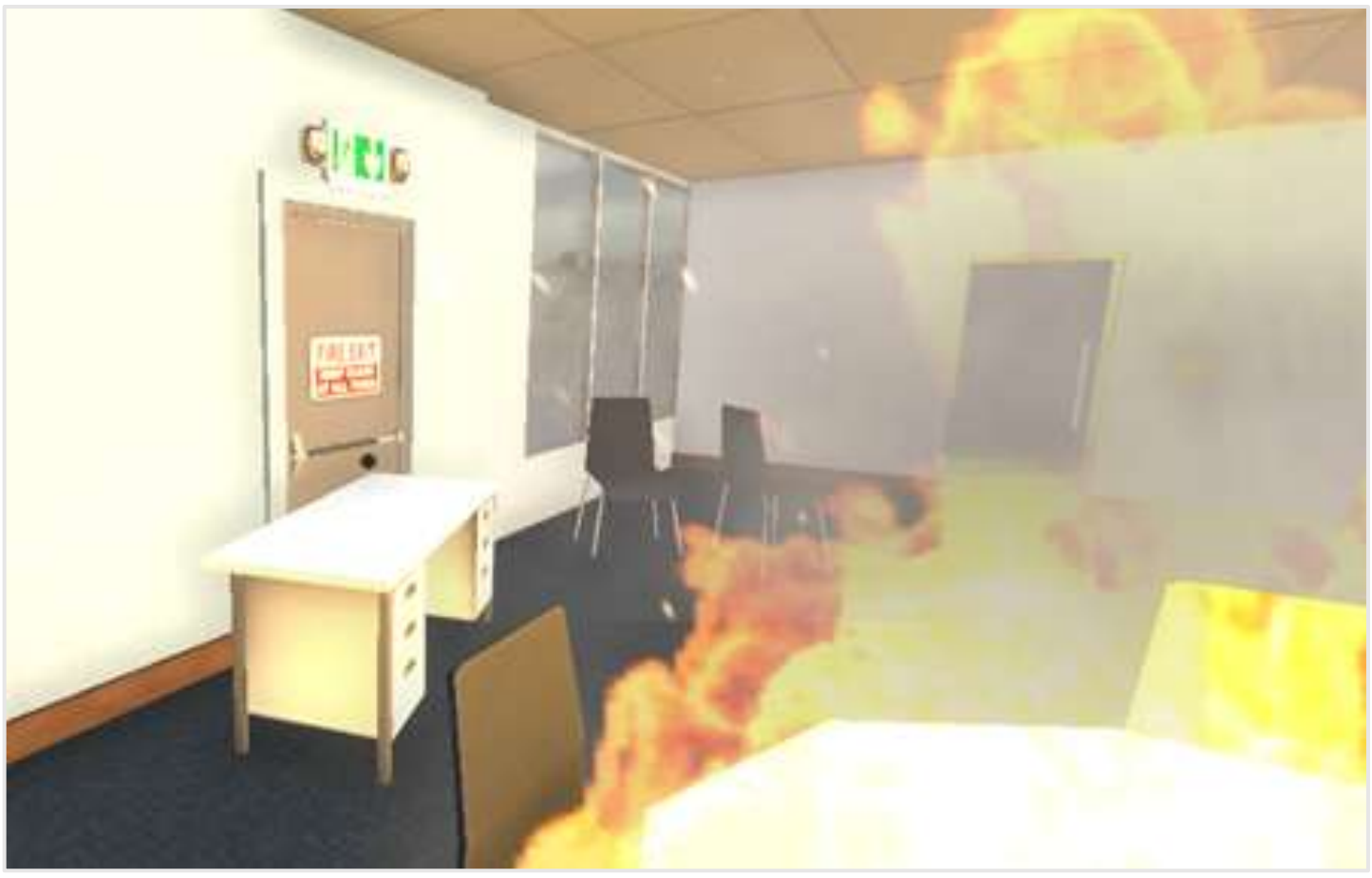

*Figure 4-61 A fire breakout and the blocked exit*

If users took actions that increased the spread of fire to the point where evacuation was no longer possible, they would receive a pop-up (see Figure 4-62) informing them that they had been trapped by the fire and to try the scenario again. The pop-up was tailored to provide the user with feedback on their actions and to reinforce what fire safety actions they should take. The pop-up then restarted the fire scenario from the receptionist desk to reduce the time it took to complete the task.

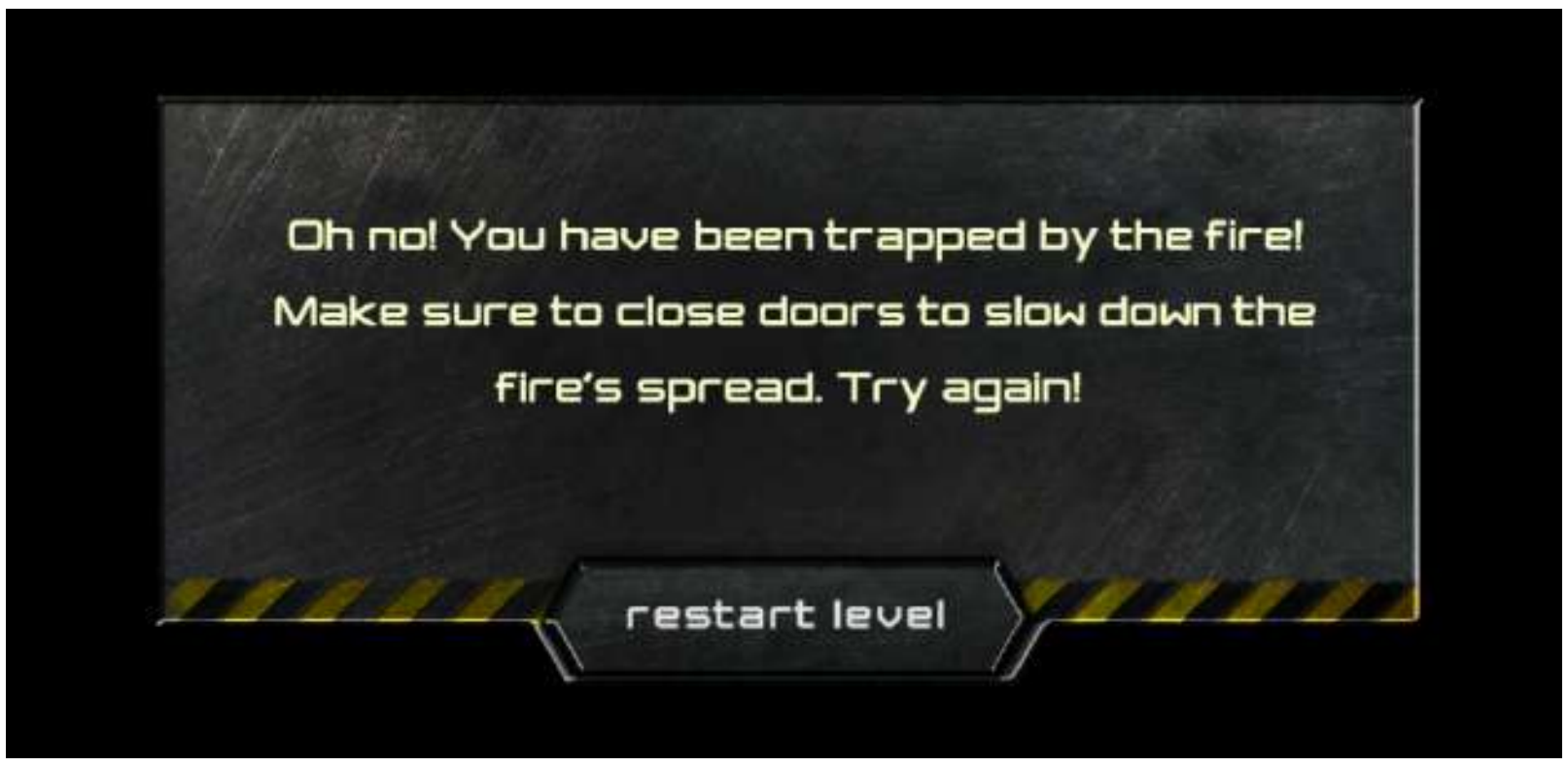

*Figure 4-62 The instructional pop-up message upon failing to evacuate*

As users exited the building, they received a pop-up containing customised feedback on the correct and incorrect actions taken during the simulation; see Figure 4-63:

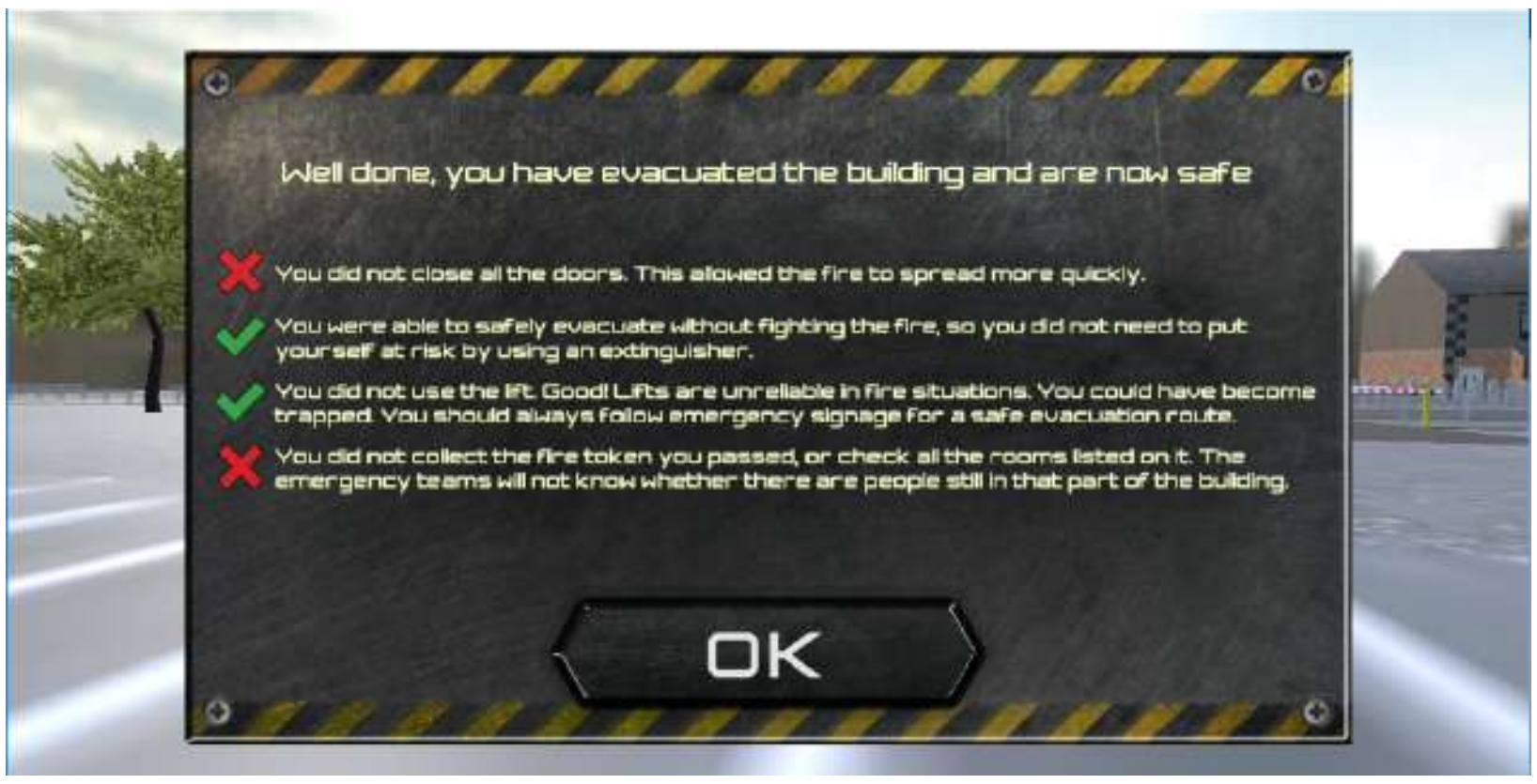

*Figure 4-63 Customised feedback showing correct and incorrect fire safety actions taken*

### 4.9.2 Engine disassembly use-case training VE

A single virtual room was produced for the purposes of the engine disassembly use case. The room was designed to resemble a typical industrial workshop and populated with corresponding features such as workbenches, soldering stations and other forms of machinery. Multiple animated workers were also distributed throughout the room to provide further context and realism. The centre of the room was occupied by a replica model car engine. Users were given a full six degrees of freedom to move around and explore the room in any manner they wished. It was thus perfectly feasible for the user to walk around the engine and view it from different angles while completing the tasks detailed below. Figure 4-64 shows an image of the workshop:

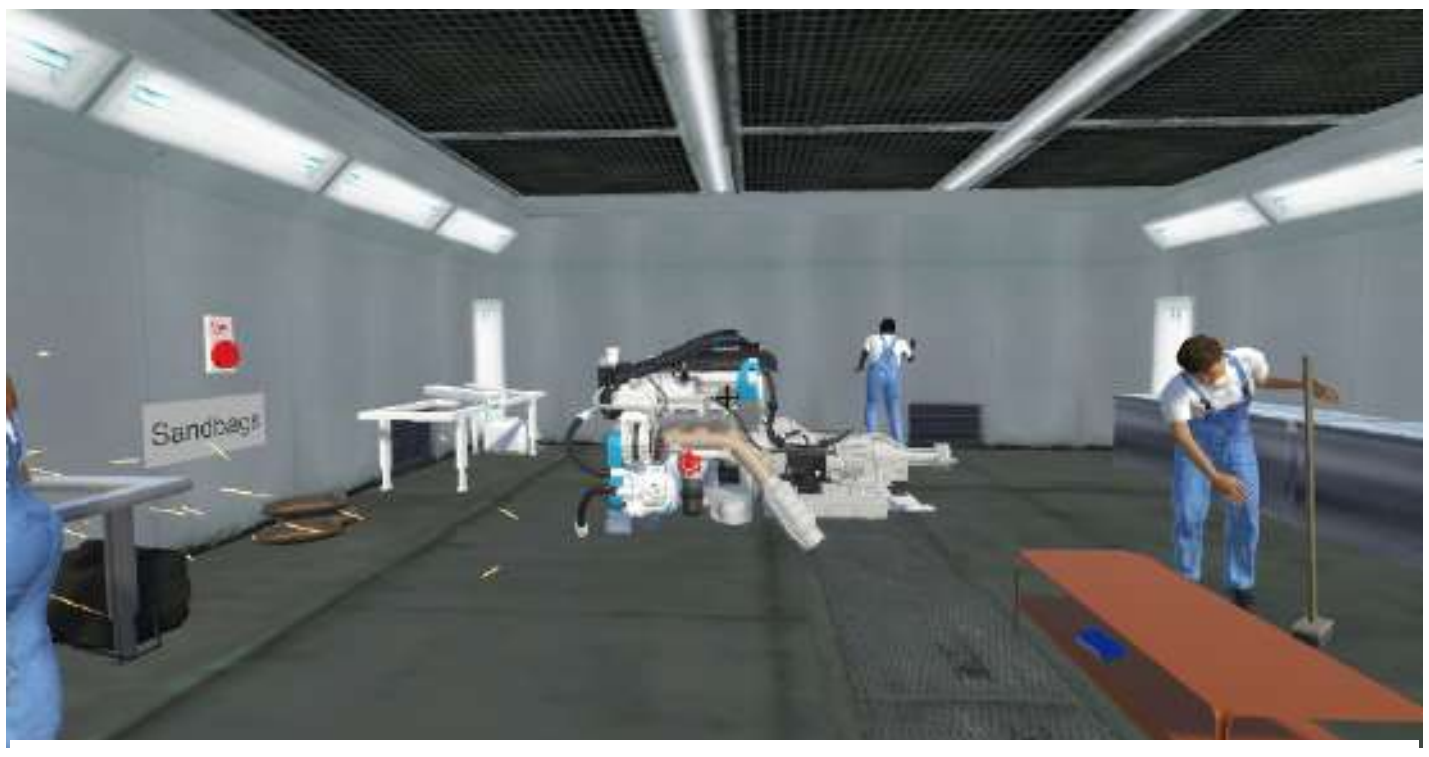

*Figure 4-64 VE workshop for engine disassembly use case*

For the engine disassembly scenario, participants were required to complete two tasks:

1. A vehicle engine disassembly task had the users fully disassemble an engine.
2. A fuel leak scenario task had the users walk through a set of safety procedures that ought to be carried out in the event of a fuel leak.

Each task was repeated twice: first with step-by-step guidance and a second time without any instructions. For both tasks, users were not just required to carry out the correct actions, but also to carry them out in a correct order.

Each training session was initiated by a visual pop-up text window that was also read aloud as a concurrent audio message. The message introduced the training module and encouraged the users to familiarise themselves with the virtual environment. The purpose of this familiarisation, as per the fire safety use case, was to allow users to experience the VE and to practise using the controls to navigate around the environment. An example of the introductory pop-ups is shown in Figure 4-65 below:

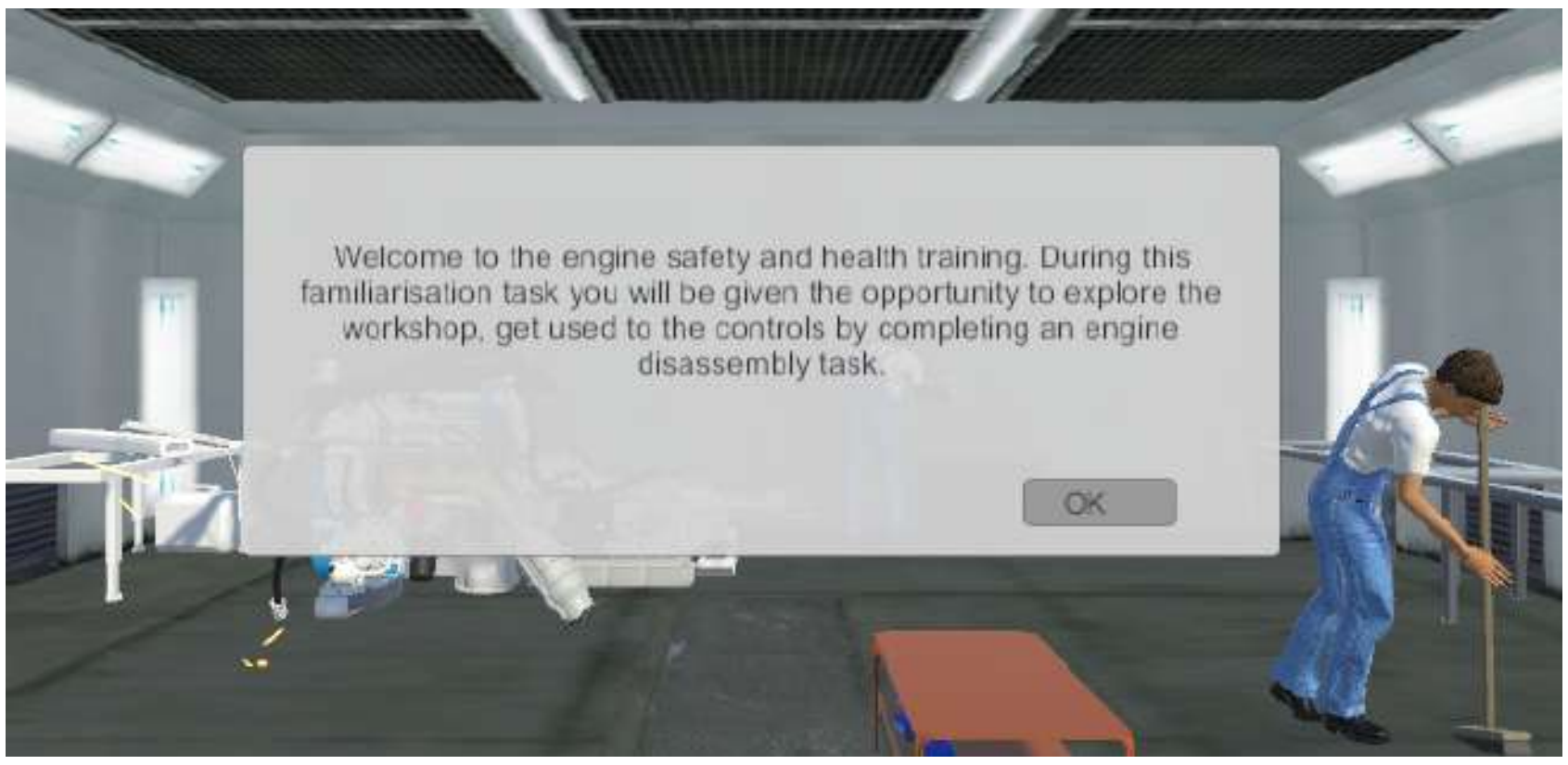

*Figure 4-65 Screenshot of the welcome message*

Once the users had familiarised themselves with the VE, a series of visual pop-ups were used to guide them step by step through the engine disassembly task (Figure 4-66). Users were required to remove components from the engine in a specific order that mirrored the training provided in the PowerPoint condition.

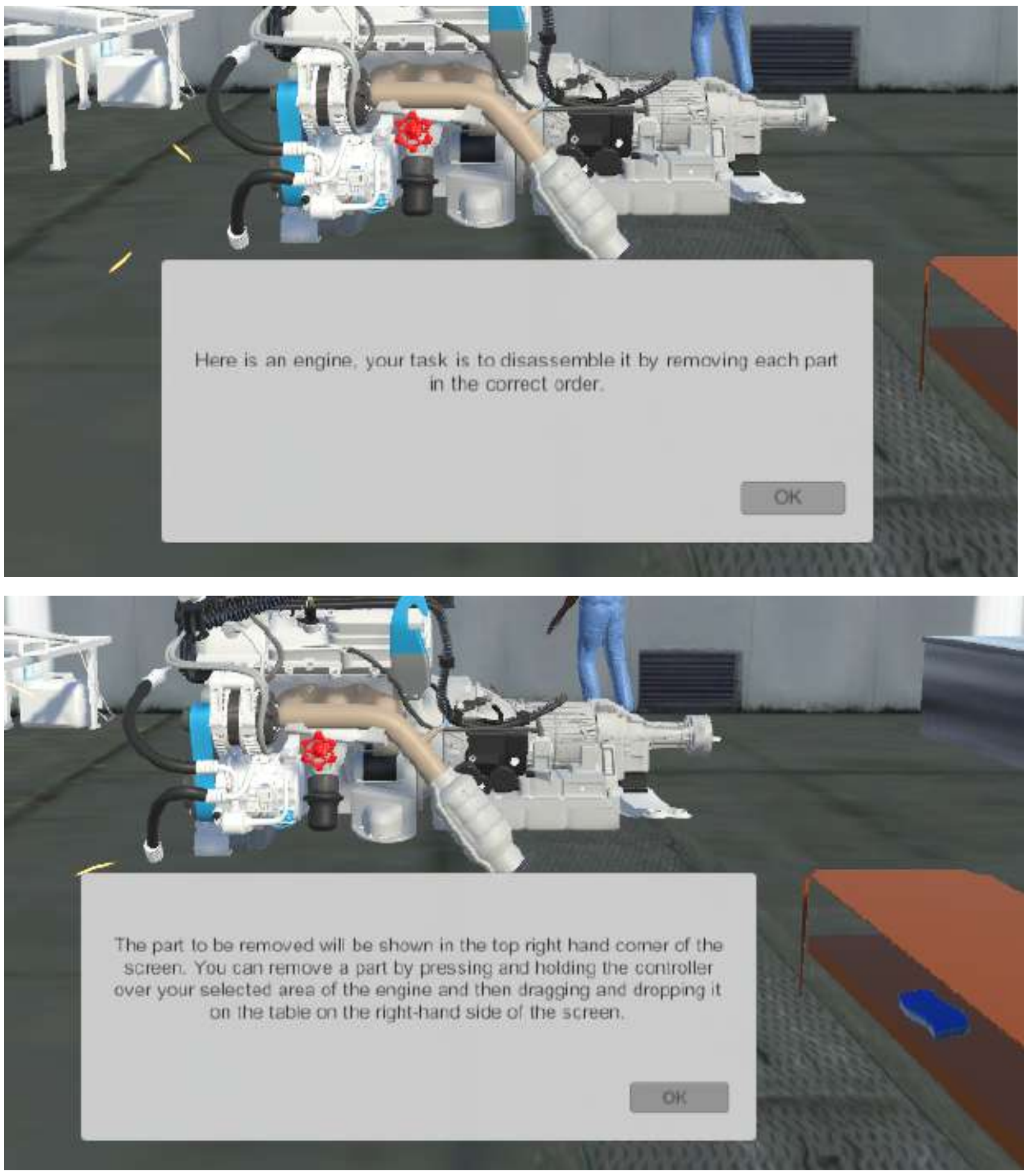

*Figure 4-66 Step-by-step instructions for engine disassembly task*

A sequence of visual pop-ups (Figure 4-67) provided the users with instructions on the specific engine components that needed to be removed as well as the specific order in which this disassembly had to be done. Throughout the task, a labelled picture of the component to be removed was presented to the user in the top right-hand corner of the screen to make the identification of the correct component easier. Users were required to use the mouse to select and drag components from the engine on to a nearby table, as seen in Figure 4-68.

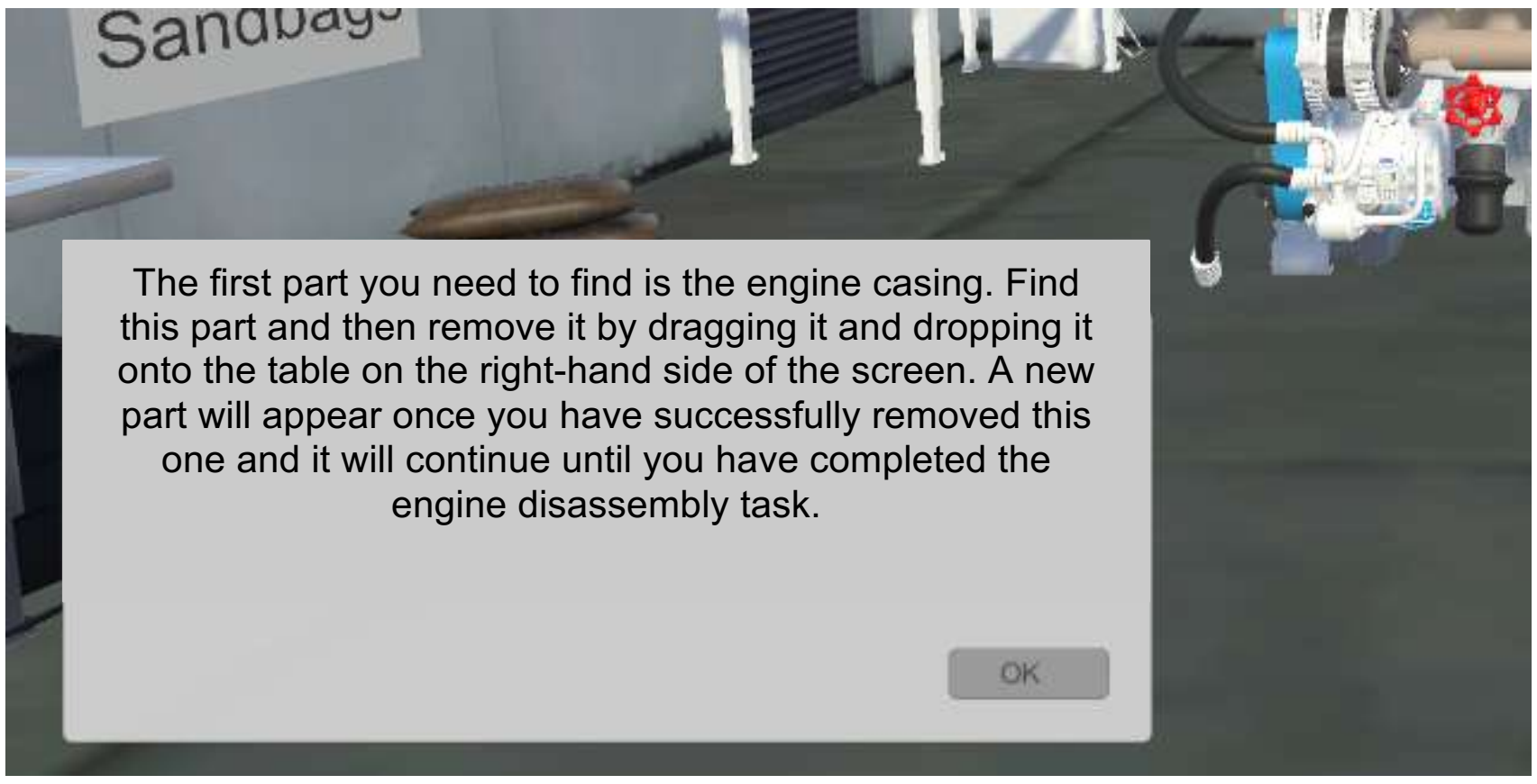

*Figure 4-67 Instructions on removing parts from the engine*

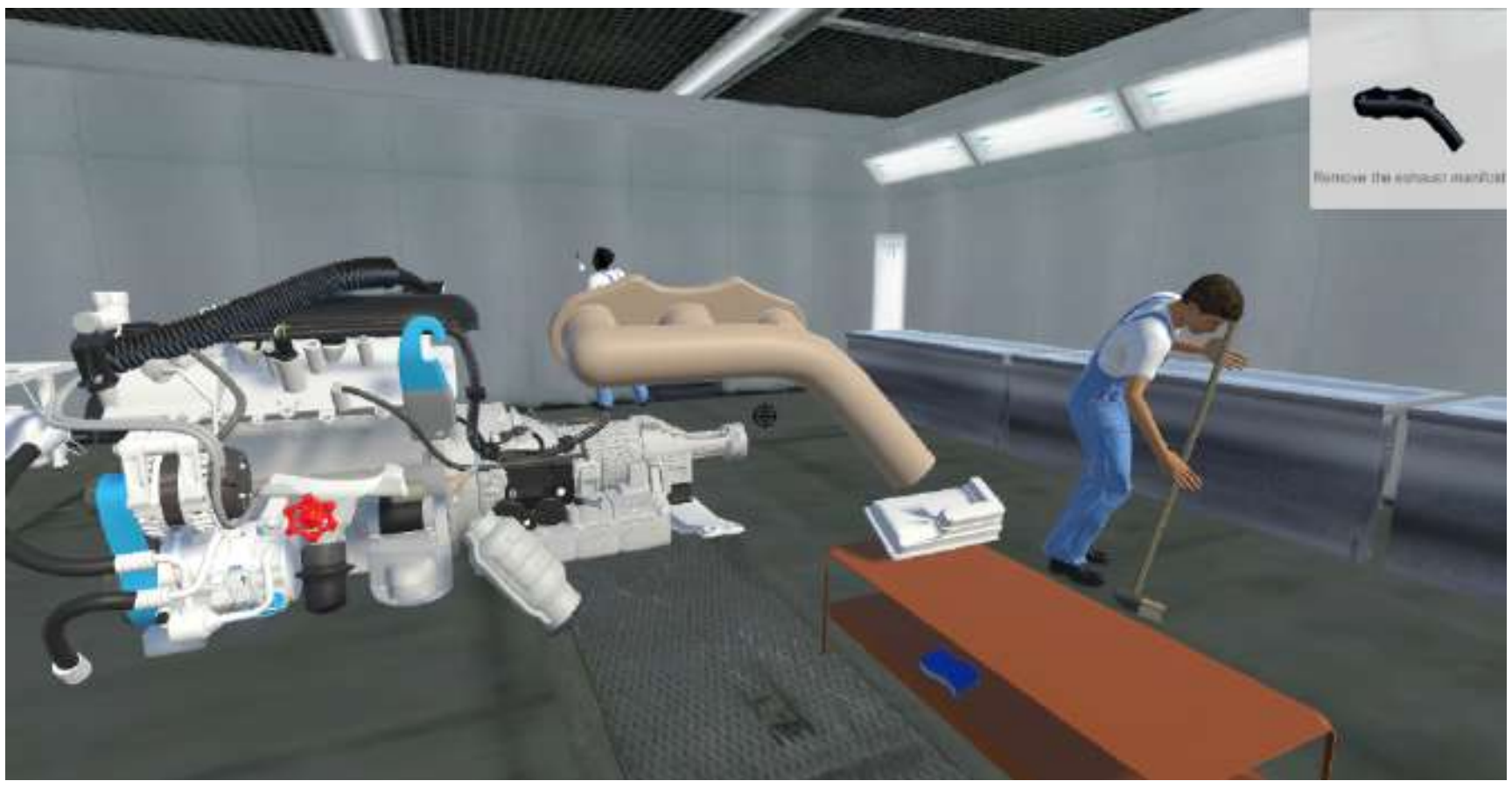

*Figure 4-68 Engine part being removed. Text reads: "Remove the exhaust manifold"*

Users were only able to complete the task successfully if they removed the components in the correct order. If the incorrect component was chosen and dragged on to the table, it would snap back to the engine as soon as the user would release the mouse button. On successful completion of the guided disassembly task, a pop-up message instructed the user to repeat the task again, but this time without any guidance (Figure 4-69):

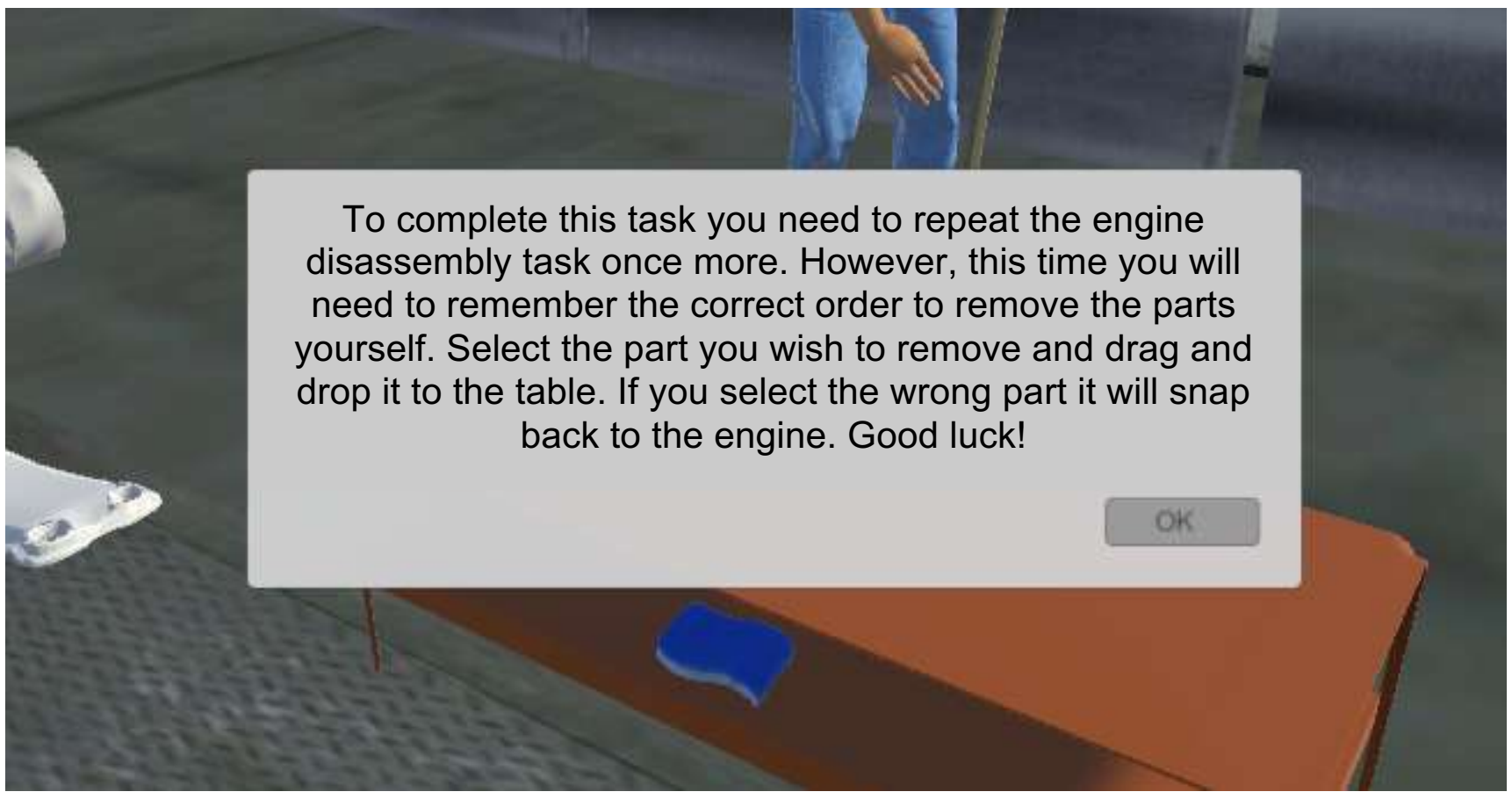

*Figure 4-69 Instructions on repeating the engine disassembly task*

A successful completion of this unguided disassembly task would automatically trigger the fuel leak scenario. The fuel leak simulation included animated visual representation of fuel dripping out from underneath the engine. In order to safely deal with the fuel leak, the user was able to interact with items placed around the workshop. The following interactive objects were placed around the VE:

- an alarm, which triggered an audible alarm on interaction with the user
- a valve, which spun towards the right as a visual representation of being 'shut off' on interaction with the user
- sandbags, which could be dragged and dropped over the fuel spillage to 'contain it'
- a sponge, which visibly removed the fuel animation on the floor of the workshop when it was moved over the image of the spilled fuel.

A visual pop-up was used to introduce the fuel leak scenario and provide the user with step-by-step guidance on what actions they needed to take and in what order to complete the task (Figure 4-70):

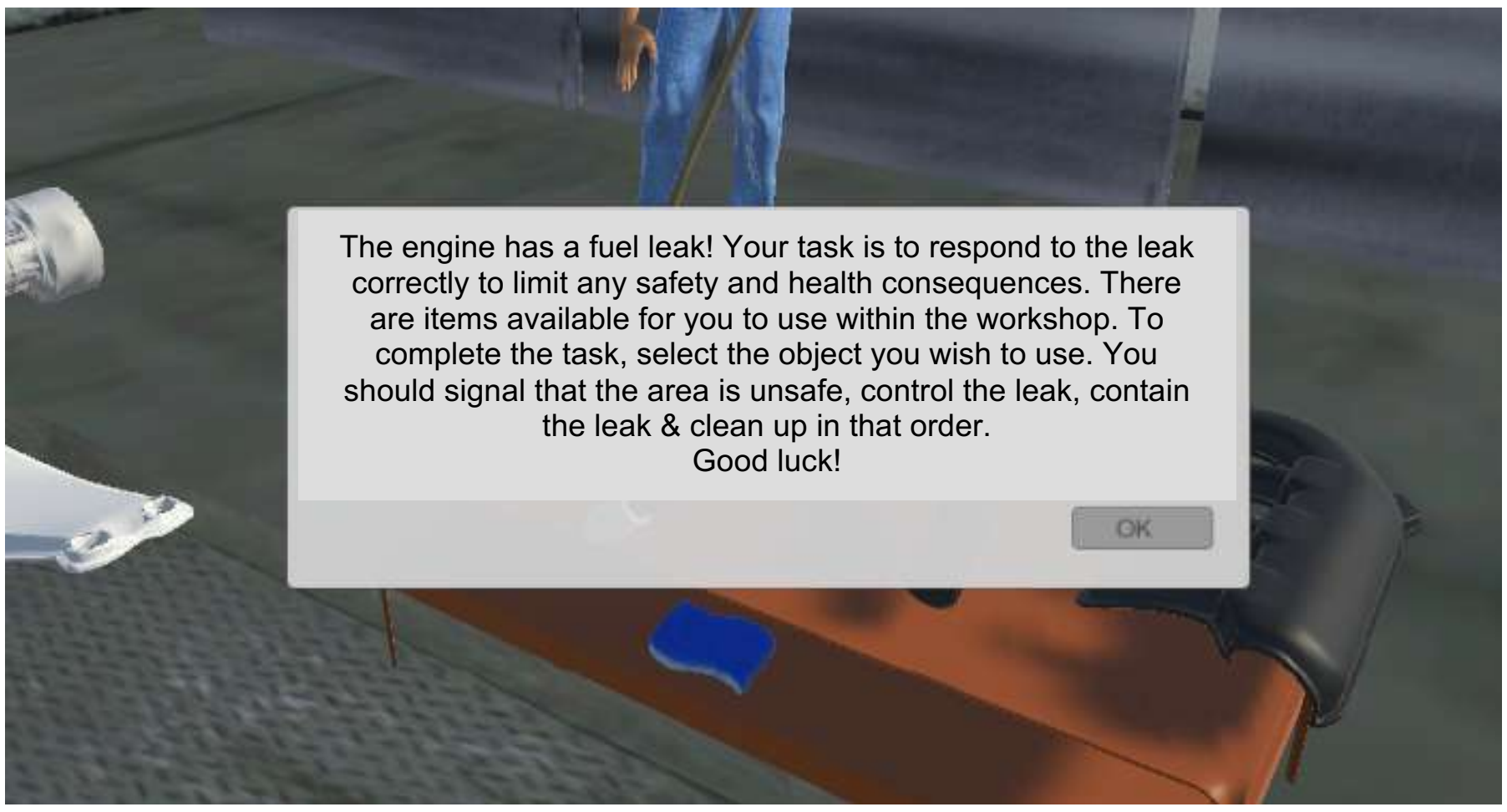

*Figure 4-70 Initial instruction for the fuel leak scenario*

To provide further assistance to the users, green arrows were used to guide them towards the location of the correct object within the VE. For example, Figure 4-71 shows an arrow guiding the user towards the alarm:

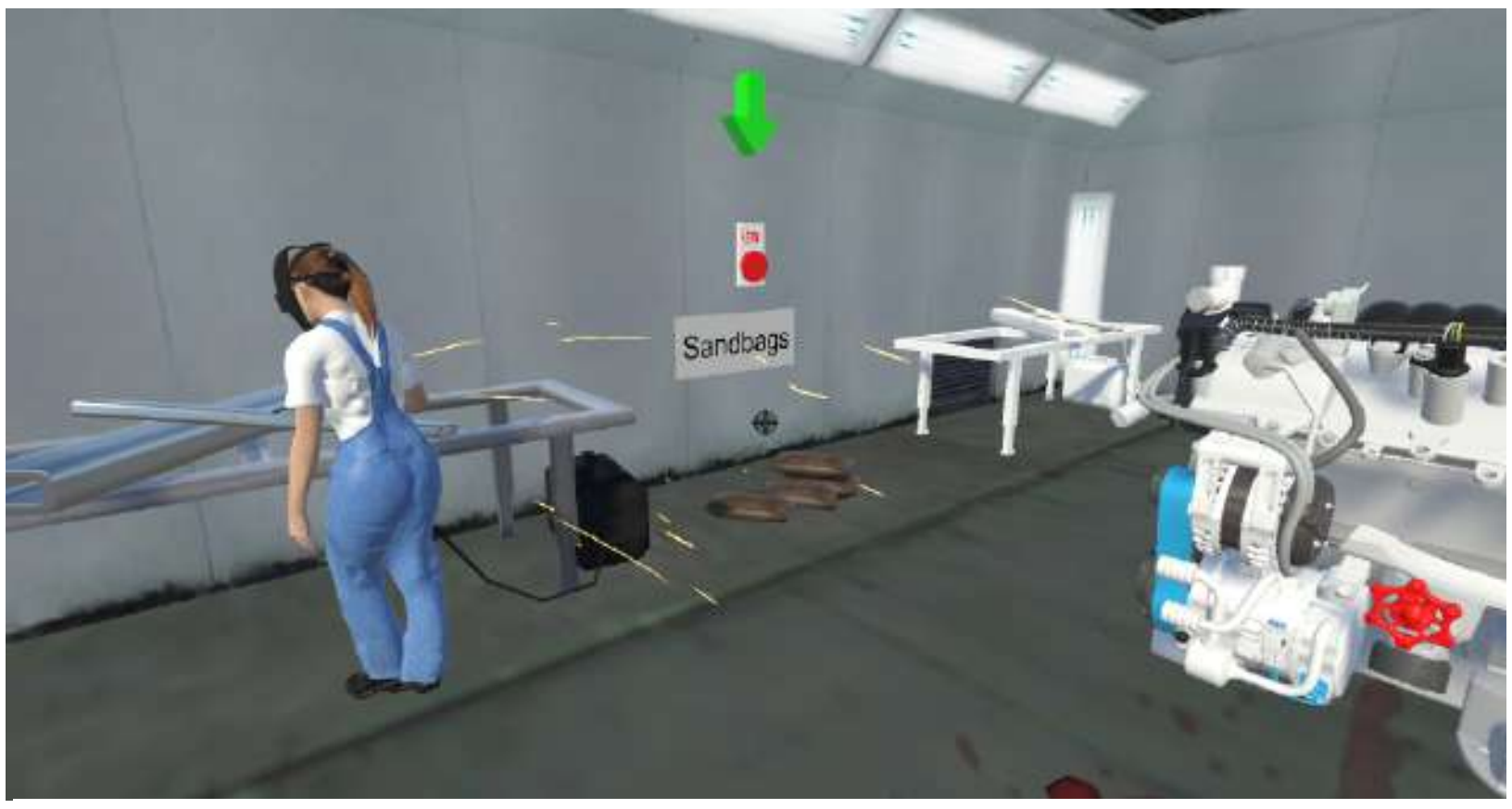

*Figure 4-71 Dynamic arrows guide the user to the correct safety procedure*

As with the engine disassembly task, users were required to repeat the fuel leak task without assistance. If users failed to complete the appropriate actions in the correct order required to effectively deal with the fuel leak, they experienced consequences within the simulation, such as a fire breaking out due to sparks from machinery igniting the spilled fuel (Figure 4-72).

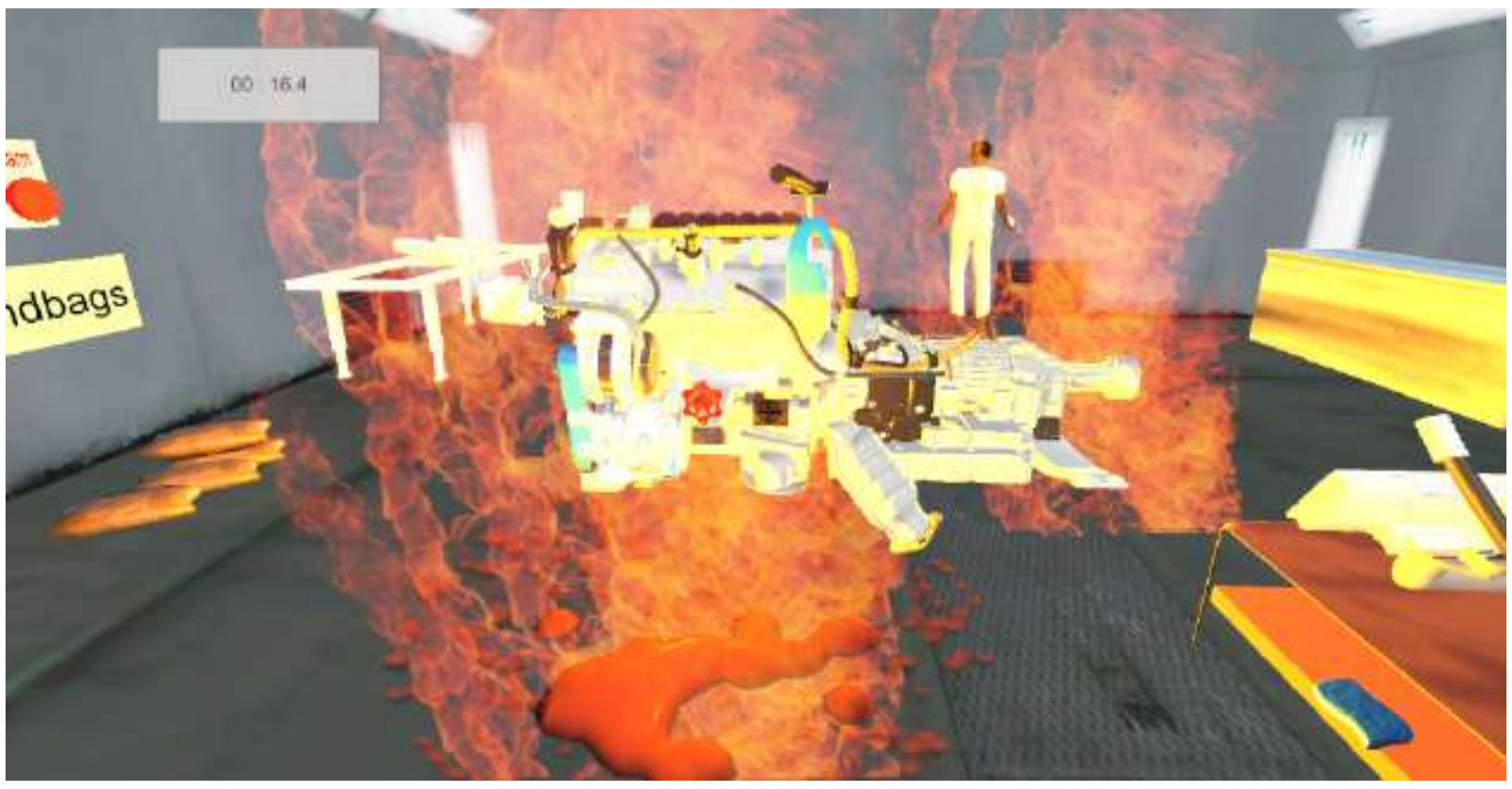

*Figure 4-72 The fire that would break out if the right sequence is not followed*

As with the fire scenario in the fire safety training use case, the fuel leak scenario would automatically reset if the user took the wrong action, allowing them to start over.

# 5 Study 1 – Behavioural validity

This study aimed to investigate how people behave in VE in a fire evacuation scenario and to see whether their perception and actions differ when thermal and olfactory feedback are added. The objective was to investigate whether the additional MS feedback encourages more realistic participant responses to the VE emergency scenario.

## 5.1 Design

The experiment used an HMD VE to run an emergency fire scenario in an office building. The display device was chosen for its highly immersive capabilities (19). A between-subjects design was employed so that participants were blind to the experimental scenario and task, as previous research reported that surprise might have a positive impact on immersion (72).

The independent variable in our main experiment was feedback modality. Participants completed the VE scenario in one of two conditions: AV and MS.

*Table 5-1 Experimental conditions*

| Condition | Description |
|---|---|
| AV (Audio-visual) | Audio and visual feedback only |
| MS (Multi-sensory) | Audio, visual, olfactory (smell) and thermal feedback |

Dependent variables were based on evacuation behaviour and measures identified from fire evacuation studies in the literature. These included (list not exhaustive):

- subjective perceptions of danger in the VE scenario (15)
- pre-evacuation, evacuation and total egress time in seconds (26)
- exit choice (6)
- changes in movement before and during the emergency (4,62).

The study received ethical approval by the Faculty of Engineering Ethics Committee at the University of Nottingham.

## 5.2 Participants

Fifty-two students, staff and associates at the University of Nottingham were recruited using poster advertisements on campus, online and circulated via university mailing lists. Participants were all over 18 and were screened to ensure they were not at increased risk of simulator sickness (SS), they had no previous traumatic experience in an emergency situation and they had normal or corrected-to-normal perception across all senses.

Participants were asked to provide their age, sex and gaming experience (self-reported on a five-point scale) using a demographics form. This information was used to balance participant assignment to the two conditions and counter potential confounding effects of individual differences reported to affect behaviour in previous studies (for example (6,10,15)).There were nine exclusions from the main data: eight due to early dropout following SS symptoms and one due to technology problems. A demographic summary is shown in Table 5-2:

*Table 5-2 Demographic summary for each condition*

| | AV | MS |
|---|---|---|
| **Sex (N)** | 10 female<br>13 male | 9 female<br>11 male |
| **Age ($\bar{x}$)** | 29.3 | 28.4 |
| **Gaming exp. ($\bar{x}$)** | 2.95 | 2.91 |

## 5.3 Equipment

### 5.3.1 Hardware

Configuration of hardware was implemented as concluded from user studies and Vive HMD and controllers were used.

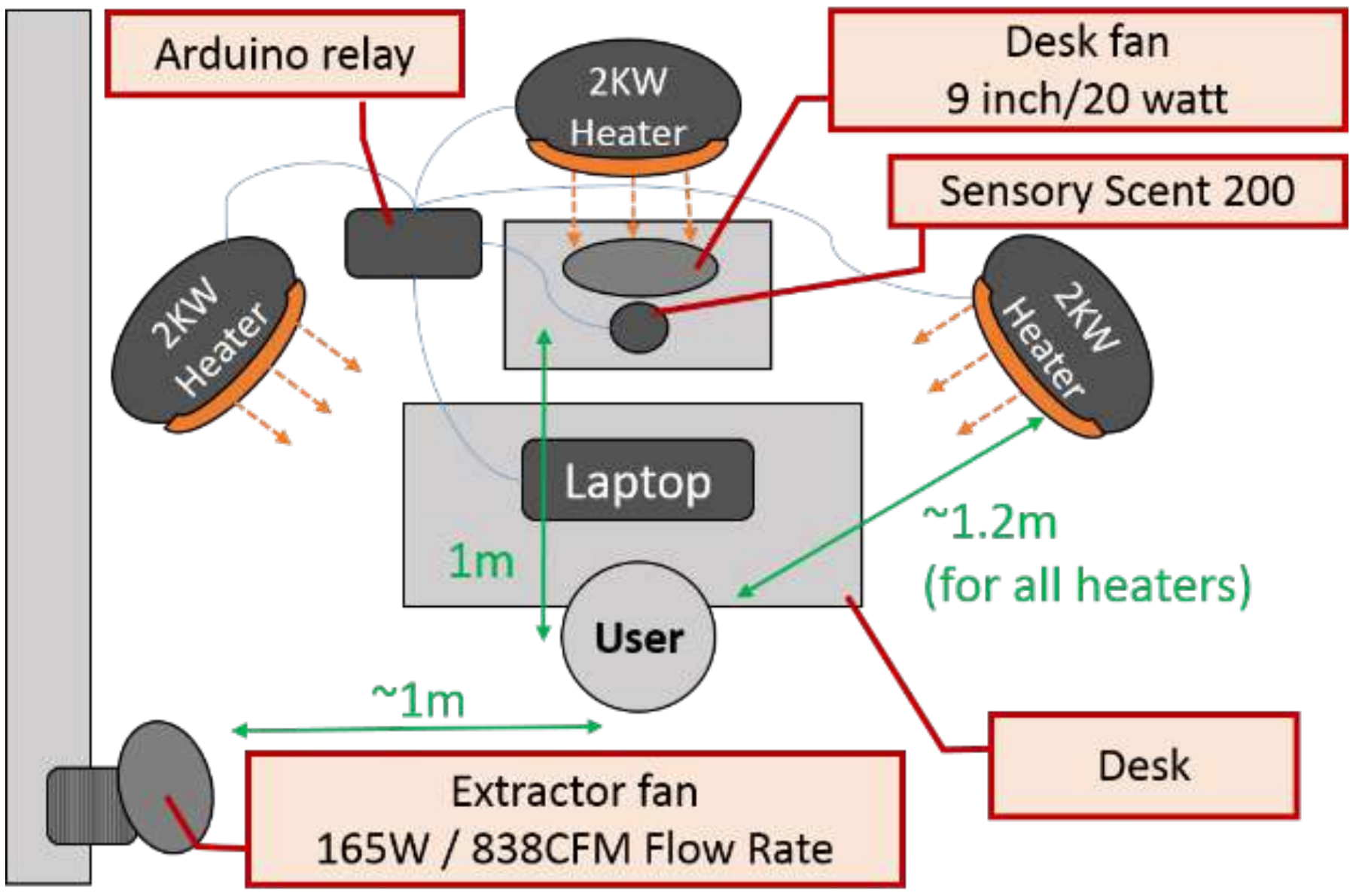

*Figure 5-1 Configuration of hardware*

#### 5.3.2 VE – Office building

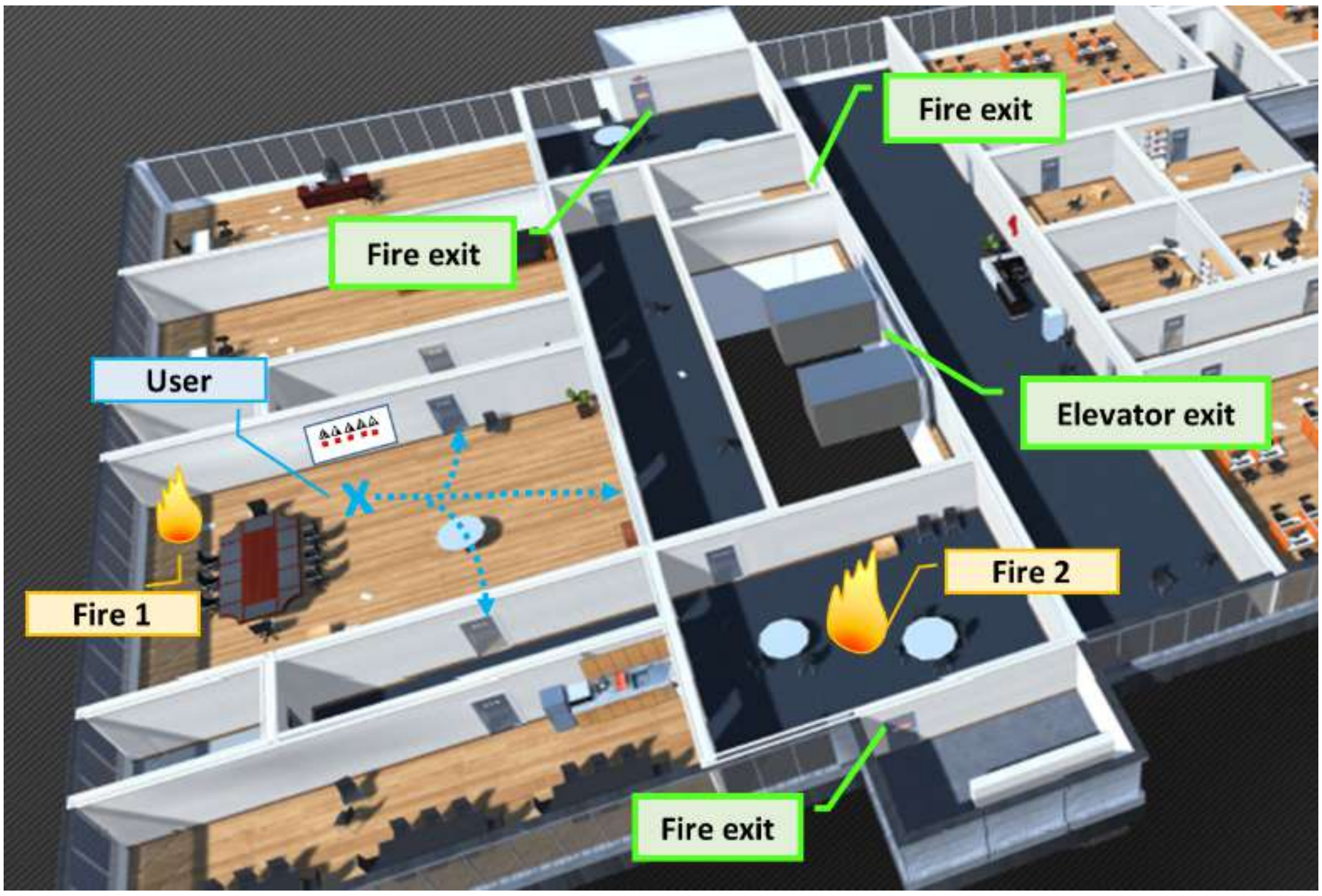

*Figure 5-2 Layout of office building interior, showing exits and fire placement, and possible exit route choices*

An office building was chosen for the VE as behavioural taxonomies in the literature exist for these structures (6,15). The structure was geo-typical as research has shown familiarity with the building can influence evacuation behaviour (28), so we did not want to recreate an actual building with which participants may have familiarity. There were four building exits: three marked fire exits (including the main stairwell) and the elevators. It also included other features that would be characteristic of office buildings, such as a reception on the ground floor and meeting and breakout rooms. Audio simulation of ambient office noise and static non-player characters (NPCs) were placed in peripheral areas to give the impression that the building was occupied but avoid encouraging interaction that was outside the scope of the design. The building layout was designed in conjunction with the placement of fires and design of the experimental scenario such that the visual simulation of the fire, e.g. smoke, could be seen through doors (Figure 5-3) and, on encountering the fire, participants could vary their proximity to it and find an alternative evacuation route if they desired (see Section 4.6). The building was designed to allow the exploration of behaviours such as the tendency to evacuate using familiar routes.

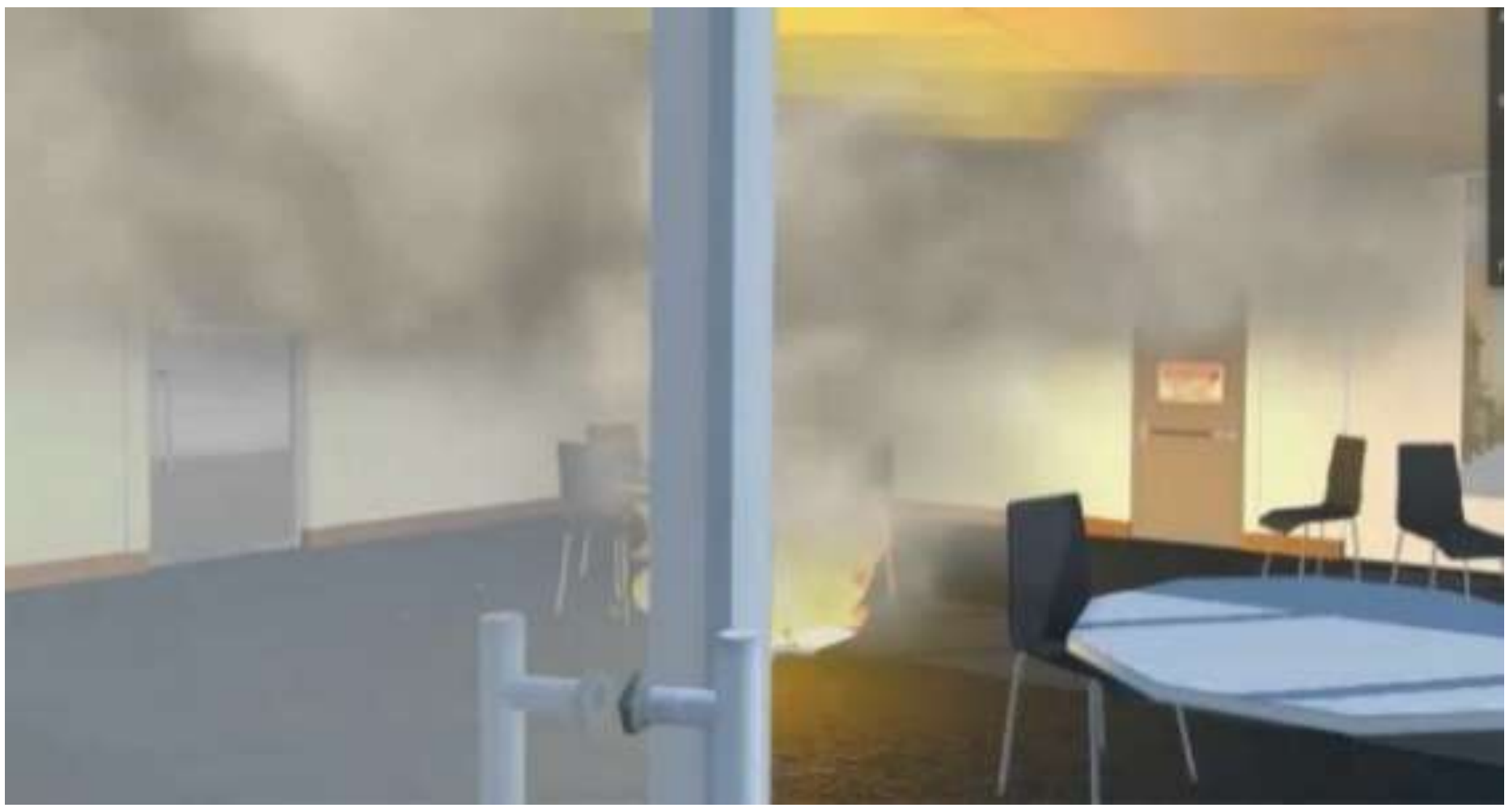

*Figure 5-3 Screenshot of fire 2 scene from the VE*

### 5.3.3 VE – Fire emergency scenario

Research suggests that if information from visual cues within the VE is contradicted by information from other sources, e.g. voice of the investigator, the immersive experience can be compromised (83,84). For this reason, the scenario included contextual purpose: attending the office building for a job interview and skills assessment and a set of related tasks, the instructions for which were all placed on signs in the VE (see Figure 5-4). The tasks included signing in, going to a specified meeting room and completing an IQ skills task ('Task 1' in Figure 5-4; also see Section 4.6).

A single emergency fire scenario was used for both conditions, providing consistency of input for behavioural response. The scenario was initiated manually by the investigator once the participant was engaged with Task 1 and consisted of an alarm and two separate fires, triggered concurrently. A small fire placed next to Task 1 was to assist with initiating evacuation as research shows that alarms are often disregarded without the addition of other fire-related cues (27,29,85). A second, larger fire blocked the familiar exit route. Fires were enhanced with visual effects, e.g. flames and smoke; audio effects, e.g. crackling; and (MS only) thermal (heat) and olfactory effects e.g. burnt wood scent effects. NPCs were removed from the building at the point the fire was triggered and placed outside the building with an NPC fire marshall. They were deliberately not involved in the fire scenario, as this has been found to cause distress in previous VE fire emergency research (e.g. (10)). The participant was not given explicit instructions to respond to the fire emergency, but the expectation was that they would begin the process of evacuation after the fire was triggered. If they did not begin evacuation after a several minutes they were prompted to respond to the scenario as if they were in the building in real life and this was noted in the data.

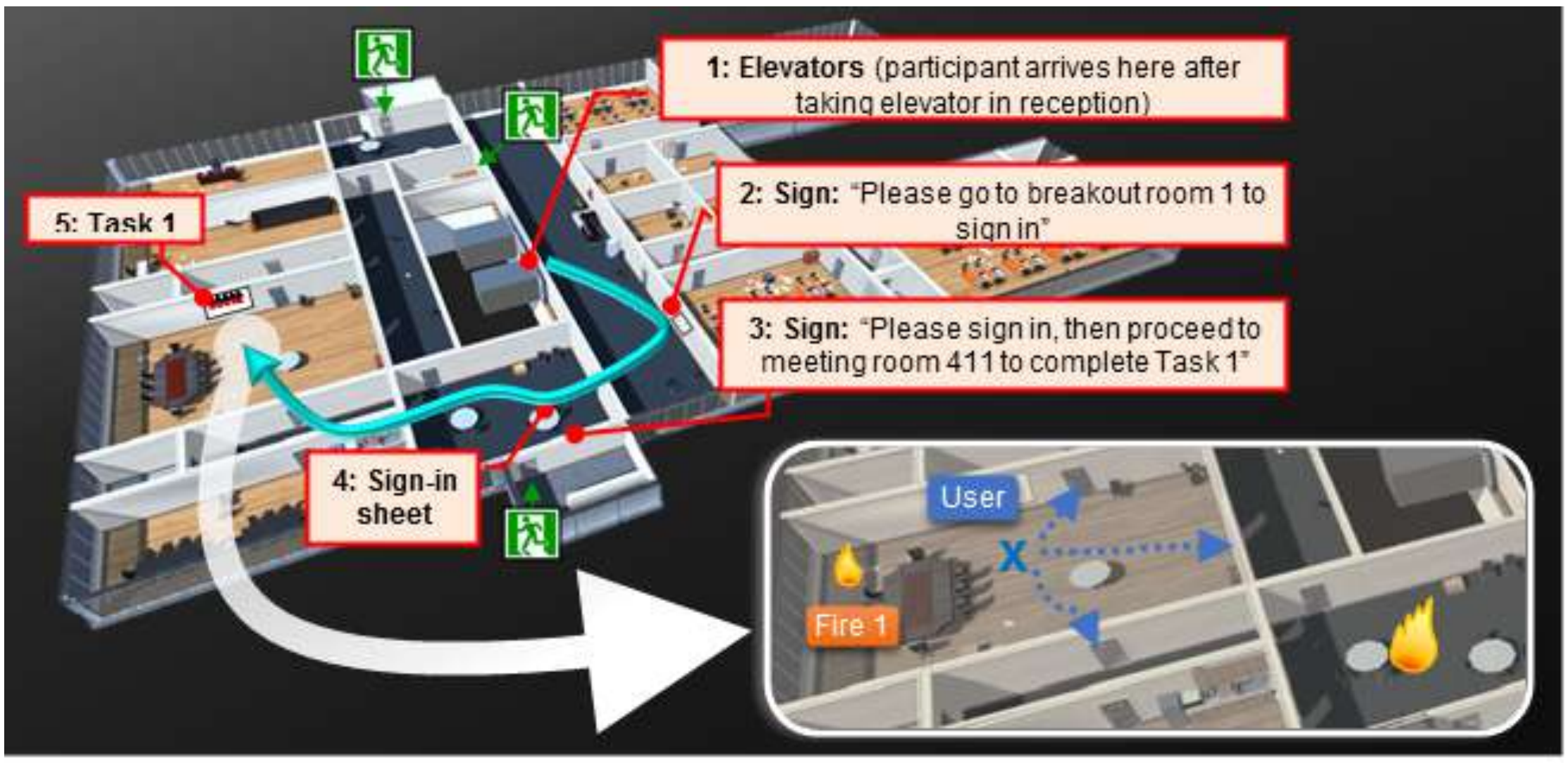

*Figure 5-4 Upper floor of the office VE showing route participants are directed along pre-fire, and potential routes post-fire (inset)*

## 5.4 Procedure

An experimental script was used for procedural consistency. After an introduction to the experiment, participants were asked to review the information sheet and sign a consent form. Participants completed a simulator sickness questionnaire (SSQ) (86) before and after the study to monitor their condition and collect information for future studies using this system. A rapid wellbeing test (87), which asked participants to verbally rate how they felt on a scale from 1 (I feel fine) to 10 (I feel awful, like I will vomit), was used as a quick and non-disruptive way to assess participants. It was used after the practice trial and whenever the investigator observed indications for concern. In cases where wellness scores worsened, sessions were terminated.

Participants were given a practice session moving around the outside of the VE building to become familiar with the VE and controllers. For the experimental session, participants were asked to imagine they had been invited for an interview and skills assessment for a new job. They were given free navigation of the VE but were asked to follow instructions within the building to complete a series of tasks. Participants were asked to pretend the investigator was not there unless they needed assistance. Participants were directed to an upper floor of the building via an elevator, to breakout room 1 (BO1) to sign in, and then, via a corridor, to 'room 411', which displayed an IQ skills task on a large whiteboard ('Task 1'); see Figure 5-4 for route. This initial task series served the purposes of familiarising participants with the building and providing a known entry route and distractor task.

Once the participant was engaged with Task 1, the investigator manually triggered the fire scenario, activating the alarm, a small fire next to the whiteboard and a more substantial fire in BO1 concurrently. In the MS condition, the fires were accompanied by the release of the fragrance, and the heater fins opened to increase temperature. Participants were not given explicit instruction on how to respond but were prompted to respond to fire cues if they remained engaged in the task for several minutes. Evacuation was considered complete after participants had exited the main building.

After evacuation, participants repeated the SSQ and were asked to complete the post-task subjective questionnaire (Appendix A). This was followed by a short unstructured interview, during which the investigator asked them to talk about their experience with the VE and explain their thoughts and actions, for example what cues they experienced that made them think there was a fire, their rationale for their evacuation route and exit choice and whether they noticed the larger fire and what their thought processes were on encountering it. Any unusual actions or behaviours were also explored, and participants were given the opportunity to comment on any aspect of the experimental session. Participants were fully debriefed on the purposes of the study and asked whether they were happy to have their data included in analysis following this disclosure.

## 5.5 Data collection

### 5.5.1 Movement data

An XSplitBroadcaster (88) was used for recording screen capture of the VE during experimental sessions and data logs showing positional data at each moment in the VE. This was used to observe, categorise and analyse movement patterns within the VE. A bespoke visualisation app was created to view the positional logs. This aided analysis through the creation of visual representation of movement that could be toggled to show trails of routes taken in and out of the building. Positions of elements present in the VE were represented so that movement patterns could be analysed in relation to key items and the position of the fires could be toggled to analyse evacuation movement patterns in relation to fire proximity. The app also allowed the collection of speed and distance data within the VE. It was beyond the scope of this project to analyse this data in full, but it was used to provide a quantitative spot check of the qualitative analysis of movement patterns relating to speed. A video camera was set up to capture participants' body language and physical actions and record post-task interviews.

### 5.5.2 Subjective questionnaire and interview

To gain a deeper understanding of participant action responses, a post-task questionnaire (Appendix A) was administered. The first three questions used a five-point ratings scale (1=very low; 5=very high) to assess participant perceptions of level of risk, stress/anxiety and time pressure. Participants were then asked to rate statements, e.g. "the building was occupied", according to how true they felt they reflected their experience while in the VE (1=not at all; 5=extremely). Some statements were included to assess the effectiveness of the VE design and, therefore, were excluded from behavioural analysis for the study. A short, unstructured questionnaire was also used to investigate participant experience with the VE and retrospective thought processes regarding actions and responses during the scenario.

## 5.6 Results

The results are presented with quantitative and qualitative analysis below. Qualitative analyses are based on review of screen capture of the VE, video capture of the user, positional data logs showing movement around the VE and participant comments in interviews.

### 5.6.1 Evacuation times and pre-evacuation behaviours

Based on the human behaviour in fire literature described in the literature review (Section 2.1), we investigated the effect of multi-sensory feedback on evacuation times. Evacuation times were analysed over three phases: pre-evacuation, evacuation and total egress (see Table 5-3). There were a number of ways the start of the evacuation phase could have been quantified, for example the point at which a participant moves away from the whiteboard. However, it was felt that leaving the room where the fire was initiated was the clearest quantifiable indicator to use.

*Table 5-3 Definitions used for quantifying times for each movement phase*

| Movement Phase | Description |
|---|---|
| **Pre-evacuation** | Point at which the fire is triggered to the point at which the user leaves room 411 (where the 'Task 1' is located) |
| **Evacuation** | Point at which the user leaves room 411 to the point at which the user goes through the chosen building exit |
| **Total egress** | Point at which the fire is triggered to the point at which the user goes through the chosen building exit |

Four participants who dropped out prematurely post fire scenario due to SS were included in the data set but had missing data entries for evacuation and total egress times. Three participants (two in MS, one AV) who did not respond to the fire cue and received a prompt from the investigator were excluded from the data set, based on this and confirmed outlier data from the statistical analysis. Pre-evacuation times for these participants were 80, 86 (MS) and 69 (AV) seconds.

*Table 5-4 Mean time and standard deviation (s) for each movement phase during evacuation*

| | Pre-evacuation | | Evacuation | | Total egress | |
|---|---|---|---|---|---|---|
| | **MS** | **AV** | **MS** | **AV** | **MS** | **AV** |
| **Mean time (s)** | 23 | 23 | 44 | 40 | 67 | 60 |
| **Standard deviation** | 14 | 10 | 24 | 21 | 35 | 24 |

To test for the assumptions of an independent *t*-test, a Shapiro-Wilk test was run to test for normality. Data for each condition in all movement phases were not normal. Therefore, the non-parametric Mann-Whitney U test was selected for analysis. Distributions of evacuation times for all three movement phases were not similar, as assessed by visual inspection. No statistically significant differences were found for any of the movement phases between MS and AV conditions.

*Table 5-5 Results of the Mann-Whitney U test using an exact sampling distribution for U*

| | Mean rank | | U score | Z score | Sig. (two tailed) | Retain/reject H0 |
|---|---|---|---|---|---|---|
| | MS | AV | | | | |
| **Pre-evacuation** | 19.63 | 21.16 | 183.50 | -0.395 | 0.70 | Retain |
| **Evacuation** | 19.62 | 17.50 | 142.50 | -0.603 | 0.56 | Retain |
| **Total egress** | 19.32 | 17.76 | 147.50 | -0.444 | 0.67 | Retain |

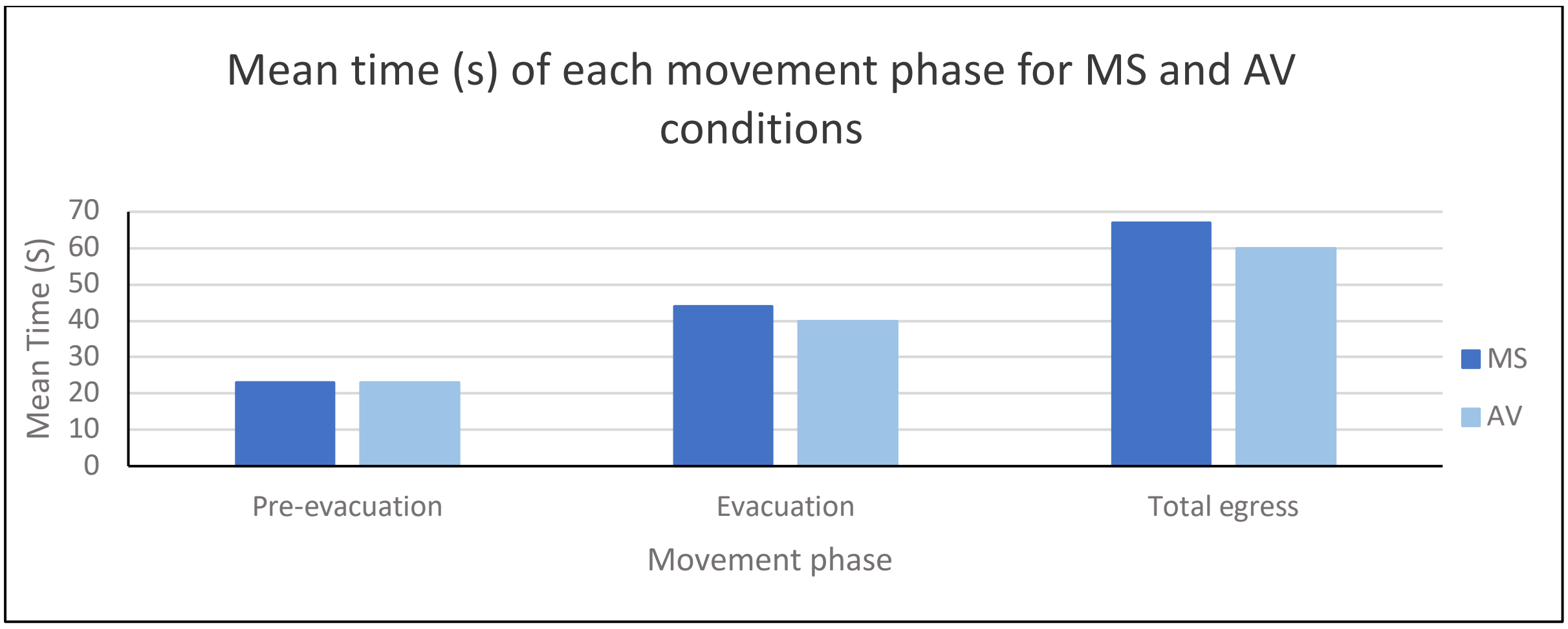

*Figure 5-5 Mean time (s) for each movement phase for each condition*

The time taken to evacuate in a fire situation is an important factor as delayed evacuation increases risks to evacuees. Results of pre-evacuation, evacuation and total egress movement phases were similar in both conditions, with slightly higher times for the MS condition. This finding contradicts the literature which suggests that additional cues would prompt faster initiation of evacuation (1,21). However, analysis of behaviour showed that times were affected by several factors that make interpretation complex.

Participants spent anything from 10 to 57 seconds before exiting the room with the first fire. The ways in which participants responded were variable. To understand the delays to evacuation, and in particular the effects of heat and smell, salient comments were extracted from the interviews and qualitative data. These are shown below:

Pre-evacuation actions noted from the qualitative data included continuing with the task, warning others and investigating or attempting to fight the fire. One notable participant from the MS condition walked towards the fire and physically shook the Vive controllers at it, in an attempt to put it out:

*"At the beginning I wanted to run away from the room and find somebody else to warn"*

- **MS participant, interview**

*“I was trying to find a fire extinguisher”*

- **AV participant, interview**

Participants reported being unsure about the meaning associated with the alarm, but reported stimuli such as visual effects, smell and heat or encountering the larger fire influenced their decision to evacuate or the route they took:

*“My very first thought was it's quite a small fire maybe I could try and put it out … but the heat and then when the alarm went off I thought oh I should probably get out.”*

- **MS participant, interview**

*“I wanted to check to see if it [fire] was a small area or a big area, so that's why I looked around a little bit and when I realised actually it's a lot of places on fire then I decided I should go.”*

- **AV participant, interview**

### 5.6.2 Analysis of movement

Movement behaviour was analysed from screen capture and data logs showing participants' movements through the VE before and after the fire was triggered. Qualitative statements from participant interviews were also used to support and elaborate on key findings from the movement data. Four participants were excluded as the data sets were incomplete due to early dropout from SS; therefore N=39.

All analysis of movement data failed the assumption of the Chi-square test, that the expected value in each cell is greater than 5, calculated using the following: row total*column total/overall total (89). Therefore, movement data was analysed descriptively, shown in Sections 5.6.2.1 and 5.6.2.2.

#### 5.6.2.1 Exit choices and retracing steps

Participants were deliberately given a known entry route by starting them outside the building and giving them a series of signs and tasks through the corridors and rooms so that this could be compared to their chosen exit route. Each building exit was defined as shown in Figure 5-6. Retracing steps was defined as the participant attempting to evacuate the building the way they entered (shown as red-dotted line).

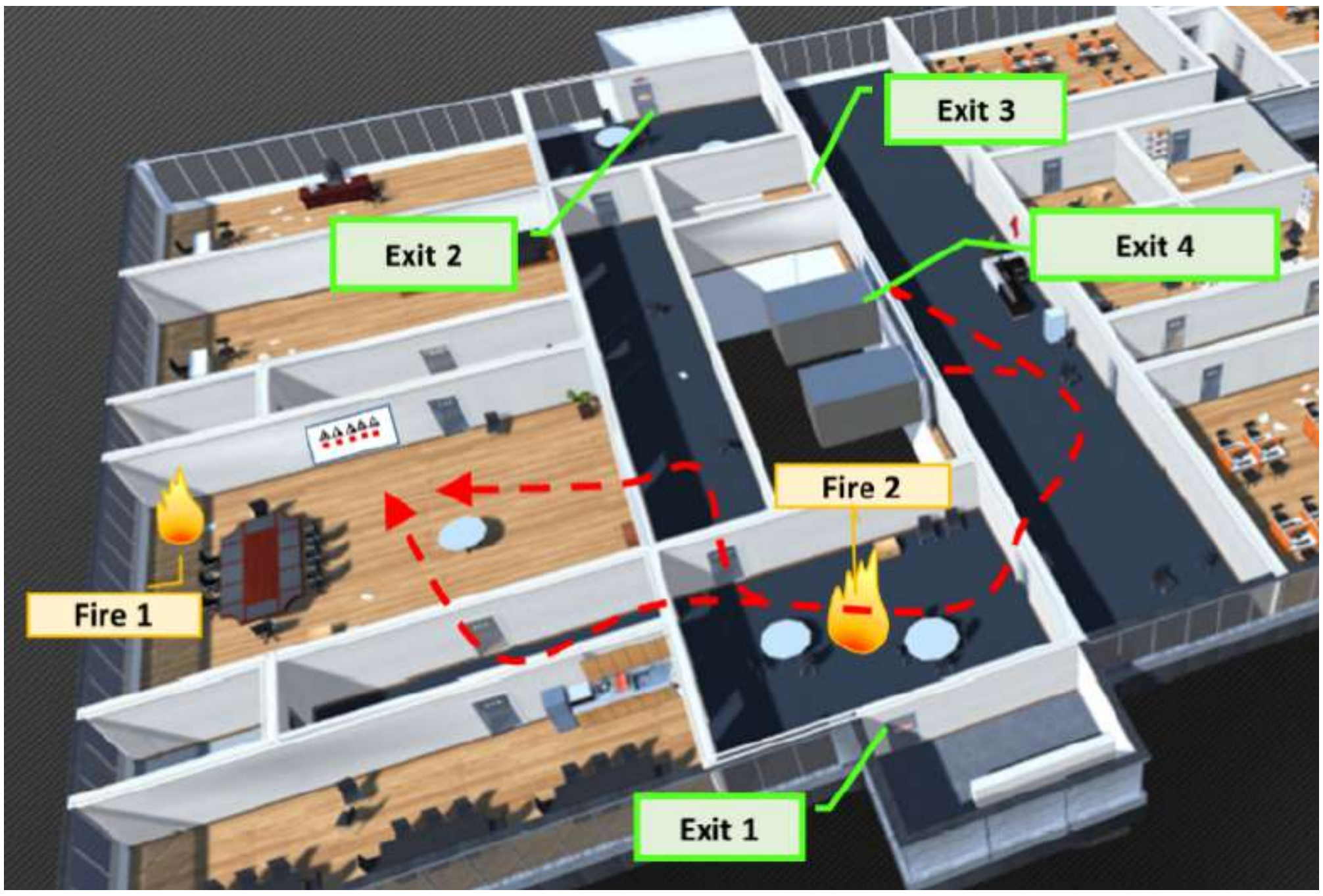

*Figure 5-6 Building exits and routes into the building plus placement of fires*

*Table 5-6 Frequency (count and %) of exit choice in each condition*

| | MS | | AV | | Totals (MS&AV) | |
|---|---|---|---|---|---|---|
| **Exit** | **Count** | **% (1 d.p.)** | **Count** | **% (1 d.p.)** | **Count** | **%** |
| **1** | 3 | 15.8 | 1 | 5.0 | 4 | 10.3 |
| **2** | 13 | 68.4 | 13 | 65.0 | 26 | 66.7 |
| **3** | 2 | 10.5 | 6 | 30.0 | 8 | 20.5 |
| **4** | 1 | 5.3 | 0 | 0.0 | 1 | 2.6 |
| **N** | 19 | | 20 | | 39 | |

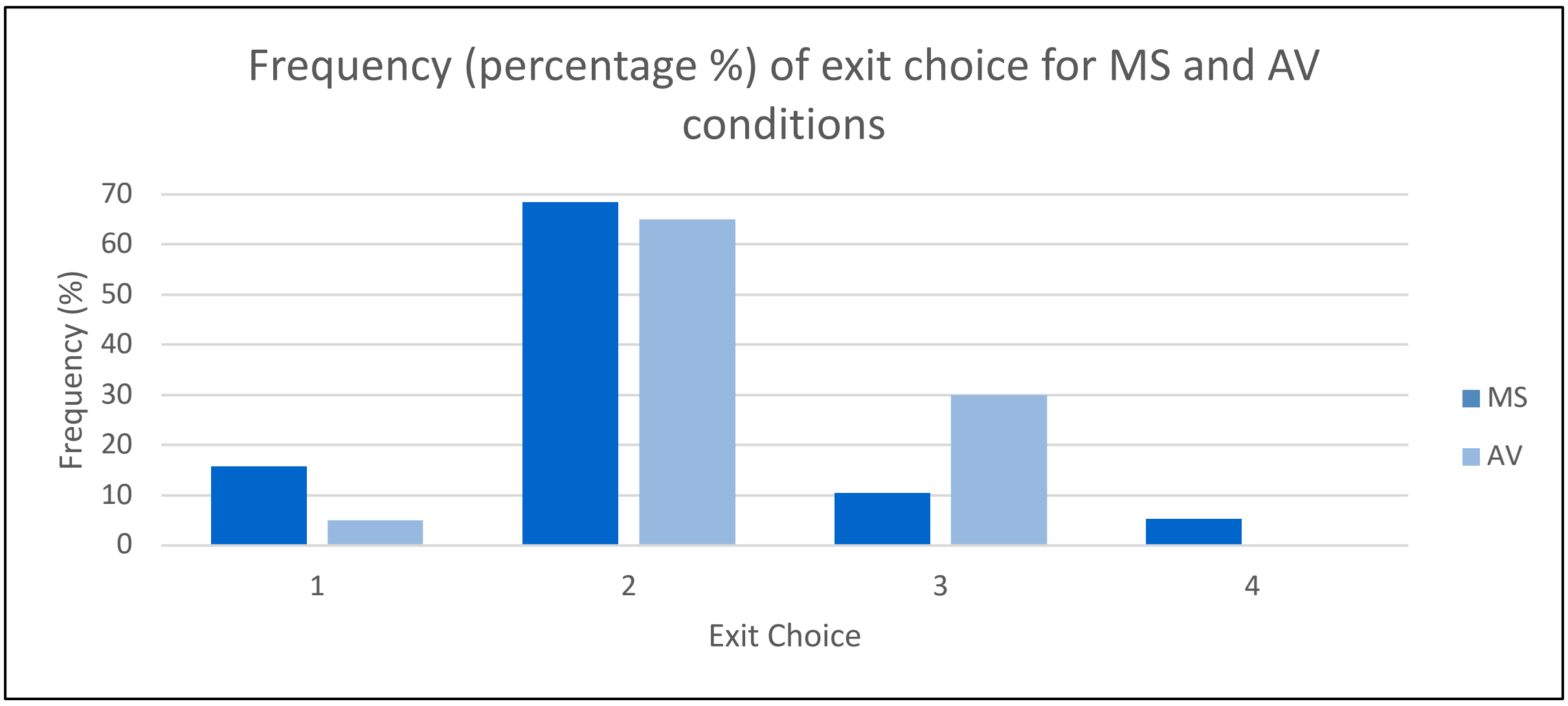

*Figure 5-7. Frequency (%) of exit choice for MS and AV conditions*

The majority (78%) of participants attempted to retrace their entry route after the fire trigger, which demonstrates an agreement with real-world behaviour. This was similar in both conditions (80% in MS, 76% in AV). However, more participants in the MS condition who retraced their steps altered their exit routes on encountering the large fire contained in Room BO1.

There were a number of qualitative statements during the interviews regarding participant exit choices and retracing steps, for example:

*"…the way I'd come in when I came out of the lifts I had registered fire exits to my left. However, the route I'd taken was clearly now blocked and not knowing the building you lose your total sense of direction."*

- **MS participant, interview**

*"I actually chose the other door and not the door I came through and after I'd done it I was like oh maybe I shouldn't have done that and like gone back the route that I knew because I'd seen that there was a fire exit right at the end of the first corridor outside the lifts, so I knew that was there already"*

- **AV participant, interview**

#### *5.6.2.2 Proximity to fire*

The second, larger fire was deliberately positioned so that it obstructed the known route participants were guided through when they entered the building, so that if participants retraced their steps during evacuation their response to the fire could be observed and analysed. Participant evacuation movement was analysed according to proximity to the large fire placed in Breakout Area 1 (BO1) and behaviour was grouped into six categories as shown in Figure 5-8:

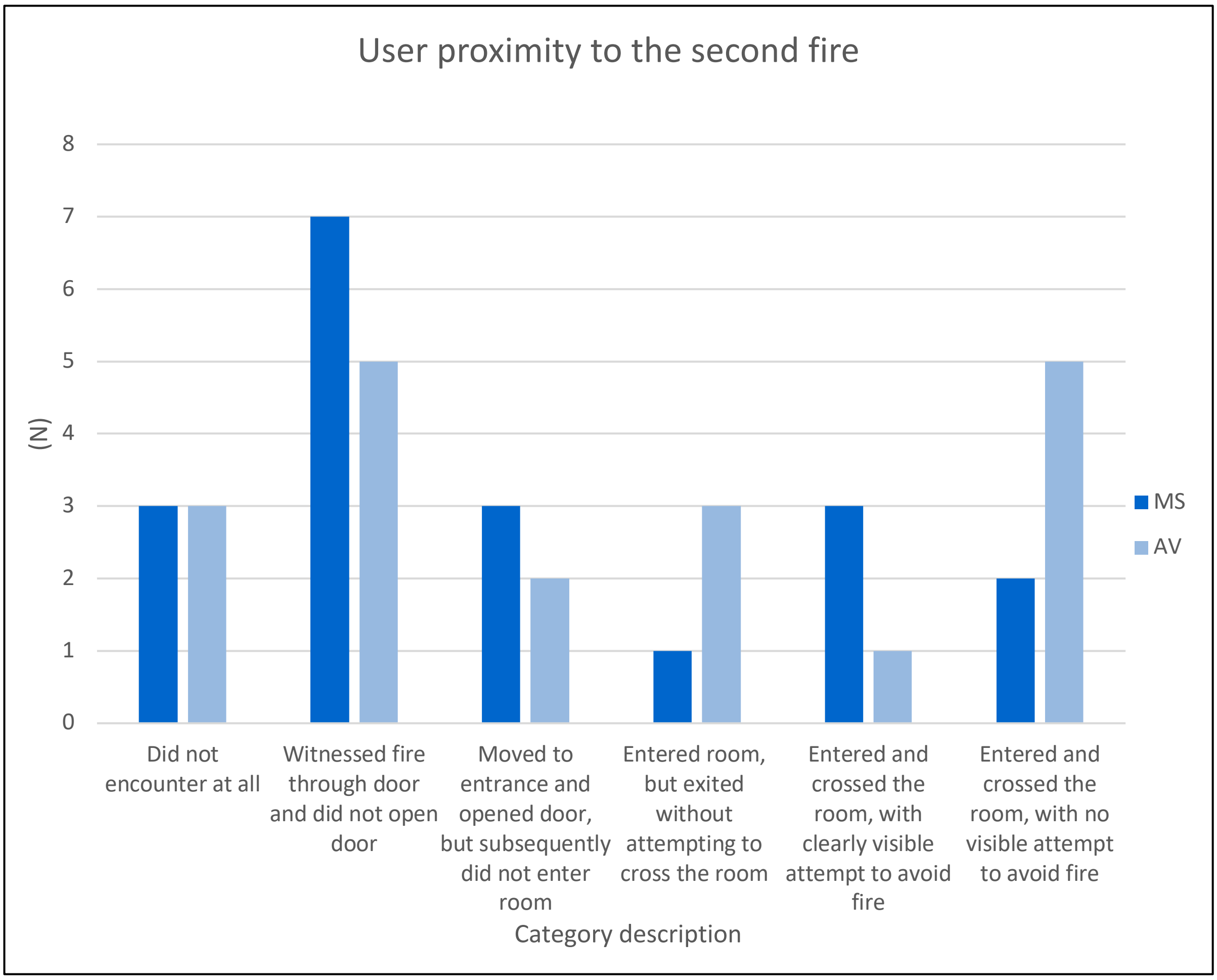

*Figure 5-8. Categorisation of user proximity to the second fire*

The majority (71%) of participants who encountered the large fire altered their evacuation route to avoid exiting via BO1. Of this population, 81% avoided entering this room altogether, having seen the fire through the transparent door. The remainder entered but exited again, finding an alternative evacuation route. Eleven participants chose an evacuation route that crossed BO1, exiting either via the main entrance (Exit 3 – see Figure 5-6), which retraced their full entry route, or using the exit located in BO1 (Exit 1). However, 86% of the AV participants who crossed the room made no visible attempt to avoid the fire, walking right next to or through the fire with minimal hesitation, whereas 60% of MS participants who crossed the room showed signs of hesitation, such as inspecting the room before entering, and made attempts to walk around the fire, for example edging around the room. Examples of participant movement across BO1 is shown in Figure 5-9 below with the red line depicting participant movement across the room. The left image shows an MS participant waiting and peering into the room at the door and skirting around the wall in order to reach the exit on the other side. The right image shows an AV participant walking directly into and across the room with smooth movement and minimal deviation from the chosen path.

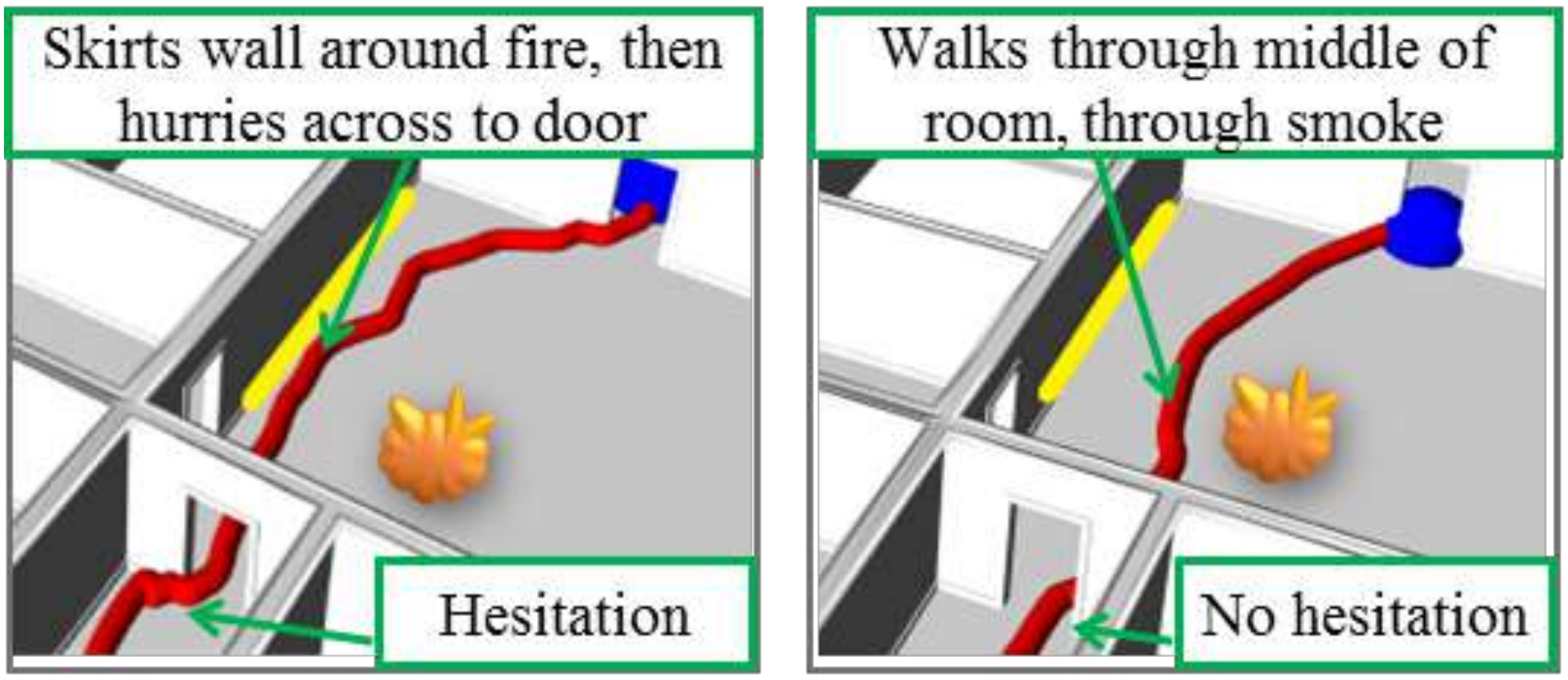

*Figure 5-9 Example of Category 5 (Left, MS participant) and 6 (Right, AV participant)*

There were a number of qualitative statements during the interviews that expanded on participant thoughts regarding their proximity to fire during evacuation:

**Category 3:** ***moved to entrance and opened door, but subsequently did not enter room***

*"I looked at it and I thought well if the room's there and the fire's there I could probably get away with it, but I thought nah, there is a lot of smoke, I probably shouldn't. If it was a game I'd have probably risked it."*

- **MS participant, interview**

**Category 4:** ***entered room, but exited without attempting to cross the room***

*"I think if the heat was there it would have made a difference because when I was walking through the fire, I saw the fire, but I didn't feel the fire. I didn't feel like I was in danger, I felt like the avatar was in danger, but there was no real urgency, so that was one of the reasons I felt I could run through it."*

- **AV participant, interview**

**Category 5:** ***entered and crossed the room, with clearly visible attempt to avoid fire***

*"I did have a sense of worry moving through the room with the fire, I was like is the floor going to collapse or something, yeah I was a bit unsure of what would happen next."*

- **MS participant, interview**

**Category 6:** ***entered and cross the room, with no visible attempt to avoid fire***

*"I felt like I was in a game, so for example there were some obstacles that I should have probably gone around, but I just went through."*

- **AV participant, interview**

## 5.7 Subjective questionnaire analysis

Subjective results were based on participant responses to a subjective questionnaire completed post-experiment. The Shapiro-Wilk test was run to check for normality. The results showed that the data for each condition were not normal. Therefore, the differences between the ordinal ratings of selected questions and statements were investigated using the rank-based non-parametric Mann-Whitney U test, results of which are shown in Table 5-7.

*Table 5-7 Descriptive results for the subjective responses; highlighted rows indicate statistically significant difference*

| | Mean rank | | Median rank | | U score | z score | p value (one tailed) |
|---|---|---|---|---|---|---|---|
| **Question** | AV | MS | AV | MS | | | $p \leq 0.05$ |
| **Max perceived level of risk** | 20.78 | 23.4 | 3 | 3 | 478 | -0.704 | 0.25 |
| **Level of stress/anxiety** | 20.54 | 23.68 | 3 | 3 | 472.5 | -0.849 | 0.20 |
| **Level of time pressure** | 19.2 | 25.23 | 3 | 4 | 441.5 | -1.646 | 0.05 |
| **Need to exit the building asap** | 20.26 | 24 | 4 | 5 | 466 | -1.053 | 0.15 |
| **The building is on fire** | 16.7 | 28.1 | 4 | 5 | 384 | -3.112 | 0.001 |
| **I need to take the nearest exit** | 19.35 | 25.1 | 5 | 5 | 445 | -1.837 | 0.04 |

There were a number of qualitative statements from MS participants who reported feeling a sense of urgency due to a combination of sensory cues they were perceiving:

*"I felt the heat and smelt the smell of burning from the fire…I think having all the senses triggered does create more of a panic state, like you did really feel that [time] pressure, whereas if it was just a video game…it doesn't feel very real, so there's no like actual pressure."*

- **MS participant, interview**

Some AV participants explicitly reported not feeling a sense of urgency and a reduced sense of reality regarding the fire itself, which they reported as having influenced the way they interacted with the fire:

*"I didn't think the fire looked really realistic. It didn't make me feel like I needed to leave quickly or under pressure … it's like a game, what happens when I walk into the trap … oh there's a fire, what happens if I walk in, I will just test it and see, but I wouldn't do that in a real fire."*

- **AV participant, interview**

MS participants reported feeling the need to move away from the fire after experiencing olfactory simulation:

*"I remember having a bit of a blast of a smell and then from that realising that I would need to leave because it was coming very close to me very quickly and I remember thinking I just need to turn around and get out this door."*

- **MS participant, during the interview**

AV participants reported using fire safety guidelines for decision making:

*[You exited immediately when you saw the fire, what was the reason for that?] "That is what you're supposed to do… I wasn't tempted to use the lift as we learnt that you must take the stairs."*

- **AV participant, interview**

## 5.8 Behavioural analysis

In addition to the a priori measures above, exploratory analysis was carried out to see what could be interpreted from behaviour in the VE. Observations were based on a combination of screen capture, data logs, video camera footage and interview responses. The basis for the behavioural analysis coding was taken from studies of behaviours from the real world: taxonomy for multiple occupancy and hospital fires in Canter et al. (17) (1980), and frequency of emerging acts identified by human behaviour in fire studies conducted by Lawson (15, p. 157), which coded acts based on the same taxonomy. The majority of identified behaviours from the VE are evident in literature on real fire situations and referenced in the taxonomy of acts. However, emergent behaviours* from the present study not accounted for under this coding, but which are useful to understanding behaviour, were added to the list of action categories:

*Table 5-8 List of observed actions, codes and explanations. Used for coding frequency (%) of participants who have been observed conducting the actions emerging from the study (action observed Y/N)*

| Action code | Action category | Explanation/notes |
|---|---|---|
| **10a** | Evasive | Stepping back from fire, changing path to avoid fire. Excluded participants that required a prompt to 'behave as they would in the real world' |
| **3c** | Note fire development/ worsening of immediate situation | Comment or response to second larger fire in the building relating to an increase in fire development i.e. the fire getting worse |
| **6a** | Seek information and investigate | Searching behaviour i.e. looking for fire extinguisher/looking around the room before evacuating |
| **1n*** | Increase in speed/amount of movement/head movement | Increase in urgency/erratic nature of movement following fire trigger |
| **24a** | Leave immediate area | Moving away from immediate location of fire when fire is initially triggered |
| **32a** | Look for/at fire signs (signage) | Explicit mention of using signage either commenting during task or during post-task interview |
| **3a** | Perception of stimulus (unambiguous) | Specific mention of a stimulus in relation to evacuation |
| **17a** | Note persistence of stimulus | Evidence of perceived persistence of fire-related stimulus in post-task interview |
| **2a** | Perception of stimulus (ambiguous) | Evidence of perceived ambiguity of stimulus during post-task interview |
| **39a** | Overshoot | Walking past a door, exit or other object, colliding with a wall |
| **33a** | Wait | Hesitation or pause |
| **19a** | Alter plan (self-initiated) | Change the direction of walking after experiencing the stimulus |
| **2n*** | Inspect escape route | For example peering around or through doors before entering |
| **22a** | Note nothing unusual/stay calm | Continuing to act calmly as before the fire scenario |
| **35a** | Get lost | Comment on getting lost during post-task interview |
| **3n*** | Skirts walls | Walk close to the walls within the virtual building post fire |
| **16c** | Experience negative feelings | Commenting on, for example, fear, anxiety, stress, panic |
| **27a** | Discuss/debate/ask | For example asking questions about whether they should leave or making statements about what they plan to do |
| **18a** | Receive verbal information | Verbal interaction with investigator during the experimental session |
| **5b** | Disregard/ignore stimulus/continue prior activity | Continue an activity unrelated to fire; walk through fire |
| **34a** | Problem with VE | Issues getting through doors; no perception of specific thermal or olfactory stimulus |
| **4n*** | Startled reaction | Verbal indications (e.g. "Oh!"); gasps; physically jumping |
| **5n*** | Attempt interaction | Attempting to physically interact with the environment or objects in it e.g. waving controllers at the fire |

***'n' used to depict new actions emerging from this study**

The raw data was analysed and coded in accordance with the action category descriptions. Four data entries were excluded from the analysis as they had dropped out of the experiment early due to SS.

*Table 5-9 Frequency of participant action observed = 'Yes' (%) for MS and AV conditions in each action category (N=39)*

| Action code | Action category | Freq (%) MS | Freq % AV |
|---|---|---|---|
| **10a** | Evasive | 90 | 90 |
| **3c** | Note fire development/worsening of immediate situation | 84 | 85 |
| **6a** | Seek information and investigate | 74 | 70 |
| **1n*** | Increase in speed/amount of movement/head movement | 84 | 60 |
| **24a** | Leave immediate area | 68 | 70 |
| **32a** | Look for/at fire signs (signage) | 68 | 75 |
| **3a** | Perception of stimulus (unambiguous) | 84 | 30 |
| **17a** | Note persistence of stimulus | 74 | 55 |
| **2a** | Perception of stimulus (ambiguous) | 63 | 60 |
| **39a** | Overshoot | 68 | 55 |
| **33a** | Wait | 68 | 55 |
| **19a** | Alter plan (self-initiated) | 53 | 55 |
| **2n*** | Inspect escape route | 63 | 45 |
| **22a** | Note nothing unusual/stay calm | 21 | 55 |
| **35a** | Get lost | 58 | 25 |
| **3n*** | Skirts walls | 53 | 20 |
| **16c** | Experience negative feelings | 63 | 5 |
| **27a** | Discuss/debate/ask | 32 | 30 |
| **18a** | Receive verbal information | 32 | 25 |
| **5b** | Disregard/ignore stimulus/continue prior activity | 16 | 35 |
| **34a** | Problem with VE | 26 | 15 |
| **4n*** | Startled reaction | 37 | 5 |
| **5n*** | Attempt interaction | 16 | 0 |
| | **Total no. of pps** | **19** | **20** |

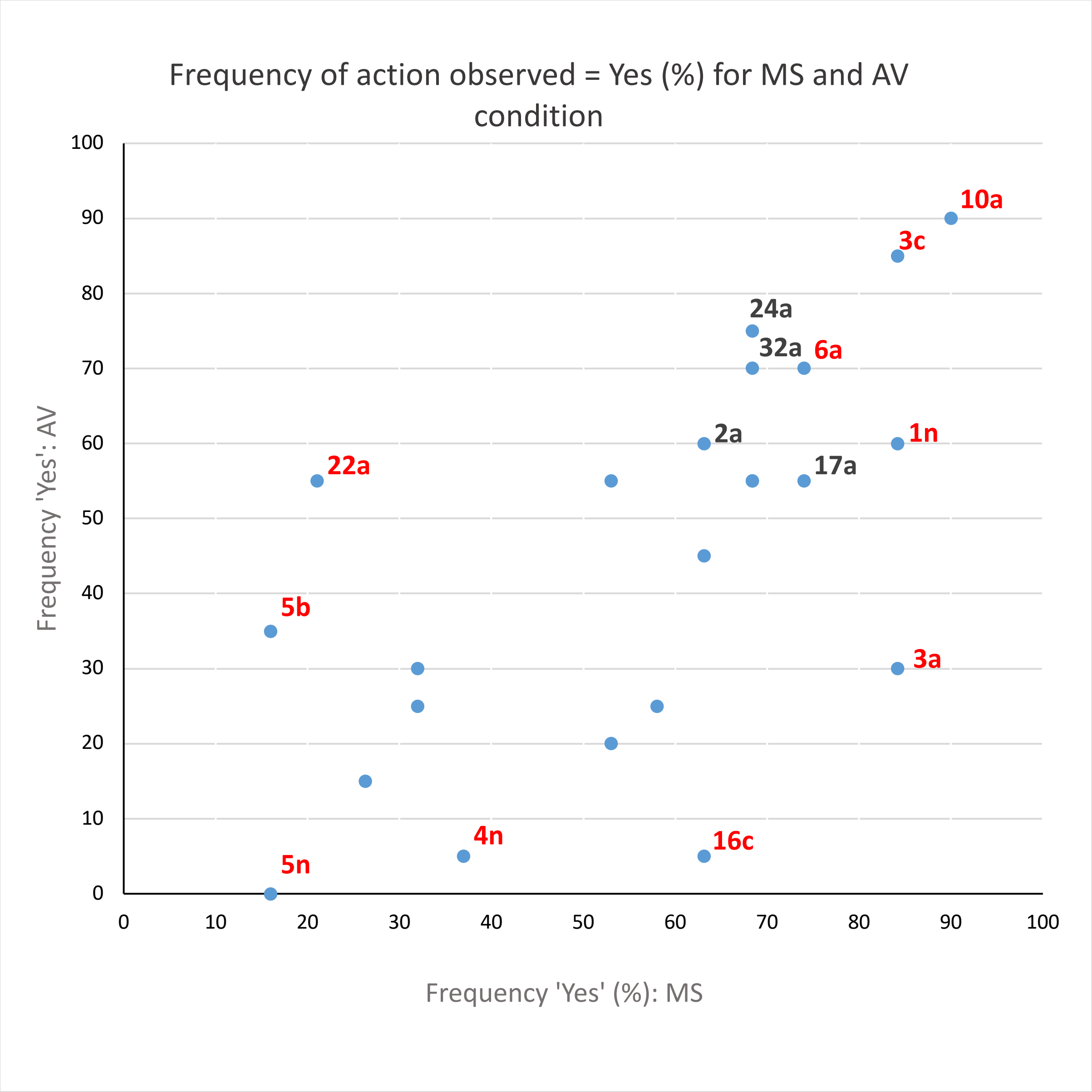

*Figure 5-10 Scatter plot showing the frequency (%) of behavioural analysis*

A higher proportion of MS than AV participants perceived stimuli that were unambiguous (0.84; 0.30) (**3a**); experienced negative feelings (0.63.5; 0.05) (**16c**); and showed a 'startled' reaction (0.37; 0.5) (**4n**) to events within the VE. A higher proportion of AV than MS participants noted nothing unusual or stayed calm (0.35; 0.16) (**22a**); and disregarded or ignored a stimulus or continued a prior activity (0.55; 0.21) (**5b**). The top three confirmed actions for MS and AV were the same: evasive (**10a**), note fire development (**3c**) and seek information and investigate (**6a**). Additionally, 84.2% MS and 60.5% AV participants showed a visible increase in frequency or speed of movement AF (**1n**). Only MS participants attempted to interact with elements of the VE (**5n***).

As well as qualitative observations from a review of screen capture of the VE, positional data logs and video capture of the user were reviewed. There were a number of qualitative quotes from interviews that supported some of the above findings:

*"I think I smelt the smell as well, but it's not as noticeable as the heat. I didn't think it would affect me as much as it did, but it made me quite anxious … Fire alarm was more like background noise rather than when you hear a real fire alarm, which is just so loud …"*

- **MS participant**

*"Not genuinely scared, but quite panic inducing and enough that it made me want to leave the building. I definitely felt the need to get out asap. Smell was the biggest cue. If the fire was on its own visually it wouldn't have felt so panic inducing. The smell really brought it to life."*

- **MS participant, interview**

*"I didn't really notice there was an alarm. I noticed smoke first, but I checked to see there was actually fire there."*

- **AV participant, interview**

*"I didn't think it was really a fire, smoke is brown and not black and I didn't see a flame. I wanted to go out that way because I knew it, but when I saw the 2nd fire I thought I would jump through the window."*

- **AV participant, interview**

## 5.9 Discussion

Following comparison of the results presented for the MS and AV conditions, a number of themes were identified: 'sense of urgency', 'evacuation behaviour', 'evacuation routes and exit choices', 'adaptive movement', 'unambiguous cues' and 'sense of immersion'. Interpretation of the results are presented in relation to these themes and the significance of the differences and similarities between the two conditions relating back to literature on human behaviour in fire and multi-sensory VEs.

### 5.9.1 Sense of urgency

Results suggest that participants in the MS condition felt a greater sense of urgency in response to the fire being triggered than the AV condition. As evidenced by the higher ratings afforded to 'perceived time pressure' and the greater agreement with the statement 'I need to find the exit nearest to me'. This was further evidenced in the adaptive movement of MS participants, which appeared to increase in frequency and direction as well as speed. For example, the head motion of MS participants increased as well as the side to side motion, causing the route patterns of participants to appear zig-zag-like rather than a smooth line as was more apparent with AV participants. Users were able to increase their speed of motion, but maximum speed was limited in order to reduce the chances of SS. Therefore, the pattern of motion seen, which would have involved the use of both controllers, may have been an attempt by users to override the speed parameters of forward motion, which are apparent when using the single controller for cross-axis locomotion. This movement modification was observed in both conditions, but less so in the AV condition and supports the findings of Gamberini et al, (4,62) who posited that these navigational changes reflect adaptive behaviour in response to the perception of a threatening situation.

### 5.9.2 Evacuation behaviour

The time taken to evacuate in a fire situation is an important factor as delayed evacuation increases risks to evacuees. Results of pre-evacuation, evacuation and total egress movement phases were similar in both conditions, with slightly higher times for the MS condition. This finding contradicts the literature which suggests that additional cues would prompt faster initiation of evacuation (1,21). However, analysis of behaviour showed that times were affected by several factors that would suggest that egress times do not provide a representative measure of validity in the VE when other behavioural measures are considered. Firstly, the parameters used to delineate each movement phase could have been chosen differently. Although the approach used a clear and tangible point to define the beginning and end of each movement phase, it did mean that times would have been inaccurate for some participants. For example, one participant reported leaving room 411 to search for a fire extinguisher; therefore, although the evacuation time would have begun at the point they exited 411, they were still carrying out actions that would be categorised as pre-evacuation. Likewise, there were participants who began the process of evacuation as soon as they turned away from the whiteboard.

Secondly, in terms of pre-evacuation behaviours, participant responses to the first fire, which was small and contained in a bin, varied, including immediate evacuation, approaching and investigating the fire, searching for something to fight it with and in one case attempting to interact with it using the controllers. On encountering the second, larger fire, no participants tried to fight it and similar numbers of participants in both conditions were observed subsequently altering their evacuation. Noting fire development was the second-highest behavioural observation recorded for both conditions and participant comments suggest that the perception of the level of seriousness increased once they saw the second fire, eliminating any ambiguity in the decision to evacuate. These findings support research by Wood et al. (6) who found that action decisions relating to fire emergencies depended on individual perception of the seriousness of the fire. They also highlight that participants attempted to find exit routes that avoided the larger fire, undoubtedly impacting evacuation times. It can be suggested from these findings that qualitative behavioural analysis of participant responses to the fire tells us much more than the egress times alone and suggests evidence of valid behaviour when compared to research in human behaviour in fire from real-world situations. From an applied perspective, these findings highlight the usefulness of VE in terms of a training tool, as Wood et al. (6) also purported that judging the fire not to be serious could reflect the layperson's misjudgement of the seriousness of the fire. Therefore, fire emergency scenarios using VE can provide an effective way to raise awareness of the dangers associated with fire and appropriate actions to take in an evacuation (7).

### 5.9.3 Exit routes

The experimental scenario used for the VE study deliberately gave participants a known entry route so that analysis could include the use of a familiar exit. Additionally, as this route was subsequently blocked by the placement of the main fire within the building, proximity to fire could also be explored. Findings showed the majority of participants in both conditions attempted to retrace their steps as a first route of choice, reflecting the literature that occupants use rational decision making in fire emergencies and prefer to evacuate using familiar routes (22,26,27). Additionally, there was evidence of the use of satisficing in the MS condition, which reported a statistically significant higher rating for time pressure (23–25), with one MS participant reporting using the fire to tell them "where not to go" rather than looking for fire exit signs. There was only one reportedly 'irrational' response to encountering the second fire. A participant in the AV condition attempted to *"jump out of the window"*; the same participant reported that they *"didn't think the fire was very realistic"* and conclusions were drawn that they were interacting with the VE in a game-like manner and their responses are unlikely to reflect valid real-world behaviour. As reported in the review of the literature, research by Edelman (27) demonstrated the importance of considering human behaviour in fire on exit choices and the intentional nature of human action. If VEs can demonstrate valid exit choice behaviour, as is suggested by the evidence here, they could be utilised for fire safety training in specific buildings to see how people behave and take corrective measures.

Both conditions showed a preference for retracing their steps to find an exit route. However, qualitative behavioural differences were identified in the way participants behaved when in close proximity to the large simulated fire in BO1. AV participants tended not to wait before entering and walked a more direct path across BO1, in some cases walking right next to the fire, whereas MS participants were observed skirting walls, waiting by doorways to inspect the escape route and reacting in a startled manner in response to elements in the VE, e.g. gasping at doors opening automatically as they walked past them. These findings suggest that participants in the MS condition were responding to stimuli within the VE as they would in the real environment, which is supported by similar behaviour reported in studies investigating overt behaviours in VEs to assess presence in threatening situations, e.g. (90). Studies by Malbos et al. (72) specifically noted behaviours that included gasping and avoiding approaching fire when walking in relation to a fire emergency scenario.

### 5.9.4 Adaptive movement

Changes in the style of navigational movement before and after the fire was triggered were observed in both AV and MS conditions. The number of participants using backwards motion increased during the emergency for both conditions. The same change was found for collision data, but there was a greater change in the number of MS participants colliding with objects during the emergency than AV. In fact, one of the AV participants noted that she had walked through furniture because she knew it was a game, whereas MS participants showed a startle reaction, such as a gasp or exclamation, when they had a collision. The adaptive movement during the emergency, in particular for backwards motion, which previous studies have associated specifically with fire scenarios (4), suggests a level of psychological response occurred for participants in both conditions. However, differences in levels would suggest a greater sense of reality of the emergency scenario for participants in the MS condition. Analysis for these behavioural indices was limited, as frequency of action data wasn't easy to collect or define using Vive controllers and frequency of participant data was too small to run inferential statistics. However, the evidence shows support for the studies by Gamberini et al. (4,62) and seem to lend support for this type of adaptive movement as a way to assess psychological response to emergency scenarios within VEs. These behavioural indices could, therefore, be considered for future studies to provide a way of validating behavioural responses within the VE. Other observations that support claims of greater immersion within the MS condition were higher levels of participants exhibiting negative emotion, e.g. fear and anxiety, reported in the observed behavioural analysis or explicitly during interviews. This supports findings by Zou et al., (91) who found that higher levels of negative emotion were associated with higher levels of realism in a fire emergency VE scenario. This is an optimistic finding as both surprise and negative emotional arousal have been purported to be particularly effective in promoting knowledge retention in VR-based safety education (8). This would again support the suggestion that the ecological validity of the VE was higher in the MS than the AV condition, thus creating a more immersive experience where participants were responding directly to simulated events within the VE (84).

### 5.9.5 Unambiguous cues

Results show that the perception that "the building is on fire" was statistically significantly greater in the MS condition that the AV condition. Additionally, following behavioural analysis coding, the number of participants who were recorded as having "perceived unambiguous fire cues" was higher in MS than AV conditions. Measure of unambiguity was taken as participants explicitly reporting a cue as being indicative of fire, for example comments such as "*I could feel warmth and the smell, so yes all the three indications of fire."* There was also a statistically significant difference in the agreement with the statement "the building is on fire" from the subjective questionnaire results. These findings can be interpreted as supporting views in the literature. For example, Chalmers et al. (71) suggested that in order for users of VE to feel sufficiently immersed in the environment they need to experience it in relation to multiple senses. Kuligowski (21) suggested that people are more likely to define the situation as a fire when there are a higher number of consistent and unambiguous sensory cues. This second point may explain some of the inconsistency in results in relation to some of the MS participants who reported that they *"did not think it was real"* as these same participants were reported as having asked for and received information from the investigator. As highlighted, our experimental procedure was designed to maximise immersion and avoid contradiction of sensory cues within the VE through receipt of inconsistent cues in the laboratory, e.g. the voice of the investigator. It is suggested that the interaction between participants in both groups and the investigator may have reduced the immersive experience for these participants.

#### 5.9.6 Sense of immersion

Immersion is said to be created through the engagement of multiple senses so that the participant feels directly engaged in the environment rather than remotely interacting with a virtual world (84,92). Evidence of direct engagement with the VE can be seen in the way some of the MS participants talked about their experience in the VE, for example an MS participant commented: "*It was coming very close to me, very quickly and I remember thinking I just need to turn around and get out this door and then thinking I need to find the nearest exit."* This was also evidenced in unprompted verbal comments MS participants made during the VE, e.g. "*Oh my god, I'm not going in there"*; *"Ah, help"*; and *"Is there water? … no … get out, out."* There were fewer examples of this within the AV condition, with participants commenting that they felt *"it was like a game";* and that they felt the avatar was in danger, but they weren't in danger. They also commented on rules associated with fire safety as the reasons for making certain decisions within the VE and this use of top down knowledge appeared to take precedence over information from cues within the VE. For example, when the investigator asked an AV participant, "What stopped you from going through the fire if you knew you weren't in danger?" they commented, *"Because I was taught never to do that."* Additionally, when looking at the subjective data in more depth, it was noted that a similar number of participants in both conditions showed a high level of agreement for "I need to take the exit nearest to me" (the highest level of agreement was statistically significant for MS participants). However, a greater number of MS participants also rated their perception of time pressure as high, whereas this was not the case for AV participants. From this we can infer that AV participants showed high levels of agreement with the need to take the nearest exit because that is a known rule associated with fire emergency scenarios, whereas MS participants showed the level of agreement because they felt under time pressure due to a greater immersive experience in the VE.

## 5.10 Limitations

#### 5.10.1 Level of risk and social interaction

NPCs were not included in the fire emergency scenario as they have been shown to cause participant distress, e.g. (10). However, certain real-world behaviours in response to a fire scenario are associated with social interaction and could affect route choice and egress times, e.g. influence on evacuation initiation and subsequent wayfinding (1). Future studies could include limited interaction of NPCs for elements such as provision of fire warnings or advice to evacuate. However, caution should be taken with respect to ethical considerations and whether participants need to be informed of what they will experience before consenting to take part.

#### 5.10.2 Development of qualitative methodology

The focus of the study was to explore observable behavioural differences in response to the simulated fire emergency scenario between the two conditions. As such, analysis concentrated on identifying observed and emergent behaviours and quantifying the number of MS and AV participants who exhibited them. This meant that the majority of behavioural data gathered did not meet the assumption threshold for inferential statistics using Chi-square, which would have allowed inferences to be made on the generalisability of the findings. Conducting a frequency of acts analysis on future studies of this nature would provide further insight into the robustness of this potential methodological tool.

#### 5.10.3 Simulator sickness and display technologies

Several participants dropped out of the study due to simulator sickness (eight prior to reaching the fire scenario and a further four before completing the evacuation task). While this is not unexpected for an immersive VE experience, practical consideration of this issue in applied contexts such as its use in OSH training is required. Future research should consider use of other display types, such as a desktop VE, and alternative methods for locomotion to reduce simulator sickness e.g. use of ‘TriggerWalking’ with the Vive controls (93).

## 5.11 Implications for VEs applied in safety contexts

The results of this study suggest potential benefits of MS simulation to OSH training, for example through increased engagement and due to the fact that we found evidence of a more direct experience with the environment, which is known to be a learning advantage as outlined in the literature review. However, evaluation of the training effectiveness is required to validate these potential benefits. We go on to empirically assess this in Study 2.

The exhibition of valid user behaviours identified in this study lead us to conclude that virtual environments would be valuable in predicting human behaviour in emergency situations. The analyses from this study highlight where such predictions are more and less likely to be valid. For example, our study shows that in VE users are likely to demonstrate more risky behaviours around fire than they would in real life, although our study suggests this was partially mitigated by the introduction of multi-sensory feedback. We also found that pre-evacuation delays were different in the VE to those we know take place in real life, because VEs do not provide effective motivation for participants to (for example) continue tasks, save work or establish whether others are safe. Therefore, interpretation of human behaviour should take these predictive limitations into account.

However, actions around assessment of the fire and decisions on firefighting appear to be valid. Exit choice (e.g. exiting via the most familiar route) is also very consistent with real-world behaviour, which may be useful when planning layout, signage and training for the workplace. Understanding human behaviour in fire and its potential impacts in emergencies is of critical importance in ensuring safe outcomes.

## 5.12 Conclusions

Both AV and MS VEs were shown to elicit behavioural responses from participants seen in real-world fires, from which we can infer some behavioural validity. However, findings suggest a differential between the two conditions in the level of immersive experience they afforded their users, with a greater level of immersion suggested in the multi-sensory VE with the added heat and smell simulation.

These results suggest that we can increase the validity of user behaviour in VEs simulating threatening emergency scenarios by using an MS interface. Future investigations could develop the behavioural tool used for analysis in the present study. Analysis using frequency of acts information gathered from similar studies could deepen the level of analysis achieved if the threshold for assumptions of inferential statistics is met, potentially providing stronger evidence of validity. Further work could also be carried out on the relative saliency of sensory information to inform prototype design for potential use as a training tool in industry.

This study has presented optimistic outputs for the applied use of multi-sensory VR in fire safety training, for two reasons. Firstly, the direct interaction with the VE is more likely to raise awareness of any contextual issues relating to human behaviour in fire that requires addressing through training scenarios. Secondly, the surprise and affective responses evidenced in this study indicate the likelihood of increasing motivation to learn by making training fun and engaging and making the lessons more likely to stick.

# 6 Study 2 – Effectiveness of training

This study aimed to evaluate the effectiveness and efficiency and participant satisfaction of different training methods for use in occupational health and safety (OSH). The primary objective for the study is to assess the knowledge uptake and retention of two OSH-related use cases, fire safety and engine disassembly, following training delivered using one of three different methods: a traditional education method (PowerPoint presentation), audio-visual virtual environment (AV VE) or multi-sensory (MS) VE.

The experimental hypothesis stated that there would be increased engagement, motivation to learn and knowledge uptake and retention shown by the participants who are trained using the MS VE method than participants trained using the traditional and AV VE methods. Proven benefits of such a system could lead to improvements in current OSH training practices. Additionally, as the developed prototype system used affordable off-the-shelf technologies, it would serve as an example of a reusable, low-cost solution for companies that require an easily customisable system for their training needs.

### 6.1.1 Fire safety training use case

The training and assessment covered the following fire safety concepts:

- **Fire alarm activation (used for familiarisation)** – all building occupants should know where fire alarms are located. Most buildings are fitted with fire safety sensors that will automatically trigger alarms. However, occupants can manually trigger fire alarms where necessary.
- **Fire token system** – this is a procedure used in high occupancy building such as universities, to assist with the evacuation of all building occupants in the event of a fire emergency. As long as it is safe to do so, if an occupant passes a fire token on their route out of the building, they are to collect it. They are then required to check the room(s) that corresponds to the number written on the fire token and encourage anybody in the room to evacuate the building.
- **Spatial awareness of fire safety objects/routes** – building occupants should know the location of their nearest safety-related objects and evacuation routes.
- **Optimal route through building** – all evacuation routes and fire exits should remain clear/unobstructed so that evacuation is as quick as possible.
- **Use of fire extinguishers** – fire extinguishers should only be used in the event that all exit routes are blocked and a path out of the building is only possible by extinguishing part of the fire. If a fire extinguisher does need to be used, the correct one should be selected depending on the type of fire that has been ignited.
- **Prevention of fire/smoke spreading** – internal doors should be closed to reduce the chances of fire and smoke spreading within the building.

### 6.1.2 Vehicle disassembly use case

The training and assessment covered the following procedural and safety concepts:

**Bimanual vehicle engine disassembly –** the correct procedure will need to be followed to safely disassemble a vehicle engine.

**Fuel leak** – the correct safety procedure will need to be followed to deal with an engine fuel leak, to safeguard against the safety risk.

For both elements, the order in which the tasks are conducting was of equal importance to the tasks themselves.

## 6.2 Design

A mixed design was employed with three factors: training medium, feedback modality and use-case scenario, each with two levels shown in Table 6-1 below:

*Table 6-1 Experimental factors and levels*

| Factor | Levels | |
|---|---|---|
| Medium | Virtual environment (VE) | PowerPoint (PP) |
| Modality | Audio-visual (AV) | Multi-sensory (MS) |

The modality factor was within subjects; all participants completed training activities for both the fire safety and engine disassembly use cases, experiencing one scenario with AV feedback and one with MS feedback. The medium factor was tested between subjects with half the participants completing both training scenarios through VE, and the other half completing both through PP. The presentation of scenario, modality and medium was fully counterbalanced so that there are eight different session types, shown in Table 6-2 below:

*Table 6-2 Counterbalanced participant groups*

| Participant group | Learning task 1 | | | Learning task 2 | | |
|---|---|---|---|---|---|---|
| | **Medium** | **Modality** | **Scenario** | **Medium** | **Modality** | **Scenario** |
| **A** | VE | AV | Fire | VE | MS | Engine |
| **B** | VE | AV | Engine | VE | MS | Fire |
| **C** | VE | MS | Fire | VE | AV | Engine |
| **D** | VE | MS | Engine | VE | AV | Fire |
| **E** | PP | AV | Fire | PP | MS | Engine |
| **F** | PP | AV | Engine | PP | MS | Fire |
| **G** | PP | MS | Fire | PP | AV | Engine |
| **H** | PP | MS | Engine | PP | AV | Fire |

## 6.3 Participants

Fifty students, staff and associates at the University of Nottingham were recruited using poster advertisements on campus, online and circulated via university mailing lists. Participants were all over 18 and were screened to ensure they were not at increased risk of simulator sickness, they had no previous traumatic experience in an emergency situation, they had normal or corrected-to-normal perception across all senses, they were not pregnant and they did not have any of the following conditions:

a) respiratory conditions e.g. asthma
b) allergies
c) odour intolerances
d) motion sickness
e) migraines
f) epilepsy
g) blurred vision
h) dizziness or vertigo
i) sleep disorders.

Participant age ranged from 19 to 43 (mean age = 27.3 years, SD = 5.4). Each participant was compensated for their time with a £10 voucher upon completion of the first part of the study and another £5 voucher once they completed a follow-up phone call one week later.

## 6.4 Equipment

### 6.4.1 Hardware

Hardware was configured as per Study 1; however, as there were a number of participants who dropped out of the previous study due to simulator sickness (SS), it was decided that for the VE in Study 2 the HMD and Vive controllers would be replaced by a keyboard and mouse to control movement through the VE. Hardware for the prototype design was as shown previously in Figure 5-1 and the desktop solution in use in Figure 6-1:

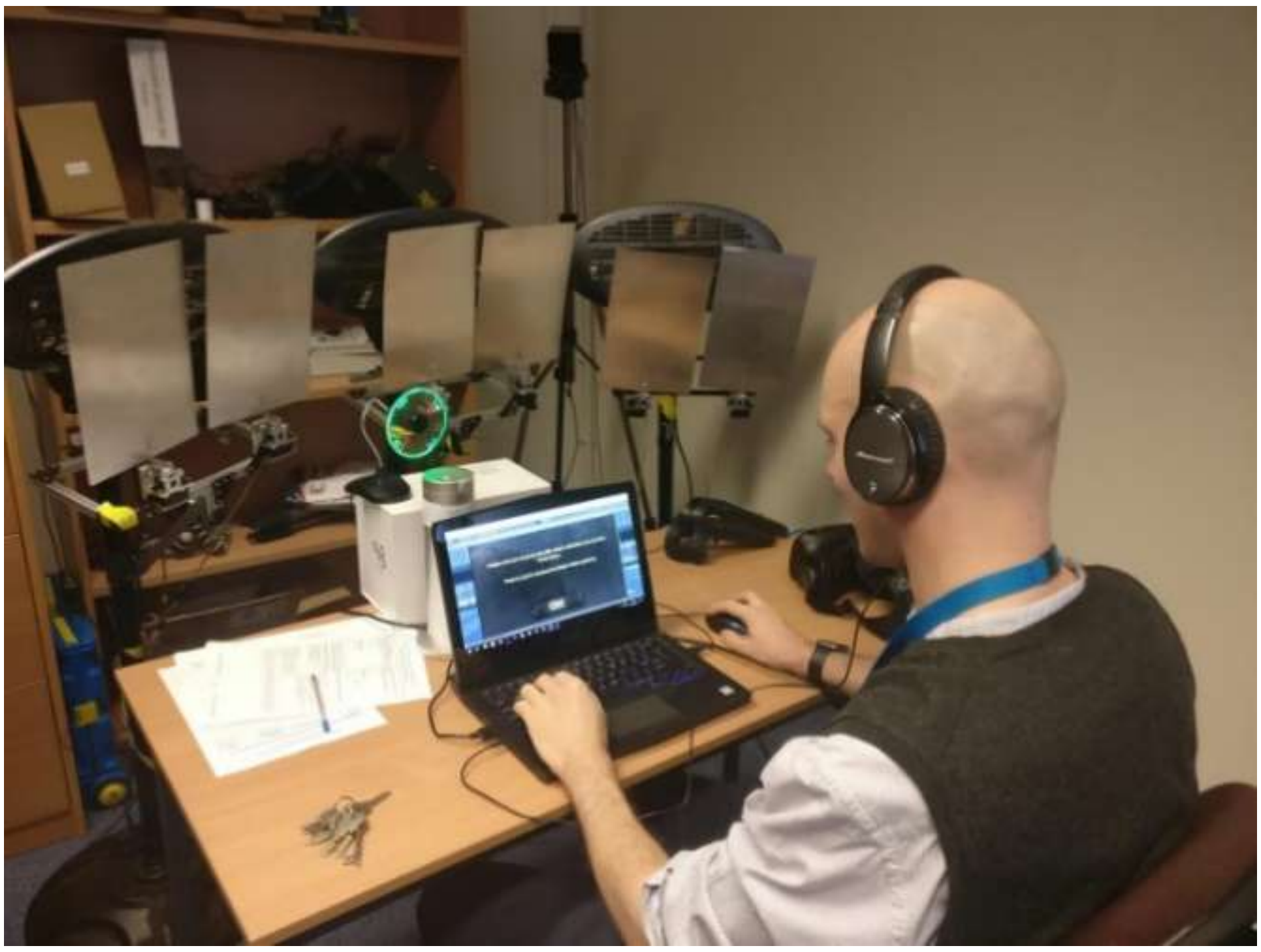

*Figure 6-1 Desktop solution in use*

:

Traditional training was provided via a PowerPoint presentation. Figure 6-2 and Figure 6-3 show sample slides from the fire safety and engine disassembly training presentations:

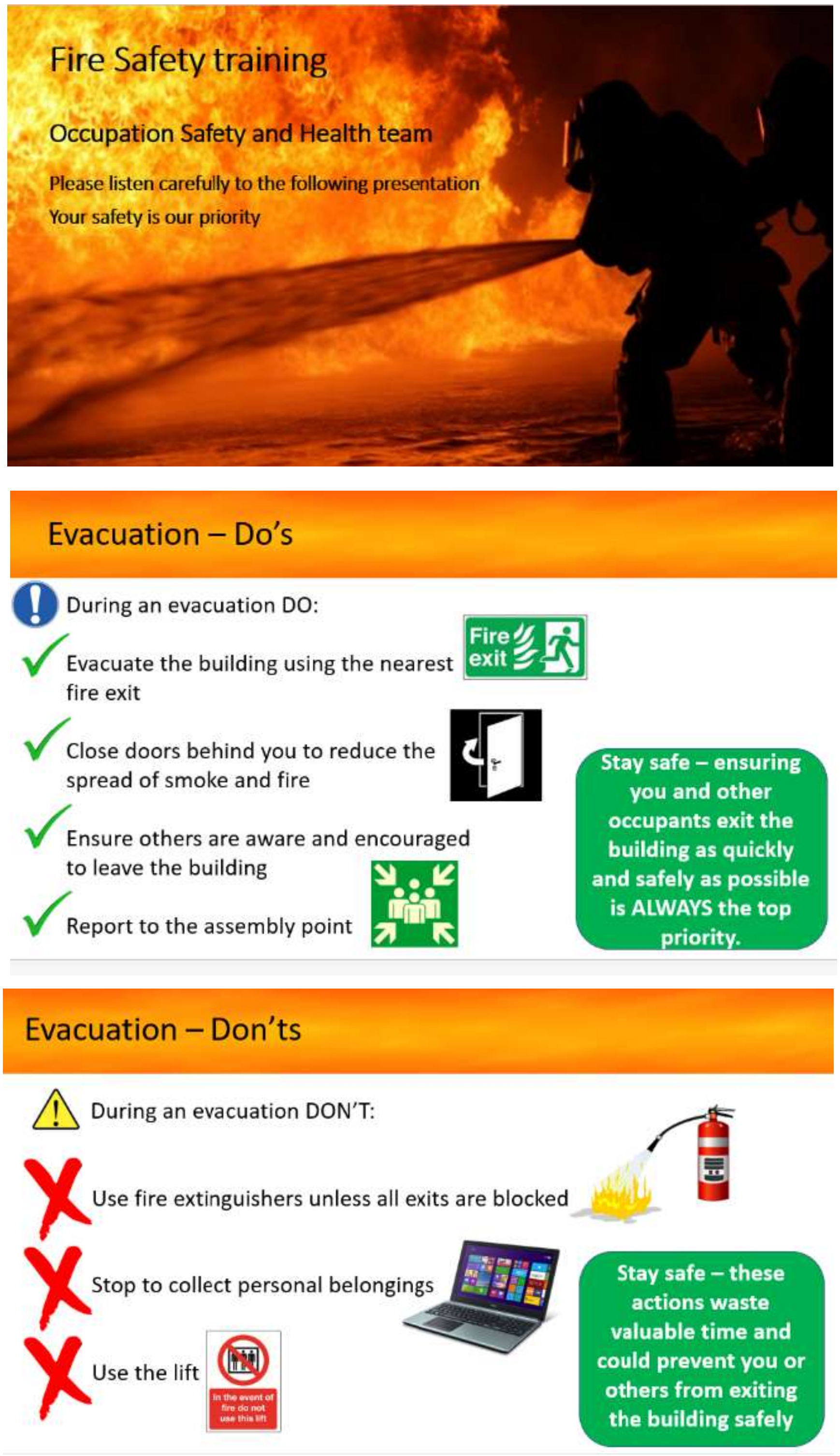

*Figure 6-2 Example slides from the traditional fire safety training presentation*

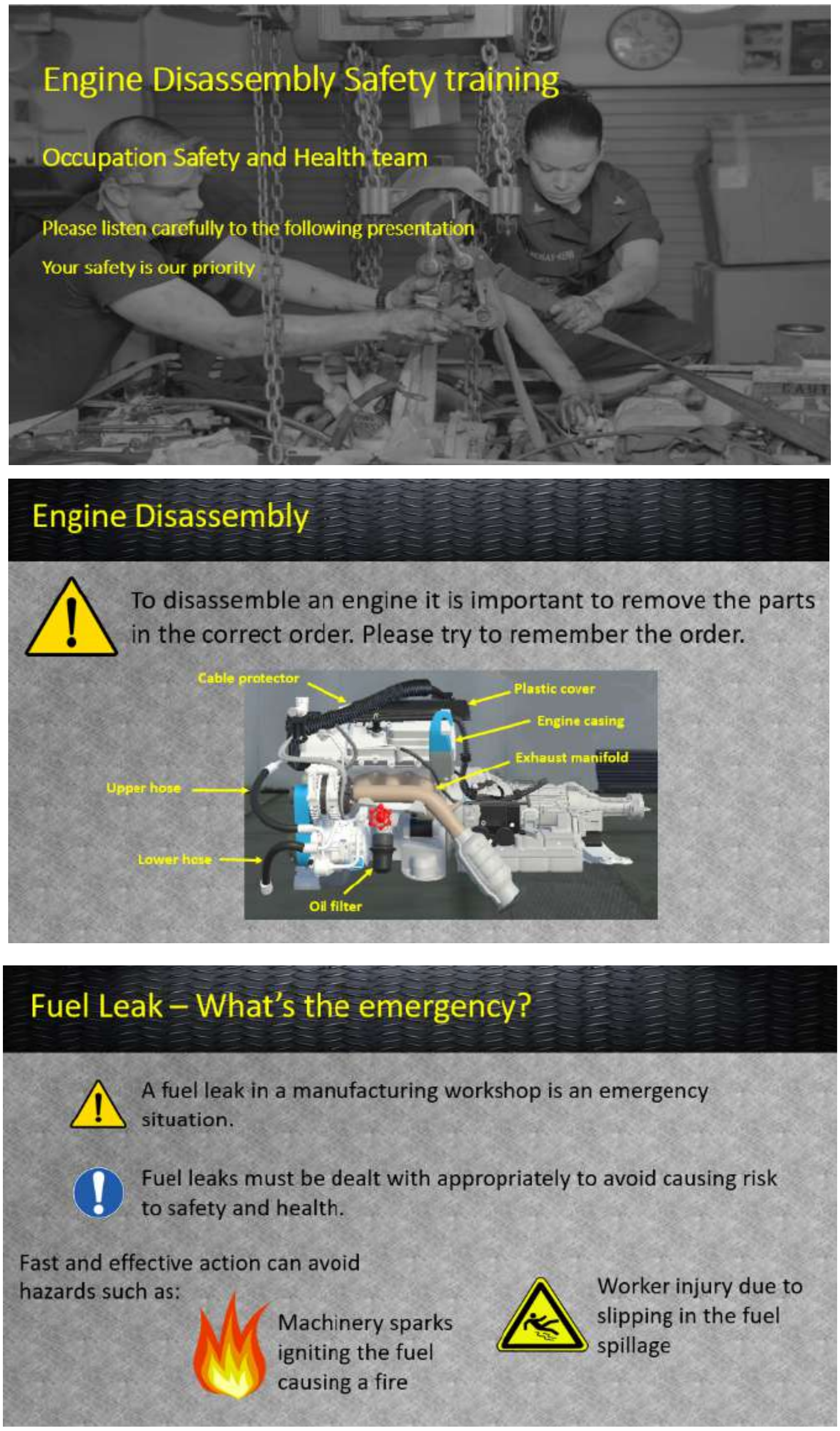

*Figure 6-3 Example slides from the traditional engine disassembly training presentation*

The VE system was designed and run using Unity on a laptop PC with external graphics card. Noise-cancelling headphones delivered audio simulation for both the PP and VE training, which masked any background noise and the noise of apparatus during use.

### 6.4.2 VE office building

The office building layout used for Study 1 was also used for Study 2 – see Figure 5-2. However, a new fire safety scenario was designed and built, which changed some of the building's previous features, for example users were able to interact with doors to open and close them within the new fire scenario design.

## 6.5 Procedure

The total running time of the study was approximately one hour. An experimental script was used for procedural consistency. Prior to any data collection participants were asked to review the participant information sheet and sign a consent form. Subsequently, their gender and age, as well as their gaming, motor vehicle maintenance and fire/fuel leak training experience was recorded on a demographics sheet for reference using a self-reported rating scale (1=no experience; 5=very experienced). Participants completed a simulator sickness questionnaire (SSQ) (86) before and after the study to monitor their condition and collect information for future studies using this system.

Participants were asked to answer a pre-training knowledge assessment (Appendix B and Appendix C) before starting the training in either the PP or VE conditions. If the participant was completing the training in the MS condition, the olfactory and thermal simulation would be triggered automatically via sets of colliders built into the PowerPoint presentations and the VE. The scent diffuser and patio heaters were disconnected for participants in the AV condition. Participants in all conditions were asked to put the noise-cancelling headphones on to begin the training.

### 6.5.1 PP condition

Participants in the PP condition were asked to read a self-paced PowerPoint presentation (see Figure 6-2 and Figure 6-3) and listen to the accompanying pre-recorded verbal instructions.

### 6.5.2 VE condition

Participants in the VE condition had to complete a variety of tasks within the VE, which varied for the fire safety and engine disassembly use cases. All tasks were carried out within the VE for each scenario and the instructions for the tasks were provided via the VE training module and not the experimenter.

#### *6.5.2.1 Fire safety use case*

Participants completed three sub-tasks during this training session: a familiarisation task, to allow participants to practise using the controls; an instructional task, where participants were given fire safety information consistent with the PowerPoint condition; and a fire emergency scenario task, where participants needed to apply the fire safety information to safely evacuate the virtual building.

Details of the procedure participants experienced when progressing through the VE training can be found in Section 4.9.1.

#### *6.5.2.2 Engine disassembly use case*

Participants completed three sub-tasks during this training session: a familiarisation task, to allow participants to practise using the controls; an engine disassembly task, where participants were required to remove parts from an engine in the correct order; and a fuel leak emergency scenario task, where participants needed to apply the correct actions to safely deal with the situation. Both the engine disassembly and fuel leak sub-tasks were repeated twice, the first time with step-by-step guidance and the second time with no assistance. All information used for the engine disassembly and fuel leak tasks was consistent with the PowerPoint presentation. For the fire leak scenario, if participants did not carry out actions in the correct order to safely respond to the fire leak scenario, the scenario would be reset so that the participant needed to restart the task. Further details of the VE training can be found in Section 4.9.2.

### 6.5.3 Post-training

Following completion of the training session, participants were asked to answer a post-training knowledge assessment (Appendix B and Appendix C) followed by a subjective questionnaire (Appendix D). Participants were fully debriefed and offered the chance to ask questions. A follow-up knowledge-retention assessment was carried out with participants one week after completing the training session.

The schematic of the experimental conditions and the procedure can be seen below:

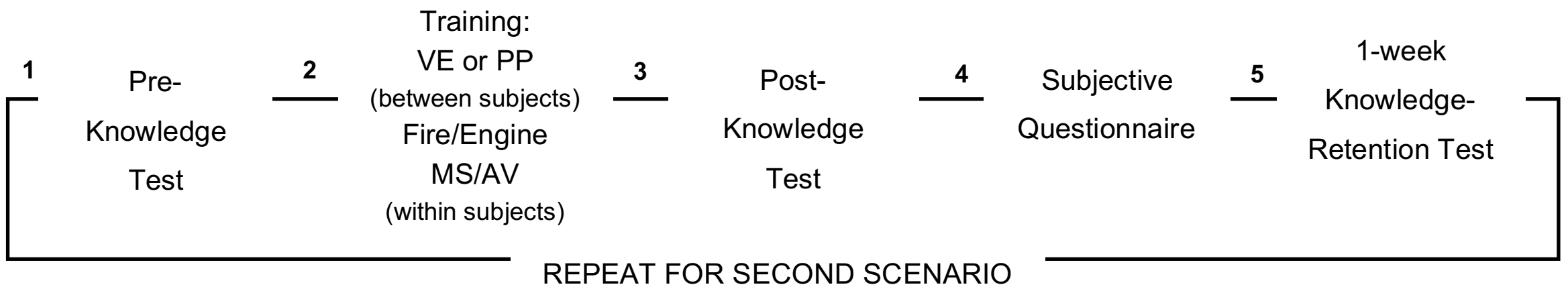

*Figure 6-4 Schematic showing experimental conditions and procedure*

## 6.6 Results

Results are presented in two separate analyses: (1) knowledge tests (scores for uptake and retention) and (2) subjective ratings (scores from participant questionnaire responses). Knowledge test scores were analysed with parametric statistical tests, while questionnaire scores were treated as non-parametric data as responses were on a five-point ordinal scale and data were not normally distributed across the range of scores.

### 6.6.1 Knowledge tests

#### *6.6.1.1 Scoring*

Model answers were produced, and two assessors were trained to score according to these model answers. To ensure that the grading was consistent throughout the whole set of data, a scoring approach was iteratively developed. First, both assessors graded five knowledge tests, assigning up to a maximum of two points per question (in 0.5 point intervals), thus up to 14 points in total for each of the two training scenarios. The answers and scoring were compared and discussed to generate a systematic scoring system.

Once the grading system had been produced, 10 answer sheets randomly sampled were graded by each marker, and the scoring was analysed for inter-rater reliability (using Cohen's kappa). The resulting score ($k=0.852$, $p<.001$) is categorised as an outstanding level of agreement between the two raters (94) (Landis and Koch, 1977), demonstrating that the scoring criteria produced reliable results and the grading approach was robust. Thus the remainder of the answer sheets (including the previous sample of five) were then graded independently according to the agreed score system.

#### *6.6.1.2 Analysis*

From the knowledge test results, scores were calculated for **knowledge uptake** (difference between pre- and post-training scores) and **knowledge retention** (difference between post-training and one-week follow-up scores). A 2x2 mixed design multivariate analysis of covariance (MANCOVA) was conducted. The two independent variables were **modality** (multi-sensory vs audio-visual; within subjects) and **medium** (PowerPoint vs VE; between subjects) and uptake and retention were treated as two separate dependent variables. Scenario, gaming experience, experience of fire safety training and engine knowledge were treated as co-variates to control for these factors when analysing potential differences between modality and medium.

There was a significant main effect of medium (PPT vs VE) overall ($F(2,41)=8.522$, $p=.001$), and on each dependent variable separately, uptake ($F(1,42)=6.124$, $p<.05$) and retention ($F(1,42)=15.065$, $p<.001$). Participants in the PPT condition had higher knowledge uptake (Mean=6.76) than participants in the VE condition (mean=5.57). However, for retention, participants in the VE condition (mean=-0.31) retained more knowledge than those in the PPT condition (Mean=-1.89). Figure 6-5 illustrates the descriptive statistics for a comparison of PPT and VE knowledge test scores before training, after training and after an interval of one week.

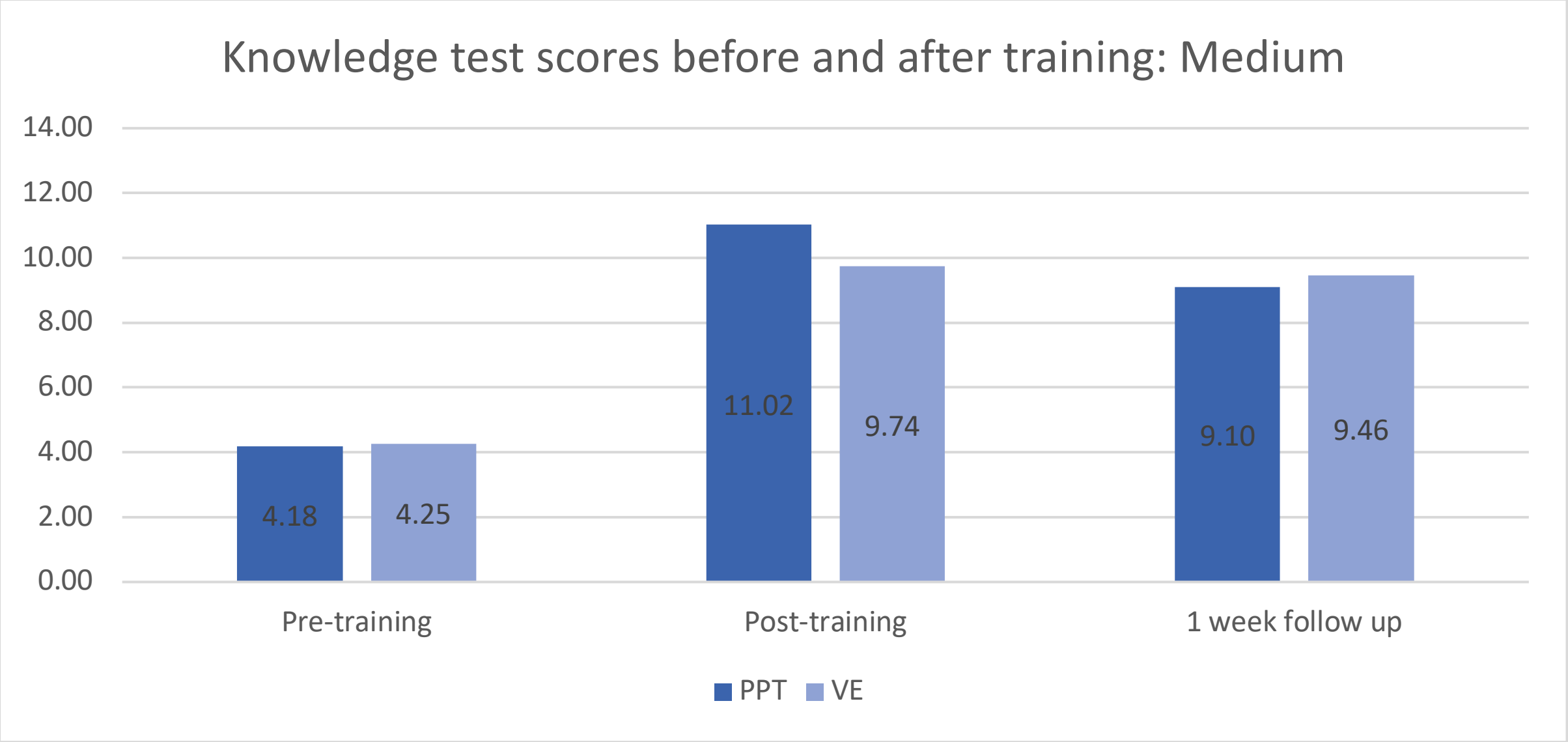

*Figure 6-5 Comparison of knowledge test scores for PPT vs VE conditions*

There was also a significant main effect of modality (MS vs AV) ($F(2,41)=20.814$, $p<.001$), with a significant difference in both uptake ($F(1,42)=42.643$, $p<.001$) and retention ($F(1,42)=13.876$, $p=.001$). MS (mean=6.21) was higher than AV (mean=6.13) for uptake. For retention, MS (mean=-1.23) performed worse (i.e. participants on average lost more knowledge) than AV (mean=-0.97). Figure 6-6 illustrates the descriptive statistics comparing MS and AV scores in the knowledge tests.

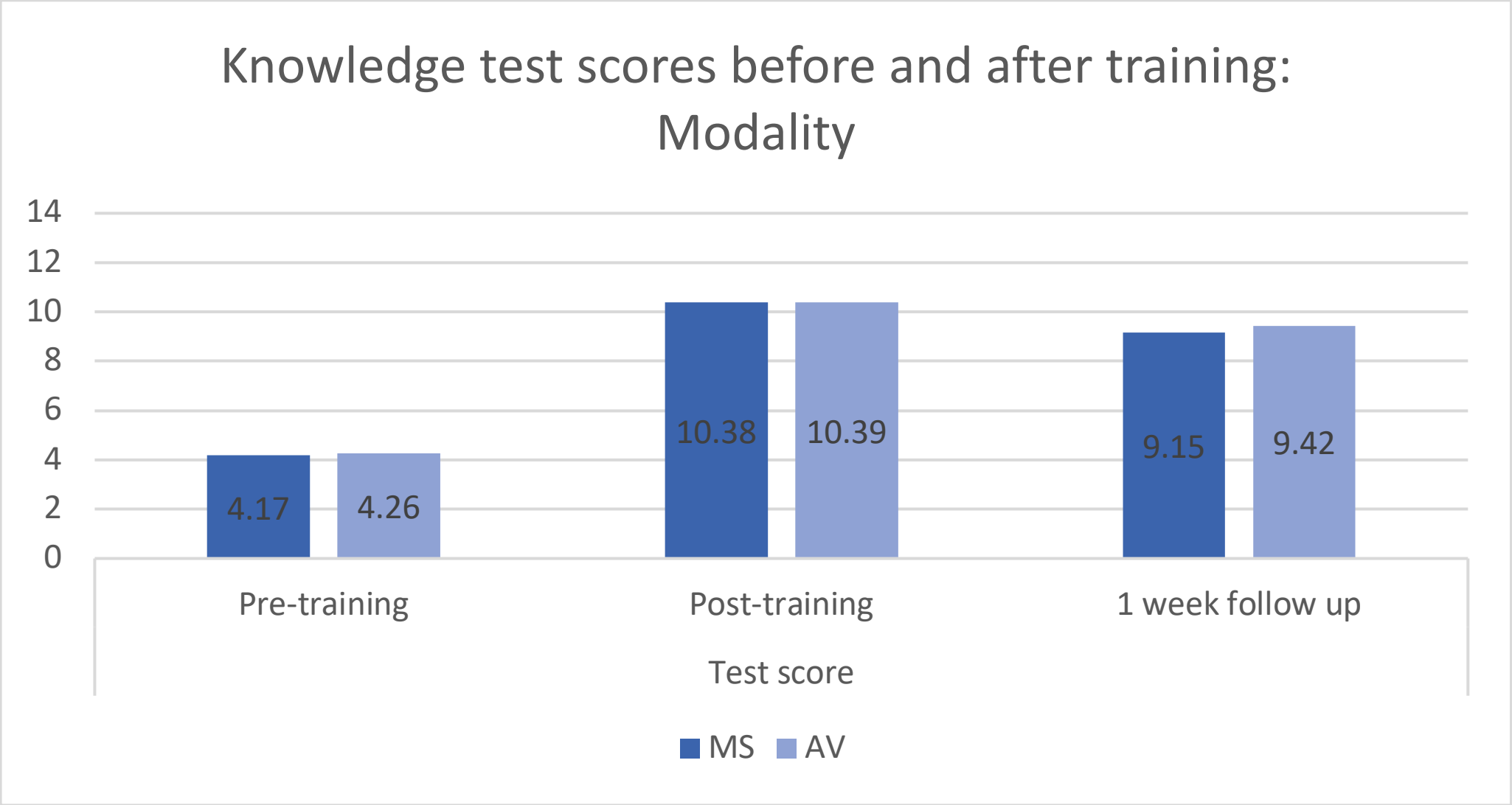

*Figure 6-6 Comparison of knowledge test scores for MS vs AV conditions*

There was no significant interaction (Figure 6-7) between the effects of modality and condition for either uptake (F1,42)=.084), p>.05) or retention ($F(1,42)=1.692$, $p>.05$).

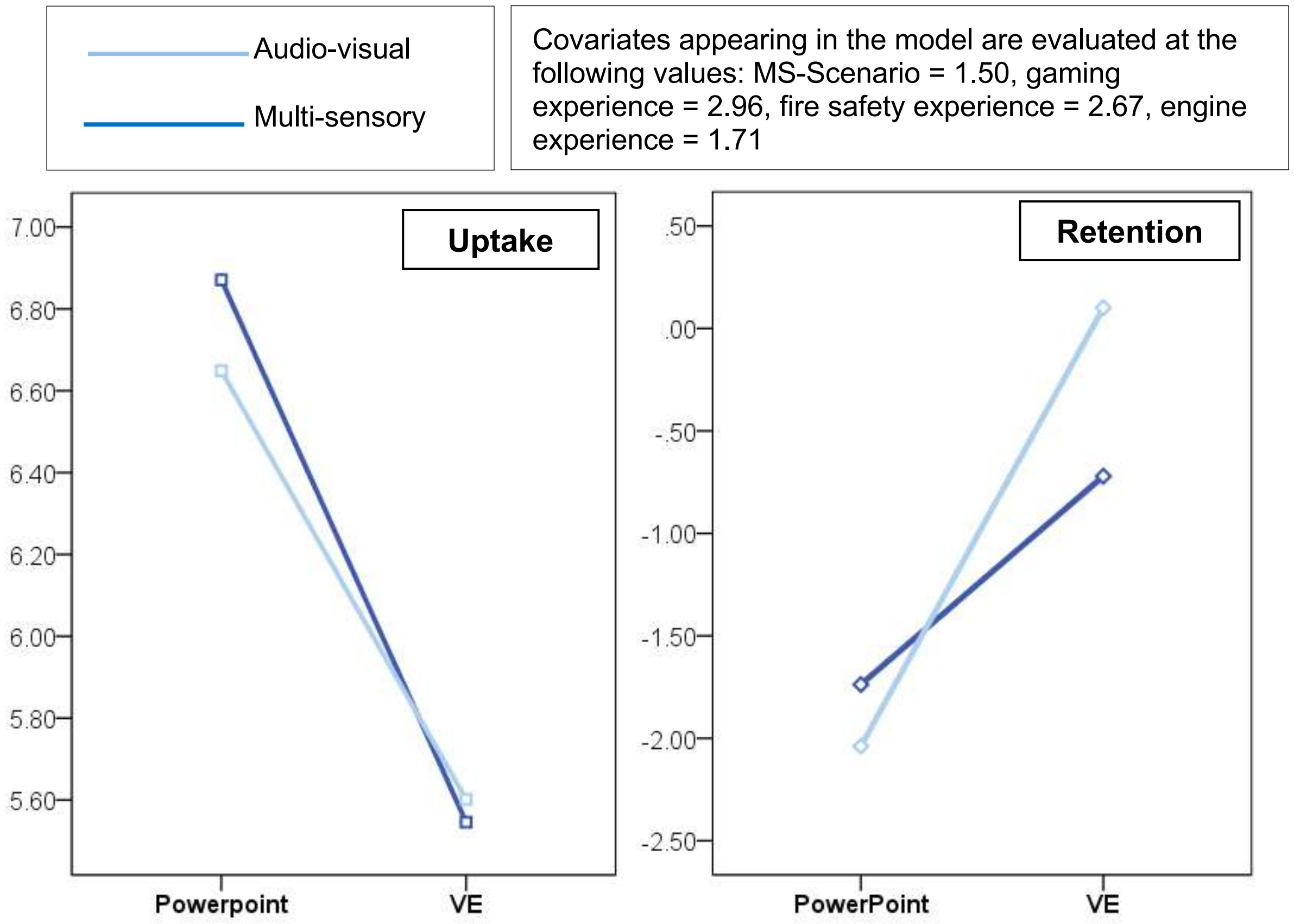

*Figure 6-7 Uptake and retention scores comparing medium and modality*

Summary descriptive statistics for each condition are shown in Table 6-3 below.

*Table 6-3 Descriptive statistics for uptake and retention by medium and modality*

| | Medium | Mean | Std. Deviation | N |
|---|---|---|---|---|
| MS Uptake | PowerPoint | 6.896 | 2.952 | 24 |
| | VE | 5.521 | 2.429 | 24 |
| | Total | 6.208 | 2.763 | 48 |
| AV Uptake | PowerPoint | 6.792 | 3.892 | 24 |
| | VE | 5.458 | 2.934 | 24 |
| | Total | 6.125 | 3.476 | 48 |
| MS Retention | PowerPoint | -1.854 | 2.402 | 24 |
| | VE | -.604 | 1.302 | 24 |
| | Total | -1.229 | 2.013 | 48 |
| AV Retention | PowerPoint | -1.979 | 2.872 | 24 |
| | VE | .042 | 1.661 | 24 |
| | Total | -.969 | 2.536 | 48 |

### 6.6.2 Subjective ratings

Questions 1–8 were combined into an overall score for ‘engagement’, with positive statements adding the respective agreement score and negative statements subtracting the agreement score for the total. Questions 9 and 10 were treated individually, as they tested distinct concepts around transfer of training (“The skills I learned during the training would be helpful in learning how to respond to this situation in real-life”) and desire to undertake this type of training again (“I would want to undertake this type of training again”) respectively. Questions 11–13 were combined into an overall ‘attitude towards health and safety’ score.

#### 6.6.2.1 Modality

The effect of modality (within subjects) on ratings was analysed with a Wilcoxon signed ranks test. Considering the differences between MS and AV conditions (Figure 6-8), scores differed significantly for engagement ($z$=-2.40, p<0.05), with a higher rating for AV (Mdn=12.5) than MS (Mdn=12.0). All other subjective ratings were not significantly different between the conditions (attitudes: z=-0.17, p=NS, MS Mdn=12.0, AV Mdn= 12.0; Q9: z=-1.18, p=NS, MS Mdn=4.0, AV Mdn= 5.0; Q10: z=-0.75, p=NS, MS Mdn=4.0, AV Mdn= 4.0).

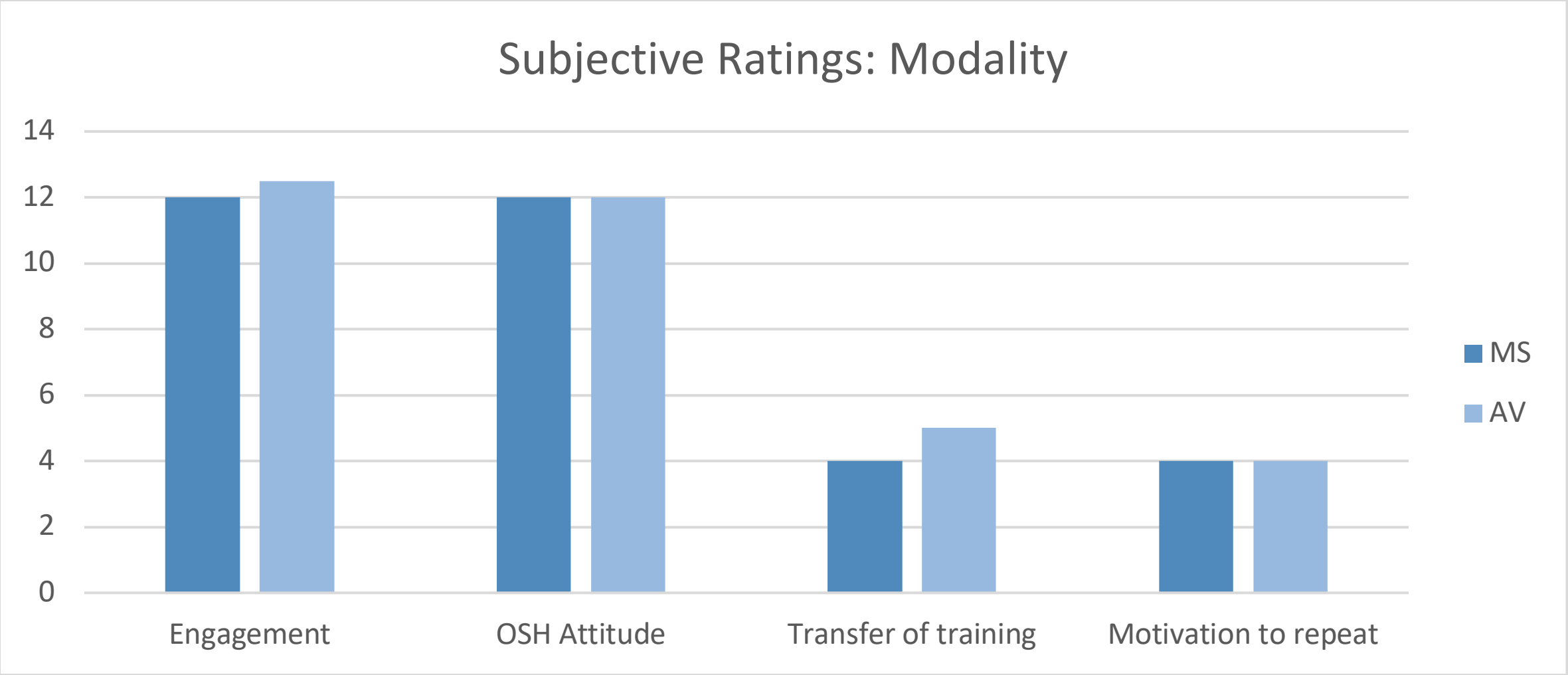

*Figure 6-8 Comparison of median subjective ratings for MS and AV conditions*

#### 6.6.2.2 Medium

The differences between PowerPoint and VE were analysed with Mann-Whitney *U* tests. Engagement differed significantly ($U$=693.50, $z$=-3.37, p=0.001) with the VE scoring higher (Mdn=14.0) than PowerPoint (Mdn=10.0). Attitudes towards health and safety also differed significantly ($U$=569.00, $z$=-4.39, p<0.001), with VE (Mdn=14.5) rated higher than PowerPoint (Mdn=10.0), as did Question 10, desire to repeat this type of training ($U$=606.00, $z$=-4.252, p<0.001; VE Mdn=5.0, Ppt Mdn=4.0). Q9 on transfer of training was not significantly different between the conditions ($U$=1070.00, z=-0.67, p=NS, VE Mdn=5.0, Ppt Mdn=4.5).

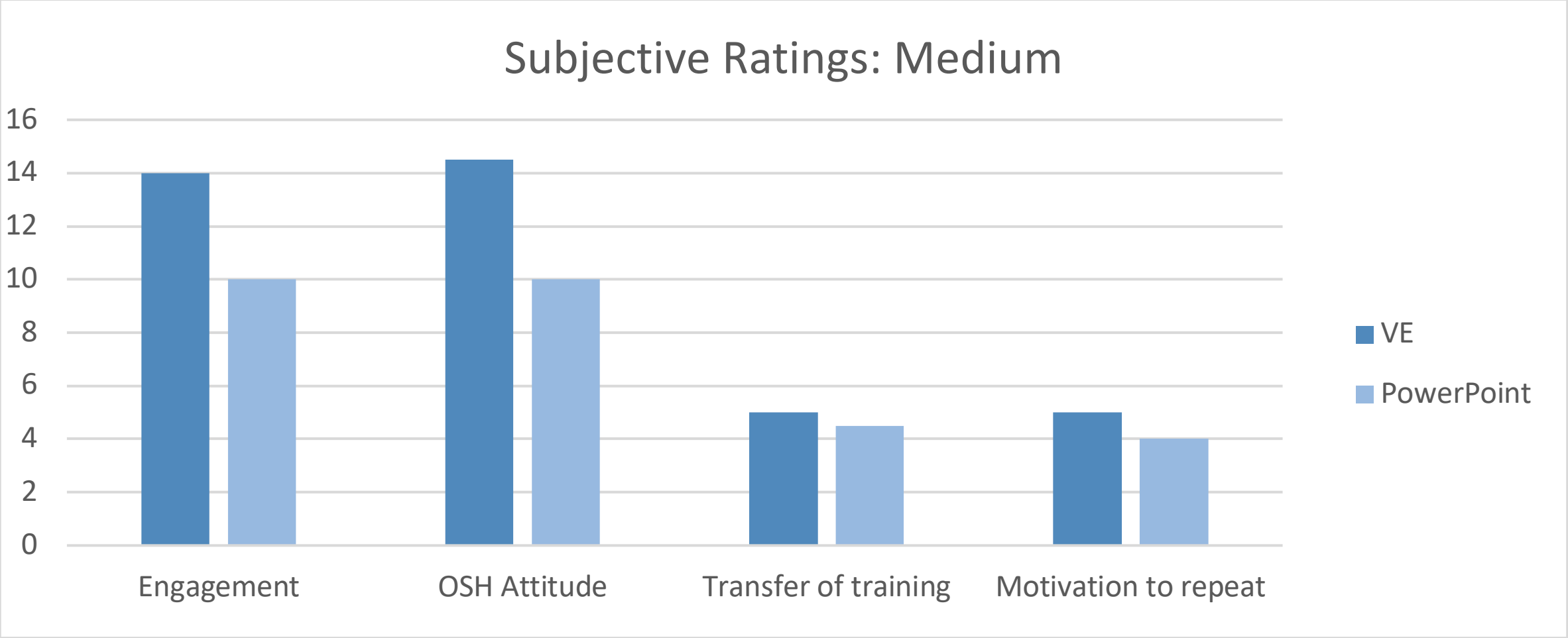

*Figure 6-9 Comparison of median subjective ratings for VE and PPT conditions*

### 6.6.3 Summary

A summary of all results, comparing PowerPoint and VE (Table 6-4) and comparing multi-sensory and audio-visual (Table 6-5) conditions for each measure (test scoring and subjective ratings) is below, showing which produced the superior results in each case.

*Table 6-4 Summary comparison of PPT and VE across all measures*

| Measure | Superiority |
|---|---|
| **Knowledge uptake** | PPT |
| **Knowledge retention** | VE |
| **Engagement** | VE |
| **OHS attitude** | VE |
| **Motivation for future training** | VE |

*Table 6-5 Summary comparison of MS and AV across all measures*

| Measure | Superiority |
|---|---|
| **Knowledge uptake** | MS |
| **Knowledge retention** | AV |
| **Engagement** | AV |
| **OHS attitude** | No significant difference observed |
| **Motivation for future training** | No significant difference observed |

## 6.7 Discussion

The VE performed comparatively well in this study. Self-reported levels of engagement, attitude to occupational safety and health, and the extent to which participants would want to undertake training in the future were all significantly higher for the VE than for PowerPoint. These attitudinal factors could play an important role in improving general health and safety culture in the workplace. If employees indicate negative attitudes towards health and safety training, VE-based training could prove a useful method to encourage engagement with the training, and more generally improve motivation.

While participants trained through PowerPoint presentations had a greater increase in knowledge test scores (from before undertaking training to immediately after), there was a significantly larger decrease in knowledge scores when participants were re-tested one week later, demonstrating that retention was better with the VE. The mean score on the final test was therefore actually higher for the VE condition, despite the short-term improved performance for PowerPoint. We hypothesise that the format of the PowerPoint training aligned more closely with the testing method (traditional fact-based verbal Q&A), and that it had the benefit of simplified delivery of direct messages. In the shorter term it was comparatively easy for participants to recall the answer, but the limited mental processing means that the knowledge, despite being superficially well learned for immediate recall in a similar form to which it was delivered, is not well retained. We speculate that increased cognitive engagement of learning in the VE, combined with the benefits of constructivist learning from the experiential affordance of VE training, creates more established and comprehensive mental models. This warrants further exploration but presents a clear implication that testing an employee's knowledge immediately following health and safety training may not be an effective means of gauging long-term knowledge of OSH. This is common in typical OSH training, for example where an employee undertakes online training that contains a test at its conclusion as a summative measure and is often a pass/fail appraisal of whether the employee should repeat the training. Our results suggest that the results may be different after an interval, and thus that either alternative or additional testing should take place in a separate later session.

The results of this study overall provide greater evidence for the body of work supporting the use of VE training in emergency situations (e.g. 7,40,69), in this case showing that VE can add value for training safe procedures in fire emergency situations and in engine disassembly.

The results of this study provided weak evidence for the use of multi-sensory (heat and smell) simulation for health and safety training, and would not justify the investment required to add the additional feedback modalities. This outcome was surprising given the theoretical importance of multi-sensory feedback for virtual experiences that are comparable to the real world reported in the academic literature (5,71,72,76) and prior research that suggests the benefits of multi-sensory virtual environments in learner motivation and engagement (8,77). The results were also unexpected given prior research in virtual fire simulations which specifically highlights the need for multi-sensory simulation (e.g. 10,39). Finally, the results seemed unexpected given the differences in experiences between multi-sensory and audio-visual VEs, identified in Study 1 of this research. The findings in Study 2 may be down to the specific implementation of the hardware, training programme, scenarios and experimental design, and should be explored further. The notion that the outcomes of virtual experiences can vary depending on the specific instances has been reported by other authors, for example, Lawson et al. (95), who comment on inconsistencies in perceptions of virtual worlds depending on the VE set-up and scenario.

A key difference in the experimental hardware for Study 2 as compared with Study 1 is that we did not use a head-mounted display. This decision was made based on ethical concerns given the levels of cybersickness demonstrated by participants in Study 1, and that in Study 2 the research focus was on training, not the behavioural validity, and it was thus less important to hide visibility of the heaters and fins operating. The sickness and practical issues experienced in use of the HMD imply that, with current mainstream technologies, use of HMD is not necessarily a suitable option for training large numbers of employees, and thus we wanted to explore whether the VE training was effective without it. This is likely to affect the overall trainee experience, as the level of immersion is lower when the VE is viewed on a monitor than when viewed through stereoscopic HMD, and this lower immersion may be incompatible with the effects of the additional sensory simulation of heat and smell. Future work could specifically compare the effects of different visual display types within multi-sensory training and the potential interaction between display type and effectiveness of additional sensory stimuli.

Moreover, there are an almost infinite number of ways to design the training programme and scenarios, and it may be that some of these reveal greater benefits of the multi-sensory simulation. The current study adapted traditional training material for VE delivery, and thus is unlikely to be exploiting the affordances of VE or MS technologies effectively. Any experimental design is ethically limited as the test of training cannot take place in real-life emergencies or physical environments that pose physical danger or psychological distress; it may be that in these situations those trained with multi-sensory virtual environments fare better than those trained in the AV conditions.

In conclusion, while Study 2 did not provide the expected evidence for multi-sensory simulation in VE-based training, this may be due to the particular set up and use in this experiment and should be explored further. Study 2 did, however, provide further evidence for the use of VE to encourage employees to engage with virtual health and safety training.

# 7 Benefits to business of a multi-sensory virtual training system

The MS virtual training system and scenarios were presented to the industrial partners to obtain their thoughts on how the system might benefit their businesses (Objective 6). This was done during two separate events: one with two employees (Health, Safety and Environmental Advisor) of Rolls Royce's nuclear submarine manufacturing facility, and the other with the Digital Manufacturing Manager of Jaguar Land Rover.

The format for each session began with the industrial partners being given the opportunity to try the MS training simulator used in Study 2 and to provide comment. The findings from Study 1 were presented, as were the preliminary results from Study 2. This was followed by open discussion, focusing on the impact and benefits the MS training system could have within their organisations. Discussions were followed up with a semi-structured interview. This included questions about how the system might be used, i.e. for which purpose and application, and in line with Objective 6, also included questions about the potential benefits the MS system could offer with regard to cost and time savings, OSH compliance, motivation to train and knowledge management.

## 7.1 Industry partner 1 – Rolls Royce, nuclear submarine manufacturer

Both representatives commented positively on the MS training simulator.

*"Really good. The senses really add to it. Felt a lot more real than I expected it to!"*

- **Industry Partner 1, demonstrator feedback**

They felt that the fire scenario would be useful for getting people used to alarms or emergency exits in their workplace. In the engine scenario, they liked the sequence of actions, including containment and raising the alarm. They liked that the scenario was linked to fire training (i.e. a consequence of inappropriate action). The partners commented that currently the safety training is in a 44-page manual, and that a lot of knowledge is held tacitly by the experience operator; the VE could allow the operator to obtain (virtual) experience and remove some of the trial and error associated with work in the real world.

However, the partners were unclear on the real-world value. That is, they questioned whether people trained with this system would actually evacuate the building faster. Moreover, their legal obligations would mean that this VE system could not reduce the time/cost associated with real-life drills. One opportunity where this system could add value to their business is addressing the challenges of training shift workers. Evacuation drills are conducted routinely and periodically, and the primary drills are predominantly conducted during core office hours to involve the most staff and capture worst-case scenarios. The VE could act as a complementary approach to the drills, giving those on alternative shift patterns or absent at the time of the primary drill the opportunity to access a wider range of emergency scenarios as immersive experiences. Furthermore, the partners noted that drills tend to be relatively routine and systematic, meaning that the drill reflects a typical task with limited variability, usually restricted only to their usual workplace, and thus it is often not applicable in less familiar situations or when faced with specific challenges that may be present during an emergency. However, they stressed that this is in fact where the VE could add most value: their impression from the system was that it would be ideal for teaching personal safety rather than a specific procedure or route, and that in fact the personal safety aspects are applicable much more broadly. These aspects are not well trained through routine drills and safety exercises.

The partners also commented on the potential of the VE system for improving engagement over current health and safety induction procedures.

Considering some of the potential future developments, the partners were interested in the sequence of an emergency. That is, releasing a smell as a pre-starter for a fire scenario and using the VE to challenge trainees to speak up and report something is not right. They also reported interest in specific environments, such as noisy areas in which employees are using hearing protection, and therefore the alarm must be supplemented with blue xenon flashing lights. One of the partners highlighted that in a building fire, the olfactory cue would not be wood (as determined through our user testing) but plastic. This could easily be integrated in the VE.

In a subsequent follow-up correspondence with the Health, Safety and Environmental Advisor, he concluded that there were two separate training opportunities:

a) a system for training personal safety and awareness

b) a system for training building familiarisation or specific fire safety procedures.

The representative was very positive about (a) and the value it offered in making safety personal, and giving staff a lifelong skill set. He felt that this was a positive approach to OSH and should be the direction safety moves towards: being *"everyone's responsibility and not something that is done to you"*. He felt that (a) holds particular value as no-one ever teaches people what fire signs look like, or the skill set to know where to look for salient objects in an emergency. Moreover, the representative mentioned the difficulty (cost, time and effort) of producing and maintaining the system if it is purposed towards (b). The representative felt that for the purposes of (a) the current prototype was already suitably configured to offer benefit to the business and could effectively be used 'as-is', while for the purposes of (b) it would require resource to tailor the environment and training objectives and then to design the learning scenario for them, which would require further cost-benefit analysis.

## 7.2 Industry partner 2 – Jaguar Land Rover, automotive manufacturer

Several of the themes reported by industry partner 1 were also mentioned by partner 2. He commented positively on the multi-sensory simulation: *"The heat and the smell definitely enhance it. The smell comes through, you can tell it's burning."* Partner 2 also mentioned the possibility for VR-based training to be used during induction processes, to increase engagement over the current video-based approach. He highlighted that this could be done remotely if desired, and while this would only have visuals and sounds, Partner 2 saw the value in this solution. Partner 2 emphasised that this would not teach you everything you need to know, but could be a foundation for in-situ training.

Similar to partner 1, partner 2 also mentioned the importance of validating the training: *"I think it would work. Needs testing against something physical."* While this was explored in Study 2, the industrial partners obviously see a need for validation in the real world, which is practically and ethically difficult, as discussed in Section 6.7.

Unlike partner 1, partner 2 did see the value in the VE as a familiarisation tool, for example: *"If there was a set procedure that was unique to the company, this is the collection point, this is who you report to, this is the fire marshal."* He also saw duel aims in the engine scenario, one being to train the disassembly and the other for the emergency events, but generally thought that *overall "it would work"*.

Partner 2's main concern was the lack of HMD. He felt that this reduced the subjective integration of the heat into the simulation, as you could see the heaters and fins move. He therefore felt that the actual heat was not associated with the scenario. Partner 2 also thought that the HMD would be more engaging, and this would improve learning.

A further concern was that the training was too game-like. This was partly due to the way the instruction was designed, but also elements like the fire progression, which partner 2 felt was too rapid. The game-like approach had been considered during the prototyping stages but, according to this partner, should be reduced in future developments. However, the feedback in this respect was more positive after the third (final) part of the training, which was the less structured experiential task section.

A final issue was the suitability of the technology for industry. Partner 2 commented that our hardware was suitable for scientific research but should be repackaged for industry. He also felt that there could be trade-offs, for example compromising the heat intensity for a more portable solution. However, this counters the findings of Wareing et al. (9), which revealed that the heat should have as a minimum three 2 kW patio heaters, and ideally more. That said, the current system is not easily portable and there is further opportunity to develop the practicality, robustness, safety and space-efficiency of the hardware.

Partner 2 felt strongly that overall virtual environment training is preferable to traditional training and should be rolled out more widely. He felt that while VR might not yet be a substitute for full in-situ training, it would be the foundation for it. In discussing specific OSH training applications, he felt that the VR would be ideal for primary health and safety inductions, noting that it could be completed in advance before arriving at a site. Currently for some of their industrial sites, visitors/employees are required to watch a video of up to 20 minutes by way of safety induction before entering, which he commented is *"not very memorable"*. He felt that this could be replaced with more engaging and effective VR training.

## 7.3 Relation to project findings

The findings from industry feedback sessions link closely to several of the project outcomes.

The investigation into custom VEs (page 34) found that there are resource and expertise requirements to creating bespoke workplace models that create barriers for their use in the real world. This corresponds with industry partner feedback on the viability of bespoke training environments and on the advantage of using VE to teach generalisable personal safety approaches rather than individual situations. Relatedly, we found that bespoke synthesised fragrances are not a viable option when it comes to training familiarisation with hazardous substances. During early meetings with the industrial partners, familiarisation with smells that would allow employees to identify hazards and dangerous substances was raised as a possible use case, but the work on this project demonstrated that obtaining, and proving the safety of, bespoke fragrances is not practically achievable.

The partners' concerns about the cost, time and effort of producing an effective training system are reflected in the processes of developing the prototype system in the current project. On this project, creating a functional, full training module was a time-consuming task and difficult to get right without both instructor participation and instructional design expertise. We did find evidence that our VE was more effective for OSH training overall than PowerPoint, but only after a heavy investment in time (~6 months) with experienced VE developers and human factors experts for two specific modules. Such investment would only be cost-effective for larger companies and those with existing in-house skills. Even in the current project, it is clear that the advantages of the VE technology were not fully utilised and that even more effective training is likely to result from increased time and resource. Thus, we would recommend on balance that with the exception of certain cases, use of 'off-the-shelf' training solutions from external providers as opposed to tailor-made individual scenarios is a more effective option. Exceptions would be for large companies for whom the economic trade-off is more efficient, and employers who regularly require training OSH topics that physically *can not* be taught through other methods. Applications of VR in training surgery and flight simulators are examples of when practical training is essential prior to conducting tasks in the real world.

The industry partners' suggestion that the MS VE would be effective for improving people's personal sense of safety awareness ties well with the outcomes of Study 1 – that people are more engaged, recognise the urgency of the situation and take it more seriously or less like a game when multi-sensory simulation is added. Essentially, this MS system speaks to an attitudinal shift rather than a knowledge one. Study 2 also found that VE-based training improved subjective attitudes, similarly supporting the concept of VE for promoting individual and team responsibility for OSH rather than simply for procedural or knowledge teaching.

## 7.4 Costs of VE training

While we have commented here on time-saving and other benefits, it is important to illustrate the scale of outlaid costs involved in the provision of VE-based training. It was an objective of the current project to investigate the feasibility of using low-cost and easily obtainable hardware and software that would be likely to be accessible to SMEs.

There are many combinations of hardware and software and while some higher- and lower-cost variants are available, for illustrative purposes this section provides an indicative cost of a complete MS VR system.

### 7.4.1 Hardware

- Laptop/PC of sufficient specification (with adequate graphics capabilities) or the addition of an external graphics card to support the VR on a lower-spec machine: **£2000**.
- Programmable scent nebuliser, maintenance and spares: **£300**.
- Specialist fragrance oil: **£60**.
- Over-ear headphones with ANC: **£50**.
- Arduino relay system: **£145**.
- IR heaters: £35 x2, **£70**.
- Servo motors and frame kit to support and control fins: **£140**.

Total hardware cost: **£2,765** (N.B. This assumes that a high-specification PC/laptop is not already available).

**Optional addition:** head-mounted display, HTC Vive including sensors and controls: **£500**.

### 7.4.2 Custom VE development

For scanning premises to create customised VEs, a Google Tango scanning device would cost below £300, but since the outset of this project the Google Tango programme has been discontinued (as of March 2018). A less powerful but more accessible version has been launched with ARCore, which works through a broader range of consumer-level devices:
https://www.blog.google/products/arcore/arcore-augmented-reality-android-scale/

The Matterport scanning device was estimated to cost £2,830 plus subscription costs at time of writing. Meanwhile, higher-end devices such as Faro laser scanners start at approximately 25,000 USD.

By comparison, Unity Pro subscription costs 125 USD per month, inclusive of three seat licences and several support benefits, and has discounted or free options for some categories of users/organisations such as educational or those with low revenue capacity. Virtual environments created in Unity can be developed from scratch or adapted from pre-purchased or more limited freely available models.

## 7.5 Summary of business benefits

Our overall conclusion from industry stakeholder work in relation to our study findings is that with the current state-of-the-art in technologies, virtual environments would be more effective for general inductions and for teaching personal safety, rather than for in-depth, detailed procedures, processes and navigation.

Based on the input from our industrial partners, study outcomes and prior research, we present the following list of benefits to business of virtual environment training in health and safety over traditional approaches:

1. Increased engagement with OSH training.
2. Improvement in attitudes towards OSH training.
3. Greater motivation to participate in future OSH training.
4. Better retention of OSH knowledge.
5. Opportunity to conduct OSH induction sessions remotely, reducing the costs associated with on-site training and with delays while waiting for new starts to be trained.
6. Opportunity to enhance the training given to those populations who were not able to participate in primary evacuation drills (for example absent workers/shift workers).
7. Related to the above, the opportunity to conduct training at a time that is convenient to the trainee.
8. Ability to customise the training to the needs of the trainee. In this way, the trainee only needs to spend the time needed to achieve competence, reducing their time away from other work activities.

The following benefits to business are suggested for the use of **MS** virtual environments specifically:

1. The ability to train employees to respond to a cue (e.g. smell of smoke) during the early stages of an emergency. In this way, employees' actions could be more likely to lead to containment of the emergency and/or lead to greater safety of themselves and others.
2. Predicting human behaviour in emergencies. Valid predictions of behaviour can be important when designing buildings, training or response procedures (15) and this research showed that MS VEs revealed behaviours that were closer to those seen in the real world than behaviours from audio-visual VEs. However, the limits of behavioural prediction must be considered in this scenario, including that even the MS VE revealed behaviours that would not be expected in real life.

The above benefits are underpinned by the association between health and safety and business benefits such as reductions in absenteeism, fewer accidents, reduced legal costs and greater productivity associated with sound health and safety practices (96). That is, if employees are adequately trained, have better attitudes towards health and safety and greater motivation to engage with it, companies can reduce the time and cost associated with accidents, damage and ill health, and consequently increase profitability.

A particularly well-supported finding, from both studies and from industry partner feedback, was that the VE may be most useful for personal, attitudinal change, rather than purely factual learning. The importance of positive influence on OSH culture and 'behavioural safety' is outlined in the IOSH guide "Promoting a positive culture – a guide to health and safety culture" (97). This report goes into further detail on the advantages of initiatives to support individuals to commit to and cooperate with health and safety objectives. Our findings suggest that VE-based training may be ideally suited to complement cultural change strategies.

# 8 Overall discussion

This research demonstrated the feasibility of creating a multi-sensory (visual, auditory, thermal and olfactory) virtual environment for health and safety applications. The system used low-cost hardware, with infrared heaters and a scent diffuser that are commercially available. A custom fin-system was developed, which proved effective in controlling the heat delivery to the user. The scent diffuser was triggered through an Arduino-controlled script to release a commercially available fragrance.

When considering the validity of behaviours demonstrated in the VE, the MS simulation resulted in behaviours that are more akin to those expected in a real fire. This was attributed to the greater immersion experienced with the addition of heat and smell. However, when tested for benefits to training outcomes, few advantages were seen for MS simulation; most of the benefits were from using VE- rather than PPT-based training. The training benefits of VE are not surprising given the previous work in this area (e.g. 13,40). However, more surprising is the lack of performance and subjective improvements as a consequence of MS simulation given the potential benefits these offer (outlined in section 2.5) and potential to address issues with previous VR systems (10,39).

Considering the practical relevance of these findings, the greatest potential for MS VEs was not seen in economic benefits related to training, for example a reduction in time taken to achieve competence. Rather, and supported by the industrial collaborators on the project, the greatest potential seems to be as a tool that gives people a higher-fidelity experience of an emergency situation, and therefore training them in personal safety and awareness. Thus, the potential benefits of the tool are closer to cultural and social outcomes, rather than economic and quantitative measures. We did not explicitly explore personal safety, awareness and effect on behavioural safety in Study 2, which was more focused on factual knowledge. Future studies could explore this and investigate valid methods of assessment rather than traditional question-and-answer tests, within the ethical constraints of not deliberately exposing participants to adverse risk. These potential benefits are, however, supported by the findings of Study 1, which showed that behaviour in the MS experience was more representative of a real emergency than in an AV VE alone. It is known that traditional training often fails to motivate and engage employees, and that cultural change is a long-term challenge for organisations; thus this is a potentially important affordance.

One of the industry partners mentioned a particular training scenario that would necessitate MS simulation, in which trainees are presented with a smell (as if in the early stages of an emergency) and are challenged to take responsive actions. For scenarios such as this, olfactory simulation is necessary, and the VE would offer the generic advantages of VET such as experiential and active learning, control and evaluation and assessment outlined in Section 2.2.

Beyond training, we found further evidence to support the already hypothesised benefit of VE as a predictive tool for human behaviour in emergencies. In particular, we found that the addition of MS simulation may improve the validity of such predictions.

The project work also revealed other opportunities for VET, which do not necessarily need MS simulation. For example, one industrial partner highlighted that failing to be part of the primary fire drill due to absence, such as on night shift, may detract from the full value of fire drills and that VET could be used to address this. The other partner was keen to see VET used to train workers in health and safety activities as part of the induction process, training that could be done remotely. In this instance, the system would be far more likely to be delivered using vision and auditory feedback alone, rather than with olfactory and heat factors.

The financial, resource and expertise costs required, and the limitations in simulation fidelity of heat and particularly smell, means that the use of the technology to create bespoke training modules for the workplace is not likely to be a suitable solution (based on current technological capabilities). We found that low-cost and low-expertise scanning technologies are not currently a viable option for creating scanned representations of actual workplaces, given the computational demands and expertise needed to translate the scans into usable VE models. While reasonable low-cost scanning technologies exist, these do not produce effective interactive models without extensive skilled processing; thus our recommendation to the technology sector is to continue efforts to facilitate *interactive* model construction of bespoke environments such that SMEs could more easily use replicas of their own workplaces for VET. The final finding on the technical development was that bespoke synthesised fragrances are not really a viable option when dealing with hazardous substances, given the difficulty of finding suitable simulated fragrances and proving their safety. The only times when this technology might be suitable for bespoke training modules is (a) where there is a very focused and specific scenario, or where the current training cost is very high **and** available training budget is very high, or where the current training is wholly ineffective. Current traditional fire safety training is specifically designed to be as effective as possible within the constraints of the learning medium, and thus it would be beneficial to explore other scenarios that are not currently taught through traditional training.

## 8.1 Limitations

The aforementioned results were likely influenced by several project limitations. Perhaps most importantly, conducting research in emergency situations poses methodological challenges given the ethical need to protect the health and wellbeing of study participants. That is, exposing participants to actual emergencies in a real-world environment, either to test the validity of their behaviours in the VE or the transfer of knowledge uptake from the VE, is ethically and practically difficult. While it is possible to use other approaches, such as reviewing behaviours against those reported in the literature, or using knowledge tests and attitudinal ratings, questions exist over how well these VE findings translate to the real world. Interestingly, this was raised by both industrial partners, who wanted validation of the results in the real world.

Related to the above, in Study 2 (training effectiveness), a verbal question-and-answer test was used for knowledge uptake, given the challenges associated with any type of real-world testing mentioned above. This type of test arguably favours the text-based PowerPoint training, given the closer link between training media and test. The virtual environment potentially offers benefits that were unable to be tested, for example recall during the state in which the knowledge was learned; that is, if someone were distressed in an actual emergency, their recall from the training session may be greater from the higher-fidelity MS virtual environment (98).

The project raised the importance of effective training design. In other words, simply creating a virtual environment does not mean that this will be effective as a training tool. The training must be designed with consideration given to learning objectives, knowledge types, feedback and progression. Overall, the time and resources needed for training design are easy to underestimate, and yet are essential for successful outcomes. Given that this was a significant challenge in a two-year research project, this is likely to be a barrier to use in practice.

Considering future directions for this research, from Study 1 (behavioural validity) the relative impact of the heat and/or smell on the behaviour in the VE was unclear. Anecdotal evidence suggested that the olfactory simulation was more impactful than the heat. If this were true, it would be far easier for companies to create (and use) MS environments with a scent diffuser, without the series of heaters and fins. However, this should be proven through further experimental work. Based on the combined results of Studies 1 and 2, there would also be significant value in exploring the interaction between different configurations of modalities and their congruence with the scenario and with each other which could, for example, reveal whether heat and smell offer greater benefit when combined with more immersive displays.

As reported in Section 6.7, the outcomes of Study 2 may be different with different training scenarios, VE set-ups (in particular using an HMD) and evaluation methods. While the latter always poses ethical challenges of using real-world scenarios for validation in a health and safety context, different training scenarios (including a more complex task structure) and hardware could be explored to identify whether different instances result in outcomes that are more favourable to MS VEs. From this, it may be possible to identify the characteristics of which scenarios are suited to which VE configurations. Related to this, it would be worth exploring the training outcomes and obtaining feedback on the VE training system in actual workplaces using company employees. In this way, the technology can be optimised for use in the workplace.

## 8.2 Recommendations for practice

Table 8-1 details the recommendations drawn from throughout the project for practical implementation in OSH. Links are provided to the corresponding sections of this report that discuss and evidence the items listed.

*Table 8-1 Practical guidance based on the findings of this project*

| | Recommendation | Source/further information |
|---|---|---|
| Applications of VE and MS VE | | |
| **1.** | Virtual environments can be used to facilitate prediction of human behaviour in emergency scenarios. | 5.11 |
| **2.** | Caution should be taken in interpreting human behaviour in VE, particularly around pre-evacuation delays, time to evacuate and where discrepancies arise from the VE interface, e.g. overshooting errors resulting from use of the controls. | 5.9 |
| **3.** | Multi-sensory simulation including heat, smell, audio and visuals should be used where viable to increase the validity of behaviour predictions, particularly in relation to risky behaviours and where emotional or psychological response is salient. | 5.9 |
| **4.** | Virtual environments are recommended to deliver OSH training for improved retention. | 6.7 |
| **5.** | Investment in multi-sensory systems for OSH training is not recommended without further exploration of appropriate implementation, or further advancement of affordable technologies. | 6.7 |
| **6.** | Head-mounted displays may improve immersion and block external influences, but pose practical and sickness issues; therefore decisions should be taken on a case-by-case basis as to the most appropriate display technology for the specific application. | 6.7 |
| **7.** | Use established techniques to reduce risks of simulator sickness, and screen potential users for increased susceptibility to adverse effects. | 5.10.3 |
| **8.** | Limit potential distress for trainees in VE, particularly where realism is high, and screen for previous traumatic experience. | 5.2 |
| Technical implementation of multi-sensory VE | | |
| **9.** | Use infrared heaters to provide an appropriate simulation of radiant heat from a fire. | 4.5.1 |
| **10.** | More than one heat source should be used for thermal simulation, positioned symmetrically around the user. The absolute position of the heaters is not important for this purpose, provided that they are symmetrical. | 4.5.1 |
| **11.** | A minimum of two 2 kw infrared heaters provides safe heat simulation and is reliably perceived by users, but using three or more heaters improves the experience and increases realism. | 4.5.1 |
| **12.** | Control thermal feedback by blocking the heating element with movable heat-deflective panels, rather than by turning the infrared heater on and off. | 4.5.1 |
| **13.** | For olfactory simulation, position a diffuser approximately 1 metre in front of the user, with a fan immediately behind it to direct the fragrance particles. | 4.5.2 |
| **14.** | Use fragrances approved by IFRA for olfactory simulation. | 4.5.2 |
| **15.** | Conduct COSHH assessment on fragrances and establish safe limits for diffusion. | 4.5.2 |
| **16.** | Use appropriate extraction equipment to prevent build-up of fragrance. | 4.5.2 |
| **17.** | Use the olfactory system in a well-ventilated area. | 4.5.2 |
| **18.** | Ensure an odour-free environment for olfactory simulation, removing or covering soft furnishings and odour-absorbing materials. | 4.5.2 |
| **19.** | Test fragrances with users to establish the lowest usable concentration that is consistently perceived and recognised. | 4.5.2 |
| **20.** | Screen users for asthma, odour intolerances, allergies and respiratory conditions prior to exposure to olfactory simulation. | 4.5.2 |
| Developing bespoke training scenarios | | |
| **21.** | Current off-the-shelf low-cost scanning technologies are unlikely to be suitable for creating interactive training environments, and scans will require significant post-processing by an experienced technician to adapt them for use. | 4.6.1 |
| **22.** | Adapt pre-made VEs or create models from existing building plans in a suitable VE development tool such as Unity, rather than investing in expensive scanning technologies. | 4.6.1 |
| **23.** | Without dedicated instructional design, customised training scenarios may not be effective. | 6.7; 7.2 |
| **24.** | In many cases, the cost of creating bespoke VE-based training may outweigh potential benefits; therefore where suitable off-the-shelf software and training modules are available, they should be used. | 7 |
| Testing OSH training | | |
| **25.** | Employees should be tested on their knowledge after an interval, rather than immediately following training, in order to evaluate their HS knowledge. | 6.7 |
| **26.** | Traditional testing is appropriate for traditional training but may not be a valid indicator of success for VE-based training, and does not address transfer of training to the real world. | 6.7 |

# 9 Overall conclusions

The overall aim of this work was to understand the potential multi-sensory (MS) virtual environments with visual, auditory, olfactory and thermal simulation offer as engaging and effective training solutions. This was to address current issues with health and safety training, such as lack of interest and engagement experienced by trainees, lack of relevance to their own workplaces, and differences in the psychological states between training environment and target scenario. The work was focused on the needs of businesses through the involvement of representatives from Rolls Royce and Jaguar Land Rover.

The project successfully developed a low-cost MS simulator. This was based on the Unity virtual environment but used Arduino-controlled actuators to invoke heat and/or scent as the user navigated through the environment. A solution was found, using fins that act as heat shields, to control the heat fluctuations that were expected based on the avatar's movement in the VE. Research was conducted to understand the level of heat needed (minimum of three 2 kW heaters) and location (symmetrically placed around the user). Scent was delivered through an off-the-shelf fragrance diffuser and standard scents. Proving the safety of uncontrolled chemicals was challenging, and to support this a protocol for the safe use of fragrances in the VR prototype was developed.

The project also investigated scanning technologies, and in particular Google Tango, to determine whether it would be feasible to scan a building, which was then imported into the virtual training simulator. However, there were several challenges, such as quality issues, and the technical knowledge required to manipulate the scan data. While scanning workplaces is likely to be impractical with current technology, a protocol was developed that would improve the scan data results. As part of this work, an example is given of how classifier objects (i.e. virtual objects with user interaction which form a salient part of the training) could be implemented in a scanned environment.

Experimental work was conducted in two main studies. The first investigated the differences in behaviour demonstrated in VE between MS and AV training simulators. The results showed that, overall, MS simulation increased the validity of user behaviour in the VE. This was observed through differences in behaviours and comments between the two conditions, with responses from MS participants indicating that they felt part of the simulated event, rather than behaving in accordance with their behavioural expectations.

The second study looked at the training outcomes, and compared virtual reality to PowerPoint-based training, giving participants either MS or AV experiences in both. The study generally showed that on the subjective measures (engagement, attitude towards health and safety, desire to repeat training) the virtual environment performed better than the PowerPoint training. The evidence for MS simulation was weak, but that may be down to the specific implementation. Importantly, this was also likely affected by the evaluation protocol, which was practically, ethically and methodologically limited to a question-and-answer test; it is likely that the AV and MS VR would yield better results in tests that more accurately represent the context of the training.

More generally, creating bespoke training modules for the workplace is not likely to be viable, given the financial, resource and expertise costs required, and the limitations in simulation fidelity of heat and particularly smell. Interestingly, though, the cost, effort and expertise of implementing the MS element specifically is comparatively low. An MS VE specifically is likely to be suitable for bespoke training modules where there is a very focused scenario, for larger companies where the benefits outweigh the costs for a large workforce and sufficient resource is available for effective implementation, or where the current training is not effective or is cost-prohibitive. Beyond that, the VE would be most useful for a cultural shift or attitudinal change in people, rather than training specific and bespoke contexts/cases. This is harder to do through traditional training. Future work could investigate whether people change their real-world behaviours as a consequence of VE training.

Finally, this project found further evidence to support the already hypothesised benefit of VE as a predictive tool for human behaviour in emergencies, and found that the addition of MS simulation may improve the validity of such predictions.

# 11 Appendix

## 11.1 Appendix A – Study 1, Post-task questionnaire

1. What was your maximum perceived level of risk during the previous scenario

   1 2 3 4 5

Very low Very high

2. How would you rate your level of stress/anxiety during the previous scenario

   1 2 3 4 5

Very low Very high

3. How would you rate the level of time pressure you felt during the previous scenario

   1 2 3 4 5

Very low Very high

4. Please indicate how true each statement was of your thoughts *while in the VE* in the previous scenario:

Not at all = 1, a little bit =2, somewhat =3, very much =4, extremely = 5

| **Questions** | **1** | **2** | **3** | **4** | **5** |
|---|---|---|---|---|---|
| The building is occupied | | | | | |
| I need to exit the building as quickly as possible | | | | | |
| The fire alarm is going off | | | | | |
| The building is on fire | | | | | |
| I know how to exit the building | | | | | |
| I need to find the exit nearest to me | | | | | |
| I can smell that the building is on fire | | | | | |
| I can feel that the building is on fire | | | | | |
| I can see that the building is on fire | | | | | |

Additional comments:

## 11.2 Appendix B – Engine knowledge test

Engine test: Pre/post/1 week (circle as appropriate)

Participant ID:.................. Experimenter:............................. Date/time: …..................

1. You are disassembling the engine. Which item would you remove after the exhaust manifold?

2. What would you do first upon identifying a leak in the workshop?

3. What item could you use to stop a leak from spreading?

4. Why is it important to raise the alarm in the event of a fuel leak?

5. What would you do after you've contained the fuel leak?

6. Which item would you remove after the lower hose when disassembling an engine?

7. Why is it important to stop the leak from a health and safety perspective?

## 11.3 Appendix C – Fire safety knowledge test

Fire knowledge test: Pre/post/1 week (circle as appropriate)

Participant ID:.................. Experimenter:............................. Date/time: …..................

1. You are evacuating the building due to a fire, you pass a fire token hanging on the wall, what do you do next?

2. In a fire emergency, what can you do to reduce the spread of fire in the building?

3. If the fire alarm goes off, what should you take with you when you are evacuating a building?

4. In what circumstances should you use a fire extinguisher?

5. What fire safety features should you know the location of within the building you're in?

6. What should you do if you see a fire, but the fire alarm has not activated?

7. What reduces the effectiveness of a fire exit?

For the **post-task** test only (and not in the session 1 week later): do the floor plan test

## 11.4 Appendix D – Study 2, Subjective questionnaire

Participant ID:

Scenario:

Please tick the box that best represents how much you agree with each of the statements below, where 1= strongly disagree and 5 = strongly agree:

| **Statement** | **1 (strongly disagree)** | **2** | **3** | **4** | **5 (strongly agree)** |
|---|---|---|---|---|---|
| I felt interested | | | | | |
| I felt bored | | | | | |
| I paid attention to the things I needed to remember | | | | | |
| I formed new questions in my mind as I participated in the training | | | | | |
| I did not want to stop at the end of the training | | | | | |
| I asked myself questions as I went along to make sure the training made sense to me | | | | | |
| I was ‘zoned out’ and not really thinking during the training | | | | | |
| I let my mind wander during the training | | | | | |
| The skills I learned during the training would be helpful in learning how to respond to this situation in real life* | | | | | |
| I would want to undertake this type of training again | | | | | |
| I better understand the importance of safety and health knowledge after completing this training | | | | | |
| I take safety and health matters more seriously after completing this training | | | | | |
| I find safety and health matters more interesting after completing this training | | | | | |

*Important: Please be aware that the training content used in this study is only for the purposes of the experiment; always familiarise yourself with, and follow, local procedures.

**IOSH**
The Grange
Highfield Drive
Wigston
Leicestershire
LE18 1NN
UK

t +44 (0)116 257 3100
www.iosh.com
twitter.com/IOSH_tweets
facebook.com/IOSHofficial
tinyurl.com/IOSH-linkedin
youtube.com/IOSHchannel
instagram.com/ioshofficial

IOSH is the Chartered body for health and safety professionals. With more than 47,000 members in over 130 countries, we're the world's largest professional health and safety organisation.

We set standards, and support, develop and connect our members with resources, guidance, events and training. We're the voice of the profession, and campaign on issues that affect millions of working people.

IOSH was founded in 1945 and is a registered charity with international NGO status.